\documentclass[11pt]{article}
\usepackage[a4paper,margin=28mm]{geometry}
\usepackage{amsmath,amssymb,amsthm,bm,booktabs,array,graphicx,xcolor,needspace,placeins,hyperref}
\definecolor{theoremgray}{gray}{0.4}
\usepackage[nameinlink,noabbrev]{cleveref}
\hypersetup{colorlinks=true,linkcolor=blue,citecolor=blue,urlcolor=blue,
pdftitle={Quantum Hypothesis Testing with Covariant Measurements},
pdfauthor={Masahito Hayashi and Hao-Chung Cheng},
pdfsubject={Revised manuscript, version 122 (pre-arXiv local corrections)}}
\newtheorem{theorem}{Theorem}[section]
\newtheorem{proposition}[theorem]{Proposition}
\newtheorem{lemma}[theorem]{Lemma}
\newtheorem{corollary}[theorem]{Corollary}
\theoremstyle{definition}\newtheorem{definition}[theorem]{Definition}
\theoremstyle{remark}\newtheorem{remark}[theorem]{Remark}
\newcommand{\Tr}{\operatorname{Tr}}
\newcommand{\one}{\mathbf 1}
\newcommand{\Dmax}{D_{\max}}

\newcommand{\Sym}{\operatorname{Sym}}

\newcommand{\cR}{\mathcal R}
\newcommand{\wt}{\mathrm{wt}}

\newcommand{\Rreg}{\overline{\operatorname{co}}}

\title{\bfseries \Large Quantum Hypothesis Testing with Covariant Measurements}
\author{\large Masahito Hayashi$^{1,2,3}$ \quad Hao-Chung Cheng$^{4,5,6}$\\[0.5em]
\parbox{0.95\textwidth}{\centering\footnotesize
$^1$School of Data Science, The Chinese University of Hong Kong, Shenzhen, China\\
$^2$International Quantum Academy, Shenzhen, China\\
$^3$Graduate School of Mathematics, Nagoya University, Nagoya, Japan\\[0.3em]
$^4$Department of Electrical Engineering and Graduate Institute of Communication Engineering,\\
National Taiwan University, Taipei, 10617, Taiwan\\
$^5$Physics/Mathematics Division, National Center for Theoretical Sciences, Taiwan\\
$^6$Hon Hai (Foxconn) Quantum Computing Center, Taiwan}}
\newcommand{\DR}{D_{\mathrm R}}
\newcommand{\dd}{\,\mathrm d}

\DeclareMathOperator{\diag}{diag}
\DeclareMathOperator{\dist}{dist}
\DeclareMathOperator{\rank}{rank}

\providecommand{\Proj}{\mathbb P^1}
\providecommand{\ucov}{\mathrm{ucov}}

\providecommand{\pauli}{\boldsymbol\sigma}
\begin{document}
\maketitle

\begin{abstract}
We study binary quantum hypothesis testing for faithful states using
unitary-covariant measurements followed by arbitrary randomized classical
decision rules. We consider two classes of tests. The first uses Keyl's
measurement and applies the decision rule to its reported state estimate.
The second allows any unitary-covariant measurement and applies the decision
rule to its outcome.

For Keyl’s measurement, we determine the exact Stein, Hoeffding, and strong-converse exponents through variational formulas involving reverse relative entropy. For the second class, we determine the Stein exponent at every fixed type-I error tolerance \(0<\epsilon<1\). Unlike the unrestricted Stein exponent, which is independent of \(\epsilon\), the covariant Stein exponent interpolates continuously between the reverse and ordinary relative entropies in the limits \(\epsilon\downarrow0\) and \(\epsilon\uparrow1\), respectively.

We bound the optimal covariant Hoeffding exponent and show that Keyl's
measurement attains it for qubits. The optimal covariant strong-converse
exponent equals the exponent under unrestricted measurements and can be
strictly smaller than Keyl's for noncommuting states.
\end{abstract}

\section{Introduction} \label{sec:introduction-reorganized}

\subsection{Covariant measurements and binary tests}

Let $\rho$ and $\sigma$ be known faithful (full-rank) states. At blocklength
$n$, we distinguish $\rho^{\otimes n}$ from $\sigma^{\otimes n}$. The
type-I error $\alpha_n$ is the probability of rejecting $\rho$ when it
is true; the type-II error $\beta_n$ is the probability of accepting
$\rho$ when $\sigma$ is true. We ask how restricting the measurement to
be unitary covariant changes the rates at which these errors can decay.

We consider three ways to constrain the errors. In the Stein regime, we
allow $\alpha_n\le\epsilon$ for a fixed $0<\epsilon<1$ and maximize the
decay exponent of $\beta_n$. In the Hoeffding regime, we require the
decay exponent of $\beta_n$ to be at least $r$ and maximize that of
$\alpha_n$. In the strong-converse regime, we keep the same type-II
constraint and minimize the decay exponent of $1-\alpha_n$, the
probability of accepting $\rho$ when it is true.

When any binary quantum measurement is allowed, the fixed-error Stein
exponent is $D(\rho\Vert\sigma)$, the quantum relative entropy
\cite{HiaiPetz1991,OgawaNagaoka2000}. The unrestricted Hoeffding and
strong-converse exponents are also known in terms of R\'enyi divergences
\cite{HayashiHoeffding2007,NagaokaHoeffding2006,AudenaertEtAl2008,
MullerLennertEtAl2013,WildeWinterYang2014,Beigi2013,MosonyiOgawa2015}.

\Needspace{8\baselineskip}
Restrictions on the measurement can change these exponents, as seen
for local operations and classical communication
\cite{BrandaoEtAl2020,ChengWinterYu2023,RippchenSreekumarBerta2025}.
We focus on unitary-covariant measurements: if the same unitary
rotates every input copy, the law of the recorded outcome rotates
accordingly~\cite{Holevo1982}. Covariance is not required of the
subsequent decision rule, which may use the known states $\rho,\sigma$.
For example, when the outcome is a state estimate, a rule may accept
$\rho$ if that estimate is close to $\rho$. Thus the question is what
information a covariant measurement can provide for decisions tailored
to the two hypotheses.

Keyl and Werner constructed a collective measurement that estimates
the spectrum of a state and determined the exponential rate of its
estimation errors~\cite{KeylWerner2001}. A spectral estimate alone cannot
distinguish states that have the same eigenvalues. Keyl extended their
construction to a covariant measurement that reports an estimate of the
full density matrix~\cite{Keyl2006}. For a faithful input state $X$,
estimates near $\tau$ have large-deviation rate $\DR(\tau\Vert X)$,
the \emph{reverse relative entropy} defined in
\eqref{eq:cov-input-SKeyl-principal-minor}. Recent work has studied
covariant tomography rates and compared measurement
protocols~\cite{GrootveldEtAl2026,GrootveldEtAlRefined2026}.

For binary hypothesis testing, estimates near $\rho$ are typical when the input is
$\rho$, while under $\sigma$ their local large-deviation rate is
$\DR(\rho\Vert\sigma)$. Accepting such estimates is one possible
decision rule. 
However, a rule can also change with $n$ and assign different
acceptance probabilities to different estimates. To find the best
tests based on Keyl's measurement, we therefore analyze the exact
distribution of its reported estimate at each blocklength.

\medskip\noindent\textbf{Two measurement classes.}
Tests in the first class use Keyl's measurement, which reports a state
estimate. Tests in the second class may use any unitary-covariant
measurement. Both classes allow an arbitrary randomized classical
decision rule on the reported outcome.
The decision rule may use the known states $\rho,\sigma$.

\subsection{Main results}

Table~\ref{tab:exponent-summary} summarizes our results for the two
measurement classes. For Keyl's measurement, we obtain exact formulas
for all three exponents. The Stein exponent is
$\DR(\rho\Vert\sigma)$ for every $0<\epsilon<1$, even after optimizing
the decision rule
(Theorem~\ref{thm:keyl-stein-exact-reorganized}).
For a type-II decay rate $r>0$, the Hoeffding and strong-converse
exponents are
$\min_{\tau:\,D_{\mathrm R}(\tau\Vert\sigma)\le r}
D_{\mathrm R}(\tau\Vert\rho)$
and
$\inf_{\tau}
\bigl\{D_{\mathrm R}(\tau\Vert\rho)
+(r-D_{\mathrm R}(\tau\Vert\sigma))_+\bigr\}$, respectively.
Both variational formulas have the same form as their classical
counterparts, with relative entropy replaced by reverse relative
entropy.\footnote{Replacing reverse relative entropy by ordinary
	relative entropy yields the corresponding variational expressions
	in terms of the log-Euclidean R\'enyi divergence
	\cite{Hayashi2006,MosonyiOgawa2017}. These expressions need not equal
	the exponents under unrestricted measurements.}

%These formulas allow us to assess what is gained by choosing the covariant measurement as well as the decision rule.

\begin{table}[!htbp]
\centering
\hypersetup{linkcolor=theoremgray}
\caption{Optimal error exponents for the two measurement classes. The Stein formulas hold for $0<\epsilon<1$, the Hoeffding formulas for $r>0$, and the strong-converse formulas for $r\ge0$. Here $H_\rho=\{h\in U(d):h\rho h^\dagger=\rho\}$ is the group of unitaries that leave $\rho$ unchanged, $\mu$ is its normalized Haar measure, and $A$ ranges over measurable subsets of $H_\rho$. The optimizations over $\tau$ range over the set $\mathcal D_d$ of density operators. We write $(x)_+=\max\{x,0\}$, and $\widetilde D_\alpha$ is the sandwiched R\'enyi divergence. In higher dimensions, bounds on the optimal covariant Hoeffding exponent appear in Theorem~\ref{thm:ucov-hoeffding-renyi-converse}. All logarithms are natural.}
\label{tab:exponent-summary}
\small
\setlength{\tabcolsep}{4pt}
\renewcommand{\arraystretch}{1.3}
\begin{tabular}{@{}>{\raggedright\arraybackslash}p{2.8cm}>{\raggedright\arraybackslash}p{5.5cm}>{\raggedright\arraybackslash}p{6.5cm}@{}}
\toprule
Exponent & Keyl's measurement & All covariant measurements\\
\midrule
Fixed-error Stein
& $\displaystyle D_{\mathrm R}(\rho\Vert\sigma)$
& $\displaystyle\sup_{\substack{A\subseteq H_\rho\\\mu(A)>1-\epsilon}}\lim_{n\to\infty}\frac{D\!\left(\rho^{\otimes n}\middle\Vert\int_A(h^\dagger\sigma h)^{\otimes n}\frac{d\mu}{\mu(A)}\right)}{n}$\\[0.5mm]
& \textcolor{theoremgray}{(Theorem~\ref{thm:keyl-stein-exact-reorganized})}
& \textcolor{theoremgray}{(Theorem~\ref{thm:merge-ucov-fixed-interior-stein})}\\[1.5mm]
Hoeffding
& $\displaystyle\min_{\tau:\,D_{\mathrm R}(\tau\Vert\sigma)\le r}D_{\mathrm R}(\tau\Vert\rho)$
& $\displaystyle\min_{\tau:\,D_{\mathrm R}(\tau\Vert\sigma)\le r}D_{\mathrm R}(\tau\Vert\rho)$ (for qubits)\\[0.5mm]
& \textcolor{theoremgray}{(Theorem~\ref{thm:keyl-hoeffding-exact-reorganized})}
& \textcolor{theoremgray}{(Theorem~\ref{thm:merge-qubit-full-hoeffding})}\\[1.5mm]
Strong converse
& $\displaystyle\inf_{\tau\in\mathcal D_d}\bigl\{D_{\mathrm R}(\tau\Vert\rho)+(r-D_{\mathrm R}(\tau\Vert\sigma))_+\bigr\}$
& $\displaystyle\sup_{\alpha>1}\frac{\alpha-1}{\alpha}\bigl\{r-\widetilde D_\alpha(\rho\Vert\sigma)\bigr\}$\\[0.5mm]
& \textcolor{theoremgray}{(Theorem~\ref{thm:keyl-strong-converse-exponent})}
& \textcolor{theoremgray}{(Theorem~\ref{thm:general-completeness})}\\
\bottomrule
\end{tabular}
\end{table}

\begin{figure}[!htbp]
\centering
\includegraphics[width=.82\linewidth]{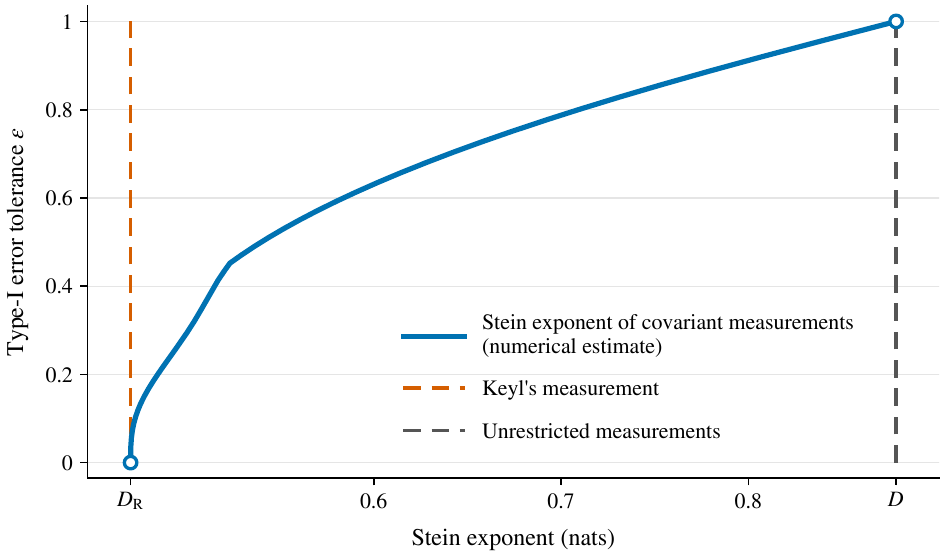}
\caption{Fixed-error Stein exponents for the faithful qubit pair $\rho=\bigl(\begin{smallmatrix}9/10&0\\0&1/10\end{smallmatrix}\bigr)$ and $\sigma=\bigl(\begin{smallmatrix}1/2&2/5\\2/5&1/2\end{smallmatrix}\bigr)$, with natural logarithms. The blue curve is a numerical estimate obtained from the two families of covariant tests described in Appendix~\ref{subsubsec:stein-qubit-arc-reduction}. The empirical estimate of the absolute numerical error for evaluating these families is about $2\times10^{-5}$ nats. Whether these families attain the optimal exponent remains open. The vertical dashed lines show the tolerance-independent Stein exponents of Keyl's measurement, $D_{\mathrm R}(\rho\Vert\sigma)=\frac45\log\frac95\simeq0.4702$, and unrestricted measurements, $D(\rho\Vert\sigma)=\frac25\log9\simeq0.8789$. Open circles mark the exact one-sided endpoint limits as $\epsilon\downarrow0$ and $\epsilon\uparrow1$, respectively (Section~\ref{sec:covariant-stein-regime}).}
\label{fig:intro-stein-boundary}
\end{figure}

The unrestricted Stein exponent $D(\rho\Vert\sigma)$ is also
independent of $\epsilon$ and is strictly larger than Keyl's for
noncommuting states
(Proposition~\ref{prop:reverse-relative-entropy-equality}).
The optimal Stein exponent over all covariant measurements connects these
two fixed benchmarks: it is continuous and nondecreasing in
$\epsilon$, approaching $\DR(\rho\Vert\sigma)$ as
$\epsilon\downarrow0$ and $D(\rho\Vert\sigma)$ as
$\epsilon\uparrow1$
(Section~\ref{sec:covariant-stein-regime}).
Figure~\ref{fig:intro-stein-boundary} illustrates this behavior for
a noncommuting qubit pair. Details of the numerical estimate are
given in Appendix~\ref{app:merge-stein-proof}.

In the Hoeffding regime, our bounds show that the optimal covariant
exponent is positive for $0<r<\DR(\rho\Vert\sigma)$ and vanishes for
$r\ge\DR(\rho\Vert\sigma)$
(Corollary~\ref{cor:ucov-hoeffding-nontight-sandwich}).
For qubits, we determine the exponent exactly: Keyl's measurement
attains the optimum at every $r>0$, so its variational formula also
gives the optimal covariant Hoeffding exponent
(Theorem~\ref{thm:merge-qubit-full-hoeffding}).

\begin{figure}[!htbp]
\centering
\includegraphics[width=\linewidth]{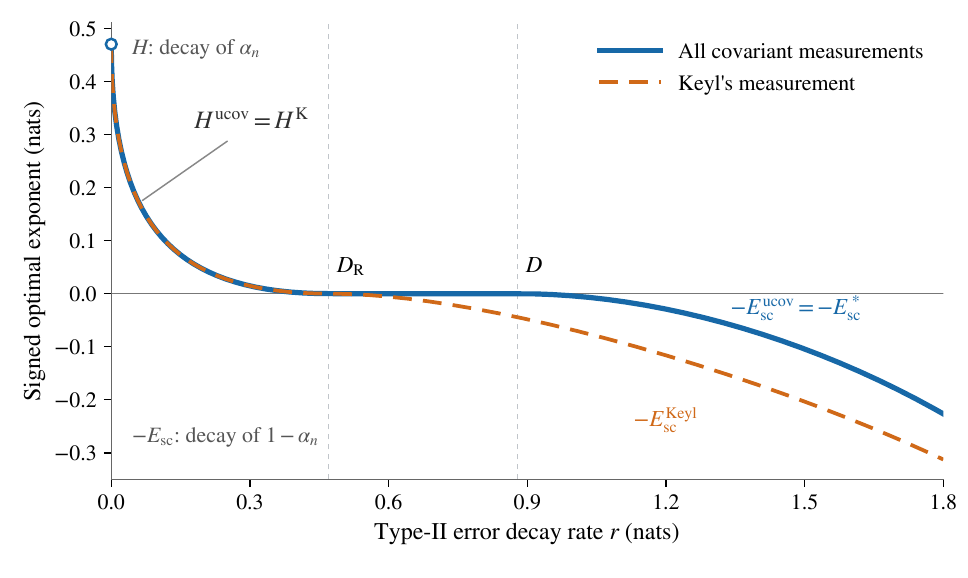}
\caption{Hoeffding and strong-converse exponents for the faithful qubit pair $\rho=\bigl(\begin{smallmatrix}9/10&0\\0&1/10\end{smallmatrix}\bigr)$ and $\sigma=\bigl(\begin{smallmatrix}1/2&2/5\\2/5&1/2\end{smallmatrix}\bigr)$, with natural logarithms. The horizontal axis is the prescribed type-II error decay rate $r$. Above zero we plot the Hoeffding exponent $H$, which describes the decay of $\alpha_n$. Below zero we plot $-E_{\mathrm{sc}}$, so the magnitude gives the decay exponent of $1-\alpha_n$. Theorems~\ref{thm:merge-qubit-full-hoeffding} and~\ref{thm:keyl-strong-converse-exponent} give the common Hoeffding curve and the Keyl strong-converse curve, respectively. Theorem~\ref{thm:general-completeness} gives the covariant strong-converse curve, which equals the unrestricted one. Here $E_{\mathrm{sc}}^*$ denotes the unrestricted strong-converse exponent. For this pair, $D_{\mathrm R}(\rho\Vert\sigma)=\frac45\log\frac95\simeq0.4702$ and $D(\rho\Vert\sigma)=\frac25\log9\simeq0.8789$. Both Hoeffding exponents vanish for $r\ge D_{\mathrm R}$, while the covariant strong-converse exponent remains zero up to $D$. The open point at $r=0$ denotes the right-hand limit $H(0+)$. Curves are numerical evaluations of the exact variational formulas.}
\label{fig:intro-qubit-exponents}
\end{figure}

In the strong-converse regime, the optimal covariant exponent equals
the unrestricted exponent
(Theorem~\ref{thm:general-completeness}).
For noncommuting states, Keyl's strong-converse exponent is strictly
larger throughout
$\DR(\rho\Vert\sigma)<r\le D(\rho\Vert\sigma)$, and the gap persists
for rates just above $D(\rho\Vert\sigma)$
(Theorem~\ref{thm:highest-weight-loss-summary}).
Figure~\ref{fig:intro-qubit-exponents} shows both the Hoeffding and strong-converse exponents of the two classes of measurements for the qubit pair.

%The improvement in Stein performance does not carry over to the
%Hoeffding regime for qubits: Keyl's measurement attains the optimal
%covariant exponent at every $r>0$
%(Theorem~\ref{thm:merge-qubit-full-hoeffding}). In higher dimensions,
%we obtain bounds rather than an exact formula. These bounds still determine
%the threshold: the optimal covariant Hoeffding exponent is positive for
%$0<r<\DR(\rho\Vert\sigma)$ and zero for
%$r\ge\DR(\rho\Vert\sigma)$
%(Corollary~\ref{cor:ucov-hoeffding-nontight-sandwich}).
%
%In the strong-converse regime, however, allowing other covariant
%measurements recovers the unrestricted exponent
%(Theorem~\ref{thm:general-completeness}). Keyl's exponent can be strictly
%larger. For noncommuting states, the strict gap holds throughout
%$\DR(\rho\Vert\sigma)<r\le D(\rho\Vert\sigma)$ and persists for
%rates just above $D(\rho\Vert\sigma)$
%(Theorem~\ref{thm:highest-weight-loss-summary}).
%Figure~\ref{fig:intro-qubit-exponents} compares the Hoeffding and
%strong-converse exponents for the qubit pair in
%Figure~\ref{fig:intro-stein-boundary}.

\FloatBarrier

\FloatBarrier

\section{Operational setting and notation}
\label{sec:covariant-test-class}

Let $\mathcal H:=\mathbb C^d$, with $d$ fixed. Unless a theorem explicitly states
otherwise, the null state $\rho$ and the alternative state $\sigma$ are faithful
(full-rank) density operators on $\mathbb C^d$. Repeated eigenvalues are allowed.
At blocklength $n$, the hypotheses are $\rho^{\otimes n}$ and
$\sigma^{\otimes n}$. All logarithms are natural. For positive operators,
$B\ll A$ means $\operatorname{supp}B\subseteq\operatorname{supp}A$.
When singular operators occur, the support convention is stated with the
corresponding information quantity or theorem.

\subsection{Covariant measurements and decision rules}

At blocklength $n$, let $M_n$ be a positive operator-valued measure
(POVM) on $\mathcal H^{\otimes n}$ with standard Borel outcome space
$\Omega_n$ and a jointly Borel measurable $U(d)$ action. The measurement
is covariant if, for every $g\in U(d)$ and every Borel set
$A\subseteq\Omega_n$,
\[
 M_n(gA)=g^{\otimes n}M_n(A)(g^\dagger)^{\otimes n}.
\]
Given outcome $\omega$, a measurable rule $f_n:\Omega_n\to[0,1]$
accepts $\rho$ with probability $f_n(\omega)$. The resulting acceptance
effect is
\[
 T_n=\int_{\Omega_n}f_n(\omega)M_n(d\omega),\qquad 0\le T_n\le I.
\]
Covariance is imposed on $M_n$. The rule may give different answers
for two rotated outcomes. For example, it may accept when the reported
state estimate is close to $\rho$.

Let $\mathfrak T_d^{\mathrm{ucov}}$ consist of all sequences $(T_n)_{n\ge1}$
obtained in this way, with a covariant $M_n$ and a measurable $f_n$
at each $n$. In the subclass $\mathfrak T_d^{\mathrm K}$, we fix
$M_n=M_n^{\mathrm K}$, Keyl's covariant measurement~\cite{Keyl2006},
and choose the decision rule.

A Keyl outcome consists of the Young label $\lambda$ and the partial
flag of cumulative eigenspaces. Write $\Omega_n^{\mathrm K}$ for this
outcome space and $P_{X,n}^{\mathrm K}$ for its law under $X^{\otimes n}$.
The outcome determines the estimate $\widehat\tau_n$, which has
ordered eigenvalues $\lambda/n$. The estimate also recovers the outcome.
Section~\ref{subsec:keyl-physical-labelled-output} describes this
correspondence. Write $P_{X,n}^{\mathrm{phys}}$ for the law of
$\widehat\tau_n$ on
$\mathcal D_d:=\{X\in\mathcal B(\mathbb C^d):X\ge0,\ \Tr X=1\}$.
We equip this compact state space with the trace-norm topology.

\subsection{Error exponents}

For an acceptance effect $T_n$, write
\[
 p_n(T_n)=\Tr\rho^{\otimes n}T_n,\qquad
 \alpha_n(T_n)=1-p_n(T_n),\qquad
 \beta_n(T_n)=\Tr\sigma^{\otimes n}T_n.
\]
For a Keyl decision rule $f_n$ inducing $T_n$, the acceptance probabilities
are $p_n(T_n)=\int f_n\,dP_{\rho,n}^{\mathrm K}$ and
$\beta_n(T_n)=\int f_n\,dP_{\sigma,n}^{\mathrm K}$.
Use $-\log0=+\infty$.

For an admissible sequence class ${\cal T}$, let ${\cal T}_n$ be its blocklength-$n$ set of obtainable effects.  Define
\[
\begin{aligned}
 \beta_{\epsilon\Vert{\cal T}}
 (\rho^{\otimes n}\Vert\sigma^{\otimes n})
 &:=\inf\left\{\beta_n(T):T\in{\cal T}_n,\ \alpha_n(T)\le\epsilon\right\},\\
 D^H_{\epsilon\Vert{\cal T}}
 (\rho^{\otimes n}\Vert\sigma^{\otimes n})
 &:=-\log\beta_{\epsilon\Vert{\cal T}}
 (\rho^{\otimes n}\Vert\sigma^{\otimes n}).
\end{aligned}
\]
For the full unitary-covariant class, define the lower and upper asymptotic Stein exponents
at fixed error tolerance $0<\epsilon<1$ by
\begin{align}
 \underline B^{\mathrm{ucov}}_\epsilon(\rho\Vert\sigma)
 &:={\liminf}_{n\to\infty}\frac1n
 D^H_{\epsilon\Vert\mathfrak T_d^{\mathrm{ucov}}}
 (\rho^{\otimes n}\Vert\sigma^{\otimes n}),\\
 \overline B^{\mathrm{ucov}}_\epsilon(\rho\Vert\sigma)
 &:={\limsup}_{n\to\infty}\frac1n
 D^H_{\epsilon\Vert\mathfrak T_d^{\mathrm{ucov}}}
 (\rho^{\otimes n}\Vert\sigma^{\otimes n}).
\end{align}
Theorem~\ref{thm:merge-ucov-fixed-interior-stein}
proves that they coincide for every $0<\epsilon<1$. Both are nondecreasing
in $\epsilon$, and their endpoint limits are examined in
Section~\ref{sec:covariant-stein-regime}.

In the Hoeffding regime, the type-II error is required to decay with exponent
at least $r$, and the type-I exponent is maximized.  For
${\cal C}\in\{\mathfrak T_d^{\mathrm K},
\mathfrak T_d^{\mathrm{ucov}}\}$ and $r>0$, define
\begin{equation}
 H^{\cal C}(r\mid\rho\mid\sigma)
 :=\sup_{(T_n)_{n\ge1}\in{\cal C}}
 \left\{
 \liminf_{n\to\infty}-\frac1n\log\alpha_n(T_n)
 \,\middle|\,
 \liminf_{n\to\infty}-\frac1n\log\beta_n(T_n)\ge r
 \right\}.
 \label{eq:operational-hoeffding-definition}
\end{equation}
Write $H^{\mathrm K}$ and $H^{\mathrm{ucov}}$ for the exponents
for Keyl's measurement and all covariant measurements, respectively.

In the strong-converse regime, the same type-II constraint is imposed, but the
quantity minimized is the exponential decay rate of the null-acceptance
probability $p_n=1-\alpha_n$. For either class ${\cal C}$ above and $r\ge0$, define
\begin{equation}
 E_{\mathrm{sc}}^{\cal C}(r\mid\rho\mid\sigma)
 :=\inf_{(T_n)_{n\ge1}\in{\cal C}}
 \left\{
 \limsup_{n\to\infty}-\frac1n\log p_n(T_n)
 \,\middle|\,
 \liminf_{n\to\infty}-\frac1n\log\beta_n(T_n)\ge r
 \right\}.
 \label{eq:operational-strong-converse-definition}
\end{equation}
The following abbreviations distinguish the two strong-converse exponents:
$E_{\mathrm{sc},d}^{\mathrm{ucov}}(r) :=E_{\mathrm{sc}}^{\mathfrak T_d^{\mathrm{ucov}}} (r\mid\rho\mid\sigma)$
and
$E_{\mathrm{sc}}^{\mathrm{Keyl}}(r\mid\rho\mid\sigma) :=E_{\mathrm{sc}}^{\mathfrak T_d^{\mathrm K}} (r\mid\rho\mid\sigma)$.

\subsection{Information quantities}

For density operators with $\operatorname{supp}\rho\subseteq\operatorname{supp}\sigma$,
the ordinary relative entropy is
$D(\rho\Vert\sigma):=\Tr\rho(\log\rho-\log\sigma)$, with $0\log0:=0$;
it is $+\infty$ otherwise.
The max-relative entropy~\cite{Datta2009} is
$D_{\max}(\rho\Vert\sigma):=\inf\{c:\rho\le e^c\sigma\}$.
For faithful states and $\alpha\in(0,1)\cup(1,\infty)$,
the Petz R\'enyi divergence~\cite{Petz1986}
and the sandwiched R\'enyi divergence~\cite{MullerLennertEtAl2013,WildeWinterYang2014}
are, respectively,
\[
 D^{\rm P}_{\alpha}(\rho\Vert\sigma)
 :=\frac1{\alpha-1}\log\Tr\rho^\alpha\sigma^{1-\alpha}
\]
and
\[
 \widetilde D_\alpha(\rho\Vert\sigma)
 :=\frac1{\alpha-1}
 \log\Tr\left(
 \sigma^{\frac{1-\alpha}{2\alpha}}
 \rho
 \sigma^{\frac{1-\alpha}{2\alpha}}
 \right)^\alpha.
\]
Both have the order-one limit $D(\rho\Vert\sigma)$. We set
$D^{\rm P}_1(\rho\Vert\sigma)=\widetilde D_1(\rho\Vert\sigma)
:=D(\rho\Vert\sigma)$.
For the sandwiched divergence,
$\lim_{\alpha\to\infty}\widetilde D_\alpha(\rho\Vert\sigma)
=D_{\max}(\rho\Vert\sigma)$, and we set
$\widetilde D_\infty(\rho\Vert\sigma):=D_{\max}(\rho\Vert\sigma)$.

For mutually absolutely continuous probability measures $P$ and $Q$
(that is, $P\ll Q$ and $Q\ll P$), the classical R\'enyi divergence of
order $\alpha\in(0,1)\cup(1,\infty)$ is
\[
 D_\alpha(P\Vert Q):=\frac{1}{\alpha-1}
 \log\int\left(\frac{dP}{dQ}\right)^\alpha dQ,
\]
where $dP/dQ$ is the likelihood ratio. At orders one and infinity, we use
$D_1(P\Vert Q)=D(P\Vert Q):=\int\log(dP/dQ)\,dP$ and
$D_\infty(P\Vert Q):=\log\operatorname*{ess\,sup}_Q(dP/dQ)$.
The reverse-sandwiched quantities needed in the later reductions are
defined at their first use.

The reverse relative entropy $D_{\mathrm R}(\tau\Vert X)$ is the state-space
rate function of Keyl's measurement~\cite[Theorem~3.2]{Keyl2006}; see also
Ref.~\cite{GrootveldEtAl2026}. It is not obtained by interchanging the
arguments of the ordinary relative entropy. For $\tau\in\mathcal D_d$ and
faithful $X$, write
$\tau=U\operatorname{diag}(t_1,\ldots,t_d)U^\dagger$, with
$t_1\ge\cdots\ge t_d$ and $t_{d+1}:=0$. Then
\begin{equation}
 D_{\mathrm R}(\tau\Vert X)
 =-H(t)-\sum_{k=1}^d(t_k-t_{k+1})
 \log\Delta_k(U^\dagger XU),
 \label{eq:cov-input-SKeyl-principal-minor}
\end{equation}
where $H(t):=-\sum_{i=1}^d t_i\log t_i$, with $0\log0:=0$, is the
Shannon entropy of the eigenvalues, and $\Delta_k(A)$ is the determinant
of the upper-left $k\times k$ submatrix of $A$. The columns of $U$ form
an ordered orthonormal eigenbasis of $\tau$. The expression is independent
of the basis chosen within any degenerate eigenspace because terms
involving an intermediate subspace have coefficient $t_k-t_{k+1}=0$. For faithful $X$,
it is finite and continuous on the compact state space $\mathcal D_d$.
When $\tau$ and $X$ commute, evaluation in a common eigenbasis gives
$D_{\mathrm R}(\tau\Vert X)=D(\tau\Vert X)$.

The same quantity arises as a limit of quantum R\'enyi divergences in
Audenaert and Datta~\cite[Theorems~2--3]{AudenaertDatta2015}.
For faithful arguments, it is also the quantity denoted by
$D^{\star}(\tau\Vert X)$ by Lipka-Bartosik, Chubb, Renes, Tomamichel,
and Korzekwa~\cite[Eqs.~(83b) and (85)]{LipkaBartosikEtAl2024}, who obtain it from the
right-pinched R\'enyi relative entropy as $\alpha\uparrow1$.
It also appears in Hayashi's quantum Sanov theorem~\cite{HayashiSanov2026}
and in the composite hypothesis-testing problem of
Hayashi and Fang~\cite[Definition~4 and Corollary~26]{HayashiFang2026}.

\section{Exact exponents for Keyl's measurement}
\label{sec:keyl-main-bounds-reorganized}
\subsection{Exponent formulas}
\label{subsec:keyl-main-results}
We fix Keyl's measurement and optimize the decision rule applied to its
state estimate. All three exponents are expressed using the rate function
$I_X(\tau):=D_{\mathrm R}(\tau\Vert X)$ of the estimate under $X^{\otimes n}$,
given in Eq.~\eqref{eq:cov-input-SKeyl-principal-minor}.

\begin{theorem}[Stein exponent for Keyl's measurement]
\label{thm:keyl-stein-exact-reorganized}
For faithful states $\rho,\sigma$ and each $0<\epsilon<1$,
\begin{equation}
 \lim_{n\to\infty}\frac1n
 D^H_{\epsilon\Vert\mathfrak T_d^{\mathrm K}}
 (\rho^{\otimes n}\Vert\sigma^{\otimes n})
 =D_{\mathrm R}(\rho\Vert\sigma).
 \label{eq:keyl-stein-exact}
\end{equation}
\end{theorem}
\begin{theorem}[Hoeffding exponent for Keyl's measurement]
\label{thm:keyl-hoeffding-exact-reorganized}
\label{cor:keyl-output-hoeffding-boundary}
For faithful $\rho$ and $\sigma$ and every $r>0$,
\begin{equation}
 H^{\mathrm K}(r\mid\rho\mid\sigma)
 =\min_{\tau:\,D_{\mathrm R}(\tau\Vert\sigma)\le r}
 D_{\mathrm R}(\tau\Vert\rho).
 \label{eq:keyl-hoeffding-exact}
\end{equation}
Replacing the minimum with an infimum over the strict sublevel
$D_{\mathrm R}(\tau\Vert\sigma)<r$ gives the same value.
\end{theorem}

If the required type-II decay rate exceeds $D_{\mathrm R}(\rho\Vert\sigma)$,
the type-I error tends to one. The next theorem gives the optimal decay
rate of $1-\alpha_n$. We write $(x)_+=\max\{x,0\}$.

\begin{theorem}[Strong-converse exponent for Keyl's measurement]
\label{thm:keyl-strong-converse-exponent}
For faithful $\rho$ and $\sigma$ and every $r\ge0$,
\begin{equation}
 E_{\mathrm{sc}}^{\mathrm{Keyl}}(r\mid\rho\mid\sigma)
 =\inf_{\tau\in\mathcal D_d}
 \left\{D_{\mathrm R}(\tau\Vert\rho)
 +(r-D_{\mathrm R}(\tau\Vert\sigma))_+\right\}.
 \label{eq:keyl-strong-converse-exact}
\end{equation}
\end{theorem}
To see the positive-part term, start with the rule that accepts estimates
near $\tau$. As the neighborhood shrinks after the blocklength limit, the
two acceptance exponents approach $I_\rho(\tau)$ and $I_\sigma(\tau)$.
If $I_\sigma(\tau)<r$, accepting these estimates with exponentially
smaller probability adds the missing rate to both exponents.

The direct bounds use large deviations on neighborhoods and sublevels
of $I_X$. For the strong-converse bound we use the exact likelihood
ratio of the reported estimate
(Lemma~\ref{lem:finite-keyl-likelihood-ratio}). Its consequence for
acceptance above the Stein threshold
(Corollary~\ref{cor:keyl-output-stein-strong-converse-implication}) gives
the Stein and Hoeffding converses. The Hoeffding direct bound uses the
closed sublevel $I_\sigma\le r$, and the converse uses the strict
sublevel $I_\sigma<r$. Section~\ref{subsec:no-positive-local-minima-DR}
shows that the strict sublevel is dense in the closed sublevel for $r>0$,
which gives equality of the two variational values.
\subsection{Direct bounds}
\label{sec:keyl-direct-reorganized}
Write $\alpha_n$ and $\beta_n$ for the errors of the rule in use, and set
$p_n=1-\alpha_n$.

\subsubsection{Large deviations of the state estimate}
\label{subsec:keyl-direct-common-ldp}
The direct constructions use the following open- and closed-set bounds.

\begin{proposition}[Large deviations of Keyl's state estimate]
\label{prop:keyl-physical-output-ldp}
Let $X$ be faithful, and let $P^{\mathrm{phys}}_{X,n}$ be the law of the
state estimate $\widehat\tau_n$ under $X^{\otimes n}$.
Then $(P^{\mathrm{phys}}_{X,n})_{n\ge1}$ satisfies a large-deviation principle (LDP) on
$\mathcal D_d$ with rate $I_X(\tau):=D_{\mathrm R}(\tau\Vert X)$.
For open $G\subseteq\mathcal D_d$ and closed $F\subseteq\mathcal D_d$,
\begin{align}
 \limsup_{n\to\infty}-\frac1n\log
 P^{\mathrm{phys}}_{X,n}(G)
 &\le \inf_{\tau\in G}I_X(\tau),
 \label{eq:keyl-physical-open-ldp}\\
 \liminf_{n\to\infty}-\frac1n\log
 P^{\mathrm{phys}}_{X,n}(F)
 &\ge \inf_{\tau\in F}I_X(\tau).
 \label{eq:keyl-physical-closed-ldp}
\end{align}
The rate $I_X$ is finite and uniformly continuous on $\mathcal D_d$, has compact sublevel sets, and has its unique zero at $X$.
\end{proposition}

\begin{proof}
The LDP and principal-minor rate follow from
\cite[Theorem~3.2]{Keyl2006}; the unique-zero property is part of the LDP
formulation in \cite[Definition~2.2]{Keyl2006}.
The identification with $D_{\mathrm R}(\cdot\Vert X)$ is also recorded in
\cite[Eq.~(6) and Sec.~1.3]{GrootveldEtAl2026}.

For completeness, continuity at degenerate reported spectra follows directly
from the principal-minor formula.  For
$\tau=U\operatorname{diag}(t)U^\dagger$,
\begin{equation}
 I_X(\tau)
 =-H(t)-\sum_{k=1}^d(t_k-t_{k+1})
 \log\Delta_k(U^\dagger XU),
 \qquad t_{d+1}:=0.
 \label{eq:keyl-ldp-regularity-formula}
\end{equation}
Faithfulness gives
$\Delta_k(U^\dagger XU)\ge\lambda_{\min}(X)^k>0$ uniformly in $U$.
The displayed function is therefore finite and continuous in $(t,U)$.
Changes of basis within a repeated-eigenvalue block affect only minors
whose coefficients $t_k-t_{k+1}$ vanish.  Thus the function is constant on
the fibres of the continuous map
$(t,U)\mapsto U\operatorname{diag}(t)U^\dagger$ from a compact space onto
$\mathcal D_d$, and descends to a continuous function of the reported state,
including at zero and repeated eigenvalues; see also
\cite[Theorem~3 and Eq.~(24)]{AudenaertDatta2015}.
Compactness of $\mathcal D_d$ gives uniform continuity and compact sublevel sets.
\end{proof}

\begin{corollary}[Local estimates]
\label{cor:keyl-local-physical-output-estimates}
Let $X$ be faithful and let $G_\delta(\tau)$ be an open neighborhood of $\tau$ contained in
$\{\eta\in\mathcal D_d:\|\eta-\tau\|_1<\delta\}$. Then
\[
 \lim_{n\to\infty}-\frac1n\log P^{\mathrm{phys}}_{X,n}(G_\delta(\tau))
 =\inf_{\eta\in G_\delta(\tau)}I_X(\eta).
\]
This infimum tends to $I_X(\tau)$ as $\delta\downarrow0$.
\end{corollary}

\begin{proof}
Continuity gives the same infimum over $G_\delta(\tau)$ and its closure, so the two large-deviation bounds agree. Continuity at $\tau$ gives the limit as $\delta\downarrow0$.
\end{proof}
\subsubsection{Stein direct bound}
\label{subsec:keyl-stein-direct}
\begin{proposition}[Stein direct bound for the Keyl measurement]
\label{prop:keyl-stein-direct-reorganized}
For faithful states $\rho,\sigma$ and each $0<\epsilon<1$,
\begin{equation}
 \liminf_{n\to\infty}\frac1n
 D^H_{\epsilon\Vert\mathfrak T_d^{\mathrm K}}
 (\rho^{\otimes n}\Vert\sigma^{\otimes n})
 \ge D_{\mathrm R}(\rho\Vert\sigma).
 \label{eq:keyl-stein-direct-target}
\end{equation}
\end{proposition}
\begin{proof}
Fix $\eta>0$. By continuity, choose an open trace-norm ball $G_\eta$
about $\rho$ such that
\begin{equation}
 \inf_{\tau\in\overline G_\eta}I_\sigma(\tau)
 \ge D_{\mathrm R}(\rho\Vert\sigma)-\eta.
 \label{eq:keyl-stein-direct-neighbourhood-rate}
\end{equation}
Accept exactly when $\widehat\tau_n\in G_\eta$.
The complement is compact and excludes the unique zero $\rho$ of $I_\rho$,
so the closed-set LDP gives
\begin{align}
 \liminf_{n\to\infty}-\frac1n\log\alpha_n
 &\ge\inf_{\tau\in G_\eta^c}I_\rho(\tau)>0,
 \label{eq:keyl-stein-direct-type-I-chain}\\
 \liminf_{n\to\infty}-\frac1n\log\beta_n
 &\ge\inf_{\tau\in\overline G_\eta}I_\sigma(\tau)
 \ge D_{\mathrm R}(\rho\Vert\sigma)-\eta.
 \label{eq:keyl-stein-direct-type-II-chain}
\end{align}
For the second inequality, enlarge $G_\eta$ to its closure before applying
the LDP under $\sigma$. Thus $\alpha_n\le\epsilon$ eventually; use the
always-accept rule at the finitely many earlier blocklengths. Letting
$\eta\downarrow0$ proves the claim.
\end{proof}

\subsubsection{Strong-converse direct bound}
\label{subsec:keyl-strong-converse-direct}
Write $\Psi_{\mathrm{sc}}^{\mathrm K}$ for the variational expression in
Eq.~\eqref{eq:keyl-strong-converse-exact},
\begin{equation}
 \Psi_{\mathrm{sc}}^{\mathrm K}(r\mid\rho\mid\sigma)
 :=\inf_{\tau\in\mathcal D_d}
 \left\{I_\rho(\tau)+(r-I_\sigma(\tau))_+\right\},
 \qquad r\ge0.
 \label{eq:keyl-Psi-definition}
\end{equation}

\begin{proposition}[Strong-converse direct bound for the Keyl measurement]
\label{prop:keyl-strong-converse-direct}
For faithful $\rho$ and $\sigma$ and $r\ge0$,
$E_{\mathrm{sc}}^{\mathrm{Keyl}}(r\mid\rho\mid\sigma) \le \Psi_{\mathrm{sc}}^{\mathrm K}(r\mid\rho\mid\sigma)$.
\end{proposition}

\begin{proof}
Continuity on the compact state space gives a minimizer $\tau_*$ in
Eq.~\eqref{eq:keyl-Psi-definition}. Let $G_\delta$ be the open trace-norm
ball of radius $\delta$ about $\tau_*$, relative to $\mathcal D_d$, and put
$b_\delta:=\inf_{\tau\in\overline G_\delta}I_\sigma(\tau)$.
Use the randomized acceptance rule
\[
 f_{n,\delta}(z)
 :=e^{-n(r-b_\delta)_+}
   \mathbf1\{\widehat\tau_n(z)\in G_\delta\}.
\]
The random acceptance factor is the same under both hypotheses. The closed-set LDP under $\sigma$ and the open-set LDP under $\rho$ give
\begin{align}
 \liminf_{n\to\infty}-\frac1n\log\beta_n
 &\ge b_\delta+(r-b_\delta)_+\ge r,
 \label{eq:keyl-direct-beta-explicit}\\
 \limsup_{n\to\infty}-\frac1n\log p_n
 &\le\inf_{\tau\in G_\delta}I_\rho(\tau)+(r-b_\delta)_+
 \le I_\rho(\tau_*)+(r-b_\delta)_+.
 \label{eq:keyl-sc-direct-success-chain}
\end{align}
Each fixed $\delta$ defines a sequence meeting the type-II constraint.
Letting $\delta\downarrow0$ gives $b_\delta\to I_\sigma(\tau_*)$.
\end{proof}

\subsubsection{Hoeffding direct bound}
\label{subsec:keyl-hoeffding-direct}
The closed-sublevel minimum in Eq.~\eqref{eq:keyl-hoeffding-exact} is
\begin{equation}
 H_{\mathrm{cl}}(r\mid\rho\mid\sigma)
 :=\min_{\tau:\,I_\sigma(\tau)\le r} I_\rho(\tau),
 \qquad r\ge0.
 \label{eq:keyl-Hcl-definition}
\end{equation}

\begin{proposition}[Hoeffding direct bound for Keyl's measurement]
\label{prop:keyl-hoeffding-direct}
Every rate $r>0$ satisfies
$H^{\mathrm K}(r\mid\rho\mid\sigma) \ge H_{\mathrm{cl}}(r\mid\rho\mid\sigma)$.
A deterministic divergence threshold attains the lower bound.
\end{proposition}

\begin{proof}
Accept when $I_\sigma(\widehat\tau_n)\ge r$. The acceptance set is closed,
and the rejection set lies in the closed set $\{I_\sigma\le r\}$.
The closed-set bounds under $\sigma$ and $\rho$ give
\begin{align}
 \liminf_{n\to\infty}-\frac1n\log\beta_n
 &\ge r,
 \label{eq:keyl-hoeffding-direct-beta-chain}\\
 \liminf_{n\to\infty}-\frac1n\log\alpha_n
 &\ge \min_{\tau:\,I_\sigma(\tau)\le r}I_\rho(\tau)
 =H_{\mathrm{cl}}(r\mid\rho\mid\sigma).
 \label{eq:keyl-hoeffding-direct-alpha-chain}
\end{align}
\end{proof}

\begin{remark}[Finite-cell implementations]
\label{rem:keyl-hoeffding-finite-cell-implementations}
Fix $a>r$ and $\eta>0$. Choose $\delta>0$ such that $I_\sigma$
varies by at most $\eta$ between states at trace-norm distance at most $\delta$. Partition the compact state space into finitely
many measurable cells of diameter at most $\delta$, and let
$A_\delta^{(a)}$ be the union of cells whose closures lie in
$\{I_\sigma\ge a\}$. Then
\[
 A_\delta^{(a)}\subseteq\{I_\sigma\ge a\},
 \qquad
 \mathcal D_d\setminus A_\delta^{(a)}
 \subseteq\{I_\sigma\le a+\eta\}.
\]
The two closed-set bounds give a type-II exponent at least $a$ and a
type-I exponent at least $H_{\mathrm{cl}}(a+\eta\mid\rho\mid\sigma)$.
The function $H_{\mathrm{cl}}$ is right-continuous. Indeed, if $b_j\downarrow b$,
a limit point of minimizers at $b_j$ is feasible at $b$, so continuity of $I_\rho$
gives $H_{\mathrm{cl}}(b\mid\rho\mid\sigma)\le\lim_j H_{\mathrm{cl}}(b_j\mid\rho\mid\sigma)$.
The opposite inequality follows by inclusion of the feasible sets.
Thus letting $\eta\downarrow0$ and $a\downarrow r$ approximates the ideal threshold exponent with finite-cell rules.
\end{remark}

\subsection{The finite-block Keyl law}
\label{sec:keyl-finite-block-reorganized}
In the outcome $(\lambda,\bar F)$, the Young label
$\lambda=(\lambda_1,\ldots,\lambda_d)$ consists of nonincreasing nonnegative
integers summing to $n$. Its entries give the estimated eigenvalues
$\lambda/n$. Set $\lambda_{d+1}=0$. A complete flag $F$ is a nested family
$V_1\subset\cdots\subset V_d=\mathbb C^d$ with $\dim V_k=k$. The reported
partial flag $\bar F$ retains only the subspaces at the boundaries between
distinct eigenvalues, together with $V_d$. Thus
\[
 (\lambda,F)
 \xrightarrow{\ \mathrm{forget\ added\ subspaces}\ }(\lambda,\bar F)
 \longleftrightarrow\widehat\tau_n.
\]
The bijection on the right holds because the eigenvalues and eigenspaces
recover the reported estimate and vice versa. We write
$\widetilde P_{X,n}^{\mathrm K}$, $P_{X,n}^{\mathrm K}$, and
$P_{X,n}^{\mathrm{phys}}$ for the three laws under $X^{\otimes n}$.

\subsubsection{The reported state and complete flags}
\label{subsec:keyl-physical-labelled-output}
For the principal-minor calculation, sample the intermediate subspaces
inside each repeated eigenspace from a conditional law independent of the
input. This completion of the flag leaves both testing probabilities unchanged.

\begin{lemma}[Equivalence of the output representations]
\label{lem:keyl-lifted-output-equivalence}
Let $P_{X,n}^{\mathrm K}$ be the physical partial-flag output law and
$\widetilde P_{X,n}^{\mathrm K}$ its complete-flag lift.  Let
$K_\lambda(dF\mid\bar F)$ be the state-independent normalized invariant
conditional probability on complete flags above the physical partial flag
$\bar F$ in the orbit belonging to $\lambda$.
Apply a rule on $(\lambda,\bar F)$ to a complete flag by ignoring the
added subspaces. This preserves the acceptance probability.
Conversely, every measurable rule $f_n^{\mathrm{lift}}(\lambda,F)\in[0,1]$ on
complete flags defines a rule on the reported partial flag by averaging,
$\overline f_n^{\mathrm{lift}}(\lambda,\bar F):=\int f_n^{\mathrm{lift}}(\lambda,F)K_\lambda(dF\mid\bar F)$,
and satisfies the following acceptance-probability identity under
every input state $X$:
\begin{equation}
 \int f_n^{\mathrm{lift}}\,d\widetilde P_{X,n}^{\mathrm K}
 =\int \overline f_n^{\mathrm{lift}}\,dP_{X,n}^{\mathrm K}.
 \label{eq:keyl-lift-acceptance-equality}
\end{equation}
At fixed $n$, the estimator map from $(\lambda,\bar F)$ to
$\widehat\tau_n$ is a Borel isomorphism onto its image.
\end{lemma}

The next identity comes from Keyl's finite-block covariant measurement
\cite[Eq.~(20) and Sec.~4.5, Eqs.~(140)--(141)]{Keyl2006}.
Here $\pi_\lambda$ is the irreducible polynomial representation with
highest weight $\lambda$, and $v_\lambda$ is a normalized highest-weight
vector.

\begin{lemma}[Finite-block fibre constancy]
\label{lem:keyl-finite-block-fibre-constancy}
Fix a Young index $\lambda$, and let $\mathcal O_\lambda$ be the physical orbit of the highest-weight ray $[v_\lambda]$.
Let $F$ be a complete flag above $\bar F\in\mathcal O_\lambda$, represented
by a unitary $U_F$.  When $X$ is full rank, the state-dependent factor in
the finite-block Keyl density satisfies
\begin{equation}
 \left\langle v_\lambda,
 \pi_\lambda(U_F^\dagger XU_F)v_\lambda\right\rangle
 =\prod_{k=1}^d
 \Delta_k(U_F^\dagger XU_F)^{\lambda_k-\lambda_{k+1}}.
 \label{eq:keyl-fibre-constancy-generalized-power}
\end{equation}
All choices of $U_F$ representing the same reported partial flag
$\bar F$ give the same value. This applies to each full-rank input state,
so the likelihood ratio is constant across all completions of the flag.
\end{lemma}

Appendix~\ref{app:keyl-finite-output-law} gives the proofs.

\subsubsection{Finite-block likelihood ratio}
\label{subsec:keyl-exact-likelihood-ratio}
The finite-block density in Proposition~\ref{prop:appendix-finite-keyl-law} gives a likelihood ratio expressed entirely in terms of the reported state.

\begin{lemma}[Finite-block Keyl likelihood ratio]
\label{lem:finite-keyl-likelihood-ratio}
For faithful states $\rho$ and $\sigma$, the Keyl laws
$P_{\rho,n}^{\mathrm K}$ and $P_{\sigma,n}^{\mathrm K}$ are mutually
absolutely continuous. On their common support,
\begin{equation}
  \frac1n\log
  \frac{dP_{\rho,n}^{\mathrm K}}
       {dP_{\sigma,n}^{\mathrm K}}(z)
  =
  D_{\mathrm R}(\widehat\tau_n(z)\Vert\sigma)
  -D_{\mathrm R}(\widehat\tau_n(z)\Vert\rho).
  \label{eq:finite-keyl-likelihood-ratio}
\end{equation}
\end{lemma}

\begin{proof}
Faithfulness makes both densities in
Eq.~\eqref{eq:appendix-finite-keyl-law} strictly positive with respect to the
same reference measure. Their ratio cancels the dimension and multiplicity
factors. For a reported outcome $z=(\lambda,\bar F)$, complete $\bar F$ to
a flag $F$ and put $t_k^{(n)}:=\lambda_k/n$. Then
\begin{align}
 \frac1n\log\frac{dP_{\rho,n}^{\mathrm K}}{dP_{\sigma,n}^{\mathrm K}}(z)
 &=\sum_{k=1}^d(t_k^{(n)}-t_{k+1}^{(n)})
 \log\frac{\Delta_k(U_F^\dagger\rho U_F)}
           {\Delta_k(U_F^\dagger\sigma U_F)}
 \notag\\
 &=D_{\mathrm R}(\widehat\tau_n(z)\Vert\sigma)
   -D_{\mathrm R}(\widehat\tau_n(z)\Vert\rho).
 \label{eq:finite-keyl-lr-identification-chain}
\end{align}
The last equality uses the principal-minor formula: the entropy terms
cancel because they depend only on the reported spectrum.
Lemma~\ref{lem:keyl-finite-block-fibre-constancy} ensures that this expression
is independent of the chosen flag above $\bar F$.
\end{proof}

\subsection{Converse bounds}
\label{sec:keyl-converse-reorganized}
\subsubsection{Strong-converse bound}
\label{sec:keyl-output-strong-converse}
We split null acceptance into two parts. The probability of a large
likelihood ratio is bounded by the LDP under $\rho$. On the rest of the
outcome space, the likelihood ratio gives an upper bound in terms of
the type-II error. For a rule $f$ write
$P_{X,n}^{\mathrm K}(f):=\int f\,dP_{X,n}^{\mathrm K}$, so
$p_n(f)=P_{\rho,n}^{\mathrm K}(f)$ and $\beta_n(f)=P_{\sigma,n}^{\mathrm K}(f)$.
\begin{proposition}[Converse for Keyl's strong-converse exponent]
\label{prop:keyl-strong-converse-converse}
Fix $r\ge0$. Let \(\rho\) and \(\sigma\) be full-rank states, and let
\(f_n^{\mathrm K}:\Omega_n^{\mathrm K}\to[0,1]\) be arbitrary measurable randomized
decision rules satisfying
$\liminf_{n\to\infty} -\frac1n\log\beta_n(f_n^{\mathrm K}) \ge r$.
Under this constraint, the null-acceptance exponent satisfies
\begin{equation}
  \liminf_{n\to\infty}
  -\frac1n\log p_n(f_n^{\mathrm K})
  \ge
  \Psi_{\mathrm{sc}}^{\mathrm K}(r\mid\rho\mid\sigma).
  \label{eq:keyl-converse-success-bound}
\end{equation}
\end{proposition}

\begin{proof}
Write $\psi:=\Psi_{\mathrm{sc}}^{\mathrm K}(r\mid\rho\mid\sigma)$.
By Lemma~\ref{lem:finite-keyl-likelihood-ratio},
$G(\tau):=I_\sigma(\tau)-I_\rho(\tau)$ is the normalized log likelihood of
$\rho$ relative to $\sigma$ at the reported state $\tau$.
If $\psi=0$, the claim follows from $p_n\le1$. For every $\tau\in\mathcal D_d$,
the variational definition gives
\[
 \psi\le I_\rho(\tau)+(r-I_\sigma(\tau))_+
 =\max\{I_\rho(\tau),r-G(\tau)\}.
\]
Choose $0<\delta<\psi$ and set $c=r-\psi+\delta$ so that on
$\{G\ge c\}$ the second term of the maximum is at most $\psi-\delta$.
Hence $I_\rho\ge\psi$ on this closed set. On $\{G<c\}$ the likelihood
ratio is bounded by $e^{nc}$. Using $f_n^{\mathrm K}\le1$ on the first set
and changing measure on the second gives the standard likelihood-ratio
bound \cite[Eq.~(14)]{NagaokaHayashi2007}
\[
 p_n(f_n^{\mathrm K})
 \le P_{\rho,n}^{\mathrm{phys}}\{\tau:G(\tau)\ge c\}
      +e^{nc}\beta_n(f_n^{\mathrm K}).
\]
The first term has exponent at least $\psi$ by the closed-set LDP,
and the second has exponent at least $r-c$ by the type-II constraint. Thus
\[
 \liminf_{n\to\infty}-\frac1n\log p_n(f_n^{\mathrm K})
 \ge\min\{\psi,r-c\}=\psi-\delta.
\]
Let $\delta\downarrow0$. The same argument applies along any subsequence.
\end{proof}

\begin{proof}[Completion of the proof of Theorem~\ref{thm:keyl-strong-converse-exponent}]
Proposition~\ref{prop:keyl-strong-converse-direct} gives
$E_{\mathrm{sc}}^{\mathrm{Keyl}}\le\Psi_{\mathrm{sc}}^{\mathrm K}$.
Proposition~\ref{prop:keyl-strong-converse-converse}, together with
$\limsup\ge\liminf$, gives the reverse inequality after taking the
infimum over the same class of admissible randomized rules.
\end{proof}
\begin{corollary}[Decay of acceptance above the Stein threshold]
\label{cor:keyl-output-stein-strong-converse-implication}
Let $\tau$ and $X$ be faithful states, and let $f_n^{\mathrm K}$ be arbitrary measurable
randomized decision rules applied to Keyl's state estimate.  If
$\liminf_{n\to\infty}-\frac1n\log P^{\mathrm K}_{X,n}(f_n^{\mathrm K}) >D_{\mathrm R}(\tau\Vert X)$,
then $P^{\mathrm K}_{\tau,n}(f_n^{\mathrm K})$ tends to zero exponentially.
The result also holds along any subsequence, and for the complementary rule
when it satisfies the same rate condition.
\end{corollary}
\begin{proof}
Choose $r$ strictly between $D_{\mathrm R}(\tau\Vert X)$ and the assumed
liminf rate. The continuous nonnegative function
\[
 \eta\longmapsto D_{\mathrm R}(\eta\Vert\tau)
       +(r-D_{\mathrm R}(\eta\Vert X))_+
\]
attains its minimum on $\mathcal D_d$. A zero minimum would force
$\eta=\tau$ by the unique-zero property and simultaneously require
$D_{\mathrm R}(\tau\Vert X)\ge r$, a contradiction.
Hence $\Psi_{\mathrm{sc}}^{\mathrm K}(r\mid\tau\mid X)>0$, and
Proposition~\ref{prop:keyl-strong-converse-converse} gives exponential decay
of $P_{\tau,n}^{\mathrm K}(f_n^{\mathrm K})$.
\end{proof}
Appendix~\ref{subsec:keyl-stein-two-converses} compares this result with the covariant-tomography converse of Grootveld et al.
\subsubsection{Stein converse}
\label{subsec:keyl-stein-converse-reorganized}
\begin{proposition}[Stein converse for post-processing the Keyl output]
\label{prop:keyl-stein-converse-reorganized}
Let $\rho$ and $\sigma$ be faithful states and fix $0<\epsilon<1$.  Then
\begin{equation}
 \limsup_{n\to\infty}\frac1n
 D^H_{\epsilon\Vert\mathfrak T_d^{\mathrm K}}
 (\rho^{\otimes n}\Vert\sigma^{\otimes n})
 \le D_{\mathrm R}(\rho\Vert\sigma).
 \label{eq:keyl-stein-converse-target}
\end{equation}
\end{proposition}

\begin{proof}
If the asserted limsup bound failed, nearly optimal tests along a
subsequence $(n_j)$ would satisfy $\alpha_{n_j}\le\epsilon$ and, for some
$\delta>0$,
\[
 \liminf_{j\to\infty}-\frac1{n_j}\log
 P_{\sigma,n_j}^{\mathrm K}(f_{n_j}^{\mathrm K})
 \ge D_{\mathrm R}(\rho\Vert\sigma)+\delta.
\]
Corollary~\ref{cor:keyl-output-stein-strong-converse-implication}
then gives $P_{\rho,n_j}^{\mathrm K}(f_{n_j}^{\mathrm K})\to0$, contradicting
$P_{\rho,n_j}^{\mathrm K}(f_{n_j}^{\mathrm K})=1-\alpha_{n_j}\ge1-\epsilon>0$.
\end{proof}

Combining this proposition with the Stein direct bound in
Proposition~\ref{prop:keyl-stein-direct-reorganized} proves
Eq.~\eqref{eq:keyl-stein-exact}, completing the proof of
Theorem~\ref{thm:keyl-stein-exact-reorganized}.

\subsubsection{Hoeffding converse}
\label{subsec:keyl-output-hoeffding-converse-expanded}
The strict rate requirement in
Corollary~\ref{cor:keyl-output-stein-strong-converse-implication} leads to
\begin{equation}
 H_{\mathrm{op}}(r\mid\rho\mid\sigma)
 :=\inf_{\tau:\,I_\sigma(\tau)<r} I_\rho(\tau),
 \qquad r>0.
 \label{eq:keyl-Hop-definition}
\end{equation}
We apply the corollary to a rule and its complement under the same
auxiliary state. Their two acceptance probabilities must sum to one.

\begin{theorem}[Hoeffding converse for arbitrary processing of the Keyl output]
\label{thm:keyl-output-hoeffding-converse}
Let $r>0$, and let $f_n^{\mathrm K}$ be arbitrary measurable randomized
decision rules applied to Keyl's state estimate satisfying
$\liminf_{n\to\infty}-\frac1n\log\beta_n\ge r$.
Then
\begin{equation}
 \limsup_{n\to\infty}-\frac1n\log\alpha_n
 \le H_{\mathrm{op}}(r\mid\rho\mid\sigma).
 \label{eq:keyl-hoeffding-converse-target}
\end{equation}
\end{theorem}

\begin{proof}
Suppose the claimed bound fails. Choose
$h>H_{\mathrm{op}}(r\mid\rho\mid\sigma)$ and a subsequence $(n_j)$ such that
$\liminf_j-n_j^{-1}\log\alpha_{n_j}\ge h$.
The type-II rate remains at least $r$ along this subsequence.
By the strict-sublevel definition, there is a state $\tau$ with
\[
 D_{\mathrm R}(\tau\Vert\sigma)<r,
 \qquad D_{\mathrm R}(\tau\Vert\rho)<h.
\]
We may take $\tau$ faithful: if necessary, replace it by
$(1-\delta)\tau+\delta I/d$ with $\delta>0$ small enough that continuity
preserves both strict inequalities.

Apply Corollary~\ref{cor:keyl-output-stein-strong-converse-implication}
along $(n_j)$ to the following two choices:
\begin{align}
 f_{n_j}^{\mathrm K}:&\quad
 \text{null }\tau,\quad\text{alternative }\sigma,
 \quad\text{type-II quantity }\beta_{n_j};\notag\\
 1-f_{n_j}^{\mathrm K}:&\quad
 \text{null }\tau,\quad\text{alternative }\rho,
 \quad\text{type-II quantity }\alpha_{n_j}.
 \label{eq:keyl-hoeffding-converse-role-map}
\end{align}
Both $P_{\tau,n_j}^{\mathrm K}(f_{n_j}^{\mathrm K})$ and
$P_{\tau,n_j}^{\mathrm K}(1-f_{n_j}^{\mathrm K})$ tend to zero, contradicting
their sum being one. This proves Eq.~\eqref{eq:keyl-hoeffding-converse-target}
for every admissible randomized sequence and hence the operational bound
$H^{\mathrm K}(r\mid\rho\mid\sigma)\le H_{\mathrm{op}}(r\mid\rho\mid\sigma)$.
\end{proof}

\subsubsection{Equality of the strict and closed variational values}
\label{subsec:no-positive-local-minima-DR}
It suffices to approximate every state with $I_\sigma(\tau)=r>0$ by states
with smaller $I_\sigma$. For a noncommuting state we move the eigenspaces
while keeping the eigenvalues fixed. For a commuting state we interpolate
toward $\sigma$ in a common eigenbasis.

\begin{lemma}[Attainment of the closed-sublevel minimum]
\label{lem:keyl-Hcl-closed-realization}
Let $\rho$ and $\sigma$ be faithful states.  Every rate $r\ge0$
admits a minimizer in the definition of
$H_{\mathrm{cl}}(r\mid\rho\mid\sigma)$.
\end{lemma}

\begin{proof}
The continuous function $I_\rho$ attains its minimum on the compact set $\{I_\sigma\le r\}$, which contains $\sigma$.
\end{proof}

\begin{lemma}[Decrease of the Keyl rate at fixed spectrum]
\label{lem:weighted-flag-descent}
Let $X>0$, and let
$\tau=\sum_{a=1}^m q_a P_a$ be the spectral decomposition into distinct
eigenvalues $q_1>\cdots>q_m\ge0$. Evaluate $D_{\mathrm R}(\tau\Vert X)$
by the principal-minor formula also when $X$ is unnormalized. If
$[\tau,X]\ne0$, there is a smooth unitary curve $U_s$ with $U_0=I$ such that,
for sufficiently small $s>0$,
\begin{equation}
 D_{\mathrm R}(U_s\tau U_s^\dagger\Vert X)
 <D_{\mathrm R}(\tau\Vert X).
 \label{eq:weighted-flag-descent}
\end{equation}
\end{lemma}

\begin{proof}
Normalizing $X$ adds a constant independent of $\tau$.
Set $Q_a:=P_1+\cdots+P_a$ and $V_a:=\operatorname{Ran}Q_a$. Grouping the
minors at the distinct eigenvalues gives
\begin{equation}
 D_{\mathrm R}(\tau\Vert X)
 =-S(\tau)-\sum_{a=1}^{m-1}(q_a-q_{a+1})
 \log\det(Q_aXQ_a\mid V_a)-q_m\log\det X,
 \label{eq:DR-partial-flag-form}
\end{equation}
where $S(\tau)=-\Tr\tau\log\tau$.
The coefficient $q_a-q_{a+1}$ weights the log determinant of the
compression to $V_a$. We keep the spectrum of $\tau$ fixed and deform
all of these subspaces by the same map $e^{sX}$, so that
$V_a(s)=e^{sX}V_a$. Apply Gram--Schmidt to the image of an orthonormal
basis adapted to this flag. The resulting unitary $U_s$ maps $V_a$ to
$V_a(s)$ for every $a$, varies smoothly in $s$, and satisfies $U_0=I$.

For a fixed $a$, take an isometry $W$ onto $V_a$ and put $A=W^\dagger XW$.
The columns of $e^{sX}W(W^\dagger e^{2sX}W)^{-1/2}$ form an orthonormal basis of $V_a(s)$. Hence the determinant of the compression of $X$ onto $V_a(s)$ is
\[
 \frac{\det(W^\dagger Xe^{2sX}W)}{\det(W^\dagger e^{2sX}W)}.
\]
To differentiate the log determinant at $s=0$, write
$W^\dagger X^2W=A^2+W^\dagger X(I-Q_a)XW$. The $A^2$ term cancels
the derivative from the denominator, leaving
\begin{align*}
 2\Tr(A^{-1}W^\dagger X^2W)-2\Tr(A)
 &=2\Tr(A^{-1}W^\dagger X(I-Q_a)XW)\\
 &=2\|(I-Q_a)XWA^{-1/2}\|_2^2,
\end{align*}
where $\|\cdot\|_2$ denotes the Hilbert--Schmidt norm.
This derivative is positive exactly when $X$ couples $V_a$ with its
orthogonal complement. Since
$\tau=q_m I+\sum_{a=1}^{m-1}(q_a-q_{a+1})Q_a$, noncommutativity of $\tau$
and $X$ forces at least one derivative to be positive. All weights in
Eq.~\eqref{eq:DR-partial-flag-form} are positive and the other terms are
constant, so $D_{\mathrm R}(U_s\tau U_s^\dagger\Vert X)$ decreases.
\end{proof}

\begin{proposition}[No positive local minimum of the Keyl rate]
\label{prop:no-positive-local-minimum-keyl-rate}
Let $\sigma$ be a faithful state. The only local minimum of
$I_\sigma(\tau)=D_{\mathrm R}(\tau\Vert\sigma)$ on $\mathcal D_d$ is
$\tau=\sigma$, where the value is zero.  In particular, the rate has no
positive local minimum.
\end{proposition}

\begin{proof}
At a state $\tau$ with $[\tau,\sigma]\ne0$, Lemma~\ref{lem:weighted-flag-descent} strictly decreases the rate along a unitary curve through $\tau$.
If $\tau$ and $\sigma$ commute, the rate reduces in a common eigenbasis to classical relative entropy. For $\tau\ne\sigma$, its convexity in the first argument gives
\[
 I_\sigma\bigl((1-\lambda)\tau+\lambda\sigma\bigr)
 \le (1-\lambda)I_\sigma(\tau)<I_\sigma(\tau)
 \qquad(0<\lambda<1).
\]
This proves the claim since $I_\sigma$ is nonnegative and vanishes only at $\sigma$.
\end{proof}

\begin{corollary}[Density of strict reverse-divergence sublevels]
\label{cor:keyl-strict-sublevel-density}
For a faithful state $\sigma$ and $r>0$,
the strict reverse-divergence sublevel is dense in the corresponding closed sublevel.
\[
 \{\tau:D_{\mathrm R}(\tau\Vert\sigma)\le r\}
 =\overline{\{\tau:D_{\mathrm R}(\tau\Vert\sigma)<r\}} .
\]
\end{corollary}

\begin{proof}
Continuity gives the inclusion of the closure in the closed sublevel. For the converse, if $I_\sigma(\tau)=r>0$, then $\tau$ is not a local minimum by Proposition~\ref{prop:no-positive-local-minimum-keyl-rate}. Every neighborhood of $\tau$ therefore contains a state with rate strictly below $r$.
\end{proof}

\begin{theorem}[Equality of the strict and closed Hoeffding values]
\label{thm:keyl-Hcl-Hop-equality}
For faithful $\rho$ and $\sigma$ and every $r>0$,
\begin{equation}
 H_{\mathrm{cl}}(r\mid\rho\mid\sigma)
 =H_{\mathrm{op}}(r\mid\rho\mid\sigma)
 =\min_{\tau:\,D_{\mathrm R}(\tau\Vert\sigma)\le r}
 D_{\mathrm R}(\tau\Vert\rho).
 \label{eq:keyl-Hcl-Hop-exact}
\end{equation}
\end{theorem}

\begin{proof}
By Lemma~\ref{lem:keyl-Hcl-closed-realization}, choose a closed-sublevel
minimizer $\tau_*$. Corollary~\ref{cor:keyl-strict-sublevel-density} gives
states $\tau_j\to\tau_*$ with $D_{\mathrm R}(\tau_j\Vert\sigma)<r$.
Continuity then yields
\[
 H_{\mathrm{op}}(r\mid\rho\mid\sigma)
 \le\lim_j D_{\mathrm R}(\tau_j\Vert\rho)
 =D_{\mathrm R}(\tau_*\Vert\rho)
 =H_{\mathrm{cl}}(r\mid\rho\mid\sigma).
\]
The reverse inequality follows from inclusion of the strict sublevel in
the closed sublevel.
\end{proof}

\begin{proof}[Completion of the proof of Theorem~\ref{thm:keyl-hoeffding-exact-reorganized}]
The direct bound in Proposition~\ref{prop:keyl-hoeffding-direct} and the
converse in Theorem~\ref{thm:keyl-output-hoeffding-converse} place the
exponent between $H_{\mathrm{cl}}$ and $H_{\mathrm{op}}$.
They coincide by Theorem~\ref{thm:keyl-Hcl-Hop-equality}.
\end{proof}
\section{Optimal exponents for covariant measurements}
\label{sec:general-ucov}
We now choose both the covariant measurement and its classical decision
rule.

\subsection{Fixed-error Stein exponent and its limits}
\label{sec:covariant-stein-regime}
We first characterize the exponent at each type-I error tolerance
$0<\epsilon<1$. We then show that it approaches $D_{\mathrm R}(\rho\Vert\sigma)$
as $\epsilon\downarrow0$ and $D(\rho\Vert\sigma)$ as $\epsilon\uparrow1$.

\subsubsection{Exact exponent at fixed error}
\label{subsubsec:merge-fixed-interior-stein}

Let $H_\rho=\{h\in U(d):h\rho h^\dagger=\rho\}$ be the stabilizer of
$\rho$ and let $\mu$ be its normalized Haar measure. This group consists
of all unitaries that act separately on the eigenspaces of $\rho$.
The construction first uses a refinement of Keyl's measurement to record
a unitary $U$. Under $\rho$, the law of $U$ converges to $\mu$. To accept
with probability at least $1-\epsilon$, it suffices to retain rotations
near a compact set with Haar measure greater than $1-\epsilon$. The test
on the remaining copies is designed against the corresponding mixture of
rotated alternatives.

For a nonempty compact $K\subset H_\rho$, let $\nu_K$ be any probability
measure with support $K$ and put
\begin{equation}
 E_\rho(K)=\lim_{n\to\infty}\frac1n
 D\left(\rho^{\otimes n}\,\middle\Vert\,
    \int_K(h^\dagger\sigma h)^{\otimes n}d\nu_K(h)\right).
 \label{eq:merge-mixture-rate}
\end{equation}
In the integral, the same rotation $h$ acts on all $n$ copies.
The limit depends only on $K$ for fixed $\rho,\sigma$, as shown in
Appendix~\ref{app:merge-stein-proof}.

\begin{theorem}[Fixed-error Stein exponent for covariant measurements]
\label{thm:merge-ucov-fixed-interior-stein}
Let $\rho,\sigma$ be faithful states on $\mathbb C^d$. For every
$0<\epsilon<1$, the lower and upper asymptotic Stein exponents coincide, and
\begin{align}
 B^{\mathrm{ucov}}_\epsilon(\rho\Vert\sigma)
 &:={\lim}_{n\to\infty}\frac1n
  D^H_{\epsilon\Vert\mathfrak T_d^{\mathrm{ucov}}}
  (\rho^{\otimes n}\Vert\sigma^{\otimes n})\nonumber\\
 &=\max_{\substack{K\subset H_\rho\;\mathrm{compact}\\
                      \mu(K)\ge1-\epsilon}} E_\rho(K)\nonumber\\
 &=\sup_{\substack{A\subset H_\rho\;\mathrm{measurable}\\
                      \mu(A)>1-\epsilon}}
  \lim_{n\to\infty}\frac1n
   D\left(\rho^{\otimes n}\,\middle\Vert\,
   \frac{1}{\mu(A)}\int_A
         (h^\dagger\sigma h)^{\otimes n}d\mu(h)\right).
 \label{eq:merge-ucov-fixed-stein}
\end{align}
The exponent is continuous and nondecreasing in $\epsilon$ on $(0,1)$.
\end{theorem}

For achievability, measure $m$ copies with the full-unitary refinement
of Keyl's measurement, with $m\to\infty$ and $m=o(n)$. Given the recorded
$U$, rotate the second-stage acceptance effect by $U^{\otimes(n-m)}$.
Its acceptance on $X^{\otimes(n-m)}$ is then that of the unrotated
effect on $(U^\dagger XU)^{\otimes(n-m)}$. For any $R<E_\rho(K)$,
Lemma~\ref{lem:stabilizer-testing}(iii) supplies one second-stage test
that accepts the null uniformly near $\rho$ and has alternative
acceptance at most $e^{-(n-m)R}$ for orientations in a neighborhood
$O$ of $K$. Choose $O$ small enough that $U^\dagger\rho U$ lies in
the null neighborhood for $U\in O$. Following a positive second-stage
outcome, accept with a continuous probability $w(U)$ equal to one on
$K$ and supported in $O$.
Under $\rho$, the law of $U$ converges to Haar measure on $H_\rho$.
Thus null acceptance tends to $\int w\,d\mu\ge\mu(K)>1-\epsilon$.
The strict margin gives the finite-block tolerance eventually, and
$m=o(n)$ preserves the type-II rate $R$ per input copy.
Appendix~\ref{app:stein-mass-optimization} uses continuity in the
mass constraint to include compact sets of mass $1-\epsilon$.

For the converse, partition $H_\rho$ into finitely many cells before
taking $n\to\infty$. At each blocklength, we extract an effect with null
acceptance bounded below and small alternative acceptance for every
rotation in a union of cells of almost the required Haar measure.
The converse for that mixture bounds the type-II exponent. Only finitely
many unions can occur at fixed partition, so the bound holds even when
the selected cells change with $n$. Appendix~\ref{app:merge-stein-proof}
gives the full argument.

\subsubsection{Phase-invariant tests for the small-error limit}
\label{subsec:covariant-stein-formulation}
Let $\rho=\sum_{a=1}^s r_a E_a$, where the $r_a$ are distinct eigenvalues
and the $E_a$ are their spectral projections. The phases act separately
on these eigenspaces through the subgroup
\[
 G_\rho=\left\{\sum_{a=1}^{s-1}e^{i\theta_a}E_a+E_s:
           \theta_a\in\mathbb R\right\}
\]
of $H_\rho$, with normalized Haar measure $\mu_\rho$. Fixing the last
phase removes the common phase, which cancels in conjugation.
The representation on $n$ copies is $g\mapsto g^{\otimes n}$.

A measurement is phase invariant if all its effects commute with
$g^{\otimes n}$ for $g\in G_\rho$. It is phase covariant if it satisfies
the covariance equation of Section~\ref{sec:covariant-test-class} for
$G_\rho$. Write \(\mathcal T_{\mathrm{phase\mbox{-}inv}}\) and
\(\mathcal T_{\mathrm{phase\mbox{-}cov}}\) for the binary-test classes
obtained by classical decision rules applied to these measurements.
An invariant measurement is covariant with the trivial action on
outcomes. Any unitary-covariant measurement is phase covariant with the
restriction of its given action to $G_\rho$. For a decision rule $f$ on
the outcome space, covariance gives
\[
 g^{\otimes n}T_f(g^{\otimes n})^\dagger
   =\int f(g^{-1}\omega)\,M(d\omega),
 \qquad T_f=\int f(\omega)\,M(d\omega).
\]
A rule constant on each $G_\rho$-orbit gives an invariant effect. A rule
that varies along an orbit can give an effect that fails to commute with
the group action.

Consider the minimax problem over all binary effects $0\le T\le I$
with $\Tr\rho^{\otimes n}(I-T)\le\epsilon$. Its objective is to minimize
$\sup_{g\in G_\rho}\Tr[(g\sigma g^\dagger)^{\otimes n}T]$, with the same
phase rotation on all $n$ copies. Averaging the conjugates of $T$ over
$G_\rho$ preserves its null acceptance and makes alternative acceptance
constant along the orbit, equal to the Haar average of the original
alternative acceptances. That average is bounded by the original
supremum. Conversely, for an invariant effect the supremum is its
acceptance on $\sigma^{\otimes n}$. The two problems therefore have
the same optimum at each blocklength. For independent phases in an
eigenbasis of $\rho$, Hayashi and Fang state the asymptotic
value~\cite[Corollary~26]{HayashiFang2026}.
Appendix~\ref{subsec:phase-invariant-stein-proof} proves the spectral-block
version below using the mixture converse and a pinching inequality at
orders below one.

\Needspace{9\baselineskip}
\begin{proposition}[Fixed-error exponent for phase-invariant observations]
\label{prop:phase-invariant-fixed-error-Stein}
For faithful $\rho,\sigma$ and every $0<\epsilon<1$,
\begin{equation}
 \lim_{n\to\infty}\frac1n
 D^H_{\epsilon\Vert{\cal T}_{\mathrm{phase\mbox{-}inv}}}
 (\rho^{\otimes n}\Vert\sigma^{\otimes n})
 =D_{\mathrm R}(\rho\Vert\sigma).
 \label{eq:phase-inv-fixed-error-exact}
\end{equation}
\end{proposition}

\begin{proposition}[Keyl achievability inside $\mathfrak T_d^{\mathrm{ucov}}$]
\label{prop:keyl-fixed-error-Stein-achievability}
For faithful $\rho,\sigma$ and $0<\epsilon<1$,
$B^{\mathrm{ucov}}_\epsilon(\rho\Vert\sigma) \ge D_{\mathrm R}(\rho\Vert\sigma)$.
\end{proposition}
\begin{proof}
Keyl's measurement is unitary covariant, so its direct bound
(Proposition~\ref{prop:keyl-stein-direct-reorganized}) applies.
\end{proof}

\subsubsection{Small-error limit}
\label{subsec:ucov-small-error-endpoint}
The small-error limit equals $D_{\mathrm R}(\rho\Vert\sigma)$. To prove
the quantitative bound, we keep entire orbits of the phase group whenever
the original rule accepts a sufficiently large fraction of their Haar
mass. The first lemma writes the measurement in orbit coordinates.
The second bounds the increase in alternative acceptance when an orbit
is kept in full. Maximally mixed null states are handled separately
in Proposition~\ref{prop:ucov-scalar-null-Stein}.

\begin{lemma}[Phase-orbit representation of a covariant POVM]
\label{lem:phase-orbit-disintegration}
Let a compact metrizable group $G$ act Borel measurably on a standard Borel
space $\Omega$, let $g\mapsto V_g$ be a finite-dimensional continuous unitary
representation, and let $M$ be a $G$-covariant POVM.  There are a standard
Borel orbit-label space $Y$, a Borel map $\pi:\Omega\to Y$, a Borel choice
$s(y)\in\pi^{-1}(y)$, a probability measure $\nu$ on $Y$, and a measurable
positive operator $S(y)$ such that, with $q(y,g):=g s(y)$, one has
$\pi(q(y,g))=y$ and, for every Borel $A\subseteq\Omega$,
\begin{equation}
 M(A)=\int_Y\int_G \mathbf 1_A(q(y,g))
 V_gS(y)V_g^\dagger\,d\mu_G(g)d\nu(y).
 \label{eq:abstract-phase-orbit-disintegration}
\end{equation}
\end{lemma}
Different group elements may represent the same point of an orbit.
Appendix~\ref{app:phase-reduction-estimates} proves the lemma using the
Borel realization theorem for Polish group actions
\cite[Theorem~1.1]{BeckerKechris1993} and the closed-class selector theorem
\cite[Theorem~12.16]{Kechris1995}.

Let $\mathbb T=\mathbb R/(2\pi\mathbb Z)$. The next bound follows by induction from the $L^1$ case of Nazarov's
estimate~\cite[Introduction, main theorem]{Nazarov2000Turan}. Its Haar
measure is normalized to one.

\begin{lemma}[Large-set estimate on a fixed-dimensional torus]
\label{lem:multivariate-large-set-turan}
Every integer $\ell\ge1$ admits constants $A_{\ell}>0$ and $\delta_{\ell}>0$
with the following property. Let $N\ge1$ and let $p\ge0$ be a trigonometric
polynomial on $\mathbb T^{\ell}$ with frequencies between $-N$ and $N$
in each coordinate. If a measurable set $E\subseteq\mathbb T^{\ell}$
has Haar measure $\mu_{\ell}(E)\ge1-\delta$ for
$0<\delta\le\delta_{\ell}$, then
\[
 \int_{\mathbb T^{\ell}}p\,d\mu_{\ell}
 \le e^{A_{\ell}N\delta^{1/\ell}}\int_E p\,d\mu_{\ell}.
\]
\end{lemma}

The proof is given in Appendix~\ref{app:phase-reduction-estimates}.

\begin{theorem}[Small-error limit of the covariant Stein exponent]
\label{thm:ucov-small-error-endpoint-qudit}
Let $\rho,\sigma$ be faithful states on $\mathbb C^d$ and suppose
$\rho$ has $s\ge2$ distinct eigenvalues. Put $d_{\mathrm{ph}}=s-1$.
There are positive constants $K_{d_{\mathrm{ph}}}$ and
$\epsilon_{d_{\mathrm{ph}}}$ depending only on $d_{\mathrm{ph}}$ such
that for $0<\epsilon\le\epsilon_{d_{\mathrm{ph}}}$,
\begin{equation}
 D_{\mathrm R}(\rho\Vert\sigma)
 \le B^{\mathrm{ucov}}_\epsilon(\rho\Vert\sigma)
 \le D_{\mathrm R}(\rho\Vert\sigma)+K_{d_{\mathrm{ph}}}\epsilon^{1/d_{\mathrm{ph}}}.
 \label{eq:ucov-qudit-small-error-sandwich}
\end{equation}
Consequently,
\begin{equation}
 \lim_{\epsilon\downarrow0}B^{\mathrm{ucov}}_\epsilon(\rho\Vert\sigma)
 =D_{\mathrm R}(\rho\Vert\sigma).
 \label{eq:ucov-small-error-endpoint-limit}
\end{equation}
If $s=2$, the upper correction is linear in $\epsilon$. If $\rho$ is
nondegenerate, then $d_{\mathrm{ph}}=d-1$ and the displayed modulus is
$O_d(\epsilon^{1/(d-1)})$.
\end{theorem}

\begin{proof}
Suppose the original rule has type-I error at most $\epsilon$.
Keep an entire orbit if its mean acceptance is at least $1-2\epsilon$.
Within a kept orbit, compare the original rule with the phases where
it accepts with probability at least $1/2$.

Apply Lemma~\ref{lem:phase-orbit-disintegration} to $G_\rho$ with
$V_g=g^{\otimes n}$. At the outcome $q(y,g)$ write $f(y,g)$ for the
acceptance probability and $w_X(y,g):=\Tr[X^{\otimes n}V_gS(y)V_g^\dagger]$
for the density under $X^{\otimes n}$. Then the acceptance probability is
\[
 \int_Y\int_{G_\rho} f(y,g)w_X(y,g)\,d\mu_\rho(g)d\nu(y).
\]
Since $\rho$ commutes with $G_\rho$, $w_\rho(y,g)$ is independent of $g$.
Write it as $w_\rho(y)$ and let $m(y)=\int f(y,g)\,d\mu_\rho(g)$.
The original rejection probability is
$\int_Y(1-m(y))w_\rho(y)\,d\nu(y)\le\epsilon$. On each discarded orbit
$1-m(y)>2\epsilon$, so the new rejection probability is at most $1/2$.
The new rule depends only on $y$. Its acceptance effect and complement
form a phase-invariant binary POVM.

For each kept orbit set $E_y=\{g:f(y,g)\ge1/2\}$. Markov's inequality
gives $\mu_\rho(E_y^c)\le4\epsilon$. In the torus coordinates
of $G_\rho$, $V_g$ has frequencies between $0$ and $n$ in each
coordinate, so $w_\sigma(y,g)$ has frequencies between $-n$ and $n$.
Choose $\epsilon_{d_{\mathrm{ph}}}=\min\{1,\delta_{d_{\mathrm{ph}}}\}/4$.
For $\epsilon\le\epsilon_{d_{\mathrm{ph}}}$,
Lemma~\ref{lem:multivariate-large-set-turan} gives
\[
 \int_{G_\rho}w_\sigma(y,g)\,d\mu_\rho(g)
 \le e^{nA_{d_{\mathrm{ph}}}(4\epsilon)^{1/d_{\mathrm{ph}}}}
      \int_{E_y}w_\sigma(y,g)\,d\mu_\rho(g).
\]
On $E_y$, $f(y,g)\ge1/2$. Integrating over the kept orbits bounds
the new type-II error by
$2e^{nA_{d_{\mathrm{ph}}}(4\epsilon)^{1/d_{\mathrm{ph}}}}\beta_n$,
where $\beta_n$ is the original type-II error.

The comparison holds for every test at tolerance $\epsilon$. Taking
the infimum of $\beta_n$ and applying
Proposition~\ref{prop:phase-invariant-fixed-error-Stein} at tolerance
$1/2$ gives the upper bound in \eqref{eq:ucov-qudit-small-error-sandwich}
with $K_{d_{\mathrm{ph}}}=4^{1/d_{\mathrm{ph}}}A_{d_{\mathrm{ph}}}$.
Proposition~\ref{prop:keyl-fixed-error-Stein-achievability} gives the
lower bound. Letting $\epsilon\downarrow0$ gives
\eqref{eq:ucov-small-error-endpoint-limit}.
\end{proof}

\begin{proposition}[Maximally mixed null state]
\label{prop:ucov-scalar-null-Stein}
If $\rho=I/d$ and $\sigma$ is full rank, then
$D_{\mathrm R}(I/d\Vert\sigma)=D(I/d\Vert\sigma)$.
For every $0<\epsilon<1$,
\begin{equation}
 B^{\mathrm{ucov}}_\epsilon(I/d\Vert\sigma)
 =D(I/d\Vert\sigma).
 \label{eq:ucov-scalar-null-Stein-exact}
\end{equation}
\end{proposition}

\begin{proof}
When the null state is maximally mixed, the principal-minor formula for the Keyl rate
\cite[Theorem~3.2]{Keyl2006}, evaluated at the output state $I/d$, reduces to
$D_{\mathrm R}(I/d\Vert\sigma) =-\log d-\frac1d\log\det\sigma =D(I/d\Vert\sigma)$.
Keyl's measurement attains this exponent for every $0<\epsilon<1$.
The unrestricted Stein converse~\cite{OgawaNagaoka2000} gives the upper
bound $D(I/d\Vert\sigma)$.
\end{proof}

\subsubsection{Large-error limit}
\label{subsec:ucov-large-error-endpoint}
\begin{theorem}[Large-error limit of the covariant Stein exponent]
\label{thm:ucov-large-error-endpoint-qudit}
Let $\rho$ and $\sigma$ be faithful states on $\mathbb C^d$. Then
\begin{equation}
 \lim_{\epsilon\uparrow1}
 B^{\mathrm{ucov}}_\epsilon(\rho\Vert\sigma)
 =D(\rho\Vert\sigma).
 \label{eq:ucov-large-error-limit}
\end{equation}
\end{theorem}

\begin{proof}
The unrestricted quantum Stein converse~\cite{OgawaNagaoka2000} gives,
for every fixed $0<\epsilon<1$,
\begin{equation}
 B^{\mathrm{ucov}}_\epsilon(\rho\Vert\sigma)
 \le D(\rho\Vert\sigma).
 \label{eq:ucov-large-error-unrestricted-upper}
\end{equation}
For $\delta>0$, let
$K_\delta:=\{h\in H_\rho:\|h-I\|_\infty\le\delta\}$.
It has positive Haar mass since it contains a neighborhood of the
identity. At fixed $\delta$, for all $\epsilon\ge1-\mu(K_\delta)$
in $(0,1)$,
Theorem~\ref{thm:merge-ucov-fixed-interior-stein} gives
\begin{equation}
 B^{\mathrm{ucov}}_\epsilon(\rho\Vert\sigma)
 \ge E_\rho(K_\delta).
 \label{eq:large-error-achievable-bound}
\end{equation}
As $\delta\downarrow0$, the sets $K_\delta$ converge in Hausdorff
distance to $\{I\}$. Lemma~\ref{lem:mixture-rate} therefore gives
\[
 E_\rho(K_\delta)\longrightarrow E_\rho(\{I\})
 =\lim_{n\to\infty}\frac1n
   D(\rho^{\otimes n}\Vert\sigma^{\otimes n})
 =D(\rho\Vert\sigma).
\]
The blocklength limit in $B^{\mathrm{ucov}}_\epsilon$ is taken for
each fixed $\epsilon$. We then let $\epsilon\uparrow1$ at fixed $\delta$,
and finally let $\delta\downarrow0$ to obtain the lower bound. With
\eqref{eq:ucov-large-error-unrestricted-upper}, this proves
\eqref{eq:ucov-large-error-limit}.
\end{proof}

\begin{remark}[Stein exponent under fixed-block repetition]
Fix a covariant measurement on $m$ copies and let $P_m,Q_m$ be its outcome
laws under $\rho^{\otimes m}$ and $\sigma^{\otimes m}$. Its independent
repetition has a Stein exponent at most $D_{\mathrm R}(\rho\Vert\sigma)$
at every fixed $0<\epsilon<1$.

Indeed, the log-likelihood ratio is bounded by
Lemma~\ref{lem:sc-uniform-likelihood-ratio-bounds}. Under $P_m^{\otimes k}$,
the log-likelihood ratio divided by $k$ converges in probability to
$D(P_m\Vert Q_m)$. Accepting above $D(P_m\Vert Q_m)-\delta$ on this scale
gives type-II probability at most $e^{-k(D(P_m\Vert Q_m)-\delta)}$
and type-I error tending to zero. Conversely, an acceptance rule with
type-I error at most $\epsilon$ has null acceptance at least
$1-\epsilon-o(1)$ on the set where the normalized log-likelihood is at
most $D(P_m\Vert Q_m)+\delta$. Its type-II probability is therefore at least
$(1-\epsilon-o(1))e^{-k(D(P_m\Vert Q_m)+\delta)}$. Letting $\delta\downarrow0$
gives the exponent $D(P_m\Vert Q_m)/m$, independent of $\epsilon$.
The product measurement is covariant, and any leftover copies may be
ignored. Hence this exponent is at most $B^{\mathrm{ucov}}_\epsilon$
for every $\epsilon$, and the small-error limit bounds it by
$D_{\mathrm R}(\rho\Vert\sigma)$. For noncommuting states this is strictly
below $D(\rho\Vert\sigma)$ by
Proposition~\ref{prop:reverse-relative-entropy-equality}.
\end{remark}

\subsection{The covariant Hoeffding exponent}
\label{sec:general-covariant-hoeffding-reorganized}
For qubits, Keyl's measurement attains the optimal covariant Hoeffding
exponent. Theorem~\ref{thm:merge-qubit-full-hoeffding} gives its value
and shows that, for each covariant test, one Keyl decision rule controls
both errors at the same blocklength up to a polynomial factor. In arbitrary dimension,
Keyl's measurement gives a lower bound and phase-invariant tests give
two R\'enyi upper bounds. Together these bounds determine the rates at
which the exponent is positive
(Corollary~\ref{cor:ucov-hoeffding-nontight-sandwich}).

\subsubsection{Reduction to phase-invariant tests}
\label{subsec:ucov-hoeffding-phase-reduction}
Let $X$ be a faithful state with $s$ distinct eigenvalues and spectral
projections $E_1,\ldots,E_s$. As in Section~\ref{sec:covariant-stein-regime},
take the relative-phase torus
\[
 G_X=\left\{\sum_{a<s}e^{i\theta_a}E_a+E_s:
                \theta_a\in\mathbb T\right\},
\]
with Haar probability measure $\mu_X$. Its $n$-copy representation has a
set of weights $\Lambda_{X,n}$ determined by the occupations of the first
$s-1$ eigenspaces. Thus $L_{X,n}:=|\Lambda_{X,n}|\le(n+1)^{s-1}$.

\begin{lemma}[Polynomial variation along a phase orbit]
\label{lem:poly-phase-var-hoeffding}
Let $X$ be faithful, let $Y\ge0$ and $A\ge0$ act on
${\cal H}^{\otimes n}$, and let $G_X$ be the relative-phase torus of $X$.
For each $g\in G_X$,
\[
 \operatorname{Tr}Yg^{\otimes n}A(g^\dagger)^{\otimes n}
 \le L_{X,n}\int_{G_X}
 \operatorname{Tr}Yh^{\otimes n}A(h^\dagger)^{\otimes n}\,d\mu_X(h).
\]
\end{lemma}

The proof is given in Appendix~\ref{app:phase-reduction-estimates}.

\begin{lemma}[Replacing a covariant test by a phase-invariant test]
\label{lem:phase-cov-to-phase-inv-hoeffding}
Let $X$ and $Y$ be faithful states.  Suppose that a $G_X$-covariant
observation, followed by a measurable randomized rule, gives a test $T_n$ that
accepts $X$.  Then there is a $G_X$-invariant test $T'_n$ such that
\begin{align}
 \operatorname{Tr}X^{\otimes n}(I-T'_n)
 &\le 2L_{X,n}\operatorname{Tr}X^{\otimes n}(I-T_n),
 \label{eq:X-error-transfer-hoeffding}\\
 \operatorname{Tr}Y^{\otimes n}T'_n
 &\le 2\operatorname{Tr}Y^{\otimes n}T_n.
 \label{eq:Y-error-transfer-hoeffding}
\end{align}
\end{lemma}
The new rule accepts each phase orbit in full if the original rule
accepts nearly all of it on average. Otherwise it rejects the entire
orbit. Appendix~\ref{app:phase-reduction-estimates} shows why this
controls both errors. Since $\log L_{X,n}=o(n)$, both decay exponents
are at least as large as before.

\subsubsection{Two R\'enyi upper bounds}
\label{subsec:ucov-hoeffding-renyi-converse}
For faithful states and $0<\alpha<1$, write
\[
 D^{\rm RS}_\alpha(\rho\Vert\sigma)
 :=\frac{\alpha}{1-\alpha}
       \widetilde D_{1-\alpha}(\sigma\Vert\rho).
\]
The divergences $\widetilde D_\alpha$ and $D^{\rm RS}_\alpha$ are the
left- and right-pinched quantities
$\overleftarrow D_\alpha$ and $\overrightarrow D_\alpha$, respectively,
in \cite[Sec.~V.B.2, Eqs.~(83a)--(85)]{LipkaBartosikEtAl2024}.
As $\alpha\uparrow1$, $D^{\rm RS}_\alpha$ tends to their
$D^\star=D_{\mathrm R}$. Define
\begin{align*}
 H_{\mathrm{RS}}(r\mid\rho\mid\sigma)
 &:=\sup_{0<\alpha<1}\frac{1-\alpha}{\alpha}
       \{D^{\rm RS}_\alpha(\rho\Vert\sigma)-r\},\\
 H_{\mathrm S}(r\mid\rho\mid\sigma)
 &:=\sup_{0<\alpha<1}\frac{1-\alpha}{\alpha}
       \{\widetilde D_\alpha(\rho\Vert\sigma)-r\}.
\end{align*}
For $r>0$, both objectives tend to $-\infty$ as $\alpha\downarrow0$
and to zero as $\alpha\uparrow1$. The suprema are therefore finite and
nonnegative. Each is a supremum of affine functions of $r$, so both
are convex and continuous on $(0,\infty)$.

Use the same phase-rotation problem as in
Section~\ref{subsec:covariant-stein-formulation}, now with a bound on
type-II error. Let $a_n^{X\mid Y}(s)$ be the minimum rejection probability
under $X^{\otimes n}$ over binary effects whose acceptance is at most
$e^{-ns}$ under every $(gYg^\dagger)^{\otimes n}$, $g\in G_X$.
For distinct faithful $X,Y$, the composite Hoeffding theorem of
Hayashi and Fang~\cite[Theorems~23--24]{HayashiFang2026} gives
\[
 \lim_{n\to\infty}-\frac1n\log a_n^{X\mid Y}(s)
 =H_{\mathrm{RS}}(s\mid X\mid Y),
 \qquad 0<s<D_{\mathrm R}(X\Vert Y).
\]
To apply it to spectral blocks of any rank, take the single-copy
projections in \cite[Lemmas~20--21]{HayashiFang2026} to be the spectral
projections $E_i$ of $X$. If $x_i$ is the eigenvalue of $X$ on $E_i$,
then $X^{\otimes n}$ equals $\prod_i x_i^{t_i}$ times the identity on
the occupation block with counts $t_i$. Both hypotheses are permutation
invariant. The null is a block-constant singleton and the alternative
is a full $G_X$-orbit. Its twirl is the singleton
$\{\mathcal E_{X,n}(Y^{\otimes n})\}$, where $\mathcal E_{X,n}$ is
pinching onto the occupation blocks. The common phase acts trivially
by conjugation.

Lemma~21 of the same reference identifies the regularized Petz
divergence against the twirled alternative with the regularized
reverse-sandwiched divergence against the orbit. Since $X$ is fixed
by $G_X$, unitary invariance and additivity make the latter
$nD^{\rm RS}_\alpha(X\Vert Y)$ before normalization. Thus
\[
 \lim_{n\to\infty}\frac1n
 D^{\rm P}_\alpha(X^{\otimes n}\Vert\mathcal E_{X,n}(Y^{\otimes n}))
 =D^{\rm RS}_\alpha(X\Vert Y),\qquad 0<\alpha<1 .
\]
The regularity condition concerns the function
\[
 (1-\alpha)D^{\rm RS}_\alpha(X\Vert Y)
 =-\log\Tr\bigl(X^{\alpha/(2(1-\alpha))}Y
                 X^{\alpha/(2(1-\alpha))}\bigr)^{1-\alpha}.
\]
The proof of \cite[Theorem~24]{HayashiFang2026} establishes its
analyticity and strict concavity for distinct faithful states, with
no condition on the projection ranks. In that argument an affine
log moment would make $\widetilde D_t(Y\Vert X)$ constant in $0<t<1$,
contrary to its limits $0$ at $t\downarrow0$ and $D(Y\Vert X)>0$
at $t\uparrow1$. Multiplying the source's base-two rate and
divergence by $\log 2$ gives the formula above.

This concavity also determines the zero region of the reverse-sandwiched
envelope. Write $F(\alpha)=(1-\alpha)D^{\rm RS}_\alpha(X\Vert Y)$
and extend it continuously to $F(1)=0$ using the finite limit of the
divergence. For $0<\alpha<\beta<1$, the chord from $\alpha$ to $1$ gives
\[
 F(\beta)\ge\frac{1-\beta}{1-\alpha}F(\alpha).
\]
Dividing by $1-\beta$ and taking $\beta\uparrow1$ yields
$D^{\rm RS}_\alpha(X\Vert Y)\le D_{\mathrm R}(X\Vert Y)$.
When $X=Y$, both divergences are zero. Thus for any faithful pair and
any $r>0$ with $r\ge D_{\mathrm R}(X\Vert Y)$, the objective in
$H_{\mathrm{RS}}(r\mid X\mid Y)$ is nonpositive at every order and
tends to zero as $\alpha\uparrow1$. The envelope therefore vanishes.

For a $G_X$-invariant effect, the acceptance under $Y^{\otimes n}$
already equals the worst-case acceptance in $a_n^{X\mid Y}(s)$.

\begin{theorem}[Upper bounds on the covariant Hoeffding exponent]
\label{thm:ucov-hoeffding-renyi-converse}
Let $\rho$ and $\sigma$ be full-rank states on $\mathbb C^d$, with $d$ fixed.
Let $(T_n)_{n\ge1}\in\mathfrak T_d^{\mathrm{ucov}}$ and suppose
$\liminf_{n\to\infty}-\frac1n\log\beta_n\ge r>0$.
Then
\begin{equation}
 \limsup_{n\to\infty}-\frac1n\log\alpha_n
 \le\min\{H_{\mathrm{RS}}(r\mid\rho\mid\sigma),
           H_{\mathrm S}(r\mid\rho\mid\sigma)\}.
 \label{eq:ucov-hoeffding-converse-target}
\end{equation}
Consequently, the same upper bound holds for
$H^{\mathrm{ucov}}(r\mid\rho\mid\sigma)$.
\end{theorem}

\begin{proof}
If $\rho=\sigma$, then $\alpha_n+\beta_n=1$ and the assumed positive
rate forces $\alpha_n\to1$; both envelopes are zero. Assume henceforth
that $\rho\ne\sigma$, and put
$h_*:=\limsup_n-n^{-1}\log\alpha_n$.
Restrict the outcome action to $G_\rho$ for the original decision
rule and to $G_\sigma$ for the complementary rule. Complementing the
rule and interchanging the hypotheses exchanges the two errors.

\noindent\textbf{Reverse-sandwiched bound.}
Lemma~\ref{lem:phase-cov-to-phase-inv-hoeffding}, with phase reference
$\rho$, gives an invariant test with errors
$\alpha'_n\le2L_{\rho,n}\alpha_n$ and $\beta'_n\le2\beta_n$.
Fix $0<s<\min\{r,D_{\mathrm R}(\rho\Vert\sigma)\}$.
The assumed liminf rate makes $\beta'_n\le e^{-ns}$ for all sufficiently
large $n$. Thus
\[
 \alpha_n\ge\frac{\alpha'_n}{2L_{\rho,n}}
             \ge\frac{a_n^{\rho\mid\sigma}(s)}{2L_{\rho,n}},
 \qquad h_*\le H_{\mathrm{RS}}(s\mid\rho\mid\sigma).
\]
Here $\log L_{\rho,n}=o(n)$, and
$-n^{-1}\log a_n^{\rho\mid\sigma}(s)$ converges,
so the bound also holds on every subsequence. If
$r<D_{\mathrm R}(\rho\Vert\sigma)$, let $s\uparrow r$.
Otherwise let $s\uparrow D_{\mathrm R}(\rho\Vert\sigma)$, where the
continuous envelope tends to zero. In both cases
$h_*\le H_{\mathrm{RS}}(r\mid\rho\mid\sigma)<\infty$.

\noindent\textbf{Sandwiched bound.}
If $h_*=0$, there is nothing more to prove. For
$0<h<\min\{h_*,D_{\mathrm R}(\sigma\Vert\rho)\}$, choose a subsequence
on which $\alpha_n\le e^{-n(h+\delta)}$ for some $\delta>0$.
Apply Lemma~\ref{lem:phase-cov-to-phase-inv-hoeffding} with reference
$\sigma$ to the complementary rule. Its rejection and acceptance errors
for the ordered pair $(\sigma,\rho)$ obey
\[
 \bar\alpha_n\le2L_{\sigma,n}\beta_n,\qquad
 \bar\beta_n\le2\alpha_n.
\]
On the chosen subsequence, $\bar\beta_n\le e^{-nh}$ eventually, while
$\bar\alpha_n$ has liminf exponent at least $r$.
Comparison with $a_n^{\sigma\mid\rho}(h)$ therefore gives
$r\le H_{\mathrm{RS}}(h\mid\sigma\mid\rho)$.
If $h_*\ge D_{\mathrm R}(\sigma\Vert\rho)$, letting $h$ approach that
endpoint would contradict $r>0$. Hence $h_*<D_{\mathrm R}(\sigma\Vert\rho)$,
and the inequality holds for every $0<h<h_*$.

The divergence identity above rewrites it as
\[
 r\le\sup_{0<\alpha<1}
       \left\{\widetilde D_\alpha(\rho\Vert\sigma)
                    -\frac{\alpha}{1-\alpha}h\right\}.
\]
For $h>0$, the expression tends to zero as $\alpha\downarrow0$ and to
$-\infty$ as $\alpha\uparrow1$. Since $r>0$, its maximum is attained
at an interior order. At that order,
\[
 h\le\frac{1-\alpha}{\alpha}
       \{\widetilde D_\alpha(\rho\Vert\sigma)-r\}
 \le H_{\mathrm S}(r\mid\rho\mid\sigma).
\]
Letting $h\uparrow h_*$ proves the second branch and hence the theorem.
\end{proof}

\begin{corollary}[Bounds and positivity of the covariant Hoeffding exponent]
\label{cor:ucov-hoeffding-nontight-sandwich}
Every rate $r>0$ satisfies
\begin{equation}
 H^{\mathrm K}(r\mid\rho\mid\sigma)
 \le H^{\mathrm{ucov}}(r\mid\rho\mid\sigma)
 \le\min\{H_{\mathrm{RS}}(r\mid\rho\mid\sigma),
           H_{\mathrm S}(r\mid\rho\mid\sigma)\}.
 \label{eq:ucov-hoeffding-full-class-sandwich}
\end{equation}
A deterministic decision rule for Keyl's measurement attains the lower
bound. In particular, $H^{\mathrm{ucov}}(r\mid\rho\mid\sigma)$ is positive
for $0<r<D_{\mathrm R}(\rho\Vert\sigma)$ and zero for
$r\ge D_{\mathrm R}(\rho\Vert\sigma)$.
\end{corollary}

\begin{proof}
Theorem~\ref{thm:keyl-hoeffding-exact-reorganized} gives the lower bound
because Keyl's measurement is covariant.
Theorem~\ref{thm:ucov-hoeffding-renyi-converse} gives the upper bound.
If $r<D_{\mathrm R}(\rho\Vert\sigma)$, the compact feasible set
in the formula for $H^{\mathrm K}$ excludes $\rho$, the unique
zero of $D_{\mathrm R}(\,\cdot\,\Vert\rho)$. The minimum is therefore
positive. For $r\ge D_{\mathrm R}(\rho\Vert\sigma)$, the
reverse-sandwiched upper bound vanishes.
\end{proof}

\subsubsection{Exact qubit Hoeffding exponent and scalar evaluation}
\label{subsubsec:merge-qubit-hoeffding-exact}

\begin{theorem}[Exact covariant Hoeffding exponent for qubits]
\label{thm:merge-qubit-full-hoeffding}
Let $\rho,\sigma$ be faithful qubit states.

\noindent\textbf{(i) Exact exponent.} For every $r>0$,
\begin{equation}
 H^{\mathrm{ucov}}(r\mid\rho\mid\sigma)
 =H^{\mathrm K}(r\mid\rho\mid\sigma)
 =\min_{\tau:\,D_{\mathrm R}(\tau\Vert\sigma)\le r}
      D_{\mathrm R}(\tau\Vert\rho).
 \label{eq:merge-qubit-H}
\end{equation}
\noindent\textbf{(ii) Finite-block comparison.}
Let $p_X\ge s_X>0$ be the eigenvalues of $X=\rho,\sigma$ and set
\begin{equation}
 L=\max_{X=\rho,\sigma}\frac{p_X-s_X}{2s_X},\qquad
 D_n=16\pi e^L(n+1)^4.
 \label{eq:covhoe-Dn}
\end{equation}
For each covariant $n$-block test with errors $(\alpha_n,\beta_n)$,
one randomized decision rule for Keyl's measurement satisfies
\begin{equation}
 \alpha_n^{\mathrm K}\le D_n\alpha_n,\qquad
 \beta_n^{\mathrm K}\le D_n\beta_n.
 \label{eq:covhoe-pair}
\end{equation}
For $\lambda>0$, let $R_n^{\mathcal C}(\lambda)$ be the minimum of
$\alpha_n+\lambda\beta_n$ over admissible $n$-block tests in the class
$\mathcal C\in\{\mathrm K,\mathrm{ucov}\}$. These minima exist by
Proposition~\ref{covhoe:prop:pure-orbit}. For every $n\ge1$ and
$\lambda>0$,
\begin{equation}
 D_n^{-1}R_n^{\mathrm K}(\lambda)
 \le R_n^{\mathrm{ucov}}(\lambda)
 \le R_n^{\mathrm K}(\lambda).
 \label{eq:covhoe-risk}
\end{equation}
The same $D_n$ applies to every weight, Schur block, and covariant
measurement.

\noindent\textbf{(iii) Scalar evaluation.} For $0<\alpha<1$ define
\begin{align}
 K_\alpha&=\max_{\|u\|=1}
  (u^\dagger\rho u)^\alpha
  (u^\dagger\sigma u)^{1-\alpha},
       \label{eq:merge-Kalpha}\\
 C_\alpha&=-\log\left(K_\alpha+
     \frac{(\det\rho)^\alpha(\det\sigma)^{1-\alpha}}
          {K_\alpha}\right).
       \label{eq:merge-Calpha}
\end{align}
Then the same exponent has the scalar representations
\begin{align}
 H^{\mathrm{ucov}}(r\mid\rho\mid\sigma)
 &=\sup_{0<\alpha<1}
    \frac{C_\alpha-(1-\alpha)r}{\alpha},
       \label{eq:merge-Hdual}\\
 K_\alpha&=\min_{s>0}\lambda_{\max}
    \left(\alpha s\rho+(1-\alpha)
          s^{-\alpha/(1-\alpha)}\sigma\right).
       \label{eq:merge-Kscalar}
\end{align}
The function minimized in~\eqref{eq:merge-Kscalar} is convex and
diverges as $s\downarrow0$ and as $s\to\infty$.
\end{theorem}

Appendix~\ref{app:merge-qubit-proof} first averages a covariant POVM over
permutations, which preserves both outcome laws on the product inputs.
The refinement records the Schur block $j$ and an index selecting a
seed ray represented by a unit vector $v\in\Sym^N(\mathbb C^2)$, where
$N=n-2j$. Conditional on the index, the measurement on the symmetric
factor has density $(N+1)$ times the projector onto $\Sym^N(g)v$
relative to Haar measure on the rotation $g$. In
Appendix~\ref{covhoe:sec:reduction}, Step~3, the refined record is
$(j,y,\ell,g)$, with $(y,\ell)$ selecting the seed and $gy$ recovering
the outcome of the averaged measurement.

In a block with $N\ge1$, a one-qubit ray represented by $u$ gives the
tensor-power seed $u^{\otimes N}$ for Keyl's measurement. Fix the block
$j$ and the error weight $\lambda>0$. Choose $u$ to maximize the
smaller of the two weighted coefficients
\[
 (\det\rho)^j\langle u|\rho|u\rangle^N,\qquad
 \lambda(\det\sigma)^j\langle u|\sigma|u\rangle^N.
\]
These coefficients omit the common block factor $m_{n,j}d_N$, where
$m_{n,j}$ is the Schur multiplicity and $d_N=N+1$ is the dimension of
the symmetric factor. Write $T_j(\lambda)$ for the maximum of their
pointwise minimum. For any seed $v$, $\Gamma_{j,v}(\lambda)$
in \eqref{covhoe:eq:Gamma} integrates the pointwise minimum of the two
weighted density coefficients against normalized Haar measure, before
multiplication by the common block factor. For a coherent seed, the
integrand is the minimum above evaluated at the rotated ray, so
$\Gamma_{j,\mathrm{coh}}(\lambda)\le T_j(\lambda)$.

For a seed $v$, let $\mu_v$ be its empirical Majorana measure and
$\Phi_{g^\dagger Xg}(\mu_v)$ its response to the rotated state, as defined
in the next subsection by \eqref{eq:majorana-phi}.
For the maximizing ray $u$, Proposition~\ref{covhoe:prop:joint} gives a
single rotation $g_0$ with $\Phi_{g_0^\dagger Xg_0}(\mu_v)\ge
\log\langle u|X|u\rangle$ for both $X=\rho$ and $X=\sigma$.
The norm bound \eqref{covhoe:eq:norm-bound} in Lemma~\ref{covhoe:lem:norm}
then gives, for each of these states,
\[
 \langle v|\Sym^N(g_0^\dagger Xg_0)|v\rangle
 \ge d_N^{-1}e^{N\Phi_{g_0^\dagger Xg_0}(\mu_v)}
 \ge d_N^{-1}\langle u|X|u\rangle^N.
\]
Multiplying by $(\det\rho)^j$ and $\lambda(\det\sigma)^j$ in the
respective terms bounds their minimum at $g_0$ below by
$d_N^{-1}T_j(\lambda)$. The rotation bound in the same lemma maintains
this estimate up to a constant factor on a neighborhood of radius
$1/(n+1)$. Integrating over its Haar mass gives the polynomial loss
in Proposition~\ref{covhoe:prop:risk-bound}. With $D_n$ from
\eqref{eq:covhoe-Dn},
\[
 \Gamma_{j,v}(\lambda)\ge D_n^{-1}T_j(\lambda)
 \ge D_n^{-1}\Gamma_{j,\mathrm{coh}}(\lambda).
\]
Minimizing over seeds in each block and summing with the block factors
gives the weighted-risk comparison. Convex separation then turns the
bounds for all weights into a single
Keyl rule satisfying~\eqref{eq:covhoe-pair}. The polynomial factors
preserve the error exponents. The scalar formulas are
proved in Appendix~\ref{app:merge-qubit-scalar}.

Equation~\eqref{eq:merge-Hdual} expresses the qubit exponent as a supremum
of affine functions of $r$, so it is convex. For $r>0$, the feasible state
$\tau=\sigma$ in \eqref{eq:merge-qubit-H} gives
$0\le H^{\mathrm{ucov}}(r\mid\rho\mid\sigma)\le D_{\mathrm R}(\sigma\Vert\rho)<\infty$,
so the exponent is continuous on $(0,\infty)$.

\subsubsection{Majorana representation of the qubit optimization}
\label{subsec:ucov-qubit-majorana}

The scalar formula evaluates the optimal qubit exponent. The Majorana
representation describes the outcome law for any pure seed in a Schur
block and gives independent upper and lower variational bounds.
Continuity of the optimal qubit exponent, obtained from the scalar
formula \eqref{eq:merge-Hdual}, then identifies both variational values
at every positive rate.

For the block $\lambda=(n-j,j)$, put $N=n-2j$. Write $\Sym^N(Y)$ for
the restriction of $Y^{\otimes N}$ to the symmetric space
$\Sym^N(\mathbb C^2)$. A pure seed is a unit vector $v_N$ in that
space. The rotated projectors of $v_N$, with their normalization,
give the measurement density in the block. Write $\mathbb P^1$ for the
projective space of one-qubit pure states, with its normalized invariant
measure $\omega$.

For $N\ge1$, the coherent amplitude is the homogeneous polynomial
$z\mapsto\langle v_N|z^{\otimes N}\rangle$ on $\mathbb C^2$. Factor it as
\[
 \langle v_N|z^{\otimes N}\rangle=c\prod_{\ell=1}^N\langle u_\ell|z\rangle,
 \qquad c\ne0.
\]
We call the rays of the unit vectors $u_\ell$ the Majorana spinors of
$v_N$. Each corresponding zero ray is the ray orthogonal to $u_\ell$
\cite{UshaDeviSudhaRajagopal2012,RomeroEtAl2024}. The empirical measure is
$\mu_{v_N}=N^{-1}\sum_{\ell=1}^N\delta_{u_\ell}$ on $\mathbb P^1$, with
one atom for each factor counted with multiplicity. The density exponent
below uses a logarithmic potential of this measure. We define that
potential for any Borel probability measure
$\mu\in\mathcal P(\mathbb P^1)$, since such measures occur as weak
limits of the empirical measures. Unit representatives are used for
rays in the formulas, and the values are independent of their phases.
Define
\begin{align}
 V_\mu(z)&=\int_{\mathbb P^1}\log|\langle u|z\rangle|^2\,d\mu(u),
 \label{eq:majorana-potential}\\
 \Phi_Y(\mu)&=\max_{z\in\mathbb P^1}
 \left\{V_\mu(z)-\log\langle z|Y^{-1}|z\rangle\right\}
 -\max_{z\in\mathbb P^1}V_\mu(z),\qquad Y>0.
 \label{eq:majorana-phi}
\end{align}
Here $\log0=-\infty$. Both maxima in Eq.~\eqref{eq:majorana-phi}
are finite.

\begin{theorem}[Pure-seed asymptotic formula for qubits]
\label{thm:majorana-asymptotic-pure-seeds}
Let $Y_N>0$ be qubit operators converging to $Y>0$, and let
$v_N\in\Sym^N(\mathbb C^2)$ be unit vectors whose empirical
Majorana measures converge weakly to $\mu$.  Then
\begin{equation}
 \lim_{N\to\infty}\frac1N\log
 \langle v_N|\Sym^N(Y_N)|v_N\rangle=\Phi_Y(\mu).
 \label{eq:majorana-seed-laplace}
\end{equation}
For every $N\ge1$, positive definite qubit operator $Y$, and unit
$v\in\Sym^N(\mathbb C^2)$, the finite-degree error obeys
\begin{equation}
 \left|\frac1N\log\langle v|\Sym^N(Y)|v\rangle
       -\Phi_Y(\mu_v)\right|
 \le\frac{\log(N+1)}N.
 \label{eq:majorana-uniform-finite-bound}
\end{equation}
The bound is independent of the seed and of the positive definite
operator $Y$.
\end{theorem}

The coherent-state resolution gives the finite-degree bound.
Continuity of both maxima in~\eqref{eq:majorana-phi} then gives the limit
under weak convergence. Appendix~\ref{app:ucov-majorana-proofs} proves
these facts and shows that every $\mu$ is a weak limit of empirical
Majorana measures.

Set $q=j/n$, the fraction of boxes in the second row of the Young diagram,
so $N/n=1-2q$. With $q\in[0,1/2]$ and $g\in U(2)$, define the
density exponent under $X^{\otimes n}$ relative to counting measure
on $j$ and normalized Haar measure on $g$.
\begin{equation}
 I_X(q,g;\mu)=-h(q)-q\log\det X
 -(1-2q)\Phi_{g^\dagger Xg}(\mu),
 \label{eq:majorana-block-rate}
\end{equation}
Here $h(q)=-q\log q-(1-q)\log(1-q)$ with $0\log0=0$.
The degree-zero block has just one seed up to phase. Assign any fixed
probability measure to its Majorana symbol, since $N/n=0$ and the
rate at $q=1/2$ is independent of $\mu$.
Equation~\eqref{eq:majorana-uniform-density-rate} gives the uniform
approximation of the normalized log density by $-I_X$. Define the variational bounds, with $\inf\varnothing=+\infty$,
\begin{align}
 \mathcal B(r)
 &=\inf_{0\le q\le 1/2}\ \sup_{\mu\in\mathcal P(\mathbb P^1)}\
 \inf_{g:\,I_\sigma(q,g;\mu)\le r}I_\rho(q,g;\mu),
 \label{eq:majorana-B-closed}\\
 \mathcal B_{<}(r)
 &=\inf_{0\le q\le 1/2}\ \sup_{\mu\in\mathcal P(\mathbb P^1)}\
 \inf_{g:\,I_\sigma(q,g;\mu)<r}I_\rho(q,g;\mu).
 \label{eq:majorana-B-open}
\end{align}
For a fixed block and seed, the direct rule accepts $\rho$ where
$I_\sigma>r+\eta$, with slack $\eta>0$. By
\eqref{eq:majorana-uniform-density-rate}, the alternative density there
is small enough to meet the type-II constraint after summing over the
polynomial number of Schur labels. On the rejected set $I_\sigma\le r+\eta$,
a lower bound on $I_\rho$ gives a lower bound on the type-I error exponent.

\Needspace{8\baselineskip}
For the converse, a rotation with $I_\sigma<r$ has a neighborhood of
positive Haar measure where the inequality holds with slack. The
type-II constraint forces the integrated acceptance weight on that
neighborhood to be exponentially small. Rejection there contributes
to the total null rejection probability, so a single such block can
limit the type-I error exponent. This explains the inner infimum over
rotations. The seed can be optimized separately for each block, and the
outer infimum takes the smallest resulting block exponent.
Appendix~\ref{app:majorana-operational-bounds} makes both bounds precise.

\begin{theorem}[Exact qubit Hoeffding variational representation]
\label{thm:ucov-majorana-hoeffding-sandwich}
For faithful qubit states $\rho,\sigma$ and every $r>0$,
\begin{equation}
 \begin{aligned}
 \mathcal B(r)=\mathcal B_{<}(r)
 &=H^{\mathrm{ucov}}(r\mid\rho\mid\sigma)\\
 &=\min_{\tau:\,D_{\mathrm R}(\tau\Vert\sigma)\le r}
      D_{\mathrm R}(\tau\Vert\rho).
 \end{aligned}
 \label{eq:ucov-majorana-continuity-equality}
\end{equation}
Independently of Theorem~\ref{thm:merge-qubit-full-hoeffding}, the
variational argument gives
\begin{equation}
 \lim_{\eta\downarrow0}\mathcal B(r+\eta)
 \le H^{\mathrm{ucov}}(r\mid\rho\mid\sigma)
 \le\mathcal B_{<}(r)
 \le\lim_{\eta\downarrow0}\mathcal B(r-\eta).
 \label{eq:ucov-majorana-hoeffding-sandwich}
\end{equation}
The function $\mathcal B$ is finite and nonincreasing on $(0,\infty)$.
In particular, $\mathcal B$ is continuous on $(0,\infty)$.
\end{theorem}

Appendix~\ref{app:ucov-majorana-proofs} proves the bounds by reducing
to one pure seed per Schur block. The variational argument gives equality
at the continuity points of $\mathcal B$. Continuity of the independently
established scalar formula for $H^{\mathrm{ucov}}$ extends the equality
to every $r>0$.

\subsection{Strong-converse exponent}
\label{sec:general-covariant-strong-converse}

We first prove equality with the unrestricted strong-converse exponent
by recording a random rotation of an unrestricted binary test. We then
construct a sequence of covariant measurements that attains the
sandwiched R\'enyi divergence asymptotically at every order above one,
including infinity. Using these measurements on the entire sample
attains the optimum asymptotically at each rate. Repeating the measurement
on sufficiently large fixed blocks approaches the optimum.

\subsubsection{Equality with unrestricted testing}

Let $\rho$ and $\sigma$ be faithful states on $\mathbb C^d$, with $d$ fixed.
The unrestricted strong-converse exponent is
\cite[Theorem~IV.10]{MosonyiOgawa2015}
\begin{equation}
 E_{\mathrm{sc}}^*(r)
 =\sup_{\alpha>1}\frac{\alpha-1}{\alpha}
 \left\{r-\widetilde D_\alpha(\rho\Vert\sigma)\right\}.
 \label{eq:sc-unrestricted-alpha-form}
\end{equation}
The objective tends to zero as $\alpha\downarrow1$, so the exponent is
nonnegative at every $r\ge0$.
If $\rho=\sigma$, the two acceptance probabilities coincide, and the
constant rule $e^{-nr}$ gives $E_{\mathrm{sc}}^*(r)=r$, also as in
\eqref{eq:sc-unrestricted-alpha-form}.

Given a binary effect $0\le T_n\le I$, choose $U$ according to normalized
Haar measure and apply the test rotated by $U$. The measurement records
$(U,b)\in U(d)\times\{0,1\}$ and has densities
\begin{equation}
\begin{aligned}
 M_n(dU,1)&:=U^{\otimes n}T_n(U^\dagger)^{\otimes n}\,dU,\\
 M_n(dU,0)&:=U^{\otimes n}(I-T_n)(U^\dagger)^{\otimes n}\,dU.
\end{aligned}
 \label{eq:covariant-embedding-binary-test}
\end{equation}
Summing over $b$ and integrating over $U$ gives $I$. Conjugation by
$g^{\otimes n}$ acts on the outcome as $(U,b)\mapsto(gU,b)$, so $M_n$
is unitary covariant. For a Borel set $S\subseteq U(d)$, the rule
$f_{n,S}(U,b):=\mathbf 1_S(U)\mathbf 1_{\{1\}}(b)$ accepts only a
positive outcome with $U\in S$. Its effect is
$T_{n,S}:=\int_S U^{\otimes n}T_n(U^\dagger)^{\otimes n}\,dU$.
Under $X^{\otimes n}$ the acceptance probability is
\begin{equation}
 \operatorname{Tr}X^{\otimes n}T_{n,S}
 =\int_S\operatorname{Tr}(U^\dagger XU)^{\otimes n}T_n\,dU.
 \label{eq:locally-twirled-probability}
\end{equation}

\begin{theorem}[Strong-converse exponent under unitary-covariant measurements]
\label{thm:general-completeness}
Every full-rank pair $\rho,\sigma$ on a fixed finite-dimensional
space $\mathbb C^d$ satisfies, at every rate $r\ge0$,
\begin{equation}
 E_{\mathrm{sc},d}^{\mathrm{ucov}}(r)
 =E_{\mathrm{sc}}^*(r).
 \label{eq:general-completeness-target}
\end{equation}
Every unrestricted test sequence has a realization by a unitary-covariant
observation followed by classical processing with the same two acceptance
exponents.
\end{theorem}

\begin{proof}
Given an unrestricted effect $T_n$, use the covariant observation in
\eqref{eq:covariant-embedding-binary-test} and accept only on
$\{(U,1):U\in C_n(I)\}$, where $C_n(I)$ is the operator-norm ball of
radius $n^{-2}$ about the identity in $U(d)$. Its normalized Haar
measure $v_n$ satisfies $-\log v_n=O(\log n)$ because small balls
in the fixed-dimensional group have volume of polynomial order in their
radius. For either hypothesis $X$, the estimate
$\|U^\dagger XU-X\|_\infty\le 2\|X\|_\infty\|U-I\|_\infty$
and $X\ge\lambda_{\min}(X)I$ give a constant $c_X$ such that, uniformly
for $U\in C_n(I)$,
\[
 e^{-c_X/n}X^{\otimes n}
 \le (U^\dagger XU)^{\otimes n}
 \le e^{c_X/n}X^{\otimes n}.
\]
The induced effect $T_{n,C_n(I)}$ therefore satisfies
\[
 v_ne^{-c_X/n}\Tr X^{\otimes n}T_n
 \le \Tr X^{\otimes n}T_{n,C_n(I)}
 \le v_ne^{c_X/n}\Tr X^{\otimes n}T_n.
\]
For each hypothesis, the two acceptance probabilities are either both
zero or have logarithms that differ by $O(\log n)$. This preserves the type-II
$\liminf$ constraint and the null-acceptance $\limsup$ objective in
\eqref{eq:operational-strong-converse-definition}. Taking the infimum over
unrestricted sequences proves
$E_{\mathrm{sc},d}^{\mathrm{ucov}}(r)\le E_{\mathrm{sc}}^*(r)$.
The reverse inequality follows from inclusion of the covariant tests
among unrestricted quantum tests.
\end{proof}

The Haar measure of the rotation ball tends to zero. Thus even when the
unrestricted test accepts the null with high probability, the covariant
construction can have a vanishing acceptance probability. The Stein
and Hoeffding exponents require the separate arguments above.

\subsubsection{R\'enyi-preserving covariant measurements}
\label{subsec:sc-renyi-preserving-covariant-measurements}

Fix an orthonormal eigenbasis of $\sigma$, choosing a basis also inside
each degenerate eigenspace, and write $\sigma=\operatorname{diag}(s_1,\ldots,s_d)$.
We first fix the measurement in randomly rotated weight bases.
Its basis contains all the vectors needed for the different R\'enyi
orders. The order will determine which outcomes we use in the proof.

For a Young diagram $\lambda\vdash m$ with at most $d$ rows, let
$\mathcal Q_\lambda$ be the corresponding polynomial Schur module and
$\pi_\lambda$ its $GL(d)$ representation. A weight space is a
joint eigenspace for the representation of diagonal matrices in the
chosen basis. For the weight $\nu=(\nu_1,\ldots,\nu_d)$, write
$\mathcal Q_\lambda[\nu]$ for this space. Every
$v\in\mathcal Q_\lambda[\nu]$ satisfies
\begin{equation}
 \pi_\lambda(D(z))v
 =\left(\prod_{i=1}^d z_i^{\nu_i}\right)v,
 \qquad
 D(z):=\operatorname{diag}(z_1,\ldots,z_d).
 \label{eq:sc-weight-space-definition}
\end{equation}
Schur--Weyl duality \cite{FultonHarris1991,Sagan2001} gives the unitary
identification
\[
 W_m:(\mathbb C^d)^{\otimes m}\longrightarrow
 \bigoplus_{\lambda\vdash m\,:,\ell(\lambda)\le d}
 \mathcal Q_\lambda\otimes\mathcal P_\lambda .
\]
The factors $\mathcal Q_\lambda$ and $\mathcal P_\lambda$ are the
representation and multiplicity spaces, denoted by $\mathcal U_\lambda$
and $\mathcal V_\lambda$ in Appendix~\ref{app:keyl-finite-output-law}.
Every positive definite $Y$ acts in the $\lambda$ block as
\begin{equation}
 \left(W_mY^{\otimes m}W_m^\dagger\right)
 \big|_{\mathcal Q_\lambda\otimes\mathcal P_\lambda}
 =\pi_\lambda(Y)\otimes I_{\mathcal P_\lambda}.
 \label{eq:sc-schur-block-action}
\end{equation}

For each $\lambda$, choose an orthonormal basis in every weight space.
Their union is an orthonormal basis
$\mathcal B_\lambda=\{v_{\lambda,a}:1\le a\le N_\lambda\}$ of
$\mathcal Q_\lambda$, where $N_\lambda:=\dim\mathcal Q_\lambda$.
On the $\lambda$ block, record $(a,U)$ with density
\[
 dM_{m,\lambda}(a,U)
 := \pi_\lambda(U)|v_{\lambda,a}\rangle\!\langle v_{\lambda,a}|
 \pi_\lambda(U)^\dagger\otimes I_{\mathcal P_\lambda}\,dU .
\]
For each $U$ the rotated basis projectors sum to the identity. Including
the Schur label $\lambda$ gives a normalized POVM $M_m$ with outcome
$(\lambda,a,U)$ and covariance under $U\mapsto gU$. A \emph{cell} fixes
the Schur label $\lambda$ and basis label $a$ and requires the
rotation $U$ to lie in a specified ball. We use balls of radius $m^{-2}$
in operator norm and denote the ball about $U_0$ by $C_m(U_0)$.
Denote the laws of $M_m$ under $\rho^{\otimes m}$ and $\sigma^{\otimes m}$
by $P_m^{M_m}$ and $Q_m^{M_m}$.
Every order $\alpha>1$ satisfies the measured-divergence bound from
quantum data processing \cite[Theorem~6]{Beigi2013}:
$D_\alpha(P_m^{M_m}\Vert Q_m^{M_m}) \le m\widetilde D_\alpha(\rho\Vert\sigma)$.
For these Borel output laws, the bound follows by applying the
finite-output inequality to every finite measurable partition and taking
the supremum over such partitions.
To prove the matching lower bound, we pass from the full outcome to
the binary indicator of a selected cell. The following lemmas estimate
its probability, combine the estimates for the two hypotheses, and
select a vector from the fixed basis.

\begin{lemma}[Probability of a small rotation ball]
\label{lem:general-local-orbit-cell}
For a faithful state $Y$ and a Borel set $C\subseteq U(d)$,
the probability assigned by $M_m$ to $\{\lambda\}\times\{a\}\times C$ is
\[
 \dim\mathcal P_\lambda
 \int_C\langle v_{\lambda,a},\pi_\lambda(U^\dagger YU)v_{\lambda,a}\rangle\,dU.
\]
We also use a measurement with only one unit vector $v$ in each Schur
block. Its density on that block is multiplied by $N_\lambda$ for
normalization. For $Y>0$ and $v\in\mathcal Q_\lambda$ of unit norm, write
\[
 P_{Y,m}(\lambda,v,C)
 :=\dim\mathcal Q_\lambda\,\dim\mathcal P_\lambda
 \int_C\langle v,\pi_\lambda(U^\dagger YU)v\rangle\,dU.
\]
This is the probability of observing the Schur label $\lambda$ and
$U\in C$ in the one-vector measurement when $Y$ is a state. If
$P_{Y,m}^{M_m}$ is the full outcome law of $M_m$ under $Y^{\otimes m}$,
then the two densities give
\[
 P_{Y,m}^{M_m}(\{\lambda\}\times\{a\}\times C)
 =\frac{P_{Y,m}(\lambda,v_{\lambda,a},C)}{N_\lambda}.
\]
For the ball $C_m(U_0)$ and every center $U_0\in U(d)$, we have
\[
 \log P_{Y,m}(\lambda,v,C_m(U_0))
 = \log\dim\mathcal P_\lambda
   + \log\langle v,\pi_\lambda(U_0^\dagger YU_0)v\rangle
   + O(\log m).
\]
The error bound is uniform in $\lambda$, $v$, and $U_0$, and one constant
suffices for $Y\in\{\rho,\sigma\}$.
\end{lemma}

This lemma and the next two are proved in
Appendix~\ref{app:sc-technical-proofs}.

\begin{lemma}[Weight-space coefficient identity]
\label{lem:sc-weight-space-coefficient-identity}
Let $0<\theta\le1$ and define
$X_\theta:=\sigma^{-\theta/2}\rho\sigma^{-\theta/2}$.
If $v$ is a unit vector in $\mathcal Q_\lambda[\nu]$, where this weight space
is defined by \eqref{eq:sc-weight-space-definition} in the fixed eigenbasis
of $\sigma$, then
$\langle v,\pi_\lambda(X_\theta)v\rangle =\left(\prod_{i=1}^d s_i^{-\theta\nu_i}\right) \langle v,\pi_\lambda(\rho)v\rangle,$
and hence
\[
 \log\langle v,\pi_\lambda(\rho)v\rangle
 -\theta\log\langle v,\pi_\lambda(\sigma)v\rangle
 =\log\langle v,\pi_\lambda(X_\theta)v\rangle.
\]
\end{lemma}

\begin{lemma}[Choosing a vector from the weight basis]
\label{lem:sc-renyi-scale-weight-seed-selection}
Fix the weight-space decompositions determined by the eigenbasis of $\sigma$.
Let
$X=U\operatorname{diag}(z_1,\ldots,z_d)U^\dagger$, where $z_1\ge\cdots\ge z_d>0$,
and let $\lambda^{(m)}\vdash m$ satisfy
$\lambda_i^{(m)}/m\to p_i$.  Every blocklength $m$ admits a unit vector
$v_m(X)\in\mathcal Q_{\lambda^{(m)}}[\nu^{(m)}]$ from one of the fixed
weight-space bases contained in the definition of $M_m$ such that
\begin{equation}
 \frac1m\log
 \langle v_m(X),\pi_{\lambda^{(m)}}(X)v_m(X)\rangle
 \longrightarrow\sum_{i=1}^d p_i\log z_i.
 \label{eq:sc-selected-seed-limit}
\end{equation}
For a fixed limiting shape $p$, the limit depends on $X$ only through its
eigenvalues. The choice of vector can also depend on its eigenvectors.
\end{lemma}

\begin{proposition}[Asymptotic recovery of the sandwiched R\'enyi divergence]
\label{prop:sc-asymptotic-renyi-recovery}
The fixed POVM sequence $(M_m)_m$ satisfies the following limits, with
the first valid for every $1<\alpha<\infty$:
\begin{align}
 \lim_{m\to\infty}\frac1m
 D_\alpha\left(P_m^{M_m}\middle\Vert Q_m^{M_m}\right)
 &=\widetilde D_\alpha(\rho\Vert\sigma),
 \label{eq:sc-renyi-recovery-finite}\\
 \lim_{m\to\infty}\frac1m
 D_\infty\left(P_m^{M_m}\middle\Vert Q_m^{M_m}\right)
 &=\widetilde D_\infty(\rho\Vert\sigma).
 \label{eq:sc-renyi-recovery-infinity}
\end{align}
\end{proposition}

\begin{proof}
\textit{Selecting a cell.}
Fix $\alpha\in(1,\infty)$ and put
$\theta:=\frac{\alpha-1}{\alpha}$ and $X_\theta:=\sigma^{-\theta/2}\rho\sigma^{-\theta/2}$.
Let $z_1\ge\cdots\ge z_d>0$ be the eigenvalues of $X_\theta$.
For a limiting Young shape $p$, the cell calculation below gives
$(1-\theta)H(p)+\sum_i p_i\log z_i$ as the limit of $-a_m+\theta b_m$,
where $a_m,b_m$ are the cell's two acceptance exponents.
To maximize this expression, use $1-\theta=1/\alpha$ and write
\[
 (1-\theta)H(p)+\sum_i p_i\log z_i
 =\frac1\alpha\left(\log\sum_i z_i^\alpha
     -D(p\Vert p^{(\alpha)})\right),
 \qquad
 p_i^{(\alpha)}=\frac{z_i^\alpha}{\sum_\ell z_\ell^\alpha}.
\]
Here $D(p\Vert p^{(\alpha)})$ is the classical relative entropy
of the two vectors. The maximizing vector $p^{(\alpha)}$ is
nonincreasing, so Young diagrams $\lambda^{(m)}\vdash m$ can be
selected with $\lambda_i^{(m)}/m\to p_i^{(\alpha)}$.

Lemma~\ref{lem:sc-renyi-scale-weight-seed-selection} selects a vector
$v_{m,\alpha}$ already in the measurement basis for this shape sequence.
The selection uses $X_\theta$ and hence can depend on $\alpha$. If
$a_{m,\alpha}$ is its basis label, take the event
\[
 C_{m,\alpha}:=\{\lambda^{(m)}\}\times\{a_{m,\alpha}\}\times C_m(I).
\]
Its probability is positive under each hypothesis, since both states are
faithful and the ball has positive Haar measure. Write
$p_m:=P_m^{M_m}(C_{m,\alpha})$, $q_m:=Q_m^{M_m}(C_{m,\alpha})$,
$a_m:=-m^{-1}\log p_m$, and $b_m:=-m^{-1}\log q_m$.

\smallskip\noindent\textit{Calculating the acceptance exponents.}
By Lemma~\ref{lem:general-local-orbit-cell}, the logarithm of each cell
probability equals $\log\dim\mathcal P_{\lambda^{(m)}}$ plus the
logarithm of its matrix coefficient, with error $O(\log m)$.
Indeed, passing from the one-vector law to the selected basis-label
event divides each probability by $N_{\lambda^{(m)}}$ and hence
subtracts $\log N_{\lambda^{(m)}}=O(\log m)$ from its logarithm.
The selected vector is an eigenvector of $\pi_\lambda(\sigma)$.
Lemma~\ref{lem:sc-weight-space-coefficient-identity} therefore combines
the two log coefficients into one for $X_\theta$. Lemma~\ref{lem:fixed-dimension-specht-multiplicity} gives
$m^{-1}\log\dim\mathcal P_{\lambda^{(m)}}\to H(p^{(\alpha)})$
along the chosen sequence of Young diagrams.
Together with \eqref{eq:sc-selected-seed-limit}, this yields
\begin{align*}
 \lim_{m\to\infty}(a_m-\theta b_m)
 & = -(1-\theta)H(p^{(\alpha)})
      -\sum_i p_i^{(\alpha)}\log z_i \\
 & = -\frac{1}{\alpha}\log\sum_i z_i^\alpha
   = -\theta\widetilde D_\alpha(\rho\Vert\sigma).
\end{align*}
The second equality uses the shape maximization above, where
$D(p^{(\alpha)}\Vert p^{(\alpha)})=0$.

\smallskip\noindent\textit{Bounding the measured divergence.}
Record whether the outcome lies in $C_{m,\alpha}$. The probabilities of
this event are $p_m$ and $q_m$, and classical data processing gives
\[
 D_\alpha(P_m^{M_m}\Vert Q_m^{M_m})
 \ge \frac{1}{\alpha-1}\log\!\left(p_m^\alpha q_m^{1-\alpha}\right)
 = -\frac{m}{\theta}(a_m-\theta b_m),
\]
where we kept only the acceptance term in the binary divergence.
This gives the lower bound in \eqref{eq:sc-renyi-recovery-finite}.
The upper bound follows from quantum data processing.

\smallskip\noindent\textit{Order infinity.}
Put $X_1:=\sigma^{-1/2}\rho\sigma^{-1/2}$ and
$z_1:=\|X_1\|_\infty$.  The endpoint construction uses the fully symmetric Young shape
$\lambda^{(m)}=(m,0,\ldots,0)$.  Lemma~\ref{lem:sc-renyi-scale-weight-seed-selection}
selects an available weight-basis seed whose normalized logarithmic
$X_1$ matrix coefficient tends to $\log z_1$.
Lemma~\ref{lem:general-local-orbit-cell}, together with the
$\theta=1$ case of
Lemma~\ref{lem:sc-weight-space-coefficient-identity}, gives a sequence of
cells $C_m$ satisfying
$\frac1m\log \frac{P_m^{M_m}(C_m)}{Q_m^{M_m}(C_m)} \longrightarrow\log z_1$.
The ratio of cell probabilities is the average of
$dP_m^{M_m}/dQ_m^{M_m}$ conditional on $C_m$ under $Q_m^{M_m}$.
It is bounded above by the essential supremum of this derivative. This
proves the lower bound in \eqref{eq:sc-renyi-recovery-infinity}.
For the upper bound, tensor
$\rho\le z_1\sigma$ and trace against each POVM effect. This gives
$dP_m^{M_m}/dQ_m^{M_m}\le z_1^m$ almost everywhere.
\end{proof}

\begin{remark}[Pinching followed by covariant embedding]
\label{rem:sc-pinching-covariant-construction}
The joint spectral measurement after pinching can also be embedded in
a covariant observation by recording its rotation. The resulting
measurement sequence is independent of the R\'enyi order and has the
limits in \eqref{eq:sc-renyi-recovery-finite} and
\eqref{eq:sc-renyi-recovery-infinity}.
Appendix~\ref{app:sc-technical-proofs} gives the construction.
\end{remark}

\subsubsection{One acceptance event at each rate}

For $0<\theta<1$, write
\[
 h(\theta)=-\theta\widetilde D_{1/(1-\theta)}(\rho\Vert\sigma),
 \qquad h(0)=0,\qquad h(1)=-D_{\max}(\rho\Vert\sigma),
\]
with the endpoints given by continuity. The cells in
Proposition~\ref{prop:sc-asymptotic-renyi-recovery} attain the lower bound
$h(\theta)$ on the limiting combination $a-\theta b$, where $a$ and $b$
are the null-acceptance and alternative-acceptance exponents.
The proof below shows that their limiting pair $(a_\theta,b_\theta)$
gives the affine upper bound $h(\eta)\le a_\theta-\eta b_\theta$ with
equality at $\eta=\theta$.
The slope gives $b_\theta=-h'(\theta)$, as shown using
\eqref{eq:sc-single-cell-support-bound} and
\eqref{eq:sc-single-cell-rate-pair}. If $\theta$ maximizes
$\theta r+h(\theta)$ in the interior, then $b_\theta=r$.
At each prescribed rate, the rule accepts only on one cell, with
additional randomized acceptance at large rates.

\begin{proposition}[Attainment by one weight-basis cell]
\label{prop:sc-single-cell-attainment}
Let $\rho,\sigma$ be faithful states in fixed finite dimension. For every
$r\ge0$, there are cells
$A_n=\{\lambda_n\}\times\{j_n\}\times C_n(I)$ of the measurements
$M_n$ and numbers $0<\gamma_n\le1$ such that the rules
$\gamma_n\mathbf1_{A_n}$ have acceptance probabilities $p_n,q_n$ satisfying
\begin{equation}
 \liminf_{n\to\infty}-\frac1n\log q_n\ge r,
 \qquad
 \lim_{n\to\infty}-\frac1n\log p_n=E_{\mathrm{sc}}^*(r).
 \label{eq:sc-single-cell-attainment}
\end{equation}
\end{proposition}

\begin{proof}
The function $h$ is continuous on $[0,1]$ by the order-one and
order-infinity limits of the sandwiched R\'enyi divergence.
Faithfulness makes the matrix powers and trace in the finite-order
formula smooth, so $h$ is differentiable on $(0,1)$. By continuity,
\eqref{eq:sc-unrestricted-alpha-form} equals the maximum of
$\theta r+h(\theta)$ over $[0,1]$.

\smallskip\noindent\textit{Exponents at an interior parameter.}
For each $\theta\in(0,1)$, use the cells constructed in
Proposition~\ref{prop:sc-asymptotic-renyi-recovery}. Their acceptance
exponents $a_{n,\theta},b_{n,\theta}$ satisfy
$a_{n,\theta}-\theta b_{n,\theta}\to h(\theta)$.
They are uniformly bounded. Indeed, if $v_n$ denotes the Haar measure
of $C_n(I)$, the cell probability under a faithful state $X$ is
\[
 \dim\mathcal P_{\lambda_n}
 \int_{C_n(I)}\langle v,\pi_{\lambda_n}(U^\dagger XU)v\rangle\,dU
 \ge v_n\lambda_{\min}(X)^n.
\]
The bound follows by restricting $X^{\otimes n}\ge\lambda_{\min}(X)^n I$
to the Schur block.
Consequently
\[
 0\le a_{n,\theta}\le-\log\lambda_{\min}(\rho)-\frac{\log v_n}{n},
 \qquad
 0\le b_{n,\theta}\le-\log\lambda_{\min}(\sigma)-\frac{\log v_n}{n},
\]
where $-\log v_n=O(\log n)$ independently of the chosen labels.

For every cell and every $0<\eta<1$, binary data processing and its
acceptance summand, as in the preceding proof, give the finite-block bound
\begin{equation}
 a_{n,\theta}-\eta b_{n,\theta}\ge h(\eta).
 \label{eq:sc-single-cell-support-bound}
\end{equation}
Subtract $a_{n,\theta}-\theta b_{n,\theta}=h(\theta)+o(1)$
from \eqref{eq:sc-single-cell-support-bound}. For fixed $\eta<\theta$ this
bounds $b_{n,\theta}$ from below, and for $\eta>\theta$ it bounds it from
above. Let $n\to\infty$, then let $\eta\to\theta$ from the two sides.
Differentiability gives
\begin{equation}
 b_{n,\theta}\longrightarrow b_\theta:=-h'(\theta),
 \qquad
 a_{n,\theta}\longrightarrow a_\theta:=h(\theta)-\theta h'(\theta).
 \label{eq:sc-single-cell-rate-pair}
\end{equation}
Taking limits in \eqref{eq:sc-single-cell-support-bound} shows that
$h(\eta)\le a_\theta-\eta b_\theta$, with equality at $\eta=\theta$.
Thus $h$ is the infimum of these affine functions and is concave.
If $\theta\in(0,1)$ maximizes $\theta r+h(\theta)$, its derivative
vanishes. Thus $r=-h'(\theta)=b_\theta$ and
$a_\theta=E_{\mathrm{sc}}^*(r)$.
The corresponding cells therefore prove the claim at such a rate.

\smallskip\noindent\textit{Endpoint cells.}
Concavity and the uniform bounds give finite one-sided limits
$b_0:=\lim_{\theta\downarrow0}b_\theta$ and
$b_1:=\lim_{\theta\uparrow1}b_\theta$, with $b_0\le b_1$.
The order-one limit $\lim_{\theta\downarrow0}h(\theta)/\theta
=-D(\rho\Vert\sigma)$ gives $b_0=D(\rho\Vert\sigma)$ by concavity.
By continuity of $h$,
\[
 (a_\theta,b_\theta)\longrightarrow(0,b_0)\quad(\theta\downarrow0),
 \qquad
 (a_\theta,b_\theta)\longrightarrow
 (b_1-D_{\max}(\rho\Vert\sigma),b_1)\quad(\theta\uparrow1).
\]
For either endpoint, choose supporting parameters $\theta_j$
approaching it. Choose strictly increasing integers $N_j$ so that
for all $n\ge N_j$ the cell for $\theta_j$ has exponents within
$1/j$ of $(a_{\theta_j},b_{\theta_j})$. At length $n$, choose the cell
for the largest $j$ with $N_j\le n$, and use any cell for the finitely many
initial lengths. This yields the endpoint pair along the full sequence.

\smallskip\noindent\textit{The rule at the prescribed rate.}
For $0\le r\le b_0$, use the first endpoint pair with $\gamma_n=1$.
It is feasible and has null-acceptance exponent $0=E_{\mathrm{sc}}^*(r)$.
For $r\ge b_1$, use the second endpoint pair and, upon observing the
cell, accept with probability $\gamma_n=e^{-n(r-b_1)}$.
The resulting pair is
$(r-D_{\max}(\rho\Vert\sigma),r)$ and attains the maximum of
$\theta r+h(\theta)$ at $\theta=1$. If $b_0<r<b_1$, the maximum is
attained in the interior because the derivatives at the two ends have
opposite signs. This covers all rates, including $b_0=b_1$.
\end{proof}

\subsubsection{Repeated finite-block measurements}
\label{subsec:sc-repeated-finite-block-measurements}

We first fix the block size $m$ and apply $M_m$ independently to $k$
blocks. We send $k$ to infinity at this fixed $m$. Choosing a sufficiently
large $m$ lets us approximate the unrestricted exponent to any accuracy.
For $n=km+\ell$ with $0\le\ell<m$, we ignore the remaining copies.
Related fixed-block arguments appear in
Refs.~\cite{Hayashi2006,Hayashi2016,MosonyiOgawa2015}.
The two output laws on the measured blocks are
$(P_m^{M_m})^{\otimes k}$ and $(Q_m^{M_m})^{\otimes k}$.
Let $M_m$ have standard Borel outcome space $\Omega_m$, let $U(d)$ act on
$\Omega_m$ by the Borel maps $\omega\mapsto g\omega$, and equip
$\Omega_m^k$ with the diagonal action
$g(\omega_1,\ldots,\omega_k) :=(g\omega_1,\ldots,g\omega_k)$.
Every $g\in U(d)$ and every Borel set $A\subseteq\Omega_m$ satisfy
the following covariance identity for $M_m$:
\begin{equation}
 M_m(gA)
 =g^{\otimes m}M_m(A)(g^\dagger)^{\otimes m}.
 \label{eq:sc-finite-block-covariance-measurable-set}
\end{equation}
For the product POVM $M_m^{\otimes k}$ the same covariance relation
holds on measurable rectangles by tensoring
\eqref{eq:sc-finite-block-covariance-measurable-set}. It extends to all
Borel sets by a monotone-class argument. Thus the product measurement is
covariant under the same unitary acting on all measured copies.

\begin{proposition}[Fixed-block realization with weight-space seeds]
\label{prop:sc-fixed-block-realization}
For every $r\ge0$ and $\varepsilon>0$, there is a fixed finite blocklength
$m$ such that repetitions of the weight-space observation $M_m$, followed
by classical post-processing, give a full covariant test sequence with
\[
 \liminf_{n\to\infty}-\frac1n\log q_n\ge r,
 \qquad
 \limsup_{n\to\infty}-\frac1n\log p_n
 \le E_{\mathrm{sc}}^*(r)+\varepsilon.
\]
\end{proposition}

\begin{proof}
By Proposition~\ref{prop:sc-single-cell-attainment}, choose $m$ and a
rule on $M_m$ with null and alternative acceptance exponents
$a_m\le E_{\mathrm{sc}}^*(r)+\varepsilon/2$ and
$b_m\ge r-\varepsilon/2$. On each block, apply this rule and keep an
acceptance with the additional probability $e^{-m(r-b_m)_+}$.
Make all randomized choices independently across blocks and accept the
null only if all blocks accept. The probabilities then multiply, giving
the exponents normalized by $km$,
$a_m+(r-b_m)_+\le E_{\mathrm{sc}}^*(r)+\varepsilon$ and
$b_m+(r-b_m)_+\ge r$. The product measurement is covariant as shown
above. At length $n=km+\ell$, use the one-outcome identity POVM on
the remaining $\ell$ copies and ignore its outcome.
The probabilities are unchanged, and $km/n\to1$ yields the stated bounds.
\end{proof}

The preceding construction uses one binary rule on each block. We now
consider the exponent obtained by optimizing the decision rule on the full
product outcome. A bounded log-likelihood ratio allows us to approximate
each fixed-block experiment by a finite one.

\begin{lemma}[Uniform likelihood-ratio bounds for measured tensor powers]
\label{lem:sc-uniform-likelihood-ratio-bounds}
Let $M_m$ be any POVM on $(\mathbb C^d)^{\otimes m}$.
Let $P_m$ and $Q_m$ be its output laws under $\rho^{\otimes m}$ and
$\sigma^{\otimes m}$, respectively. Define
$c_-:=\lambda_{\min}(\sigma^{-1/2}\rho\sigma^{-1/2})$ and
$c_+:=\lambda_{\max}(\sigma^{-1/2}\rho\sigma^{-1/2})$.
If $\rho$ and $\sigma$ are full rank, then $P_m$ and $Q_m$ are mutually
absolutely continuous: $P_m\ll Q_m$ and $Q_m\ll P_m$.
Moreover, $m\log c_-\le\log(dP_m/dQ_m)\le m\log c_+$ holds
$Q_m$-almost everywhere.
\end{lemma}
\begin{proof}
Tensor $c_-\sigma\le\rho\le c_+\sigma$ and take the trace against
$M_m(A)$ for any measurable set $A$. This gives
$c_-^mQ_m(A)\le P_m(A)\le c_+^mQ_m(A)$. Since $c_->0$, the measures
are equivalent, and these inequalities give the asserted bounds on their
likelihood ratio almost everywhere.
\end{proof}

\begin{lemma}[Strong-converse exponent from repetition of a bounded classical experiment]
\label{lem:sc-classical-repetition}
Let $P$ and $Q$ be mutually absolutely continuous probability measures, and suppose that the log-likelihood ratio
$Z:=\log(dP/dQ)$ is essentially bounded.  Define
\begin{equation}
 e_{P,Q}(R)
 :=\sup_{\alpha>1}\frac{\alpha-1}{\alpha}
 \{R-D_\alpha(P\Vert Q)\},\qquad R\ge0.
 \label{eq:sc-classical-repetition-exponent}
\end{equation}
The exponent $e_{P,Q}(R)$ is nonnegative because the expression in the supremum tends to zero
as $\alpha\downarrow1$.  Every target rate $R\ge0$ and slack $\delta>0$ admit a full sequence
of measurable randomized acceptance rules $f_k$ on the product experiment.
Under the two product laws $P^{\otimes k}$ and $Q^{\otimes k}$, the
respective acceptance probabilities $p_k:=\int f_k\,dP^{\otimes k}$ and
$q_k:=\int f_k\,dQ^{\otimes k}$ satisfy
$\liminf_{k\to\infty}-k^{-1}\log q_k\ge R$ and
$\limsup_{k\to\infty}-k^{-1}\log p_k\le e_{P,Q}(R)+\delta$.
Every sequence satisfying the constraint on $q_k$ has null-acceptance
exponent at least $e_{P,Q}(R)$. Thus $e_{P,Q}(R)$ is the infimum of the
null-acceptance $\limsup$ exponents under that constraint.
\end{lemma}

\begin{proof}
Fix $0<\eta<\delta/2$. Partition the essential range of $Z$ into finitely
many intervals of length at most $\eta$, and let $\pi$ be the resulting
measurable cell map, defined arbitrarily on the common null set.
Write $P^\pi_j=P(\pi=j)$ and $Q^\pi_j=Q(\pi=j)$, omitting null cells.
Both probabilities are positive. Since $P^\pi_j/Q^\pi_j$ is the
$Q$-conditional average of $e^Z$ on the cell,
$z_j:=\log(P^\pi_j/Q^\pi_j)$ lies in the same interval as $Z$.
Thus $|Z-z_{\pi}|\le\eta$ almost everywhere, and every $\alpha>1$ satisfies
\[
 \sum_jP^\pi_j e^{(\alpha-1)z_j}
 \ge e^{-(\alpha-1)\eta}\int e^{(\alpha-1)Z}\,dP.
\]
Taking logarithms, dividing by $\alpha-1$, and using classical data
processing for the opposite inequality gives the uniform estimate
\[
 0\le D_\alpha(P\Vert Q)-D_\alpha(P^\pi\Vert Q^\pi)\le\eta,
 \qquad \alpha>1.
\]
Consequently
$0\le e_{P^\pi,Q^\pi}(R)-e_{P,Q}(R)\le\eta$ for every $R\ge0$.

Apply the commuting case of the strong-converse theorem
\cite[Theorem~IV.10, \mbox{Eq.~(88)}]{MosonyiOgawa2015} to the finite laws
$P^\pi,Q^\pi$. For every $R\ge0$ it gives a full sequence of randomized
rules $g_k$ with $Q^\pi$-acceptance exponent at least $R$ and
$P^\pi$-acceptance exponent at most $e_{P^\pi,Q^\pi}(R)+\delta/2$,
in the respective liminf and limsup senses of the statement.
If $P^\pi=Q^\pi$, the constant rule $g_k=e^{-kR}$ has both exponents
$R=e_{P^\pi,Q^\pi}(R)$ and gives the same conclusion directly.
Pull these rules back by $f_k=g_k\circ\pi^{\otimes k}$.
Their acceptance probabilities are unchanged, so their $P$-exponent is
at most $e_{P,Q}(R)+\eta+\delta/2<e_{P,Q}(R)+\delta$, while the
$Q$-exponent is at least $R$, as required.

For the converse put $Z_k=\log(dP^{\otimes k}/dQ^{\otimes k})$.
Applying H\"older's inequality with the measure $f_k\,dQ^{\otimes k}$
and then using $0\le f_k\le1$ gives, for each $\alpha>1$,
\[
 p_k\le
 \left(\int e^{\alpha Z_k}\,dQ^{\otimes k}\right)^{1/\alpha}
 q_k^{(\alpha-1)/\alpha}
 =e^{k(\alpha-1)D_\alpha(P\Vert Q)/\alpha}
   q_k^{(\alpha-1)/\alpha}.
\]
If $q_k=0$, equivalence of the measures gives $p_k=0$. Taking
the normalized logarithms gives a $\liminf$ null-acceptance exponent
of at least $(\alpha-1)(R-D_\alpha(P\Vert Q))/\alpha$.
The supremum over $\alpha>1$ bounds the $\limsup$ objective from
below as well. The arbitrary slack in the direct bound proves the
infimum formula.
\end{proof}

The next result compares the optimized classical exponent with the
unrestricted quantum exponent. Convergence at order infinity gives
control at the upper endpoint of the optimization over orders.

\begin{corollary}[Convergence of the induced classical strong-converse exponents]
\label{cor:sc-classical-exponent-convergence}
For the measurements in
Proposition~\ref{prop:sc-asymptotic-renyi-recovery}, put
$P_m=P_m^{M_m}$ and $Q_m=Q_m^{M_m}$. Every rate $r\ge0$ satisfies
\begin{equation}
 \lim_{m\to\infty}\frac1m e_{P_m,Q_m}(mr)
 =E_{\mathrm{sc}}^*(r),
 \label{eq:sc-classical-exponent-convergence-target}
\end{equation}
The convergence is uniform for $r\ge0$. Here $e_{P,Q}$ is given by
\eqref{eq:sc-classical-repetition-exponent}.
\end{corollary}

\begin{proof}
For $0<\theta<1$, set
\[
 f_m(\theta):=\frac{\theta}{m}D_{1/(1-\theta)}(P_m\Vert Q_m),
 \qquad
 f(\theta):=\theta\widetilde D_{1/(1-\theta)}(\rho\Vert\sigma),
\]
and extend them by
$f_m(0)=f(0)=0$,
$f_m(1)=m^{-1}D_\infty(P_m\Vert Q_m)$, and
$f(1)=\widetilde D_\infty(\rho\Vert\sigma)$.
Nonnegativity and order monotonicity of R\'enyi divergence make each
$f_m$ nondecreasing. Proposition~\ref{prop:sc-asymptotic-renyi-recovery}
gives pointwise convergence to $f$ on $[0,1]$, including both endpoints.
Continuity in the order and the standard limits at $1$ and infinity make
$f$ continuous on $[0,1]$.
A finite-grid squeeze now gives uniform convergence: choose a grid on
which the oscillation of $f$ between adjacent points is at most $\varepsilon$;
for all sufficiently large $m$, the errors at the grid points are at most
$\varepsilon$, and monotonicity bounds the error everywhere by
$2\varepsilon$.

The endpoint limits leave the suprema defining both exponents unchanged
when $0<\theta<1$ is replaced by $[0,1]$. Hence
\[
 \begin{aligned}
 \left|\frac1m e_{P_m,Q_m}(mr)-E_{\mathrm{sc}}^*(r)\right|
 &=\left|\sup_{0\le\theta\le1}\{\theta r-f_m(\theta)\}
       -\sup_{0\le\theta\le1}\{\theta r-f(\theta)\}\right|\\
 &\le\|f_m-f\|_\infty\longrightarrow0.
 \end{aligned}
\]
The bound is independent of $r$.
\end{proof}

At fixed $m$, applying Lemma~\ref{lem:sc-classical-repetition} at
$R=mr$ makes $e_{P_m,Q_m}(mr)/m$ the optimal exponent per input copy
for processing the full product outcome. The direct rule at $n=km+\ell$
ignores the last $\ell$ copies, so the block-normalized exponents are
multiplied by $k/n\to1/m$. The converse along $n=km$ gives the same
lower bound for any full sequence of such rules. Thus
\eqref{eq:sc-classical-exponent-convergence-target} takes $m\to\infty$
after the repetition limit at fixed $m$.

\section{Consequences for highest-weight measurements}
\label{sec:highest-weight-consequences}\label{sec:general-highest-weight-loss}
\subsection{Two highest-weight procedures}
Keyl's measurement on all $n$ copies gives a state estimate, to which
we may apply any decision rule. For the other procedure, split the $n$
copies into a fixed finite number $L$ of blocks of lengths $m_i(n)$,
with $m_i(n)/n\to t_i>0$ and $\sum_{i=1}^L t_i=1$. This includes a
single block of all $n$ copies. On each block of size $m$ use the
full-unitary refinement of Keyl's measurement from
Appendix~\ref{app:stein-achievability}, Step~1. It records the Young label
and the full rotation $U$. Set $V_k=\operatorname{span}\{Ue_1,\ldots,Ue_k\}$.
The reported partial flag retains $V_k$ for $k<d$ with
$\lambda_k>\lambda_{k+1}$, together with $V_d=\mathbb C^d$. This flag and
$\lambda$ determine the state estimate. The Born density is constant on
each fibre of the map to the reported flag, so the conditional Haar law
on that fibre is independent of the input
(Appendix~\ref{app:keyl-flag-stabilizer}). A highest-weight cell
selects one Young label and an operator-norm ball of radius $m^{-2}$
about a prescribed rotation in $U(d)$. Select one such cell per block.
The joint event is that every block yields its selected cell.

For a sequence of cells $C_m$, write $(a,b)$ for the rates defined by
\[
 P_{\rho,m}(C_m)=e^{-ma+o(m)},\qquad
 P_{\sigma,m}(C_m)=e^{-mb+o(m)}.
\]
The first rate measures decay of correct null acceptance, and the
second measures decay of the type-II error. Write
$\mathcal R_{\mathrm{hw}}^{(1)}$ for the compact set of limiting
one-block rate pairs, including subsequential limits. If block $i$
has rates $(a_i,b_i)$ per copy, independence gives the rates of the
joint event per copy as $\sum_{i=1}^L t_i(a_i,b_i)$.
Appendix~\ref{subsec:general-highest-weight-region} proves that every
point in the convex hull can be attained with at most three growing
blocks. Section~\ref{subsec:sc-repeated-finite-block-measurements}
treats repetitions of a fixed-size block.

If a joint event has rates $(a,b)$, keep an acceptance with probability
$e^{-n(r-b)_+}$. This adds $(r-b)_+$ to both rates, making the type-II
rate at least $r$. The region and the optimum for these concatenated
cell tests with thinning are
\[
 \mathcal R_{\mathrm{hw}}
 :=\overline{\operatorname{co}}\mathcal R_{\mathrm{hw}}^{(1)},
 \qquad
 E_{\mathrm{hw}}(r)
 :=\min_{(a,b)\in\mathcal R_{\mathrm{hw}}}\{a+(r-b)_+\}.
\]
Appendix~\ref{subsec:general-highest-weight-region} proves attainment
and the converse, even when the selected cells vary with blocklength
and the joint event rates do not converge.
The two exponent formulas give
\begin{equation}
 E_{\mathrm{sc}}^*(r)\le E_{\mathrm{hw}}(r)
 \le E_{\mathrm{sc}}^{\mathrm{Keyl}}(r\mid\rho\mid\sigma),\qquad r\ge0.
 \label{eq:highest-weight-exponent-hierarchy}
\end{equation}
Here $E_{\mathrm{sc}}^*(r)$ is the unrestricted strong-converse exponent.
The concatenated tests are covariant, which gives the first inequality.
For the second, Lemma~\ref{lem:highest-weight-rate-image} identifies
$\mathcal R_{\mathrm{hw}}^{(1)}$ with the pairs
$(D_{\mathrm R}(\tau\Vert\rho),D_{\mathrm R}(\tau\Vert\sigma))$. The
Keyl formula in Theorem~\ref{thm:keyl-strong-converse-exponent} minimizes
$a+(r-b)_+$ over these pairs. The convex hull can lower this minimum.

\subsection{When the exponents differ}
At a fixed type-II rate, a larger strong-converse exponent means that
the probability of accepting the null when it is true decays faster.
\begin{theorem}[Strict gap for Keyl's measurement and highest-weight cells]
\label{thm:highest-weight-loss-summary}
Let $\rho$ and $\sigma$ be full-rank states. Suppose that
$D_{\mathrm R}(\rho\Vert\sigma)<D(\rho\Vert\sigma)$, or equivalently,
$[\rho,\sigma]\ne0$ by Proposition~\ref{prop:reverse-relative-entropy-equality}.
For $D_{\mathrm R}(\rho\Vert\sigma)<r\le D(\rho\Vert\sigma)$,
\begin{equation}
 E_{\mathrm{sc}}^*(r)=0<E_{\mathrm{hw}}(r)
 \le E_{\mathrm{sc}}^{\mathrm{Keyl}}(r\mid\rho\mid\sigma).
 \label{eq:highest-weight-loss-summary-target}
\end{equation}
The strict gap for both highest-weight procedures persists for rates just
above $D(\rho\Vert\sigma)$.
\end{theorem}

\begin{proof}
Equation~\eqref{eq:sc-unrestricted-alpha-form} gives $D(\rho\Vert\sigma)$
as the largest rate at which $E_{\mathrm{sc}}^*$ vanishes.
Lemma~\ref{lem:general-highest-weight-zero-cost} gives
$D_{\mathrm R}(\rho\Vert\sigma)$ as the corresponding threshold for
$E_{\mathrm{hw}}$. Indeed, the compact region $\mathcal R_{\mathrm{hw}}$
contains exactly one point with first coordinate zero, namely
$(0,D_{\mathrm R}(\rho\Vert\sigma))$. For $r$ above that threshold,
the minimum of $a+(r-b)_+$ is positive. The comparison with Keyl's
measurement follows from \eqref{eq:highest-weight-exponent-hierarchy}.

At $r=D(\rho\Vert\sigma)$, $E_{\mathrm{hw}}$ is already positive, and it
is nondecreasing. Continuity of $E_{\mathrm{sc}}^*$ therefore extends the
gap to the right. Theorem~\ref{thm:general-right-persistence} gives an
explicit interval.
\end{proof}

\subsection{The divergence gap and qubit cells}

\begin{proposition}[Equality of reverse and ordinary relative entropy]
\label{rem:qubit-gap-summary}\label{prop:reverse-relative-entropy-equality}
For any full-rank states $\rho$ and $\sigma$,
\begin{equation}
 D_{\mathrm R}(\rho\Vert\sigma)\le D(\rho\Vert\sigma),
 \qquad
 D_{\mathrm R}(\rho\Vert\sigma)=D(\rho\Vert\sigma)
 \iff [\rho,\sigma]=0.
 \label{eq:qubit-gap-summary-bound}
\end{equation}
\end{proposition}

\begin{proof}
Faithfulness of $\sigma$ gives $D(\rho\Vert\sigma)<\infty$.
The conclusion follows from \cite[Proposition~4.6]{GrootveldEtAl2026},
which also covers degenerate spectra.
\end{proof}

For a qubit null state
$\rho=p|0\rangle\!\langle0|+(1-p)|1\rangle\!\langle1|$ with $1/2<p<1$,
the principal-minor formula \eqref{eq:cov-input-SKeyl-principal-minor}
gives the explicit gap
\[
 D(\rho\Vert\sigma)-D_{\mathrm R}(\rho\Vert\sigma)
 =(2p-1)\bigl(\log\langle0|\sigma|0\rangle
             -\langle0|\log\sigma|0\rangle\bigr).
\]
By strict concavity of the logarithm, the parenthesized term is
positive exactly when $|0\rangle$ has nonzero components in two
eigenspaces of $\sigma$ with different eigenvalues.

To attain the unrestricted strong-converse exponent even at rates where
highest-weight cells are suboptimal, we use the weight-basis construction
of Proposition~\ref{prop:sc-single-cell-attainment}. For qubits, the corresponding family uses
Dicke vectors, which are normalized symmetric vectors with fixed numbers
of occurrences of the two basis vectors. The cell center is a rotation
that sets the oriented axis, which can be arbitrary. The acceptance
event selects one Young label and a shrinking neighborhood of the center. Its rate
pair includes the probability of the Young-label outcome.

Let $\mathcal R_{\mathrm{wt}}^{(1)}$ be the compact set of asymptotic rate
pairs over all Young shapes, occupations, and axes in
Definition~\ref{def:weight-seed-family}. Define
\[
 \mathcal R_{\mathrm{wt}}
 :=\overline{\operatorname{co}}\mathcal R_{\mathrm{wt}}^{(1)},
 \qquad
 E_{\mathrm{wt}}(r)
 :=\min_{(a,b)\in\mathcal R_{\mathrm{wt}}}\{a+(r-b)_+\}.
\]
The superscript $(1)$ denotes one collective block. The convex hull
represents concatenation on groups of copies.

\Needspace{12\baselineskip}
\begin{corollary}[Dicke-vector realization for qubits]
\label{cor:qubit-one-global-block-summary}
When $\rho$ and $\sigma$ are full-rank qubit states, for every
$0\le\theta\le1$ the minimum of $a-\theta b$ over
$\mathcal R_{\mathrm{wt}}^{(1)}$ equals $h(\theta)$ as defined before
Proposition~\ref{prop:sc-single-cell-attainment}. This is also the
infimum of $a-\theta b$ over all finite rate pairs obtained from
sequences of arbitrary binary effects $0\le T_n\le I$ with
$\Tr(\rho^{\otimes n}T_n)=e^{-na+o(n)}$ and
$\Tr(\sigma^{\otimes n}T_n)=e^{-nb+o(n)}$.
A minimizing pair is realized using one Schur decomposition of the
entire sample, one selected Young block, one Dicke vector in an
eigenbasis of $\sigma$, and one shrinking orbit cell. For every
prescribed rate $r\ge0$, one such cell attains the unrestricted exponent
with randomized acceptance if needed. In particular,
\begin{equation}
\begin{aligned}
 \min_{(a,b)\in\mathcal R_{\mathrm{wt}}^{(1)}}\{a+(r-b)_+\}
 &=E_{\mathrm{wt}}(r)\\
 &=E_{\mathrm{sc},2}^{\mathrm{ucov}}(r)
 =E_{\mathrm{sc}}^*(r),\qquad r\ge0.
\end{aligned}
 \label{eq:qubit-one-global-block-summary-target}
\end{equation}
\end{corollary}

\begin{proof}
For $0<\theta<1$, binary sandwiched R\'enyi data processing gives
$a-\theta b\ge h(\theta)$ for any such sequence of effects, and
Theorem~\ref{thm:exact-weight-support} constructs the Dicke cells that
attain equality. At $\theta=0$ we have $a\ge0=h(0)$. At $\theta=1$,
$\rho\le e^{D_{\max}(\rho\Vert\sigma)}\sigma$ gives
$a-b\ge-D_{\max}(\rho\Vert\sigma)=h(1)$ by tensoring and tracing against
$T_n$. The endpoint values are attained by Dicke cells on an eigenaxis
of $\sigma$ by compactness of their parameters and continuity of the
support minimum.
Corollary~\ref{thm:qubit-weight-completeness} shows how these cells attain
the strong-converse exponent. When an interior support parameter
maximizes the dual objective at rate $r$, the minimizing cell has $b=r$.
An endpoint cell suffices outside this range. Above the upper
endpoint, the rule also randomizes acceptance.
\end{proof}

\subsection{An explicit noncommuting qubit example}
Consider the full-rank pair whose Bloch vectors are perpendicular,
\begin{equation}
 \rho=\begin{pmatrix}9/10&0\\0&1/10\end{pmatrix},\qquad
 \sigma=\begin{pmatrix}1/2&2/5\\2/5&1/2\end{pmatrix}.
 \label{eq:pair-main-summary}
\end{equation}
For this pair,
$D_{\mathrm R}(\rho\Vert\sigma)=\frac45\log\frac95 <D(\rho\Vert\sigma)=\frac25\log9$.
Hence both highest-weight restrictions are strictly suboptimal throughout
$\frac45\log\frac95<r\le\frac25\log9$.
A highest-weight cell realizes
$(a,b)=(0,D_{\mathrm R}(\rho\Vert\sigma))$, whereas a non-highest-weight
Dicke cell realizes $(a,b)=(0,D(\rho\Vert\sigma))$.  Here $a=0$ means that
null acceptance has zero exponential decay rate. A fixed type-I error
tolerance requires separate control of the acceptance probability.
Figure~\ref{fig:intro-qubit-exponents} shows the Keyl and unrestricted
exponents for this pair. Appendix~\ref{app:qubit-quantitative-example}
gives the cell parameters, the local expansion above the Stein point, and
the numerical comparison.

\section{Discussion and open problems}
\label{sec:discussion-open-problems}

For qubits, Keyl's measurement attains the optimal Hoeffding exponent.
Its strong-converse exponent becomes positive above
$\DR(\rho\Vert\sigma)$, while the optimal covariant exponent stays
zero up to $D(\rho\Vert\sigma)$. For noncommuting states these
thresholds differ. At rates between them, both the optimal covariant
Hoeffding and strong-converse exponents vanish. The fixed-error Stein
theorem describes which type-I error tolerances permit each type-II rate
within that interval.

In any finite dimension, the covariant Hoeffding exponent is positive for
$0<r<\DR(\rho\Vert\sigma)$ and zero for $r\ge\DR(\rho\Vert\sigma)$
(Corollary~\ref{cor:ucov-hoeffding-nontight-sandwich}). For $d>2$ it
remains open whether Keyl's measurement attains the optimal exponent
in the positive range. The qubit proof gives one decision rule for
Keyl's measurement that controls both errors, by comparing the two
states under the same rotation on the Bloch sphere.

Proposition~\ref{prop:sc-single-cell-attainment} attains the unrestricted
strong-converse exponent by accepting at one Schur label, one
weight-basis vector, and a shrinking neighborhood of rotations.
For qubits the selected vector is a Dicke state in an eigenbasis of
$\sigma$ (Corollary~\ref{cor:qubit-one-global-block-summary}). Explicit
formulas for the selected weight spaces and vectors in higher dimension
would make this construction easier to evaluate.

\subsection*{Acknowledgements}
MH was supported in part by
the General R\&D Projects of 1+1+1 CUHK-CUHK(SZ)-GDST Joint Collaboration Fund (Grant No. GRDP2025-022), 
and the Guangdong Provincial Quantum Science Strategic Initiative (Grant No. 
GDZX2505003).
HC was supported by the National Science and Technology Council
under grants NSTC 115-2628-E-002-005, NSTC 114-2119-M-001-002, and
NSTC 115-2124-M-002-014, and by the Ministry of Education under grants
NTU-115V2016-1, NTU-CC115L893705, and MOE-114-YSFEE-0003-006-P2
(Yushan Fellow Program).

AI (Microsoft Copilot Enterprise and ChatGPT Pro)
was extensively used in the development and
writing of this manuscript. The research questions,
the overall research direction, and many of the
proof strategies were formulated by the authors.
AI assisted in developing the mathematical arguments
and drafting substantial portions of the manuscript.
The authors critically examined the mathematical
formulations and arguments, made revisions where
necessary, and carefully verified the mathematical
content and proofs.
The authors take full responsibility for the final
manuscript.

\appendix
\section{The finite-block Keyl law}\label{app:keyl-finite-output-law}
\subsection{Spectrum and flags}
Let $\mathcal F_d$ denote the space of complete flags on $\mathcal H$.
Each flag has subspaces $V_k$ of dimension $k$, with $V_0:=\{0\}$ and
$V_d=\mathcal H$. An orthonormal frame $(e_1,\ldots,e_d)$ represents the flag
when $V_k=\operatorname{span}\{e_1,\ldots,e_k\}$. Define
\[
 \Delta_d^\downarrow:=\{t\in[0,1]^d:t_1\ge\cdots\ge t_d,\ \textstyle\sum_i t_i=1\},\qquad
 \mathcal Z_d:=\Delta_d^\downarrow\times\mathcal F_d.
\]
The map $q(t,F):=\sum_i t_i|e_i\rangle\!\langle e_i|$ gives the state
with spectrum $t$ in this frame. The state determines the partial flag at
the boundaries between distinct eigenvalues. A completion chooses the
intermediate subspaces inside the repeated eigenspaces.

For a faithful input state $X$, the principal-minor formula on $\mathcal Z_d$ is
\begin{equation}
 D_{\mathrm R}((t,F)\Vert X)
 :=-H(t)-\sum_{k=1}^d(t_k-t_{k+1})
 \log\det(P_{V_k}XP_{V_k}\mid V_k),
 \qquad t_{d+1}:=0.
 \label{eq:labelled-reverse-divergence-paper1}
\end{equation}
Inside a block of equal eigenvalues, the coefficients of the intermediate subspaces vanish. The rate therefore depends only on the spectrum and its partial flag, and equals $D_{\mathrm R}(q(t,F)\Vert X)$.

At blocklength $n$, complete $\bar F$ by sampling the added subspaces
from the normalized invariant measure on the flags above $\bar F$.
This conditional measure $K_\lambda(dF\mid\bar F)$ is independent of the
input. Every such completion gives the same state
$\widehat\tau_n=q(\lambda/n,F)$.

\subsection{Decision rules on the reported and completed flags}
\begin{proof}[Proof of Lemma~\ref{lem:keyl-lifted-output-equivalence}]
The marginal law of $(\lambda,\bar F)$ under the completed-flag law
$\widetilde P_{X,n}^{\mathrm K}$ is $P_{X,n}^{\mathrm K}$. This gives the
same acceptance probability for any rule that ignores the added
subspaces. Conversely, integrating a rule on $(\lambda,F)$ against
$K_\lambda(dF\mid\bar F)$ and then $P_{X,n}^{\mathrm K}$ gives
Eq.~\eqref{eq:keyl-lift-acceptance-equality}.

For the state-valued representation, the ordered eigenvalues of
$\widehat\tau_n$ recover $\lambda/n$, and the subspaces corresponding to
its distinct eigenvalues recover the physical partial flag.  Conversely,
these data determine $\widehat\tau_n$.  On each Young orbit this map is a
continuous bijection from a compact space into the Hausdorff state space,
and hence a homeomorphism onto its image.  There are finitely many Young
orbits at fixed $n$, with disjoint estimator images.  The map and its
inverse are therefore Borel measurable on their union.  This proves the
stated Borel isomorphism and equivalence of the state-valued and
partial-flag decision rules.
\end{proof}
\subsection{Independence of the completed flag}\label{app:keyl-flag-stabilizer}
\begin{proof}[Proof of Lemma~\ref{lem:keyl-finite-block-fibre-constancy}]
Let $H_\lambda$ be the stabilizer of the highest-weight ray
$[v_\lambda]$. To identify this stabilizer, put $m_k=\lambda_k-\lambda_{k+1}$.
In the exterior-power construction of the irreducible representation,
the highest-weight ray is represented by the tensor of
$(e_1\wedge\cdots\wedge e_k)^{\otimes m_k}$ over $k$ with $m_k>0$.
The absolute overlap of this unit tensor with its rotation is the
product of the corresponding absolute overlaps of the unit wedge
vectors, each raised to $m_k$. It equals one precisely when the rotation
preserves every participating wedge ray. The $k$th wedge ray determines
$V_k$, and $V_d=\mathcal H$ is preserved by every unitary. Thus
$H_\lambda$ consists of the unitaries preserving the subspaces at the
row jumps of $\lambda$. A full unitary coordinate $U$ gives a complete
flag by taking the successive spans of its columns. Keeping the
row-jump subspaces and $V_d$ gives the reported partial flag.

For $h\in H_\lambda$, the vector
$\pi_\lambda(h)v_\lambda=\chi_\lambda(h)v_\lambda$ differs from $v_\lambda$
by a phase. Two representatives of $\bar F$ have the form $U_F$ and
$U_Fh$ for some $h\in H_\lambda$.
By multiplicativity of $\pi_\lambda$ and unitarity on $U(d)$,
\begin{align*}
 &\left\langle v_\lambda,
 \pi_\lambda((U_Fh)^\dagger X(U_Fh))v_\lambda\right\rangle \\
 &\quad=
 \left\langle \pi_\lambda(h)v_\lambda,
 \pi_\lambda(U_F^\dagger XU_F)
 \pi_\lambda(h)v_\lambda\right\rangle
 =\left\langle v_\lambda,
 \pi_\lambda(U_F^\dagger XU_F)v_\lambda\right\rangle.
\end{align*}
The coefficient therefore depends only on the reported partial flag.
Lemma~\ref{lem:appendix-highest-weight-coefficient} identifies it with
the principal-minor product.

In a block of equal rows $a,\ldots,b$, the reported flag fixes
$V_{a-1}$ and $V_b$. The added subspaces lie between them, with
$V_k\ominus V_{a-1}$ forming a flag inside $V_b\ominus V_{a-1}$.
For $a\le k<b$, the power $\lambda_k-\lambda_{k+1}$ is zero.
The minors for those subspaces therefore cancel out of the likelihood ratio.
\end{proof}

\subsection{Density and normalization}
\label{app:finite-keyl-output-law}
Use the Schur--Weyl and highest-weight construction
of Refs.~\cite{KeylWerner2001,Keyl2006}.

Fix a Schur--Weyl unitary $W_n:\mathcal H^{\otimes n} \longrightarrow \bigoplus_{\lambda\vdash n,\,\ell(\lambda)\le d} \mathcal U_\lambda\otimes\mathcal V_\lambda$,
where \(\mathcal U_\lambda\) carries the irreducible polynomial
\(GL(d)\)-representation \(\pi_\lambda\) and \(\mathcal V_\lambda\) is the
symmetric-group multiplicity space.  Put
\(m_\lambda:=\dim\mathcal V_\lambda\) and
\(d_\lambda:=\dim\mathcal U_\lambda\).  With all block operators written
after conjugation by \(W_n\),
$W_nX^{\otimes n}W_n^\dagger = \bigoplus_\lambda \pi_\lambda(X)\otimes I_{\mathcal V_\lambda}$,
for every positive definite \(X\).

Choose a unit highest-weight vector \(v_\lambda\in\mathcal U_\lambda\).
The stabilizer calculation in
Appendix~\ref{app:keyl-flag-stabilizer} identifies its ray orbit with the
reported partial-flag manifold. Write $\bar F$ for a flag on this orbit
and choose a unitary representative $U_{\bar F}$. Its choice changes
$\pi_\lambda(U_{\bar F})v_\lambda$ by a phase, so the corresponding
projector is well defined. Let \(\mu_\lambda\) be the normalized invariant
measure on this flag space. Schur's lemma gives
\begin{equation}
  d_\lambda
  \int
  |\bar F,\lambda\rangle\langle\bar F,\lambda|\,d\mu_\lambda(\bar F)
  =I_{\mathcal U_\lambda},
  \qquad
  |\bar F,\lambda\rangle:=\pi_\lambda(U_{\bar F})v_\lambda.
  \label{eq:appendix-orbit-resolution}
\end{equation}
Consequently, after the Schur projection, the covariant orbit POVM in block
\(\lambda\) is the operator-valued measure
\begin{equation}
  dM_\lambda(\bar F)
  :=d_\lambda
  |\bar F,\lambda\rangle\langle\bar F,\lambda|\otimes
  I_{\mathcal V_\lambda}\,d\mu_\lambda(\bar F).
  \label{eq:appendix-keyl-povm}
\end{equation}
The block effects in Eq.~\eqref{eq:appendix-keyl-povm} are obtained by
rotating the projector onto $v_\lambda$. See
Ref.~\cite{CassinelliDeVitoToigo2003} for a general treatment of covariant
POVMs. The block measures sum to a normalized POVM on
\(\mathcal H^{\otimes n}\).

\begin{lemma}[Highest-weight matrix coefficient]
\label{lem:appendix-highest-weight-coefficient}
Let \(\lambda_{d+1}:=0\). Every positive definite matrix \(Y\) satisfies the following
highest-weight matrix-coefficient identity:
\begin{equation}
  \langle v_\lambda,\pi_\lambda(Y)v_\lambda\rangle
  =
  \prod_{k=1}^d
  \Delta_k(Y)^{\lambda_k-\lambda_{k+1}},
  \label{eq:appendix-generalized-power}
\end{equation}
\end{lemma}

\begin{proof}
Apply the generalized-power identity in \cite[Eq.~(141)]{Keyl2006}
to $Y/\Tr Y$ and use degree-$n$ homogeneity on both sides. The principal
minors denoted there by $\mathrm{pm}_k$ are our $\Delta_k$.
\end{proof}

\begin{proposition}[Finite-block Keyl output law]
\label{prop:appendix-finite-keyl-law}
For a faithful state \(X\), use counting measure on the Young labels
and $\mu_\lambda$ on the partial flags for each $\lambda$. The joint law is
\begin{equation}
  dP_{X,n}^{\mathrm K}(\lambda,\bar F)
  =m_\lambda d_\lambda
  \prod_{k=1}^d
  \Delta_k(U_{\bar F}^\dagger XU_{\bar F})^{\lambda_k-\lambda_{k+1}}
  \,d\mu_\lambda(\bar F).
  \label{eq:appendix-finite-keyl-law}
\end{equation}
The density is independent of the representative $U_{\bar F}$ and is
constant across all completions of $\bar F$. The conditional measure
$K_\lambda(dF\mid\bar F)$ gives the state-independent lift.
\end{proposition}

\begin{proof}
By the block form of \(X^{\otimes n}\), the Born density of
\eqref{eq:appendix-keyl-povm} is
\[
 m_\lambda d_\lambda
 \langle\bar F,\lambda|\pi_\lambda(X)|\bar F,\lambda\rangle
 \,d\mu_\lambda(\bar F).
\]
Unitary covariance rewrites the matrix coefficient as
\(\langle v_\lambda,\pi_\lambda(U_{\bar F}^\dagger XU_{\bar F})v_\lambda\rangle\).
Lemma~\ref{lem:appendix-highest-weight-coefficient} gives the principal-minor
product. The POVM normalization makes this a probability law.
Lemma~\ref{lem:keyl-finite-block-fibre-constancy} shows that the coefficient
is independent of the chosen representative. Disintegrating invariant
measure over the possible completions of $\bar F$ gives
$K_\lambda(dF\mid\bar F)$.
\end{proof}

\section{Auxiliary estimates for covariant phase reductions}\label{app:phase-reduction-estimates}
\subsection{Covariant POVMs in orbit coordinates}
\begin{proof}[Proof of Lemma~\ref{lem:phase-orbit-disintegration}]

We seek an orbit integral that recovers $M(A)$ for every Borel set $A$,
as in \eqref{eq:abstract-phase-orbit-disintegration}. We first express
the scalar outcome law of the maximally mixed state in orbit coordinates
to obtain the measure on orbit labels. We then use these coordinates
to average the conjugates of the operator density and verify that the
resulting orbit operators recover $M(A)$ by covariance.

The Borel realization theorem \cite[Theorem~1.1]{BeckerKechris1993}
allows us to work on a Polish space with a continuous action. Each orbit
is compact, and the saturation of any open set is open. The closed-class
selector theorem \cite[Theorem~12.16]{Kechris1995} therefore gives a
Borel set $Y$ containing one representative per orbit and a Borel map
$\pi:\Omega\to Y$ sending each outcome to its representative. Write
$s:Y\hookrightarrow\Omega$ for the inclusion and $q(y,g)=g s(y)$.

Let $D$ be the dimension of the representation space. The outcome law
$\lambda(A)=D^{-1}\operatorname{Tr}M(A)$ for the maximally mixed state
is $G$-invariant by covariance and unitary invariance of the trace.
Positivity gives $M\ll\lambda$, so the
Radon--Nikodym theorem applied to matrix entries gives a Borel density
$R_0$ with $M(d\omega)=R_0(\omega)\lambda(d\omega)$. We may set $R_0=I$
on a Borel null set so that $R_0\ge0$ and $\operatorname{Tr}R_0=D$
everywhere. Put $\nu=\pi_*\lambda$. For any bounded Borel function $f$,
the average $\int_G f(g\omega)\,d\mu_G(g)$ depends only on $\pi(\omega)$.
Invariance of $\lambda$ and Fubini's theorem then give
\[
 \int_\Omega f(\omega)\,d\lambda(\omega)
 =\int_Y\int_G f(q(y,g))\,d\mu_G(g)d\nu(y).
\]

The pushforward $\nu=\pi_*\lambda$ is the outer measure in
\eqref{eq:abstract-phase-orbit-disintegration}, and the normalized Haar
measure $\mu_G$ is the inner measure. In the same coordinates $q(y,h)$,
the operator for each orbit is
\[
 S(y)=\int_G V_h^\dagger R_0(q(y,h))V_h\,d\mu_G(h).
\]
The Borel density $R_0$ makes $S$ measurable, and averaging its positive
conjugates gives $S(y)\ge0$. To verify
\eqref{eq:abstract-phase-orbit-disintegration}, insert this integral
and set $h=kg$. Haar right invariance preserves the integral, while
$q(y,kg)=kq(y,g)$ and $V_hV_g^\dagger=V_k$. The right side becomes
\begin{align*}
 &\int_G V_k^\dagger
 \left[\int_Y\int_G \mathbf 1_{kA}(q(y,h))R_0(q(y,h))
                  \,d\mu_G(h)d\nu(y)\right]V_k\,d\mu_G(k)\\
 &\qquad=\int_G V_k^\dagger M(kA)V_k\,d\mu_G(k)=M(A).
\end{align*}
For the first equality, apply the scalar identity entrywise to the
matrix field $\mathbf 1_{kA}R_0$. The bracket is then $M(kA)$.
Covariance and normalization of Haar measure give the last equality.
By construction $\pi(q(y,g))=y$.
\end{proof}  
\subsection{Multivariate large-set estimate on the torus}
\begin{proof}[Proof of Lemma~\ref{lem:multivariate-large-set-turan}]

\noindent\textbf{One-dimensional estimate.}
Let $A>0$ be the absolute constant in the $L^1$ case of Nazarov's
near-full-set theorem~\cite[Introduction, main theorem]{Nazarov2000Turan}.
For a nonnegative trigonometric polynomial $p$ with $m$ nonzero Fourier
characters and a measurable set $E\subseteq\mathbb T$ with
$\mu_1(E)\ge1/3$, its $L^1(E)$ norm is taken with the measure normalized
\emph{on $E$}.  Write $u=\mu_1(E^c)$. In the ambient Haar measure the inequality is
\begin{equation}
 \int_{\mathbb T}p\,d\mu_1
 \le \frac{\exp\{A(m-1)u\}}{1-u}
       \int_Ep\,d\mu_1.
 \label{eq:turan-one-dimensional-unnormalized}
\end{equation}
If $p$ is zero the conclusion is immediate. Otherwise, for frequencies in
$[-N,N]$ we have $m\le2N+1$. If $0\le u\le\delta\le1/2$ and $N\ge1$,
then $-\log(1-u)\le2u$, so
\begin{equation}
 \log\frac{\exp\{A(m-1)u\}}{1-u}
 \le(2AN+2)u\le(2A+2)N\delta.
 \label{eq:turan-one-dimensional-absorption}
\end{equation}
This proves the base case with $A_1=2A+2$ and $\delta_1=1/2$. The factor
$(1-u)^{-1}$ also covers a nonzero constant polynomial, for which it
gives the exact ratio of the two integrals.

\noindent\textbf{Slicing in dimension $\ell$.}
Assume the assertion in dimension $\ell-1$ for some $\ell\ge2$ and fix
$0<\delta\le\delta_{\ell}$, where $\delta_{\ell}$ is chosen below.  Write
$\bm\theta=(t,\bm x)\in\mathbb T\times\mathbb T^{\ell-1}$ and set
$E_{\bm x}:=\{t:(t,\bm x)\in E\}$. By Fubini's theorem the sections
are measurable for $\mu_{\ell-1}$-almost every $\bm x$. On a measurable
full-measure set of such coordinates, the missing mass
$u(\bm x):=1-\mu_1(E_{\bm x})$ is measurable and its integral is at most
$\delta$. Choose the slice threshold $\eta=\delta^{1/\ell}$ and let $G$
consist of the coordinates in this full-measure set with
$u(\bm x)\le\eta$. The excluded null set contributes zero mass, so
Markov's inequality gives
\begin{equation}
 \mu_{\ell-1}(G^c)\le\delta/\eta
 =\delta^{(\ell-1)/\ell}.
 \label{eq:good-slice-complement}
\end{equation}
A positive $\delta_{\ell}$ satisfying
$\delta_{\ell}^{1/\ell}\le\delta_1$ and
$\delta_{\ell}^{(\ell-1)/\ell}\le\delta_{\ell-1}$ exists; for example,
$\delta_{\ell}:=\min\{2^{-\ell},\delta_{\ell-1}^{\ell/(\ell-1)}\}$.
Then the one-dimensional estimate applies to each section with
$\bm x\in G$, and the induction hypothesis applies to the set $G$.

\noindent\textbf{Induction.}
The function $P(\bm x)=\int_{\mathbb T}p(t,\bm x)\,d\mu_1(t)$
is a nonnegative trigonometric polynomial of coordinate degree at most
$N$. On $G$, the one-dimensional estimate bounds $P(\bm x)$ by
$e^{A_1N\eta}\int_{E_{\bm x}}p(t,\bm x)\,d\mu_1(t)$.
By~\eqref{eq:good-slice-complement} the induction hypothesis gives
\begin{equation}
 \int_{\mathbb T^{\ell}}p\,d\mu_{\ell}
 \le e^{A_{\ell-1}N\eta}\int_GP\,d\mu_{\ell-1}
 \le e^{(A_{\ell-1}+A_1)N\eta}\int_Ep\,d\mu_{\ell}.
 \label{eq:turan-multivariate-induction-chain}
\end{equation}
Here $(\delta^{(\ell-1)/\ell})^{1/(\ell-1)}=\eta$, and the last
inequality integrates the slice bound and uses $p\ge0$. This proves
the claim with $A_\ell=\ell(2A+2)$.
\end{proof}
\subsection{Polynomial variation along a phase orbit}
\begin{proof}[Proof of Lemma~\ref{lem:poly-phase-var-hoeffding}]
Decompose the $n$-copy representation of $G_X$ into its $L_{X,n}$ weight
spaces and let ${\cal E}_{X,n}$ be the associated pinching.  For
$Y_g=(g^\dagger)^{\otimes n}Yg^{\otimes n}\ge0$, the pinching inequality gives
$Y_g\le L_{X,n}{\cal E}_{X,n}(Y_g)$.  The pinched operator is independent of
$g$ and equals the Haar average of the phase conjugates of $Y$.  Taking the
trace against $A\ge0$ proves the claim.
\end{proof}
\subsection{Replacing a covariant test by a phase-invariant test}
\begin{proof}[Proof of Lemma~\ref{lem:phase-cov-to-phase-inv-hoeffding}]
Use the representation of Lemma~\ref{lem:phase-orbit-disintegration}.
We base acceptance on the Haar average of the original rule along
each orbit. Write $R(y,g)=g^{\otimes n}S(y)(g^\dagger)^{\otimes n}$ for
the measurement density and $f(y,g)\in[0,1]$ for the original decision
rule at $q(y,g)$. Put $m(y)=\int_{G_X}f(y,g)\,d\mu_X(g)$ and
$L=L_{X,n}$. The new rule accepts every outcome with orbit label $y$
if $1-m(y)<1/(2L)$ and rejects every outcome otherwise. The rule
is Borel and depends only on the orbit, so its effect $T'_n$ is invariant.

The density under $X^{\otimes n}$ is constant in $g$. On a discarded
orbit the original rule rejects at least a fraction $1/(2L)$ of the
null probability in that orbit. Discarding the entire orbit therefore
increases its contribution to the type-I error by at most a factor $2L$.
Integrating proves~\eqref{eq:X-error-transfer-hoeffding}.

On a kept orbit, write $w(g)=\operatorname{Tr}Y^{\otimes n}R(y,g)$
and $W=\int_{G_X}w(g)\,d\mu_X(g)$.
Lemma~\ref{lem:poly-phase-var-hoeffding} gives $w(g)\le LW$, so
\[
 \int_{G_X}(1-f(y,g))w(g)\,d\mu_X(g)
 \le L(1-m(y))W\le W/2.
\]
The old rule already accepted at least half the $Y$-mass of this orbit.
Integrating gives~\eqref{eq:Y-error-transfer-hoeffding}.
\end{proof}

\section{Technical proofs for the covariant strong-converse construction}\label{app:sc-technical-proofs}
\begin{proof}[Proof of Lemma~\ref{lem:general-local-orbit-cell}]
Write $U=U_0V$ for $U\in C_m(U_0)$, so that
$\|V-I\|_\infty\le m^{-2}$. Apply the conjugation estimate used in
Theorem~\ref{thm:general-completeness} to the reference matrix
$U_0^\dagger YU_0>0$. Its norm and smallest eigenvalue are independent
of $U_0$, so there is a constant $c_Y$, independent of $m,\lambda,v,U_0$,
such that
$e^{-c_Ym^{-2}}U_0^\dagger YU_0 \le U^\dagger YU \le e^{c_Ym^{-2}}U_0^\dagger YU_0$.
Under Schur--Weyl duality, the $\lambda$ block of $Y^{\otimes m}$ is
$\pi_\lambda(Y)\otimes I_{\mathcal P_\lambda}$. Therefore the
preceding inequalities tensorize to
$e^{-c_Y/m}(U_0^\dagger YU_0)^{\otimes m} \le (U^\dagger YU)^{\otimes m} \le e^{c_Y/m}(U_0^\dagger YU_0)^{\otimes m}$.
Restricting to the Schur block gives the same comparison for
$\pi_\lambda(U^\dagger YU)$ and hence for its matrix coefficient at $v$.
The log coefficient varies by at most $c_Y/m$ across the ball. The
logarithms of the Haar measure of the ball and of $\dim\mathcal Q_\lambda$
are $O(\log m)$, the latter by the Weyl dimension formula at fixed $d$.
Thus
\[
 \log P_{Y,m}(\lambda,v,C_m(U_0))
 =\log\dim\mathcal P_\lambda
  +\log\langle v,\pi_\lambda(U_0^\dagger YU_0)v\rangle
  +O(\log m)
\]
uniformly in $\lambda,v,U_0$. Take the larger constant for $Y=\rho$ and
$Y=\sigma$.
\end{proof}

\begin{lemma}[Specht multiplicity at fixed dimension]
\label{lem:fixed-dimension-specht-multiplicity}
For $m\ge1$ and a Young diagram $\lambda\vdash m$ with at most $d$ rows,
pad $\lambda$ with zeros to length $d$. At fixed $d$,
\[
 \log\dim\mathcal P_\lambda=mH(\lambda/m)+O_d(\log(m+1)),
\]
uniformly in $\lambda$, where $H$ is the Shannon entropy.
\end{lemma}
\begin{proof}
The hook-length formula~\cite{Sagan2001} gives
\[
 \dim\mathcal P_\lambda
 =\frac{m!\prod_{1\le i<j\le d}(\lambda_i-\lambda_j+j-i)}
        {\prod_{i=1}^d(\lambda_i+d-i)!}.
\]
Divide by the multinomial coefficient $m!/\prod_i\lambda_i!$.
In this ratio the numerator is the product of the shifted row
differences and the denominator is $\prod_{i=1}^d\prod_{s=1}^{d-i}(\lambda_i+s)$.
Each product contains $d(d-1)/2$ factors in $[1,m+d]$, including when
some rows are equal or zero. The logarithm of the ratio therefore has
absolute value at most $\tfrac12d(d-1)\log(m+d)$.
Stirling's estimate $\log k!=k\log k-k+O(\log(k+1))$, with
$0\log0=0$, gives $mH(\lambda/m)+O_d(\log(m+1))$ for the
logarithm of the multinomial coefficient. This proves the estimate.
\end{proof}

\begin{proof}[Proof of Lemma~\ref{lem:sc-weight-space-coefficient-identity}]
Equation \eqref{eq:sc-weight-space-definition} gives
$\pi_\lambda(\sigma^{-\theta/2})v
=(\prod_i s_i^{-\theta\nu_i/2})v$.
By multiplicativity and self-adjointness of
$\pi_\lambda(\sigma^{-\theta/2})$,
\[
 \langle v,\pi_\lambda(X_\theta)v\rangle
 =\langle \pi_\lambda(\sigma^{-\theta/2})v,
          \pi_\lambda(\rho)\pi_\lambda(\sigma^{-\theta/2})v\rangle .
\]
Substituting the scalar action gives the first identity. Since
$\pi_\lambda(\sigma)v=(\prod_i s_i^{\nu_i})v$, taking logarithms gives
the second.
\end{proof}

\begin{proof}[Proof of Lemma~\ref{lem:sc-renyi-scale-weight-seed-selection}]
An orthonormal basis of $\mathcal Q_\lambda$ adapted to the fixed
weight-space decomposition resolves $\Tr\pi_\lambda(X)$ into diagonal
coefficients.  Positivity makes every coefficient nonnegative, so one
weight-basis vector $v$, which may depend on $X$, satisfies
$\langle v,\pi_\lambda(X)v\rangle
\ge\Tr\pi_\lambda(X)/\dim\mathcal Q_\lambda$.
After diagonalizing $X$, the eigenvalues of the polynomial representation
are weight monomials in the eigenvalues $z_i$ of $X$. Every weight $\nu$
satisfies $\sum_{i=1}^k\nu_i\le\sum_{i=1}^k\lambda_i$ for $k<d$, with
equality at $k=d$. Summation by parts against the decreasing $\log z_i$
shows that the largest eigenvalue is the highest-weight monomial
$\prod_i z_i^{\lambda_i}$. It is bounded by the trace and bounds every
diagonal coefficient from above. Hence
\[
 \frac{\prod_i z_i^{\lambda_i}}{\dim\mathcal Q_\lambda}
 \le\langle v,\pi_\lambda(X)v\rangle
 \le\prod_i z_i^{\lambda_i}.
\]
The Weyl dimension bound $\log\dim\mathcal Q_\lambda=O(\log m)$
shows that division of the logarithms by $m$ gives the same limit on both
sides, namely $\sum_i p_i\log z_i$.
\end{proof}

\begin{proof}[Construction for Remark~\ref{rem:sc-pinching-covariant-construction}]
Let $\widehat\rho_m$ be the pinching of $\rho^{\otimes m}$ in the spectral
projections of $\sigma^{\otimes m}$, and take a joint spectral measurement
$\{F_{m,j}\}_j$ of $\widehat\rho_m$ and $\sigma^{\otimes m}$, omitting zero
projections. Write $p_{m,j}=\Tr\rho^{\otimes m}F_{m,j}$ and
$q_{m,j}=\Tr\sigma^{\otimes m}F_{m,j}$.
The standard pinching construction gives
\[
 \lim_{m\to\infty}\frac1mD_\alpha(p_m\Vert q_m)
 =\widetilde D_\alpha(\rho\Vert\sigma),\qquad 1<\alpha<\infty,
\]
with the same spectral measurement for every order
\cite[Theorem~III.7, its proof, and Corollary~III.8]{MosonyiOgawa2015}.

Covariantize this measurement by recording $(U,j)$ with POVM density
\[
 d\widehat M_m(U,j)
 =U^{\otimes m}F_{m,j}(U^\dagger)^{\otimes m}\,dU.
\]
It is normalized and covariant, and no choice in it depends on $\alpha$.
Let $\widehat P_m,\widehat Q_m$ be its complete output laws.
On the radius-$m^{-2}$ ball $C_m(I)$, the two outcome densities lie
between $e^{-c/m}$ and $e^{c/m}$ times $p_{m,j}$ and $q_{m,j}$,
respectively, with one constant $c$ independent of $m,j$.
Writing $v_m$ for the Haar volume of this ball and restricting the
R\'enyi-moment integral to it gives
\[
 \begin{split}
 D_\alpha(\widehat P_m\Vert\widehat Q_m)
 \ge{}&D_\alpha(p_m\Vert q_m)
       -\frac{\log(1/v_m)}{\alpha-1}
       -\frac{(2\alpha-1)c}{m(\alpha-1)}.
 \end{split}
\]
Since $-\log v_m=O(\log m)$, division by $m$ and quantum data processing
prove the finite-order limits. For every fixed $\alpha>1$,
\[
 \frac1mD_\alpha(\widehat P_m\Vert\widehat Q_m)
 \le\frac1mD_\infty(\widehat P_m\Vert\widehat Q_m)
 \le\widetilde D_\infty(\rho\Vert\sigma).
\]
Taking the lower limit in $m$ and then letting $\alpha\to\infty$ proves
the order-infinity limit for the same observation sequence.
\end{proof}

\section{Relation to the covariant-tomography converse}
\label{subsec:keyl-stein-two-converses}
Grootveld, Maloney, Pollack, and Wu
\cite[Theorems~1.2 and~6.1]{GrootveldEtAl2026} consider consistent
unitary-covariant tomography protocols whose estimator laws satisfy a
large-deviation principle. They bound the exponential rate of estimating
a fixed state $\tau$ under $X$ by
$I(\tau\Vert X)\le D_{\mathrm R}(\tau\Vert X)$. The rate is determined by
the large-deviation bounds for neighborhoods of $\tau$. In their
local-to-pointwise argument \cite[Lemma~6.5]{GrootveldEtAl2026}, one takes
the blocklength limits with a fixed neighborhood radius and then lets
the radius decrease to zero.

A randomized binary rule can change at every blocklength, even within
one neighborhood. For Keyl's measurement,
Proposition~\ref{prop:keyl-strong-converse-converse} uses the exact
likelihood ratio to split the acceptance integral. The probability of
the large-ratio event under $\rho$ bounds one part, and the prescribed
type-II rate bounds the other. The split allows arbitrary measurable
decision rules.

\section{Highest-weight rate regions and strict-loss proofs}\label{app:highest-weight-loss-proofs}
\label{sec:general-highest-weight-loss-detailed}
\subsection{The two thresholds}
The unrestricted exponent is zero up to $D(\rho\Vert\sigma)$,
while the two highest-weight exponents are zero up to
$D_{\mathrm R}(\rho\Vert\sigma)$. To bound how far the gap persists beyond
$D(\rho\Vert\sigma)$, we compare $E_{\mathrm{hw}}$ with an upper bound
$F_D$ on the unrestricted exponent.

\subsection{An ordinary-relative-entropy upper envelope}
\label{subsec:ordinary-relative-entropy-envelope}

Define the ordinary-relative-entropy upper bound by
\begin{equation}
 F_D(r\mid\rho\mid\sigma)
 :=\inf_{\tau\in\mathcal D_d}
 \left\{
 D(\tau\Vert\rho)
 +\bigl(r-D(\tau\Vert\sigma)\bigr)_+
 \right\}.
 \label{eq:ordinary-relative-entropy-envelope}
\end{equation}
For $\alpha>1$, define the balanced, or log-Euclidean, R\'enyi divergence by
$D_\alpha^{\mathrm b}(\rho\Vert\sigma):=\frac{1}{\alpha-1}\log\Tr\exp\!\left(\alpha\log\rho+(1-\alpha)\log\sigma\right)$.

\begin{proposition}[Ordinary-relative-entropy support formula]
\label{prop:ordinary-relative-entropy-envelope}
When $\rho$ and $\sigma$ are full rank, the auxiliary envelope satisfies
\begin{equation}
 F_D(r\mid\rho\mid\sigma)
 =\sup_{0\le\theta<1}
 \left\{
 \theta r-
 \theta D_{1/(1-\theta)}^{\mathrm b}(\rho\Vert\sigma)
 \right\},
 \label{eq:balanced-envelope}
\end{equation}
where the support term $-\theta D_{1/(1-\theta)}^{\mathrm b}(\rho\Vert\sigma)$ is
defined to be zero at $\theta=0$.  Its continuous extension at
$\theta=1$ is $-\lambda_{\max}(\log\rho-\log\sigma)$. The supremum is unchanged if it includes this endpoint.  Moreover,
\begin{equation}
 E_{\mathrm{sc}}^*(r)
 \le F_D(r\mid\rho\mid\sigma)
 \qquad(r\ge0).
 \label{eq:unrestricted-below-ordinary-envelope}
\end{equation}
If $\rho$ and $\sigma$ commute, equality holds in
\eqref{eq:unrestricted-below-ordinary-envelope} for every $r$.
\end{proposition}

\begin{proof}
Using $(r-D(\tau\Vert\sigma))_+
=\max_{0\le\theta\le1}\theta\{r-D(\tau\Vert\sigma)\}$, we apply
Sion's minimax theorem~\cite[Theorem~3.4]{Sion1958} on the compact convex
domains $\mathcal D_d$ and $[0,1]$. The objective is continuous and
affine in $\theta$. Its dependence on $\tau$ is convex, since
\begin{equation}
 D(\tau\Vert\rho)-\theta D(\tau\Vert\sigma)
 =(1-\theta)\Tr\tau\log\tau-\Tr\tau(\log\rho-\theta\log\sigma).
\end{equation}
It follows that
\[
 F_D(r\mid\rho\mid\sigma)=
 \max_{0\le\theta\le1}\left\{\theta r+
 \min_{\tau\in\mathcal D_d}[D(\tau\Vert\rho)-\theta D(\tau\Vert\sigma)]\right\}.
\]
For $\theta<1$, the Gibbs variational principle gives the inner minimum
\[
 -(1-\theta)\log\Tr\exp\!\left(\frac{\log\rho-\theta\log\sigma}{1-\theta}\right)
 =-\theta D_{1/(1-\theta)}^{\mathrm b}(\rho\Vert\sigma),
\]
with value zero at $\theta=0$. Since the objective is continuous on a
compact set, its minimum is continuous in $\theta$.
At $\theta=1$, the state optimization is
$\inf_\tau\{-\Tr\tau(\log\rho-\log\sigma)\}
=-\lambda_{\max}(\log\rho-\log\sigma)$, which also gives the stated endpoint limit.

For $\alpha>1$, the comparison between balanced and sandwiched
R\'enyi divergences~\cite[Proposition~III.20, Eq.~(69)]{MosonyiOgawa2017} gives
$D_\alpha^{\mathrm b}(\rho\Vert\sigma) \le\widetilde D_\alpha(\rho\Vert\sigma)$.
Comparing \eqref{eq:balanced-envelope} with the unrestricted support
formula \eqref{eq:sc-unrestricted-alpha-form} proves
\eqref{eq:unrestricted-below-ordinary-envelope}.  In the commuting case,
both quantum R\'enyi divergences reduce to the same classical R\'enyi
divergence, so the two envelopes coincide.
\end{proof}

We next record the rates at which this upper bound vanishes and its
Lipschitz bound above the threshold.
\begin{lemma}[Zero-cost region and rate regularity of the ordinary envelope]
\label{lem:ordinary-envelope-zero-cost}
$F_D(r\mid\rho\mid\sigma)=0$ if and only if $0\le r\le D(\rho\Vert\sigma)$
for full-rank states $\rho$ and $\sigma$.
Moreover, $r\mapsto F_D(r\mid\rho\mid\sigma)$ is nondecreasing and
$1$-Lipschitz on $[0,\infty)$.
\end{lemma}

\begin{proof}
Choosing $\tau=\rho$ in Eq.~\eqref{eq:ordinary-relative-entropy-envelope}
gives zero for $r\le D(\rho\Vert\sigma)$.  Conversely, if this
infimum vanishes, the continuous function
$\tau\mapsto D(\tau\Vert\rho)+(r-D(\tau\Vert\sigma))_+$
attains its zero minimum on compact $\mathcal D_d$.  Both terms are
nonnegative, so
$D(\tau\Vert\rho)=0$, hence $\tau=\rho$, and the positive part can vanish only
when $r\le D(\rho\Vert\sigma)$.  For each fixed $\tau$, the function of the rate $r$ in
Eq.~\eqref{eq:ordinary-relative-entropy-envelope}, namely
$r\mapsto D(\tau\Vert\rho)+(r-D(\tau\Vert\sigma))_+$,
is nondecreasing and $1$-Lipschitz. Taking the infimum over $\tau$
preserves both properties.
\end{proof}

\begin{remark}[Regularization of the ordinary-relative-entropy bound]
\label{rem:ordinary-envelope-regularization}
Evaluating the comparison envelope on $n$ copies at the total rate $nr$
gives the same bound per copy:
\[
 \frac1n F_D(nr\mid\rho^{\otimes n}\mid\sigma^{\otimes n})
 = F_D(r\mid\rho\mid\sigma).
\]
For $\theta<1$, the Gibbs optimizer of the $n$-copy support term is
proportional to
\[
 \exp\!\left(\frac{\log\rho^{\otimes n}-\theta\log\sigma^{\otimes n}}{1-\theta}\right).
\]
The exponent is a sum of copies of $\log\rho-\theta\log\sigma$ on
separate factors. These terms commute, so the optimizer factorizes
and the support term is $n$ times its one-copy value. At $\theta=1$,
the largest eigenvalue of the sum of copies of $\log\rho-\log\sigma$
is $n$ times the one-copy largest eigenvalue. The rate term is also
multiplied by $n$, so maximizing over $\theta\in[0,1]$ and dividing
by $n$ gives the displayed equality, with the same balanced
R\'enyi support as in \eqref{eq:balanced-envelope}.
The gap proofs below use the single-copy envelope directly. Its zero
region gives the bound up to the Stein threshold, and its value at that
threshold and its Lipschitz bound control the increase just above it.
\end{remark}

\subsection{Rate region of concatenated highest-weight cells}
\label{subsec:general-highest-weight-region}
A highest-weight cell $C$ at block length $m$ selects one Young label
and an operator-norm ball of radius $m^{-2}$ about a prescribed rotation
in $U(d)$. As in Section~\ref{sec:highest-weight-consequences}, the
measurement records the full rotation. The probability $P_{X,m}(C)$ is
the one-vector law of Lemma~\ref{lem:general-local-orbit-cell}, with
$v=v_\lambda$ and the rotation ball of the cell as its set argument.
Full rank makes both entries of its rate pair
\[
 \bigl(a_m(C),b_m(C)\bigr)
 =\bigl(-m^{-1}\log P_{\rho,m}(C),-m^{-1}\log P_{\sigma,m}(C)\bigr)
\]
finite. For fixed $m$, these pairs form a compact set because the Young
labels are finite and the probabilities depend continuously on the
center of the neighborhood. The one-block region
$\mathcal R_{\mathrm{hw}}^{(1)}$ is the
closure of all limits of such pairs along block lengths tending to
infinity. The next lemma identifies this region and realizes each
point along the full sequence of block lengths.

\begin{lemma}[Asymptotic rate image of one-block highest-weight cells]
\label{lem:highest-weight-rate-image}
When $\rho$ and $\sigma$ are full rank, the one-block asymptotic highest-weight cell region is the compact continuous image
\begin{equation}
 \mathcal R_{\mathrm{hw}}^{(1)}
 =\left\{
 \bigl(D_{\mathrm R}(\tau\Vert\rho),
       D_{\mathrm R}(\tau\Vert\sigma)\bigr):
 \tau\in\mathcal D_d
 \right\}.
 \label{eq:highest-weight-rate-image}
\end{equation}
Consequently, if $(a_j,b_j)\in\mathcal R_{\mathrm{hw}}^{(1)}$ and
$a_j\to0$, then
\begin{equation}
 b_j\longrightarrow D_{\mathrm R}(\rho\Vert\sigma).
 \label{eq:highest-weight-zero-cost-stability}
\end{equation}
\end{lemma}
\begin{proof}
Consider a sequence of canonical highest-weight cells with block lengths
$m_j\to\infty$, Young diagrams $\lambda^{(j)}$, and orbit centers $U_j$.
After passing to a subsequence, compactness of the ordered simplex and of
$U(d)$ gives
$\lambda^{(j)}/m_j\longrightarrow t\in\Delta_d^\downarrow$ and $U_j\longrightarrow U$.
Write $C_{m_j}$ for each selected cell and set
$\tau:=U\operatorname{diag}(t)U^\dagger$. Apply the cell estimate
of Lemma~\ref{lem:general-local-orbit-cell} and the highest-weight
coefficient identity \eqref{eq:appendix-generalized-power}.
Lemma~\ref{lem:fixed-dimension-specht-multiplicity} gives the
multiplicity term $m_jH(\lambda^{(j)}/m_j)+O_d(\log(m_j+1))$.
Thus, uniformly for $X\in\{\rho,\sigma\}$,
\begin{align*}
 -\frac1{m_j}\log P_{X,m_j}(C_{m_j})
 &=-H(\lambda^{(j)}/m_j)
   -\sum_{k=1}^d
     \frac{\lambda_k^{(j)}-\lambda_{k+1}^{(j)}}{m_j}
     \log\Delta_k(U_j^\dagger XU_j)+o(1) \\
 &\longrightarrow D_{\mathrm R}(\tau\Vert X),
\end{align*}
where $\lambda_{d+1}^{(j)}:=0$.  Thus every subsequential one-block rate pair
belongs to the right-hand side of \eqref{eq:highest-weight-rate-image}.

Conversely, let $\tau=U\operatorname{diag}(t)U^\dagger$ with
$t\in\Delta_d^\downarrow$.  Young diagrams $\lambda^{(m)}\vdash m$ with
$\lambda^{(m)}/m\to t$ and the highest-weight rays in those blocks
determine canonical radius-$m^{-2}$ orbit cells centered at $U$.  The same
calculation yields the rate pair
$\bigl(D_{\mathrm R}(\tau\Vert\rho),
D_{\mathrm R}(\tau\Vert\sigma)\bigr)$.  This proves the reverse inclusion.
Continuity of the two rate functions on the compact state space makes the
image compact, so the closure in the definition of
$\mathcal R_{\mathrm{hw}}^{(1)}$ adds no points.

Finally, if $(a_j,b_j)$ has $a_j\to0$, represent it by states $\tau_j$ using
\eqref{eq:highest-weight-rate-image}.  Every convergent subsequence of
$\tau_j$ has a limit $\tau_*$ satisfying
$D_{\mathrm R}(\tau_*\Vert\rho)=0$.  Since the Keyl rate function has the
unique zero $\tau_*=\rho$, compactness implies $\tau_j\to\rho$.  Continuity
then gives \eqref{eq:highest-weight-zero-cost-stability}.
\end{proof}

By Carath\'eodory's theorem in $\mathbb R^2$, any point in the convex
hull of $\mathcal R_{\mathrm{hw}}^{(1)}$ is a convex combination of at
most three points. The combinations of at most three points form a
compact set, so this convex hull is already closed. For a prescribed
combination, omit any zero weights and round the group lengths in the
remaining proportions so that they sum to $n$. Use the cell sequence
for each pair on its group.
The product measurement is covariant under the same unitary on all
copies. Independence makes the rates add with the specified proportions.

For a product event $B_n$ with rate pair $(a,b)$, the rule
$f_n(\omega)=e^{-n(r-b)_+}\mathbf 1_{B_n}(\omega)$ achieves a type-II
rate of at least $r$ and a null-acceptance rate of $a+(r-b)_+$.
Choosing a minimizing pair in the compact region $\mathcal R_{\mathrm{hw}}$
therefore attains $E_{\mathrm{hw}}(r)$.

Conversely, take any feasible sequence of these tests with a fixed finite
number $L$ of blocks, $m_i(n)/n\to t_i>0$, and a selected cell $C_{i,n}$
in each block. Write $a_{i,n}=-m_i(n)^{-1}\log P_{\rho,m_i(n)}(C_{i,n})$
for the null-acceptance rate of that cell, and $b_{i,n}$ for its rate
under $\sigma$. Independence gives the joint event rates per input copy
\[
 (a_n,b_n)=\sum_{i=1}^L \frac{m_i(n)}{n}\,(a_{i,n},b_{i,n}).
\]
Let $c_n\in[0,1]$ be the probability of retaining an acceptance on the
joint event. If $c_n=0$ at infinitely many blocklengths, the null-acceptance
$\limsup$ exponent is infinite. Otherwise $c_n>0$ eventually, and
$s_n=-n^{-1}\log c_n\ge0$ adds the same amount to both rates. The constraint
$\liminf_n(b_n+s_n)\ge r$ gives, for every $\delta>0$ and all sufficiently
large $n$,
\[
 a_n+s_n\ge a_n+(r-\delta-b_n)_+.
\]

Since each $m_i(n)\to\infty$, the compactness and cell estimates in the
proof of Lemma~\ref{lem:highest-weight-rate-image} give a common subsequence
on which all $L$ pairs $(a_{i,n},b_{i,n})$ converge to points of
$\mathcal R_{\mathrm{hw}}^{(1)}$. The joint pair then converges to a point
$(a,b)\in\mathcal R_{\mathrm{hw}}$, with weights $t_i$. The limsup of
$a_n+s_n$ over the full sequence is at least $a+(r-\delta-b)_+$.
Letting $\delta\downarrow0$ gives the lower bound
$a+(r-b)_+\ge E_{\mathrm{hw}}(r)$.

Thus $E_{\mathrm{hw}}(r)$ is the optimum for the concatenated cell
procedure. Its measurements are covariant, so
\begin{equation}
 E_{\mathrm{sc}}^*(r)\le E_{\mathrm{hw}}(r).
 \label{eq:general-highest-weight-scope}
\end{equation}

\begin{definition}[Zero-cost threshold of a compact rate region]
For a nonempty compact rate region $\mathcal R\subset[0,\infty)^2$, put
$B_0(\mathcal R):=\sup\{b:(0,b)\in\mathcal R\}$,
with $B_0(\mathcal R):=-\infty$ if the displayed set is empty.
\end{definition}
\begin{lemma}[General highest-weight zero-cost threshold]
\label{lem:general-highest-weight-zero-cost}
Every full-rank pair determines the same zero-cost threshold for the
one-block and block-concatenated highest-weight regions:
$B_0(\mathcal R_{\mathrm{hw}}^{(1)}) =B_0(\mathcal R_{\mathrm{hw}}) =D_{\mathrm R}(\rho\Vert\sigma)$.
The exponent of the classical experiment generated by Keyl's measurement
vanishes exactly for
$r\le D_{\mathrm R}(\rho\Vert\sigma)$.
\end{lemma}
\begin{proof}
Lemma~\ref{lem:highest-weight-rate-image} identifies $\bigl(0,D_{\mathrm R}(\rho\Vert\sigma)\bigr)$ as the unique vertical-axis point of $\mathcal R_{\mathrm{hw}}^{(1)}$.
If a point in the convex hull has first coordinate zero, nonnegativity of the first coordinate forces every component with
positive weight to have first coordinate zero.  Each such component is the
unique point displayed above, so the second coordinate is
$D_{\mathrm R}(\rho\Vert\sigma)$.  This gives the same threshold for the convex hull.
In the exact formula \eqref{eq:keyl-strong-converse-exact}, both
nonnegative terms must vanish. The first forces $\tau=\rho$. At that
state, the second vanishes exactly when $r\le D_{\mathrm R}(\rho\Vert\sigma)$.
\end{proof}

For the support calculations below, put
$h_{\mathrm{hw}}(\theta):=\min_{(a,b)\in\mathcal R_{\mathrm{hw}}}(a-\theta b)$
for $0\le\theta\le1$.

\begin{lemma}[Highest-weight max-relative-entropy endpoint]
\label{lem:highest-weight-dmax-endpoint}
The support at $\theta=1$ is the negative max-relative entropy,
\[
 h_{\mathrm{hw}}(1)=\min_{(a,b)\in\mathcal R_{\mathrm{hw}}}(a-b)
 =-\Dmax(\rho\Vert\sigma).
\]
\end{lemma}
\begin{proof}
The operator inequality
$\rho\le e^{\Dmax(\rho\Vert\sigma)}\sigma$ tensorizes.  Hence every
positive test effect $T_m$ satisfies
$\Tr\rho^{\otimes m}T_m \le e^{m\Dmax(\rho\Vert\sigma)} \Tr\sigma^{\otimes m}T_m$.
For an asymptotic rate pair $(a,b)$ this gives
$b-a\le\Dmax(\rho\Vert\sigma)$, or equivalently
$a-b\ge-\Dmax(\rho\Vert\sigma)$.  The inequality passes to convex
combinations and closure, so it holds throughout
$\mathcal R_{\mathrm{hw}}$.

To prove the reverse inequality, choose a unit vector $v$ attaining the
largest generalized Rayleigh quotient, so that
$\log\frac{\langle v,\rho v\rangle} {\langle v,\sigma v\rangle} =\Dmax(\rho\Vert\sigma)$.
The fully symmetric Young block $(m,0,\ldots,0)$ and the coherent
highest-weight ray centered at $[v^{\otimes m}]$ define a canonical
radius-$m^{-2}$ orbit cell. Measure the normalized highest-weight orbit
POVM, record its Young label and orbit coordinate, and accept $\rho$
exactly if the label is $(m,0,\ldots,0)$ and the coordinate belongs
to that cell; reject on all other outcomes.  For the selected cell $C_m$, Lemma~\ref{lem:general-local-orbit-cell}
and the block probability formula give
$\log P_{Y,m}(C_m)=m\log\langle v,Yv\rangle+o(m)$ for both $Y=\rho,\sigma$.
Indeed, the multiplicity of the fully symmetric block is one, while the orbit
normalization and the shrinking-cell volume have logarithm $o(m)$.  The
resulting highest-weight rate pair therefore satisfies
$a-b=-\log\langle v,\rho v\rangle +\log\langle v,\sigma v\rangle =-\Dmax(\rho\Vert\sigma)$.
This pair belongs to the one-block asymptotic highest-weight region and hence
to $\mathcal R_{\mathrm{hw}}$, proving the opposite inequality.
\end{proof}

\begin{lemma}[One-sided derivatives and fixed-rate slices of a compact rate region]
\label{lem:compact-rate-envelope-derivatives}
Let $\mathcal R\subset\mathbb R^2$ be nonempty, compact, and convex, and put
$E_{\mathcal R}(r):=\min_{(a,b)\in\mathcal R} \{a+(r-b)_+\}$ and $h_{\mathcal R}(\theta):=\min_{(a,b)\in\mathcal R}(a-\theta b)$.
Then the exponent admits the support representation
$E_{\mathcal R}(r)=\max_{0\le\theta\le1} \{\theta r+h_{\mathcal R}(\theta)\}$.
The active set
$A_{\mathcal R}(r):=\operatorname*{argmax}_{0\le\theta\le1} \{\theta r+h_{\mathcal R}(\theta)\}$
is nonempty and compact, and
$E_{\mathcal R,+}'(r)=\max A_{\mathcal R}(r)$ and $E_{\mathcal R,-}'(r)=\min A_{\mathcal R}(r)$.
The slice at rate $r$ consists of the pairs $(a,b)\in\mathcal R$ with
$b=r$. Every active parameter $\theta_*\in A_{\mathcal R}(r)$ admits a
saddle pair $((a_*,b_*),\theta_*)$. If $0<\theta_*<1$, then
$b_*=r$ and
$E_{\mathcal R}(r)=a_* \ge \min\{a:(a,r)\in\mathcal R\}$.
Conversely, every $(a,r)\in\mathcal R$ gives
$E_{\mathcal R}(r)\le a$. Thus an interior active parameter implies
$E_{\mathcal R}(r)=\min\{a:(a,r)\in\mathcal R\}$.
\end{lemma}
\begin{proof}
The identity $(r-b)_+=\max_{0\le\theta\le1}\theta(r-b)$ and Sion's
minimax theorem~\cite[Theorem~3.4]{Sion1958} give the support representation.
The objective is affine in both variables on compact convex sets.
Continuity makes the active set nonempty and compact.
Danskin's theorem~\cite{Danskin1967,Bertsekas1999}, applied to
$\phi(r,\theta)=\theta r+h_{\mathcal R}(\theta)$, gives
$E_{\mathcal R}'(r;d)=\max_{\theta\in A_{\mathcal R}(r)}d\theta$.
Taking $d=1$ gives the right derivative.  Taking $d=-1$ and using
$E_{\mathcal R}'(r;-1)=-E_{\mathcal R,-}'(r)$ gives the left derivative.

Choose a primal minimizer $(a_*,b_*)$ and any dual maximizer
$\theta_*\in A_{\mathcal R}(r)$. The two optimal values coincide, so
$a_*+\theta_*(r-b_*)$ is bounded below by the dual value and above by
the primal value, with equality on both sides. Thus
$((a_*,b_*),\theta_*)$ is a saddle pair, so $\theta_*$ maximizes
$a_*+\theta(r-b_*)$ over $[0,1]$.  If $\theta_*$ is interior, this affine
function can have an interior maximum only when its coefficient vanishes,
so $b_*=r$.  The saddle value is then $a_*$.  The reverse comparison follows
by substituting any point $(a,r)$ into the primal objective.  Compactness gives
attainment of the slice minimum whenever the slice is nonempty, as it is here
because the saddle point lies on it.
\end{proof}

\subsection{The strict gap}
\label{subsec:general-strict-loss}
Write $D=D(\rho\Vert\sigma)$ and $D_{\mathrm R}=D_{\mathrm R}(\rho\Vert\sigma)$.

\begin{theorem}[Strict inequality up to the unrestricted Stein threshold]
\label{thm:general-strict-Keyl-loss}
Let $\rho$ and $\sigma$ be full-rank states on $\mathbb C^d$, and assume that
$D_{\mathrm R}<D$.
Every rate $r$ in the interval $D_{\mathrm R}<r\le D$ satisfies
$E_{\mathrm{sc}}^*(r) \le F_D(r\mid\rho\mid\sigma) <E_{\mathrm{hw}}(r)$.
Moreover, Eq.~\eqref{eq:highest-weight-exponent-hierarchy} extends the chain to
\[
 E_{\mathrm{sc}}^*(r)=F_D(r\mid\rho\mid\sigma)=0
 <E_{\mathrm{hw}}(r)
 \le E_{\mathrm{sc}}^{\mathrm{Keyl}}(r\mid\rho\mid\sigma).
\]
\end{theorem}

\begin{proof}
The comparison has two parts. The first inequality follows from Proposition~\ref{prop:ordinary-relative-entropy-envelope}. The auxiliary envelope vanishes throughout the stated interval by Lemma~\ref{lem:ordinary-envelope-zero-cost}. By
Lemma~\ref{lem:general-highest-weight-zero-cost}, the compact
highest-weight region contains no point with first coordinate zero and second
coordinate at least $r$ once $r>D_{\mathrm R}$.
On the compact region $\mathcal R_{\mathrm{hw}}$, the function
$(a,b)\mapsto a+(r-b)_+$ is continuous and nonnegative.  It attains a
minimum, which must be strictly positive because it has no zero-cost
point at this rate. The final comparison with the Keyl-output exponent
follows from Eq.~\eqref{eq:highest-weight-exponent-hierarchy}.
\end{proof}

We will also use the strict inequality between relative entropy and
max-relative entropy for distinct faithful states.
\begin{lemma}[Strict max-relative-entropy bound]
\label{lem:strict-relative-max}
If $\rho$ and $\sigma$ are full-rank states, then
$D(\rho\Vert\sigma)\le\Dmax(\rho\Vert\sigma)$,
with equality if and only if $\rho=\sigma$.
\end{lemma}
\begin{proof}
Put $c=\Dmax(\rho\Vert\sigma)$.  The operator inequality
$\rho\le e^c\sigma$ and operator monotonicity of the logarithm give
$\log\rho\le cI+\log\sigma$.  Taking the $\rho$-expectation yields
$D(\rho\Vert\sigma)\le c$.  If equality holds, the positive operator
$cI+\log\sigma-\log\rho$ has zero expectation in the faithful state $\rho$
and must therefore vanish. Thus $\rho=e^c\sigma$, and normalization
gives $c=0$ and $\rho=\sigma$.  The converse is immediate.
\end{proof}

To bound the gap above $D$, put $m_{\mathrm{hw}}:=E_{\mathrm{hw}}(D)$
and use the function $h_{\mathrm{hw}}$ defined before
Lemma~\ref{lem:highest-weight-dmax-endpoint}.
The active support set and its maximal element are $\Theta_D^{\mathrm{hw}} :=\operatorname*{argmax}_{0\le\theta\le1} \{\theta D+h_{\mathrm{hw}}(\theta)\}$ and $\theta_+:=\max\Theta_D^{\mathrm{hw}}$.

\begin{theorem}[Strict inequality above the unrestricted Stein threshold]
\label{thm:general-right-persistence}
Under the assumptions of
Theorem~\ref{thm:general-strict-Keyl-loss}, the highest-weight exponent is
strictly positive at the unrestricted Stein threshold $D$, and its largest
active support parameter is an interior point:
$m_{\mathrm{hw}}>0$ and $0<\theta_+<1$.
On the Stein slice, where $b=D(\rho\Vert\sigma)$, the exponent is the
minimum null-acceptance rate $a$.
\begin{equation}
 m_{\mathrm{hw}}
 =\min\{a:(a,D)\in\mathcal R_{\mathrm{hw}}\}.
 \label{eq:general-exact-Stein-slice}
\end{equation}
Moreover, the right-hand expansion at the Stein point is
$E_{\mathrm{hw}}(D+\varepsilon) =m_{\mathrm{hw}}+\theta_+\varepsilon+o(\varepsilon)$,
and the strict chain
\begin{equation}
 E_{\mathrm{sc}}^*(r)
 \le F_D(r\mid\rho\mid\sigma)
 <E_{\mathrm{hw}}(r),
 \label{eq:general-right-strict-chain}
\end{equation}
holds throughout
$D<r<D+\frac{m_{\mathrm{hw}}}{1-\theta_+}$.
\end{theorem}
\begin{proof}
Theorem~\ref{thm:general-strict-Keyl-loss} at $r=D$ gives
$m_{\mathrm{hw}}>0$.  Lemma~\ref{lem:compact-rate-envelope-derivatives} applies to
$\mathcal R_{\mathrm{hw}}$.  At $\theta=0$ the support value is zero, so
the dual value is zero and this endpoint is inactive because
$m_{\mathrm{hw}}>0$.  By
Lemma~\ref{lem:highest-weight-dmax-endpoint}, the dual value at
$\theta=1$ is
$D+h_{\mathrm{hw}}(1) =D(\rho\Vert\sigma)-\Dmax(\rho\Vert\sigma)<0$,
where strictness follows from Lemma~\ref{lem:strict-relative-max}, because the
divergence-gap assumption implies $\rho\ne\sigma$.  Every active parameter at $D$ thus lies in $(0,1)$, so $0<\theta_+<1$.

The interior-active-parameter conclusion of
Lemma~\ref{lem:compact-rate-envelope-derivatives} now gives the minimum
over the Stein slice
$m_{\mathrm{hw}} =\min\{a:(a,D)\in\mathcal R_{\mathrm{hw}}\}$,
which is~\eqref{eq:general-exact-Stein-slice}.  The same lemma gives
$E_{\mathrm{hw},+}'(D)=\theta_+$.
The right-derivative identity yields the stated
first-order expansion.  Moreover, keeping an active affine support at $D$ gives the global one-sided
lower bound
$E_{\mathrm{hw}}(D+\varepsilon) \ge m_{\mathrm{hw}}+\theta_+\varepsilon$ for $\varepsilon\ge0$.

Lemma~\ref{lem:ordinary-envelope-zero-cost} gives $F_D(D)=0$ and says that
$F_D$ is $1$-Lipschitz.  Hence the auxiliary envelope obeys
$F_D(D+\varepsilon\mid\rho\mid\sigma)\le\varepsilon$.
The upper bound and the active-support lower bound give
$E_{\mathrm{hw}}(D+\varepsilon) -F_D(D+\varepsilon\mid\rho\mid\sigma) \ge m_{\mathrm{hw}}-(1-\theta_+)\varepsilon$,
which is strictly positive whenever
$0<\varepsilon<m_{\mathrm{hw}}/(1-\theta_+)$.  Proposition~\ref{prop:ordinary-relative-entropy-envelope} gives the first
inequality in~\eqref{eq:general-right-strict-chain}, and the preceding
estimate gives the second.
\end{proof}

Monotonicity of the Keyl exponent also gives a direct bound on the
interval where it exceeds $F_D$.

\begin{corollary}[Strict gap above $D$ for Keyl's measurement]
\label{cor:Keyl-measurement-right-neighborhood}
Under the assumptions of Theorem~\ref{thm:general-strict-Keyl-loss}, define $m_{\mathrm{fix}} :=E_{\mathrm{sc}}^{\mathrm{Keyl}}(D\mid\rho\mid\sigma)$.
Then $m_{\mathrm{fix}}>0$, and
every rate $r$ with $D<r<D+m_{\mathrm{fix}}$ satisfies
$E_{\mathrm{sc}}^*(r) \le F_D(r\mid\rho\mid\sigma) <E_{\mathrm{sc}}^{\mathrm{Keyl}}(r\mid\rho\mid\sigma)$.
\end{corollary}

\begin{proof}
The Keyl-measurement zero-cost threshold is $D_{\mathrm R}<D$, so
$m_{\mathrm{fix}}>0$.  At $r=D+\varepsilon$, the ordinary envelope is at most
$\varepsilon$, whereas monotonicity of the Keyl-measurement exponent gives the lower
bound $m_{\mathrm{fix}}$.  This proves the claim when
$0<\varepsilon<m_{\mathrm{fix}}$.
\end{proof}

For faithful states, Proposition~\ref{prop:reverse-relative-entropy-equality}
gives $D_{\mathrm R}<D$ exactly when $[\rho,\sigma]\ne0$.

\section{Qubit Dicke-weight cells and exact support values}\label{app:qubit-one-block-proofs}\label{sec:qubit-operational}
\label{sec:qubit-operational-detailed}
\subsection{Attainable rate pairs}
For qubits, one Young block and one Dicke vector in an eigenbasis of
$\sigma$ suffice to attain each linear minimum that determines the
unrestricted strong-converse exponent. We first define the cell rates
and then choose the occupation and the Young shape.

For each $n$, take a unitary-covariant POVM $M_n$ on
$(\mathbb C^2)^{\otimes n}$ with outcome space $\Omega_n$ as in
Section~\ref{sec:covariant-test-class}, and a measurable cell
$B_n\subseteq\Omega_n$. Write $P_{X,n}(B_n)=\Tr[X^{\otimes n}M_n(B_n)]$
for its probability under $X\in\{\rho,\sigma\}$.
A pair $(a,b)\in\mathbb R^2$ is attainable for the qubit covariant
class if a sequence of such measurements and cells satisfies
$P_{\rho,n}(B_n)=e^{-na+o(n)}$ and $P_{\sigma,n}(B_n)=e^{-nb+o(n)}$.
The indicator rule $\mathbf 1_{B_n}$ gives the acceptance effect
$M_n(B_n)$, so the sequence of these effects belongs to
$\mathfrak T_2^{\mathrm{ucov}}$.
An attainable region is a nonempty set $\cR\subset[0,\infty)^2$ of
such pairs. For any nonempty region $\cR\subseteq\mathbb R^2$ put
\begin{equation}\label{eq:region-bound}
 E_\cR(r):=\inf_{(a,b)\in\cR}\{a+(r-b)_+\}.
\end{equation}

\begin{lemma}[Randomly retaining accepted outcomes of a cell test]\label{lem:thinning}
Every attainable region $\cR$ gives the operational upper bound
$E_{\mathrm{sc},2}^{\mathrm{ucov}}(r)\le E_\cR(r)$.
\end{lemma}
\begin{proof}
For a pair $(a,b)$ in $\cR$, take the POVMs $M_n$ and cells $B_n$
witnessing its attainability. To enforce a target rate $r$, replace
the indicator rule by the randomized rule
$f_n(\omega):=e^{-n(r-b)_+}\mathbf 1_{B_n}(\omega)$.
Equivalently, independently retain an acceptance on $B_n$ with probability
$e^{-n(r-b)_+}$. The induced effect is
$e^{-n(r-b)_+}M_n(B_n)$; under either state the cell probability is
multiplied by the same factor.
The type-II exponent is at least $r$, while the acceptance-probability exponent under $\rho$ is $a+(r-b)_+$.
The infimum over attainable rate pairs proves the claim.
\end{proof}

For block concatenation, independently measure the prescribed covariant
observation on each subblock and retain the tuple of complete outcomes.
Accept $\rho$ exactly if every subblock outcome lies in its chosen cell;
the induced effect is the tensor product of the corresponding cell effects.
The product observation is covariant under the simultaneous unitary action.
Products of cell events on finitely many subblocks with prescribed asymptotic proportions produce convex combinations of exponent pairs.
More precisely, if construction $j$ has rate pair $(a_j,b_j)$ and occupies a proportion $\theta_j$ of the total block length, then the product cell has rate pair $(a,b)=\sum_{j=1}^J\theta_j(a_j,b_j)$, where $\theta_j\ge0$ and $\sum_{j=1}^J\theta_j=1$.  Block concatenation is therefore the operational source of convexification.
For a single-block cell region $\cR^{(1)}$, call
$\Rreg\cR^{(1)}$ its block-concatenation regularization.
The two single-block regions used below are compact. Their convex
hulls are compact by Lemma~\ref{lem:compact-region}, and every point
in either hull is a combination of at most three single-block pairs
by Carath\'eodory's theorem. Concatenating the blocks with positive
proportions realizes these points. Throughout the comparison the
superscript $(1)$ denotes the single-block region, and the undecorated
region denotes its regularization.
As in Section~\ref{sec:highest-weight-consequences}, write
$\cR_{\mathrm{hw}}:=\Rreg\cR_{\mathrm{hw}}^{(1)}$ and
$\cR_\wt:=\Rreg\cR_\wt^{(1)}$ for the concatenated rate regions, with
exponents $E_{\mathrm{hw}}=E_{\cR_{\mathrm{hw}}}$ and
$E_\wt=E_{\cR_\wt}$. The exponent $E_{\mathrm{sc}}^{\mathrm{Keyl}}$
describes arbitrary decision rules applied to a Keyl measurement of the
whole sample.

\subsection{Compactness and support representation}
\begin{lemma}[Compactness and elementary bounds for rate regions]
\label{lem:compact-region}
A continuous map from a compact parameter space into $\mathbb R^2$
has compact image and compact convex hull. If
$\cR_1\subseteq\cR_2$, then $E_{\cR_2}(r)\le E_{\cR_1}(r)$, also after
closed convexification. For every nonempty region
$\cR\subseteq[0,\infty)^2$, $E_\cR$ is nondecreasing and
$1$-Lipschitz on $[0,\infty)$.
\end{lemma}

\begin{proof}
By Carath\'eodory's theorem, the convex hull is the image of a compact
parameter space of convex combinations of at most three image points.
Thus it is compact and already closed. Enlarging a region can only
lower the infimum in \eqref{eq:region-bound}, and convexification
preserves inclusion. Each function $r\mapsto a+(r-b)_+$ is nondecreasing
and $1$-Lipschitz, as is their infimum.
\end{proof}
\begin{lemma}[Linear-support representation of the exponent]\label{lem:support-envelope}
Let $\cR\subset\mathbb R^2$ be nonempty, compact, and convex.  Then
\begin{equation}\label{eq:support-envelope}
 E_\cR(r)=\sup_{0\le\theta\le1}
 \left\{\theta r+\inf_{(a,b)\in\cR}(a-\theta b)\right\}.
\end{equation}
If $\cR=\Rreg\cR^{(1)}$ for a compact single-block region
$\cR^{(1)}$, then the inner infimum in \eqref{eq:support-envelope} may be
taken over $\cR^{(1)}$.
\end{lemma}
\begin{proof}
The support representation is the first conclusion of
Lemma~\ref{lem:compact-rate-envelope-derivatives}.  A linear functional
has the same infimum on a compact set and on its closed convex hull, so
$\inf_{(a,b)\in\Rreg\cR^{(1)}}(a-\theta b) =\inf_{(a,b)\in\cR^{(1)}}(a-\theta b)$.
\end{proof}
\subsection{Qubit single-block support and explicit Dicke seeds}
\label{sec:qubit}
We calculate the exponential rate of a Dicke matrix coefficient, then
convert it into the probability rate of a measurable cell. Optimizing
the occupation and Young shape gives the desired linear minimum.

For a partition $\lambda=(n-j,j)$, write $q=j/n\in[0,1/2]$ and $N=n-2j$.  The corresponding irreducible factor is
$\mathcal Q_\lambda\simeq (\det)^j\otimes\Sym^N(\mathbb C^2)$.
The Young fraction $q$ controls the determinant and multiplicity
contributions. In the symmetric power of degree $N$, the number $k$
counts the occurrences of the second basis vector in the Dicke seed.
The axis $\bm n$ specifies the basis.
\begin{definition}[Qubit weight-seed family]\label{def:weight-seed-family}
In the Schur block $\lambda=(n-j,j)$, put $N=n-2j$. Fix a reference
orthonormal basis and let $|D_{N,k}\rangle$ be its normalized Dicke seed
with $k\in\{0,\ldots,N\}$ occurrences of the second basis vector
\cite[Sec.~2, Eq.~(5)]{UshaDeviSudhaRajagopal2012}. At rotation $U$,
the Haar density is $N+1$ times the projector onto
$\Sym^N(U)|D_{N,k}\rangle$, tensored with the multiplicity identity.
Integration gives the identity on this block. Choose any normalized
covariant POVM on each remaining block.

To specify the cell center, choose $U_0$ taking the first reference
basis ray to the oriented axis $\bm n$. Write $|D_{N,k}^{\bm n}\rangle$
for $\Sym^N(U_0)|D_{N,k}\rangle$ up to phase. The phases of the two
basis vectors have no effect on its projector. A \emph{weight-seed
cell} selects the indicated Young label and a shrinking neighborhood
of $U_0$ of Haar volume $e^{o(n)}$. The rule accepts exactly on this
event. Thus $k$ selects the seed in the reference basis and $\bm n$
selects its orientation at the cell center.
An asymptotic cell sequence is parameterized by
$j/n\to q\in[0,1/2]$ and, when $q<1/2$,
$k/(n-2j)\to u\in[0,1]$, together with the limiting axis $\bm n$.
When $q=1/2$, only $N=n-2j=o(n)$ is required, and the seed parameters
do not affect the limiting rate.  The exactly balanced block at even $n$
has $N=0$ and a one-dimensional symmetric factor, but this special choice
is not required.
The finite-block highest-weight seed has $k=0$ in its oriented basis.
The asymptotic rate pair obtained at $u=0$ is the highest-weight pair.
\end{definition}

The Dicke seed has the following exponential matrix coefficient.  For a positive definite qubit matrix $X$, let
$L_X(u,\bm n):=\lim_{N\to\infty}\frac1N \log\langle D_{N,\lfloor uN\rfloor}^{\bm n}|X^{\otimes N}|D_{N,\lfloor uN\rfloor}^{\bm n}\rangle$.  Write $X$ in the $\bm n$ basis as
\[
 X=\begin{pmatrix}A&B\\ \overline B&D\end{pmatrix}.
\]
Equivalently, calculate the reference-seed coefficient with $X$
expressed in the $\bm n$ basis. In that basis the Dicke vector is the
sum of all bit strings of weight $k$, divided by $\sqrt{\binom Nk}$.
For a fixed first string, a second string of the same weight differs
by changing $\ell$ zeros to ones and $\ell$ ones to zeros. The choices
give $\binom{N-k}{\ell}\binom{k}{\ell}$ partners. Each contributes
$A^{N-k-\ell}D^{k-\ell}B^\ell\overline B^\ell$.
The number of first strings, $\binom Nk$, cancels the normalization
in the matrix element, giving
\begin{equation}\label{eq:dicke-coeff}
 \langle D_{N,k}^{\bm n}|X^{\otimes N}|D_{N,k}^{\bm n}\rangle
 =\sum_{\ell=0}^{\min(k,N-k)}
 \binom{N-k}{\ell}\binom{k}{\ell}
 A^{N-k-\ell}D^{k-\ell}|B|^{2\ell}.
\end{equation}
For $N\ge1$, put $u_N=k/N$ and $s=\ell/N$. The fraction $s$ is measured
among all $N$ positions and is the same for the two directions of change.
The feasible values are multiples of $1/N$ in $[0,\min(u_N,1-u_N)]$.
With binary entropy $h(x)=-x\log x-(1-x)\log(1-x)$ and $0\log0=0$,
the two binomial factors yield $(1-u_N)h(s/(1-u_N))$ and $u_N h(s/u_N)$
in the normalized log summand up to $O(\log(N+1)/N)$. The remaining
terms come from the powers of $A$, $D$, and $|B|^2$.
Lemma~\ref{lem:dicke-varying-parameters} justifies taking the maximum
over the limiting interval as $u_N\to u$, including for a convergent
sequence of axes. This gives
\begin{align}\label{eq:L-variational}
 L_X(u,\bm n)=\max_{0\le s\le\min(u,1-u)}\{&
 (1-u)h(s/(1-u))+u h(s/u)\\
 &+(1-u-s)\log A+(u-s)\log D+s\log|B|^2\}.
\end{align}
\begin{lemma}[Boundary convention and continuity]\label{lem:L-boundary}
When $X$ is positive definite, the variational expression in \eqref{eq:L-variational},
with $0\log0:=0$, $s\log0:=-\infty$ for $s>0$, and the continuous
extension $t h(s/t)=t\log t-s\log s-(t-s)\log(t-s)$ for $0\le s\le t$,
defines a continuous function of $(u,\bm n)$. If $B=0$, only $s=0$
contributes. If $u=0$ or
$u=1$, the feasible interval consists only of $s=0$.  At $q=1/2$, the cell
exponent is independent of $L_X$ because its coefficient in \eqref{eq:Gamma}
is zero.
\end{lemma}
\begin{proof}
Positive definiteness bounds $A$ and $D$ away from zero uniformly in the
axis.  For $B\ne0$, the objective in \eqref{eq:L-variational} is continuous
on its compact feasible set, including $u=0,1$ with the stated conventions,
so its maximum depends continuously on the parameters.  It remains to
check continuity as $B\to0$.
Let $s_B$ be a maximizer.  For every fixed $\delta>0$, the term
$s\log|B|^2$ tends uniformly to $-\infty$ on $s\ge\delta$, while the other
terms remain bounded.  Comparison with $s=0$ therefore gives
$s_B\to0$ uniformly in $u$.
The entropy bound $t h(s/t)\le s(1-\log s)$ for $0\le s\le t\le1$
now yields, for small $|B|$,
\[
 0\le L_X(u,\bm n)-\bigl((1-u)\log A+u\log D\bigr)
 \le 2s_B(1-\log s_B)+Cs_B\longrightarrow0,
\]
where $C$ is uniform in the axis and we dropped the nonpositive term
$s_B\log|B|^2$.  This also covers simultaneous limits in $u$ and the
axis.  If $u\to0$ or $u\to1$ with $B$ bounded away from zero, the feasible
interval shrinks to $\{0\}$ and continuity follows directly.
Finally, at $q=1/2$ the coefficient of $L_X$ in \eqref{eq:Gamma} vanishes.
\end{proof}
For $0<u<1$ and $A,D,|B|>0$, the derivative of the objective in
\eqref{eq:L-variational} on $0<s<\min(u,1-u)$ is
\[
 \log\frac{|B|^2(1-u-s)(u-s)}{ADs^2}.
\]
It decreases strictly from $+\infty$ at the left endpoint to
$-\infty$ at the right endpoint. Thus the unique maximizer is interior
and satisfies
\begin{equation}\label{eq:saddle-equation}
 |B|^2(1-u-s)(u-s)=ADs^2.
\end{equation}
For $u=0$ or $u=1$, the feasible interval consists only of $s=0$.
When $B=0$, only $s=0$ contributes.

\begin{lemma}[Convergence for varying Dicke parameters]
\label{lem:dicke-varying-parameters}
Let $X$ be a fixed positive definite qubit matrix.  If
$N\to\infty$, $k_N/N\to u$, and $\bm n_N\to\bm n$, then
\[
 \frac1N\log\langle D_{N,k_N}^{\bm n_N}|X^{\otimes N}
                   |D_{N,k_N}^{\bm n_N}\rangle
 \longrightarrow L_X(u,\bm n).
\]
\end{lemma}
\begin{proof}
First fix the axis and write its matrix as
$Y=\left(\begin{smallmatrix}A&B\\\overline B&D\end{smallmatrix}\right)$.
Write $L_Y(u)$ for the variational value in this fixed basis.
If $B=0$, the coefficient is exactly $A^{N-k}D^k$, so the assertion
follows immediately.  If $B\ne0$, the binomial bounds
\[
 \frac{e^{m h(\ell/m)}}{m+1}\le\binom m\ell
 \le e^{m h(\ell/m)}
\]
(with the evident convention for $m=0$) apply uniformly to every summand
in \eqref{eq:dicke-coeff}.  Taking the maximum summand and then summing
at most $N+1$ terms changes the normalized logarithm by
$O(\log(N+1)/N)$, uniformly in $k$.  The two entropy terms in
\eqref{eq:L-variational} are sums of functions $x\log x$ on $[0,1]$.
At $u_N=k/N$, rounding a maximizing $s$ down to the grid
$s=\ell/N$ with $0\le\ell\le\min(k,N-k)$ therefore changes the variational
value by $O_Y(\log(N+1)/N)$, uniformly in $k$.
Consequently, the normalized logarithm of the coefficient differs from
$L_Y(k/N)$ by this same uniform order.  Continuity from
Lemma~\ref{lem:L-boundary} proves the claim for every sequence $k_N/N\to u$.

For varying axes, choose unitary representatives $U_N\to U$ of
$\bm n_N\to\bm n$ in a local coordinate neighborhood.  The phases of
the basis vectors do not change the Dicke projector.  Set
$Y_N=U_N^\dagger XU_N$ and $Y=U^\dagger XU$.  Positive definiteness and
$\|Y_N-Y\|\to0$ give numbers $\delta_N\to0$ such that
\[
 e^{-\delta_N}Y\le Y_N\le e^{\delta_N}Y.
\]
Tensoring and taking the same Dicke-vector expectation shows that the
normalized logarithms of the two coefficients differ by at most
$\delta_N$, uniformly in $k_N$.  The fixed-axis argument applies to $Y$,
including a diagonal limiting $Y$, and completes the proof.
\end{proof}

The symmetric-factor contribution now combines with the other Schur-block
contributions. For $\lambda=(n-j,j)$, Lemma~\ref{lem:fixed-dimension-specht-multiplicity}
gives $\log\dim\mathcal P_\lambda=nh(j/n)+O(\log(n+1))$, since the
normalized row lengths are $(1-j/n,j/n)$. Thus the multiplicity
contributes $h(q)$ in the limit, and the logarithmic cell weight has
the form
\begin{equation}\label{eq:Gamma}
 \Gamma_X(q,u,\bm n)
 =h(q)+q\log\det X+(1-2q)L_X(u,\bm n),
\end{equation}
and the probability exponent is
\begin{equation}\label{eq:IX}
 I_X^{\wt}(q,u,\bm n)=-\Gamma_X(q,u,\bm n).
\end{equation}
The orbit normalization is polynomial in $n$ and does not change the exponent.
At $q=1/2$, the degree $N_n=n-2j_n$ is $o(n)$.  For every unit seed
$w_n\in\Sym^{N_n}(\mathbb C^2)$,
\[
 \lambda_{\min}(X)^{N_n}
 \le \langle w_n|X^{\otimes N_n}|w_n\rangle
 \le \lambda_{\max}(X)^{N_n}.
\]
Its normalized logarithm therefore vanishes uniformly in the occupation
and axis.  This proves the seed independence of \eqref{eq:Gamma} at the
balanced limiting shape, even when $N_n$ does not vanish.

The next lemma gives the probability of a shrinking neighborhood of
the specified orientation under both hypotheses.

\begin{lemma}[Local orbit cells]\label{lem:local-cells}
Fix an admissible triple $(q,u,\bm n)$ and the full-rank pair
$(\rho,\sigma)$.  Integer sequences $j_n,k_n$ can be selected with $j_n/n\to q$ and,
when $q<1/2$, $k_n/(n-2j_n)\to u$. There are measurable rotation
neighborhoods $C_n$ for the axis $\bm n$, each of Haar volume $e^{o(n)}$,
such that for $X\in\{\rho,\sigma\}$,
$P_{X,n}(j_n,k_n,C_n)=\exp\{n\Gamma_X(q,u,\bm n)+o(n)\}$.
The same neighborhood sequence works for both states.
\end{lemma}
\begin{proof}
Choose rounded integers $j_n,k_n$ with the stated limits and put
$N_n=n-2j_n$ and $\lambda_n=(n-j_n,j_n)$. Fix the Dicke seed in a
reference basis and take a radius-$n^{-2}$ orbit cell about the
center $U_0$ from Definition~\ref{def:weight-seed-family}.
The coefficient at $U_0$ on the reference seed has input
$U_0^\dagger X U_0$. Equivalently it is the coefficient of $X$
on $|D_{N_n,k_n}^{\bm n}\rangle$ in the symmetric factor. Apply
Lemma~\ref{lem:general-local-orbit-cell} to the normalized POVM from
Definition~\ref{def:weight-seed-family}. Accept on the selected Young
label and orbit cell. The occupation $k_n$ specifies the seed.
Since $\pi_{\lambda_n}(X)=(\det X)^{j_n}\Sym^{N_n}(X)$, for each
$X\in\{\rho,\sigma\}$ the resulting probability satisfies
\[
\begin{aligned}
 \log P_{X,n}(j_n,k_n,C_n)
 &=\log\dim\mathcal P_{\lambda_n}+j_n\log\det X\\
 &\quad+\log\langle D_{N_n,k_n}^{\bm n}|X^{\otimes N_n}
                          |D_{N_n,k_n}^{\bm n}\rangle+o(n).
\end{aligned}
\]
The multiplicity estimate just given yields
$\log\dim\mathcal P_{\lambda_n}=nh(q)+o(n)$. For $q<1/2$,
Lemma~\ref{lem:dicke-varying-parameters} evaluates the last term as
$n(1-2q)L_X(u,\bm n)+o(n)$; for $q=1/2$, the spectral bound following
\eqref{eq:IX} makes it $o(n)$ uniformly in the seed.  Dividing by $n$
gives \eqref{eq:Gamma}. Thus one cell sequence realizes the rate pair
$\bigl(I_\rho^{\wt}(q,u,\bm n),I_\sigma^{\wt}(q,u,\bm n)\bigr)$
by \eqref{eq:IX}.
\end{proof}

For the highest-weight vector,
$L_X(0,\bm n)=\log\langle \bm n,+|X|\bm n,+\rangle$.
Define the single-block rate regions by
\begin{align*}
 \cR_\wt^{(1)}
 &:={\left\{
 \bigl(I_\rho^{\wt}(q,u,\bm n),I_\sigma^{\wt}(q,u,\bm n)\bigr):
 q\in[0,1/2],\ u\in[0,1],\ \bm n\in S^2
 \right\}},\\
 \cR_{\mathrm{hw}}^{(1)}
 &:={\left\{
 \bigl(I_\rho^{\wt}(q,0,\bm n),I_\sigma^{\wt}(q,0,\bm n)\bigr):
 q\in[0,1/2],\ \bm n\in S^2
 \right\}}.
\end{align*}
The balanced-shape convention in Definition~\ref{def:weight-seed-family}
keeps these parameter domains compact. To identify the two rate regions,
we check both inclusions. First, Lemma~\ref{lem:local-cells} realizes
every displayed pair by a sequence of radius-$n^{-2}$ cells with rounded
parameters.

For the reverse inclusion, consider any sequence of cells from
Definition~\ref{def:weight-seed-family}. Write $U_n$ for the center,
$C_n$ for the rotation neighborhood, and $v_n=|D_{N_n,k_n}\rangle$ for the
reference seed in the block $\lambda_n=(n-j_n,j_n)$ with $N_n=n-2j_n$.
The shrinkage and volume conditions give
$\delta_n:=\sup_{U\in C_n}\|U-U_n\|_\infty\to0$ and
$\log\nu_n=o(n)$, where $\nu_n=\int_{C_n}dU$ is the normalized Haar
volume. For each $X\in\{\rho,\sigma\}$ the conjugation estimate used in
the proof of Lemma~\ref{lem:general-local-orbit-cell} gives
\[
 e^{-\kappa_X(\delta_n)}U_n^\dagger XU_n
 \le U^\dagger XU\le
 e^{\kappa_X(\delta_n)}U_n^\dagger XU_n,\qquad
 \kappa_X(t)=\log\!\left(1+\frac{2\|X\|_\infty t}{\lambda_{\min}(X)}\right).
\]
Indeed, the difference has norm at most $2\|X\|_\infty\delta_n$
and each matrix is bounded below by $\lambda_{\min}(X)I$.
Tensoring to $n$ copies and restricting by \eqref{eq:sc-schur-block-action}
bounds each block coefficient in $C_n$ by its value
$c_{X,n}=\langle v_n,\pi_{\lambda_n}(U_n^\dagger X U_n)v_n\rangle$
times $e^{\pm n\kappa_X(\delta_n)}$. Integrating with the one-vector
normalization in Lemma~\ref{lem:general-local-orbit-cell} gives
\begin{align*}
 \log P_{X,n}(j_n,k_n,C_n)
 &=\log\dim\mathcal P_{\lambda_n}+\log\dim\mathcal Q_{\lambda_n}
   +\log\nu_n+\log c_{X,n}+\varepsilon_{X,n},\\
 |\varepsilon_{X,n}|&\le n\kappa_X(\delta_n).
\end{align*}
The representation dimension is $N_n+1$, and its logarithm,
$\log\nu_n$ and $\varepsilon_{X,n}$ are all $o(n)$.

Along a convergent parameter subsequence with $j_n/n\to q$, the
same multiplicity estimate gives $n^{-1}\log\dim\mathcal P_{\lambda_n}\to h(q)$.
The determinant power in $c_{X,n}$ contributes $q\log\det X$. If $q<1/2$,
Lemma~\ref{lem:dicke-varying-parameters} gives the remaining term
$(1-2q)L_X(u,\bm n)$ of \eqref{eq:Gamma}. When $q=1/2$, the uniform
spectral bound following \eqref{eq:IX} makes the log of the symmetric
coefficient $o(n)$ even without a limit for the seed parameters. Thus every subsequential rate limit is the image of
limiting parameters, with the same neighborhoods under both states.
The rate-pair map is continuous, so its images are compact by
Lemma~\ref{lem:compact-region}.  Hence the displayed sets are exactly the
asymptotic rate regions for cells on a single block. For the
highest-weight family, this is the qubit case of
Lemma~\ref{lem:highest-weight-rate-image}.

\subsubsection{Optimizing the occupation and Young shape}
\label{subsec:qubit-completeness}
Fix $0<\theta<1$ and choose an eigenaxis $\bm n_\sigma$ of $\sigma$.
Set $X_\theta:=\sigma^{-\theta/2}\rho\sigma^{-\theta/2}$.

\begin{proposition}[Optimal Dicke occupation in an eigenbasis of $\sigma$]
\label{prop:eigenaxis-occupation-support}
With the preceding choice of $\bm n_\sigma$, the maximum over $0\le u\le1$
of $L_\rho(u,\bm n_\sigma)-\theta L_\sigma(u,\bm n_\sigma)$
equals $\log\lambda_{\max}(X_\theta)$.
For each $N$ and $0\le k\le N$, the axis and occupation
$(\bm n_\sigma,k)$ and $(-\bm n_\sigma,N-k)$ describe the same Dicke ray.
When $N\to\infty$ and $k_N/N\to u$, the corresponding limiting
occupations $u$ and $1-u$ give the same rate pair at each fixed Young
fraction $q$. At $q=1/2$ the rates are independent of the seed.
Thus the axis can be oriented so that a maximizing occupation lies
in $[0,1/2]$.
\end{proposition}

\begin{proof}
Write $\sigma=\operatorname{diag}(s_0,s_1)$ in the selected basis.  Since
$|D_{N,k}^{\bm n_\sigma}\rangle$ belongs to the type subspace with $N-k$
copies of the first eigenvector and $k$ copies of the second,
\[
\langle D_{N,k}^{\bm n_\sigma}|X_\theta^{\otimes N}
 |D_{N,k}^{\bm n_\sigma}\rangle 
 =s_0^{-\theta(N-k)}s_1^{-\theta k}
 \langle D_{N,k}^{\bm n_\sigma}|\rho^{\otimes N}
 |D_{N,k}^{\bm n_\sigma}\rangle .
\]
Thus, for $k_N/N\to u$,
\begin{equation}
 L_{X_\theta}(u,\bm n_\sigma)
 =L_\rho(u,\bm n_\sigma)-\theta L_\sigma(u,\bm n_\sigma).
 \label{eq:X-coefficient-identity}
\end{equation}
Let $v$ be a top eigenvector of $X_\theta$ with eigenvalue $z_+$, and
let $u_*$ be its squared overlap with the second basis vector. Choose $k_N$
to be a mode of the binomial law with parameters $N,u_*$. Then
$k_N/N\to u_*$ and
$|\langle D_{N,k_N}^{\bm n_\sigma}|v^{\otimes N}\rangle|^2\ge 1/(N+1)$.
The contribution of the top eigenvector shows that
\[
 \frac{z_+^N}{N+1}
 \le \langle D_{N,k_N}^{\bm n_\sigma}|X_\theta^{\otimes N}
              |D_{N,k_N}^{\bm n_\sigma}\rangle \le z_+^N .
\]
Lemma~\ref{lem:dicke-varying-parameters} gives
$L_{X_\theta}(u_*,\bm n_\sigma)=\log z_+$.
The upper bound by $z_+^N$ holds for all occupations, so
\eqref{eq:X-coefficient-identity} gives the claimed maximum.

To reverse the axis, exchange the two basis vectors $(e_0,e_1)$.
The Dicke vector of occupation $N-k$ in the basis $(e_1,e_0)$ is the
same normalized symmetrization of $N-k$ copies of $e_0$ and $k$ copies
of $e_1$. Thus
\[
 [D_{N,k}^{\bm n_\sigma}]=[D_{N,N-k}^{-\bm n_\sigma}],
\]
where brackets denote rays, so arbitrary phases of the basis vectors
give the same identity. For $k_N/N\to u$, the exchanged occupation
satisfies $(N-k_N)/N\to 1-u$. The finite-$N$ coefficients under both
hypotheses coincide, so Lemma~\ref{lem:dicke-varying-parameters} gives
$L_X(u,\bm n_\sigma)=L_X(1-u,-\bm n_\sigma)$ for
$X\in\{\rho,\sigma\}$. With $q$ fixed, the entropy and determinant
contributions in \eqref{eq:Gamma} are unchanged, giving the same rate
pair by \eqref{eq:IX}, including at $q=1/2$ where the seed term vanishes.
The reversed axis is again an eigenaxis of $\sigma$. Applying the
exchange when $u_*>1/2$ places a maximizing occupation in $[0,1/2]$.
\end{proof}

\begin{theorem}[Exact single-block weight support for qubits]
\label{thm:exact-weight-support}
For $0\le\theta<1$,
\begin{equation}
 \inf_{(a,b)\in\cR_{\wt}^{(1)}}(a-\theta b)
 =-\theta\widetilde D_{1/(1-\theta)}(\rho\Vert\sigma).
 \label{eq:exact-weight-support}
\end{equation}
For $0<\theta<1$, cells with axis in an eigenbasis of $\sigma$ attain
the minimum.
\end{theorem}

\begin{proof}
Proposition~\ref{prop:eigenaxis-occupation-support} performs the
occupation optimization.  For $0<\theta<1$, put $c=1-\theta$ and
$M(u):=L_\rho(u,\bm n_\sigma)-\theta L_\sigma(u,\bm n_\sigma)$.
Equations~\eqref{eq:IX} and \eqref{eq:Gamma} express the support objective of a weight cell on this axis as
\begin{equation}
 \Phi_\theta(q,u)
 =-M(u)-c h(q)+q\{-\log\det X_\theta+2M(u)\}.
 \label{eq:weight-support-fixed-axis}
\end{equation}
The coefficient of $M(u)$ is $-(1-2q)\le0$, so maximizing $M(u)$
minimizes this objective for every fixed $q$.  Let $z_+\ge z_->0$ be the
eigenvalues of $X_\theta$.  Proposition~\ref{prop:eigenaxis-occupation-support}
supplies an occupation $u_\theta$
with $M(u_\theta)=\log z_+$.  At this occupation, differentiation of
\eqref{eq:weight-support-fixed-axis} gives
$c\log\frac{q}{1-q}+\log\frac{z_+}{z_-}=0$,
because $\det X_\theta=z_+z_-$.  Strict convexity in $q$ then gives the
unique minimizing Young fraction
$q_\theta= \left[1+\exp\left\{\frac{\log(z_+/z_-)}{c}\right\}\right]^{-1} \in(0,1/2]$.
Substitution into \eqref{eq:weight-support-fixed-axis} gives
\begin{equation*}
 \Phi_\theta(q_\theta,u_\theta)
 =-c\log\left(z_+^{1/c}+z_-^{1/c}\right)
 =-\theta\widetilde D_{1/(1-\theta)}(\rho\Vert\sigma).
\end{equation*}
Lemma~\ref{lem:local-cells} realizes this pair in
$\cR_{\wt}^{(1)}$.  This completes the second optimization and proves that
the left-hand side of~\eqref{eq:exact-weight-support} is at most the stated
sandwiched-R\'enyi support value.

For the converse, take any sequence of quantum effects $T_n$ with
acceptance probabilities
$p_n=\Tr\rho^{\otimes n}T_n=e^{-na+o(n)}$ and $q_n=\Tr\sigma^{\otimes n}T_n=e^{-nb+o(n)}$.
Data processing of the sandwiched R\'enyi divergence under the binary
measurement $(T_n,\one-T_n)$ gives, for $\alpha>1$,
\begin{equation*}
 e^{(\alpha-1)n\widetilde D_\alpha(\rho\Vert\sigma)}
 \ge p_n^\alpha q_n^{1-\alpha}
 +(1-p_n)^\alpha(1-q_n)^{1-\alpha}
 \ge p_n^\alpha q_n^{1-\alpha}.
\end{equation*}
With $\theta=(\alpha-1)/\alpha$, taking normalized logarithms yields
$a-\theta b\ge-\theta\widetilde D_\alpha(\rho\Vert\sigma)$.
This proves the reverse inequality. The case $\theta=0$ follows by
continuity and nonnegativity of the first rate coordinate.
\end{proof}

We now choose a cell for each prescribed rate. At an interior
maximizer of the dual objective, its type-II rate is $b=r$. The
endpoint pairs cover the remaining rates with randomization when needed.
\begin{corollary}[Exact exponent from one qubit weight-seed cell]\label{thm:qubit-weight-completeness}
For every full-rank qubit pair and every rate $r\ge0$,
\begin{equation}
\begin{aligned}
 \min_{(a,b)\in\cR_{\wt}^{(1)}}\{a+(r-b)_+\}
 &=E_{\wt}(r)\\
 &=E_{\mathrm{sc},2}^{\mathrm{ucov}}(r)=E_{\mathrm{sc}}^*(r).
\end{aligned}
 \label{eq:qubit-single-cell-exponent}
\end{equation}
The minimum is realized by a cell on an eigenaxis of $\sigma$, with randomized acceptance if needed as in Lemma~\ref{lem:thinning}.
\end{corollary}
\begin{proof}
Because $\cR_{\wt}^{(1)}$ is compact, its support function
$h_{\wt}(\theta):=\min_{(a,b)\in\cR_{\wt}^{(1)}}(a-\theta b)$ is continuous
on $[0,1]$.  Theorem~\ref{thm:exact-weight-support} and the standard endpoint
limit of the sandwiched R\'enyi divergence therefore give
$h_{\wt}(1) =\lim_{\theta\uparrow1}h_{\wt}(\theta) =-\Dmax(\rho\Vert\sigma)$.
Thus the support identity extends from $0\le\theta<1$ to the closed interval
$0\le\theta\le1$. Its value is the function $h$ in the proof of
Proposition~\ref{prop:sc-single-cell-attainment}. For $0<\theta<1$,
choose a support-minimizing cell on an eigenaxis of $\sigma$, as supplied
by Theorem~\ref{thm:exact-weight-support}. Every such pair satisfies
$a-\eta b\ge h(\eta)$ for all $\eta\in[0,1]$, with equality at
$\eta=\theta$. Differentiating these supporting inequalities at
$\theta$ gives
$(a,b)=(h(\theta)-\theta h'(\theta),-h'(\theta))$.
If $\theta$ maximizes $\theta r+h(\theta)$ in the interior, this is
$(E_{\mathrm{sc}}^*(r),r)$.

The endpoint limits of these pairs are $(0,b_0)$ and
$(b_1-D_{\max}(\rho\Vert\sigma),b_1)$, where
$b_0=D(\rho\Vert\sigma)$ and $b_1=-h'_-(1)$, as in the same proposition.
Compactness of the cell parameters on the fixed eigenaxis puts both
limits in $\cR_{\wt}^{(1)}$. The first pair attains the exponent for
$r\le b_0$. Randomized acceptance of the second attains it for $r\ge b_1$.
Together with the interior case, this gives the upper bound
$\min_{\cR_{\wt}^{(1)}}\{a+(r-b)_+\}\le E_{\mathrm{sc}}^*(r)$.
Conversely, every pair in $\cR_{\wt}^{(1)}$ is operationally attainable,
so Lemma~\ref{lem:thinning} and the unrestricted converse give the
reverse inequality.

Finally, Lemma~\ref{lem:support-envelope} applies to the closed convex
hull $\cR_{\wt}$ and the same support identity, giving
$E_{\wt}(r)=E_{\mathrm{sc}}^*(r)$. The $d=2$ specialization of
Theorem~\ref{thm:general-completeness} gives
$E_{\mathrm{sc},2}^{\mathrm{ucov}}(r)=E_{\mathrm{sc}}^*(r)$, completing the
chain.
\end{proof}

In the operational optimization, the type-II constraint applies to
the rate of the final randomized test. Replacing a constraint that
this rate be at least $r$ by the requirement that it exceed $r$
gives the same infimum. Indeed, multiplying the final acceptance
effect by $e^{-n\delta}$ raises both acceptance exponents by $\delta$
while retaining an admissible rule. Letting $\delta\downarrow0$ gives
the equality of infima. In the cell formula, $b$ is the type-II rate
before this randomization.

\section{Majorana formulas for covariant qubit measurements}
\label{app:ucov-majorana-proofs}
We bound the matrix element of a pure seed by its maximum coherent
overlap and use that estimate in the operational Hoeffding bounds.
Appendix~\ref{app:majorana-integral-route} derives the same exponential
rate from an exact integral representation.

\subsection{Coherent overlaps and the matrix-element bound}

Let $d\omega$ denote the invariant probability measure on
$\mathbb P^1$.  For $\|v\|=1$ in $\Sym^N(\mathbb C^2)$, define its
normalized Husimi density by
\begin{equation}
 Q_v(z)=(N+1)|\langle v|z^{\otimes N}\rangle|^2.
 \label{eq:majorana-Q}
\end{equation}
Schur orthogonality gives $\int Q_v(z)\,d\omega(z)=1$.

For $N\ge1$, homogeneous polynomial factorization supplies rays
$u_1,\ldots,u_N$ (including multiplicities) and a nonzero scalar $c$
such that
$\langle v|z^{\otimes N}\rangle=c\prod_{\ell=1}^N\langle u_\ell|z\rangle$.
With $\mu_v=N^{-1}\sum_\ell\delta_{u_\ell}$, normalization of
$Q_v$ becomes
\begin{equation}
 Q_v(z)=\frac{\exp\{NV_{\mu_v}(z)\}}
 {\int\exp\{NV_{\mu_v}(w)\}\,d\omega(w)}.
 \label{eq:majorana-Q-products}
\end{equation}
\begin{lemma}[Continuity of the maximum Majorana potential]
\label{lem:majorana-max-continuity}
If $\mu_i\Rightarrow\mu$ on $\mathbb P^1$ and $h_i\to h$ uniformly,
with $h_i,h$ real continuous functions, then
\[
 \max_z\{V_{\mu_i}(z)+h_i(z)\}
 \longrightarrow \max_z\{V_\mu(z)+h(z)\}.
\]
All these maxima are finite.
\end{lemma}
\begin{proof}
The kernel $k(u,z)=\log|\langle u|z\rangle|^2$ is upper
semicontinuous, bounded above by zero, and integrates to $-1$
against $d\omega(z)$ for every $u$. This gives finite maxima.
The map $u\mapsto k(u,\cdot)$ is continuous in $L^1(\omega)$.
In a projective chart the singularity is an integrable logarithm.
The integral near the moving singularity can be made uniformly small,
and away from it the kernels converge uniformly. Approximating this
continuous $L^1$-valued map on the compact projective line by a finite
partition-of-unity sum gives $V_{\mu_i}\to V_\mu$ in $L^1(\omega)$
under weak convergence.

For the upper bound, take a convergent subsequence of maximizing
rays $z_i$ along which the upper limit is achieved. The upper
semicontinuous Portmanteau bound applied to $\mu_i\otimes\delta_{z_i}$
bounds this limit by $V_\mu(z)+h(z)$ at the limiting ray $z$.
For the lower bound, center a complex chart at a maximizing ray of
$V_\mu+h$. In this chart $V_\mu$ is an average of functions
$\log|a+bw|^2$ minus $\log(1+|w|^2)$. The disk submean inequality for
the logarithms and continuity of $h$ show that, for any $\eta>0$, on
a sufficiently small fixed disk the mean of $V_\mu+h$ is at least
its maximum minus $\eta$. Write $d\lambda$ for normalized planar
area on this disk. There is a constant $c>0$ with
$d\omega\ge c\,d\lambda$ there, so $L^1(\omega)$ convergence
gives convergence of the disk means of $V_{\mu_i}+h_i$.
Each maximum bounds the corresponding mean from above.
Letting $\eta\downarrow0$ gives the lower bound.
\end{proof}

\begin{proof}[Proof of Theorem~\ref{thm:majorana-asymptotic-pure-seeds}]
For the finite-degree estimate, put
$w=\Sym^N(Y^{1/2})v$ and define
$M_a=\max_z|\langle a|z^{\otimes N}\rangle|^2$ for any
$a\in\Sym^N(\mathbb C^2)$. The coherent-state resolution gives
\[
 M_a\le\|a\|^2\le(N+1)M_a.
\]
Under the change of ray
$z=Y^{1/2}y/\|Y^{1/2}y\|$, polynomial factorization and
Eq.~\eqref{eq:majorana-phi} give
\[
 M_w=\max_z\frac{|\langle v|z^{\otimes N}\rangle|^2}
                         {\langle z|Y^{-1}|z\rangle^N}
     =M_v e^{N\Phi_Y(\mu_v)}.
\]
Since $1/(N+1)\le M_v\le1$ and
$\|w\|^2=\langle v|\Sym^N(Y)|v\rangle$, it follows that
\[
 \frac{e^{N\Phi_Y(\mu_v)}}{N+1}
 \le\langle v|\Sym^N(Y)|v\rangle
 \le(N+1)e^{N\Phi_Y(\mu_v)}.
\]
Taking logarithms proves
Eq.~\eqref{eq:majorana-uniform-finite-bound}, uniformly in both the
seed and the positive definite operator.

Since $Y_N\to Y>0$, the perturbations $-\log\langle z|Y_N^{-1}|z\rangle$
converge uniformly. Apply Lemma~\ref{lem:majorana-max-continuity}
to these perturbations and to zero to get $\Phi_{Y_N}(\mu_{v_N})\to\Phi_Y(\mu)$.
The finite-degree bound then gives~\eqref{eq:majorana-seed-laplace}.
The same maximum-continuity argument gives joint continuity of
$\Phi_Y(\mu)$ in $\mu$ and $Y>0$.

Finally, equal-weight empirical measures approximate any
$\mu\in\mathcal P(\mathbb P^1)$; the corresponding products define
nonzero binary homogeneous polynomials and thus normalized pure
symmetric tensors.
\end{proof}

\subsection{Choosing one pure seed per block}

\begin{lemma}[Constant-loss pure-orbit reduction]
\label{lem:majorana-pure-orbit-reduction}
Let $n$ be fixed.  Every binary test obtained by post-processing a
$U(2)$-covariant observation on $(\mathbb C^2)^{\otimes n}$ can be
replaced by a test that first records the Young label and then uses
one normalized rank-one covariant orbit observation in each Schur
block, with a measurable randomized decision rule, such that
\begin{equation}
 \alpha_n^{\mathrm{new}}\le3\alpha_n^{\mathrm{old}},
 \qquad
 \beta_n^{\mathrm{new}}\le3\beta_n^{\mathrm{old}}.
 \label{eq:majorana-pruning-errors}
\end{equation}
The chosen pure seed may depend on the Young label and on $n$.
\end{lemma}
\begin{proof}
The refinement in the proof of Proposition~\ref{covhoe:prop:pure-orbit}
preserves both outcome laws. In block $j$, the branch index is
$\xi=(y,\ell)$, where $y$ is the original outcome coordinate and
$\ell$ indexes the spectral resolution of the seed density $S_j(y)$.
Its normalized, state-independent law is
\[
 \eta_j(\dd y,\ell)=\nu_j(\dd y)\frac{s_\ell(y)}{d_N},
\]
where $\nu_j$ is the normalized block trace measure, $s_\ell(y)$ is
the eigenvalue of that branch, and $d_N=N+1$. On branch $\xi$ the normalized
pure-orbit measurement has seed $v_\ell(y)$ and inherits the rule $f(gy)$
at group element $g$, where $f$ is the original acceptance rule.

Let $A_j(\xi)$ be the block probability under $\rho^{\otimes n}$ times
the conditional type-I error of that branch. Define $B_j(\xi)$ by
the block probability under $\sigma^{\otimes n}$ and the conditional
type-II error of the same branch. The original block contributions
are $a_j=\int A_j\,\dd\eta_j$ and $b_j=\int B_j\,\dd\eta_j$.
Markov's inequality gives $\eta_j$-probability at least $2/3$ to
each of $A_j(\xi)\le3a_j$ and $B_j(\xi)\le3b_j$. If either average
is zero, its contribution vanishes almost surely. The intersection
has positive probability, so we can select one branch from it and
keep its rule. Summing the resulting bounds over all Schur blocks
proves \eqref{eq:majorana-pruning-errors}.
\end{proof}

For a pure orbit seed $v_{n,j}\in\Sym^N(\mathbb C^2)$ the
Haar-coordinate output density under $X^{\otimes n}$ is
\begin{equation}
 p^v_{X,n}(j,g)
 =m_{n,j}(N+1)(\det X)^j
 \langle v_{n,j}|\Sym^N(g^\dagger Xg)|v_{n,j}\rangle,
 \label{eq:majorana-Schur-density}
\end{equation}
where $m_{n,j}$ is the multiplicity of the Young block. For $N=0$,
assign any fixed probability measure to $\mu_{v_{n,j}}$; its
coefficient $1-2j/n$ in $I_X$ is zero. The logarithm has the uniform
asymptotic form
\begin{equation}
 \sup_{j,g,v_{n,j},\,X\in\{\rho,\sigma\}}
 \left|\frac1n\log p^v_{X,n}(j,g)
 +I_X(j/n,g;\mu_{v_{n,j}})\right|
 =O\!\left(\frac{\log(n+1)}n\right).
 \label{eq:majorana-uniform-density-rate}
\end{equation}
The multiplicity accompanying \eqref{eq:covhoe-product-Schur} is
\[
 m_{n,j}=\binom nj-\binom n{j-1}
        =\binom nj\frac{n-2j+1}{n-j+1}.
\]
The ratio is between $1/(n+1)$ and $1$ for every
$0\le j\le\lfloor n/2\rfloor$, including $j=0$.
Using the binomial bounds from the proof of
Lemma~\ref{lem:dicke-varying-parameters} gives
\[
 nh(j/n)-2\log(n+1)\le\log m_{n,j}\le nh(j/n).
\]
For $N\ge1$, Eq.~\eqref{eq:majorana-uniform-finite-bound} bounds
$\left|\log\langle v|\Sym^N(g^\dagger Xg)|v\rangle
-N\Phi_{g^\dagger Xg}(\mu_v)\right|$ by $\log(N+1)$.
The dimension factor in Eq.~\eqref{eq:majorana-Schur-density}
contributes another $\log(N+1)$. For $N=0$ both contributions
vanish. This proves the uniform estimate for every Schur block.
The functions $I_X$ are continuous in $q,g,\mu$.

\subsection{Operational Hoeffding variational bounds}
\label{app:majorana-operational-bounds}

\begin{proof}[Proof of Theorem~\ref{thm:ucov-majorana-hoeffding-sandwich}]
First note that, for every $\mu$,
\begin{equation}
 \min_g I_X(q,g;\mu)
 =-h(q)-q\log\det X-(1-2q)\log p_X
 =D_{\mathrm{bin}}(q\Vert s_X),
 \label{eq:majorana-min-orbit-rate}
\end{equation}
where $p_X\ge s_X$ are the eigenvalues of $X$ and $D_{\mathrm{bin}}$
is Bernoulli relative entropy.
Indeed, $\langle z|(g^\dagger Xg)^{-1}|z\rangle\ge p_X^{-1}$ gives
$\Phi_{g^\dagger Xg}(\mu)\le\log p_X$ for every $g$.
Choose a maximizer $z_0$ of $V_\mu$ and a unitary $g$ for which $z_0$
is a top eigenvector of $g^\dagger Xg$. Evaluating the defining maximum
at $z_0$ gives the reverse inequality and hence equality.

To bound $\mathcal B(r)$ for $r>0$, fix $q=s_\sigma$ in
\eqref{eq:majorana-B-closed}. For every $\mu$,
\eqref{eq:majorana-min-orbit-rate} with $X=\sigma$ gives a minimum of
$D_{\mathrm{bin}}(s_\sigma\Vert s_\sigma)=0$. A rotation attaining this
minimum is feasible and may depend on $\mu$. By joint continuity of
$I_\rho$ and compactness of $U(2)\times\mathcal P(\mathbb P^1)$
with the weak topology on $\mathcal P(\mathbb P^1)$, the bound
\[
 C:=\max_{g\in U(2),\,\mu\in\mathcal P(\mathbb P^1)}
 I_\rho(s_\sigma,g;\mu)<\infty
\]
holds uniformly over the rotations and seed measures. Thus the
constrained infimum at $q=s_\sigma$ is at most $C$ for every $\mu$,
and so is its supremum over $\mu$. Taking the infimum over $q$ gives
$\mathcal B(r)\le C$. Since \eqref{eq:majorana-min-orbit-rate} with
$X=\rho$ gives $I_\rho(q,g;\mu)\ge D_{\mathrm{bin}}(q\Vert s_\rho)\ge0$,
we have $0\le\mathcal B(r)\le C$. Monotonicity follows by inclusion of
the constrained orbit sets.

For the direct bound, fix $\eta>0$ and
$h_0<\mathcal B(r+\eta)$.  For every $q\in[0,1/2]$, the definition
of $\mathcal B$ permits a measure $\mu_q$ for which no $g$ has
both $I_\sigma(q,g;\mu_q)\le r+\eta$ and
$I_\rho(q,g;\mu_q)\le h_0$. By continuity and compactness of the
rotation group,
\[
 \min_g\max\{I_\sigma(q,g;\mu_q)-(r+\eta),\
                I_\rho(q,g;\mu_q)-h_0\}>0 .
\]
The same measure keeps this minimum positive, with at least half
its value at $q$, throughout a neighborhood of $q$. A finite cover
of $[0,1/2]$ supplies finitely many measures and a common positive
margin. For each $n$ and $j$, choose a measure from this cover for
$q=j/n$. Approximate it by an empirical Majorana measure of
$N=n-2j$ points. By joint continuity on compact parameter sets, this
approximation preserves a positive margin for all $N\ge N_0$ for
some $N_0$ independent of the chosen measure. For $N<N_0$, the
uniformly bounded seed-dependent term $\Phi_{g^\dagger Xg}$ is
multiplied by $N/n$ in~\eqref{eq:majorana-block-rate}. An arbitrary
seed then changes the minimum by at most $O(N_0/n)$. When $N=0$,
any measure can be used in the rate formula because its coefficient
is zero.

For the resulting pure-seed orbit measurement, accept $\rho$
when $I_\sigma(j/n,g;\mu_{v_{n,j}})>r+\eta$.
By Eq.~\eqref{eq:majorana-uniform-density-rate}, the
$\sigma$-probability of acceptance is at most
$e^{-n(r+\eta-o(1))}$ times a polynomial in $n$.
On the complementary outcomes the common margin gives
$I_\rho>h_0$ with a fixed margin, so the $\rho$-probability
of rejection is at most $e^{-n(h_0-o(1))}$ times a polynomial.
Letting $h_0\uparrow\mathcal B(r+\eta)$ and then
$\eta\downarrow0$ proves the first inequality in
Eq.~\eqref{eq:ucov-majorana-hoeffding-sandwich}.

For the converse, Lemma~\ref{lem:majorana-pure-orbit-reduction}
reduces any admissible sequence to one pure seed per block
without decreasing either error exponent.  Fix
$h>\mathcal B_{<}(r)$.  By
Eq.~\eqref{eq:majorana-B-open}, there is a fixed
$q\in[0,1/2]$ such that, for every $\mu$, some $g$ satisfies
$I_\sigma(q,g;\mu)<r$ and $I_\rho(q,g;\mu)<h$.
Choose labels $j_n$ with $j_n/n\to q$ and $N_n=n-2j_n\ge1$, also
possible when $q=1/2$. Along a subsequence the empirical measures of
the corresponding seeds converge to some $\mu$. Take a rotation $g_0$
that gives both strict inequalities for this $\mu$. By continuity,
there are $\delta>0$ and an open neighborhood $U$ of $g_0$ with positive
Haar measure such that, on its closure,
$I_\sigma(q,g;\mu)\le r-3\delta$ and $I_\rho(q,g;\mu)\le h-3\delta$.
The functions $I_X$ are uniformly continuous on the compact product
$[0,1/2]\times U(2)\times\mathcal P(\mathbb P^1)$, with the weak
topology for measures. Thus along the subsequence, for both states,
$I_X(j_n/n,g;\mu_{v_{n,j_n}})$ differs from $I_X(q,g;\mu)$
by at most $\delta$ uniformly in $g$ for all sufficiently large $n$.
On $U$ the actual rate functions are therefore bounded by
$r-2\delta$ and $h-2\delta$, respectively. Applying
\eqref{eq:majorana-uniform-density-rate} with error at most $\delta$
gives density lower bounds $e^{-n(r-\delta)}$ under $\sigma$ and
$e^{-n(h-\delta)}$ under $\rho$ throughout this same $U$. The rate constraint gives
$\beta_n\le e^{-n(r-\delta/2)}$ eventually. Hence, if $f_n(j,g)$
is the acceptance rule, $\int_U f_n(j_n,g)\,dg\le e^{-n\delta/2}$,
so
\[
 \alpha_n\ge e^{-n(h-\delta)}
 \left(\int_U dg-e^{-n\delta/2}\right).
\]
The type-I liminf exponent is at most $h$. Letting
$h\downarrow\mathcal B_{<}(r)$ proves the converse.

Finally, for $0<\eta<r$, the set where $I_\sigma<r$ contains
the set where $I_\sigma\le r-\eta$.  Thus
$\mathcal B_{<}(r)\le\mathcal B(r-\eta)$, yielding the last
inequality in Eq.~\eqref{eq:ucov-majorana-hoeffding-sandwich}.
A finite, monotone function has at most countably many
discontinuities, and its two one-sided limits agree at every
continuity point. The sandwich therefore proves
$H^{\mathrm{ucov}}(r\mid\rho\mid\sigma)=\mathcal B(r)$ at every
continuity point of $\mathcal B$, using only the variational argument.

To extend this identity to every $r>0$, use continuity of
$H^{\mathrm{ucov}}$ on $(0,\infty)$, as deduced from the independently
proved scalar formula \eqref{eq:merge-Hdual} in
Section~\ref{subsubsec:merge-qubit-hoeffding-exact}.
For any $r>0$, choose continuity points $r_k^-<r<r_k^+$ of
$\mathcal B$ with $r_k^-\uparrow r$ and $r_k^+\downarrow r$.
Monotonicity gives
\[
 H^{\mathrm{ucov}}(r_k^+\mid\rho\mid\sigma)
 =\mathcal B(r_k^+)\le\mathcal B(r)
 \le\mathcal B(r_k^-)
 =H^{\mathrm{ucov}}(r_k^-\mid\rho\mid\sigma).
\]
Continuity of $H^{\mathrm{ucov}}$ proves
$\mathcal B(r)=H^{\mathrm{ucov}}(r\mid\rho\mid\sigma)$ for every
$r>0$ and, consequently, continuity of $\mathcal B$.
The sandwich then gives the same equality for $\mathcal B_{<}(r)$.
Combining with Eq.~\eqref{eq:merge-qubit-H} proves
Eq.~\eqref{eq:ucov-majorana-continuity-equality}.
\end{proof}

\subsection{Integral representation and Laplace limit}
\label{app:majorana-integral-route}

The angular integral gives an exact expression for the matrix element.
Its Laplace limit recovers the exponential rate in
Theorem~\ref{thm:majorana-asymptotic-pure-seeds}. The uniform estimate
compares each logarithmic integral with the maximum for its current
seed measure, even as the seed and matrix vary with the degree.

\begin{lemma}[Integral representation of the matrix element]
\label{lem:majorana-inverse-husimi}
For every $N\ge0$, positive definite qubit operator $Y$, and unit
$v\in\Sym^N(\mathbb C^2)$,
\begin{equation}
 \langle v|\Sym^N(Y)|v\rangle
 =\frac1{\det Y}\int_{\mathbb P^1}
 Q_v(z)\langle z|Y^{-1}|z\rangle^{-N-2}\,d\omega(z).
 \label{eq:majorana-inverse-husimi}
\end{equation}
\end{lemma}
\begin{proof}
The coherent-state resolution is
$(N+1)\int|y^{\otimes N}\rangle\langle y^{\otimes N}|\,d\omega(y)
=I_{\Sym^N}$. Conjugate it by $\Sym^N(Y^{1/2})$ and change variables
to the ray of $z=Y^{1/2}y/\|Y^{1/2}y\|$. To compute the angular
measure, take a standard isotropic complex Gaussian $\xi$ in
$\mathbb C^2$. Its ray has law $\omega$ and $Y^{1/2}\xi$ has density
$(\pi^2\det Y)^{-1}\exp(-\zeta^\dagger Y^{-1}\zeta)$ at $\zeta$.
Writing $\zeta=r z$ with $\|z\|=1$, the radial integral is
\[
 \int_0^\infty r^3e^{-r^2\langle z|Y^{-1}|z\rangle}\,dr
 =\frac{1}{2\langle z|Y^{-1}|z\rangle^2}.
\]
Dividing by the radial integral for $Y=I$ gives the ray density
$(\det Y)^{-1}\langle z|Y^{-1}|z\rangle^{-2}$ relative to $\omega$.
The squared norm of the transformed
tensor contributes $\|Y^{1/2}y\|^{2N}=\langle z|Y^{-1}|z\rangle^{-N}$.
Taking the $v$ matrix element gives~\eqref{eq:majorana-inverse-husimi},
including $N=0$.
\end{proof}

Combining Eqs.~\eqref{eq:majorana-inverse-husimi} and
\eqref{eq:majorana-Q-products} yields the exact positive ratio
\begin{equation}
 \langle v|\Sym^N(Y)|v\rangle
 =\frac1{\det Y}\,
 \frac{\displaystyle\int e^{NV_{\mu_v}(z)}
  \langle z|Y^{-1}|z\rangle^{-N-2}\,d\omega(z)}
 {\displaystyle\int e^{NV_{\mu_v}(z)}\,d\omega(z)}.
 \label{eq:majorana-exact-ratio}
\end{equation}

\begin{lemma}[Laplace principle for Majorana potentials]
\label{lem:majorana-uniform-laplace}
Suppose $\mu_N\Rightarrow\mu$ weakly on $\mathbb P^1$ and
$h_N\to h$ uniformly, where the $h_N$ and $h$ are continuous real
functions.  Then
\begin{equation}
 \lim_{N\to\infty}\frac1N\log\int_{\mathbb P^1}
 \exp\{N(V_{\mu_N}(z)+h_N(z))\}\,d\omega(z)
 =\max_{z\in\mathbb P^1}\{V_\mu(z)+h(z)\}.
 \label{eq:majorana-laplace-principle}
\end{equation}
The maximum is finite.  More precisely, for every compact family
$\mathcal H\subset C(\mathbb P^1;\mathbb R)$ in the uniform norm,
\[
 \sup_{\substack{\nu\in\mathcal P(\mathbb P^1)\\h\in\mathcal H}}
 \left|\frac1N\log\int e^{N(V_\nu+h)}\,d\omega
       -\max_{z\in\mathbb P^1}\{V_\nu(z)+h(z)\}\right|
 \longrightarrow0.
\]
In particular, this uniform estimate includes all empirical Majorana
measures, without a restriction on their atoms.
\end{lemma}
\begin{proof}
The integral is at most the exponential of $N$ times the current
maximum, since $\omega$ is a probability measure. The maxima converge
by Lemma~\ref{lem:majorana-max-continuity}.
For the lower bound, take the fixed disk from the proof of that lemma
with mean at least the limiting maximum minus $\eta$.
The means of $V_{\mu_N}+h_N$ on the disk converge to that of $V_\mu+h$.
Since $d\omega\ge c\,d\lambda$ on the disk, Jensen's inequality bounds
the integral below by
$c\exp\{N\int(V_{\mu_N}+h_N)\,d\lambda\}$. Taking the degree limit
and then $\eta\downarrow0$ proves \eqref{eq:majorana-laplace-principle}.

If the uniform estimate failed, compactness of
$\mathcal P(\mathbb P^1)$ and $\mathcal H$ would give a subsequence with
the measures converging weakly and the perturbations converging
uniformly. Along it the maxima and the logarithmic integrals have
the same limit by the two preceding results, a contradiction.
\end{proof}

In \eqref{eq:majorana-exact-ratio}, for $Y_N\to Y>0$ the numerator
uses $h_N=-(1+2/N)\log\langle z|Y_N^{-1}|z\rangle$, which converges
uniformly to $-\log\langle z|Y^{-1}|z\rangle$. The denominator
uses $h_N=0$, and $(\log\det Y_N)/N\to0$. Subtracting the two
limits gives $\Phi_Y(\mu)$ when $\mu_{v_N}\Rightarrow\mu$.

For a compact family $\mathcal Y$ of positive definite qubit matrices,
put $h_Y(z)=-\log\langle z|Y^{-1}|z\rangle$. The functions $c h_Y$ with
$1\le c\le3$ and $Y\in\mathcal Y$, together with zero, form a compact
family in the uniform norm. This family contains the perturbations
$(1+2/N)h_Y$ and $0$ in the numerator and denominator for every $N\ge1$.
Replacing $(1+2/N)h_Y$ by $h_Y$ changes the numerator maximum by at
most $2\|h_Y\|_\infty/N$, and $(\log\det Y)/N$ tends to zero uniformly
on $\mathcal Y$. Applying Lemma~\ref{lem:majorana-uniform-laplace} to both
integrals in \eqref{eq:majorana-exact-ratio} therefore gives
\[
 \sup_{\substack{Y\in\mathcal Y,\ \|v\|=1\\ v\in\Sym^N(\mathbb C^2)}}
 \left|\frac1N\log\langle v|\Sym^N(Y)|v\rangle-\Phi_Y(\mu_v)\right|
 \longrightarrow 0.
\]
The operational proof uses \eqref{eq:majorana-uniform-finite-bound}
for its explicit error at finite degree.

\section{Quantitative qubit example and numerical reproducibility}\label{app:qubit-quantitative-example}
\subsection{Explicit cell rates}
\label{sec:comparison}
We give two explicit cells for the pair in
\eqref{eq:pair-main-summary}. Its Pauli representation is
\begin{equation}\label{eq:pair}
 \rho=\frac12(\one+\tfrac45\sigma_z),\qquad
 \sigma=\frac12(\one+\tfrac45\sigma_x).
\end{equation}
For this pair, the highest-weight and unrestricted zero-exponent
boundaries are, respectively,
$D_{\mathrm R}(\rho\Vert\sigma)=\frac45\log\frac95 =0.4702293319\ldots$ and $D(\rho\Vert\sigma)=\frac25\log9 =0.8788898309\ldots$.
For the highest-weight cell at $q=1/10$ and $\bm n=\bm e_z$,
the two symmetric-factor exponents are
$L_\rho(0,\bm e_z)=\log(9/10)$ and $L_\sigma(0,\bm e_z)=\log(1/2)$.
Both determinants are $9/100$. Equation~\eqref{eq:Gamma} therefore
gives null cost
\[
 -h(1/10)-\tfrac1{10}\log(9/100)-\tfrac45\log(9/10)=0.
\]
Replacing the last logarithm by $\log(1/2)$ gives alternative cost
$\tfrac45\log(9/5)=D_{\mathrm R}(\rho\Vert\sigma)$.

Now fix the same $q$ and take a Dicke cell with $u=1/2$ on
$\bm n=-\bm e_x$. In this basis $\sigma$ is diagonal with entries
$1/10,9/10$, so $L_\sigma(1/2,-\bm e_x)=\tfrac12\log(9/100)$.
For $\rho$, the two diagonal entries are $1/2$ and the off-diagonal
modulus is $2/5$. With $u=1/2$ the objective in
\eqref{eq:L-variational} is $\log(1/2)+h(2s)+2s\log(4/5)$.
The saddle equation gives $s=2/9$, and the value is $\log(9/10)$.
Substituting these values in \eqref{eq:Gamma} gives the same zero
null cost, while the alternative cost is
$-h(1/10)-\tfrac12\log(9/100)=\tfrac25\log9$. Thus
\begin{equation}
 I_\rho^{\wt}=0,
 \qquad
 I_\sigma^{\wt}=\frac25\log9=D(\rho\Vert\sigma).
 \label{eq:balanced-pair}
\end{equation}
\subsubsection{Local behavior above the Stein point}
\label{subsec:local-Stein}
We exhibit Dicke cells on a fixed axis that attain the quadratic term
of the strong-converse exponent above $D$. For the pair~\eqref{eq:pair},
write
$D:=D(\rho\Vert\sigma)=\frac25\log9$.
The Pauli representation gives
\[
 \log\rho-\log\sigma=\frac{\log9}{2}(\sigma_z-\sigma_x),
 \qquad (\sigma_z-\sigma_x)^2=2\one.
\]
Consequently, its relative entropy variance is
\[
 V(\rho\Vert\sigma)
 :=\Tr\rho(\log\rho-\log\sigma-D\one)^2
 =\frac12(\log9)^2-D^2
 =\frac{17}{50}(\log9)^2>0.
\]
This variance determines the quadratic coefficient in
Proposition~\ref{prop:quadratic-onset}.

\begin{proposition}[Exact quadratic onset realized by fixed-axis cells]\label{prop:quadratic-onset}
As $\varepsilon\downarrow0$,
$E_{\wt}(D+\varepsilon) = \frac{25}{17(\log9)^2}\varepsilon^2+O(\varepsilon^3)$.
Moreover, the matching quadratic upper bound is realized by cells with axis
$\bm n=-\bm e_x$ and parameters
\[
 q_\varepsilon
 =\frac1{10}-\frac{9}{34\log9}\varepsilon+O(\varepsilon^2),
 \qquad
 u_\varepsilon
 =\frac12-\frac{125}{136\log9}\varepsilon+O(\varepsilon^2),
\]
so that
$I_\sigma^{\wt}(q_\varepsilon,u_\varepsilon,-\bm e_x)=D+\varepsilon$.
\end{proposition}
\begin{proof}
For faithful states, the function
\[
 \theta\longmapsto(1-\theta)\log\Tr
 \bigl(\sigma^{-\theta/2}\rho\sigma^{-\theta/2}\bigr)^{1/(1-\theta)}
 =\theta\widetilde D_{1/(1-\theta)}(\rho\Vert\sigma)
\]
is analytic near zero because the matrix inside the power remains
positive definite. Its logarithm and the ensuing exponential are analytic.
The order-one derivative formula of
\cite[Theorem~5]{LinTomamichel2015} therefore gives
$\theta\widetilde D_{1/(1-\theta)}=D\theta+\frac V2\theta^2+O(\theta^3)$,
where $V=\frac{17}{50}(\log9)^2>0$.
Evaluating \eqref{eq:sc-unrestricted-alpha-form} at
$\theta=\varepsilon/V$ yields
\[
 E_{\mathrm{sc}}^*(D+\varepsilon)
 \ge\frac{\varepsilon^2}{2V}-C\varepsilon^3
\]
for some constant $C$ and all sufficiently small $\varepsilon>0$.
By Corollary~\ref{thm:qubit-weight-completeness}, the same lower bound holds
for $E_{\wt}$.

Fixed-axis cells attain the same coefficient.  Fix the axis $\bm n=-\bm e_x$.  In this basis,
$\sigma$ is diagonal and
$L_\sigma(u,-\bm e_x) =(1-u)\log\frac1{10}+u\log\frac9{10}$.
For $\rho$, the same basis gives diagonal entries $1/2$ and off-diagonal modulus $2/5$.
At $u=1/2$, the saddle equation has the interior solution $s=2/9$.
Its derivative with respect to $s$ is nonzero there, so the implicit-function theorem applied to \eqref{eq:saddle-equation} makes the saddle, and hence $L_\rho(u,-\bm e_x)$, smooth near $u=1/2$.
Let $F(u,s)$ denote the objective in \eqref{eq:L-variational} for
this matrix of $\rho$. For the smooth maximizing saddle $s(u)$,
$F_s(u,s(u))=0$ and $F_{ss}<0$. Hence
\[
 \partial_u L_\rho=F_u,\qquad
 \partial_u^2 L_\rho=F_{uu}-\frac{F_{us}^2}{F_{ss}}.
\]
Symmetry under $u\mapsto1-u$ gives $F_u=F_{us}=0$ at $u=1/2$.
Differentiating the two entropy terms, using
$t h(s/t)=t\log t-s\log s-(t-s)\log(t-s)$, gives
$F_{uu}=2(2-1/(1/2-2/9))=-16/5$ at the saddle.
The value $L_\rho=\log(9/10)$ was obtained above.

Set $(q_0,u_0)=(1/10,1/2)$.  The explicit realization in
\eqref{eq:balanced-pair} gives
$I_\rho^{\wt}(q_0,u_0,-\bm e_x)=0$ and $I_\sigma^{\wt}(q_0,u_0,-\bm e_x)=D$.
Suppress the fixed axis $-\bm e_x$ in $I_X^{\wt}$ and $L_X$.
By \eqref{eq:Gamma} and \eqref{eq:IX},
\[
 I_X^{\wt}(q,u)=-h(q)-q\log\det X-(1-2q)L_X(u).
\]
Its first derivatives are
\[
 \partial_q I_X^{\wt}=-h'(q)-\log\det X+2L_X(u),
 \qquad
 \partial_u I_X^{\wt}=-(1-2q)L_X'(u).
\]
The $qq$, $qu$, and $uu$ derivatives are
$1/[q(1-q)]$, $2L_X'(u)$, and $-(1-2q)L_X''(u)$, respectively.

At the reference point, $h'(q_0)=\log9$ and $1-2q_0=4/5$.
For $\rho$, the $q$ derivative is
$-\log9-\log(9/100)+2\log(9/10)=0$.
Since $L_\rho'(u_0)=0$, both the $u$ derivative and the mixed
derivative vanish. The two diagonal second derivatives are $100/9$
and $(4/5)(16/5)=64/25$. For $\sigma$, the identities
$2L_\sigma(u_0)=\log\det\sigma$ and $L_\sigma'(u_0)=\log9$
give the two first derivatives $-\log9$ and $-(4/5)\log9$.

Thus, with $x=q-q_0$ and $y=u-u_0$, the expansions are
\begin{align*}
 I_\rho^{\wt}
 &=\frac12\left(\frac{100}{9}x^2+\frac{64}{25}y^2\right)
    +O((|x|+|y|)^3),\\
 I_\sigma^{\wt}-D
 &=-(\log9)x-\tfrac45(\log9)y+O((|x|+|y|)^2).
\end{align*}
Let $H$ be the null-rate Hessian and $g$ the alternative-rate gradient
at the reference point.
\[
 H=\begin{pmatrix}100/9&0\\0&64/25\end{pmatrix},
 \qquad
 g=\begin{pmatrix}-\log9\\-(4/5)\log9\end{pmatrix}.
\]
To increase the alternative rate by $t$ to first order, a direction
$v$ must satisfy $g^{\mathsf T}v=1$. The quadratic null cost in this
direction is $\tfrac12 t^2 v^{\mathsf T}Hv$. Cauchy--Schwarz in the
$H$ inner product gives
$1=(g^{\mathsf T}v)^2\le(g^{\mathsf T}H^{-1}g)(v^{\mathsf T}Hv)$,
with equality for $v$ proportional to $H^{-1}g$.
Here $g^{\mathsf T}H^{-1}g=\tfrac{17}{50}(\log9)^2$.
The minimizing direction is
\[
 v:=\frac{H^{-1}g}{g^{\mathsf T}H^{-1}g}
 =\begin{pmatrix}-9/(34\log9)\\-125/(136\log9)\end{pmatrix}.
\]
It satisfies $g^{\mathsf T}v=1$.
Along $(q,u)=(q_0,u_0)+tv$, the derivative of $I_\sigma^{\wt}$ at
$t=0$ is one. We can therefore invert $I_\sigma^{\wt}-D=\varepsilon$
to get $t=\varepsilon+O(\varepsilon^2)$. This gives the stated
$(q_\varepsilon,u_\varepsilon)$ and the exact rate $D+\varepsilon$.
The parameters remain in the interior for all sufficiently small
$\varepsilon>0$. The resulting null-acceptance rate is
\[
 I_\rho^{\wt}
 =\tfrac12\varepsilon^2v^{\mathsf T}Hv+O(\varepsilon^3)
 =\frac{25}{17(\log9)^2}\varepsilon^2+O(\varepsilon^3).
\]
This single-block cell meets the type-II rate constraint on its own.
The infimum in \eqref{eq:region-bound} gives the matching upper bound
$E_{\wt}(D+\varepsilon)\le\varepsilon^2/(2V)+O(\varepsilon^3)$.
Combining it with the lower bound proves the claimed expansion.
\end{proof}

\subsection{The common large-rate formula}
The proof of Lemma~\ref{lem:highest-weight-dmax-endpoint} gives
$b-a\le\Dmax(\rho\Vert\sigma)$ for every attainable rate pair.
The coherent cell in that proof saturates this bound. It has $q=u=0$,
with its axis determined by a unit vector $v$ maximizing
$\langle v,\rho v\rangle/\langle v,\sigma v\rangle$.
Its rate pair is
$(a_{\rm coh},b_{\rm coh})=
(-\log\langle v,\rho v\rangle,-\log\langle v,\sigma v\rangle)$,
so $b_{\rm coh}-a_{\rm coh}=\Dmax(\rho\Vert\sigma)$.
For any attainable pair,
$a+(r-b)_+=\max\{a,r-(b-a)\}\ge r-\Dmax$.
The finite-block likelihood bound in the same lemma gives the
unrestricted lower bound for any test sequence. For $r\ge b_{\rm coh}$,
the coherent cell and randomized acceptance attain it.
\[
 r-\Dmax
 \le E_{\mathrm{sc}}^*(r)\le E_{\wt}(r)\le E_{\mathrm{hw}}(r)
 \le a_{\rm coh}+r-b_{\rm coh}=r-\Dmax.
\]

\begin{remark}[Role of full rank]
Full rank is used in the local Loewner comparisons, the finiteness of
$\Dmax$, and the smooth local R\'enyi expansion at the Stein point.
It also ensures $X_\theta=\sigma^{-\theta/2}\rho\sigma^{-\theta/2}>0$
for the selection in Lemma~\ref{lem:sc-renyi-scale-weight-seed-selection}.
The proof in Appendix~\ref{app:sc-technical-proofs} chooses a weight-basis
vector whose diagonal coefficient of $\pi_\lambda(X_\theta)$ is at
least the average $\Tr\pi_\lambda(X_\theta)/\dim\mathcal Q_\lambda$.
Singular pairs require separate support and endpoint analysis.
\end{remark}

\subsection{Numerical envelope comparison and reproducibility}\label{app:numerical-reproducibility}

We compare the exponent for concatenated highest-weight cells with
the unrestricted exponent. By
Corollary~\ref{thm:qubit-weight-completeness}, the unrestricted curve is
\begin{equation}\label{eq:exact-weight-seed-curve}
 E_{\wt}(r)=E_{\mathrm{sc}}^*(r)
 =\sup_{0\le\theta<1}\left\{\theta r-
 \theta\widetilde D_{1/(1-\theta)}(\rho\Vert\sigma)\right\}.
\end{equation}

For the other curve, use the support formula
\[
 E_{\mathrm{hw}}(r)=\sup_{0\le\theta\le1}\{\theta r+h_{\mathrm{hw}}(\theta)\}.
\]
The support $h_{\mathrm{hw}}$ is the minimum of
$I_\rho^{\wt}(q,0,\bm n)-\theta I_\sigma^{\wt}(q,0,\bm n)$ over
$q\in[0,1/2]$ and $\bm n\in S^2$. We minimize first over the axis and
then over the Young fraction. For $0\le\theta<1$, put $c=1-\theta$ and
\[
 M_\theta(\bm n)
 =\log\frac{1+\frac45 n_z}{2}
  -\theta\log\frac{1+\frac45 n_x}{2}.
\]
Since $\det\rho=\det\sigma=9/100$, the highest-weight cell rates give
\[
 I_\rho^{\wt}(q,0,\bm n)-\theta I_\sigma^{\wt}(q,0,\bm n)
 =-c h(q)-cq\log\frac9{100}-(1-2q)M_\theta(\bm n).
\]
For $0\le q<1/2$, the minimizing axes are exactly the maximizers
of $M_\theta$. At $q=1/2$ the coefficient of $M_\theta$ vanishes,
so every axis minimizes. A single maximizing axis of $M_\theta$
therefore suffices throughout the minimization over $q$.
The monotonicity of $M_\theta$ in the two components allows
$n_x\le0$ and $n_z\ge0$.  At fixed $n_x$, increasing $n_z$ then
sets $n_y=0$.  Thus we may write
$n_x=-x$, $n_z=\sqrt{1-x^2}$ with $0\le x\le1$, and denote the
restricted function by $M_\theta(x)$.  For $0<x<1$,
\[
 M_\theta'(x)=\frac{4/5}{1-\frac45x}\{\theta-\theta(x)\},
 \qquad
 \theta(x)=\frac{x(1-\frac45x)}
 {\sqrt{1-x^2}\,[1+\frac45\sqrt{1-x^2}]}.
\]
The function $\theta(x)$ is strictly increasing.  Indeed, put
$a=4/5$ and $s=\sqrt{1-x^2}$. Direct differentiation gives
\begin{equation*}
 \theta'(x)=
 \frac{1-2ax+ax^3+as(1-2ax+x^2)}
      {s^3(1+as)^2}.
\end{equation*}
The first numerator term has minimum
$1-16\sqrt6/45>0$ on $[0,1]$, while
$1-2ax+x^2=(x-a)^2+1-a^2>0$. Hence $\theta'(x)>0$.
Moreover, $\theta(0)=0$ and $\theta(x)\to\infty$ as $x\uparrow1$.
Consequently, for every $0<\theta<1$, the unique solution
$x_\theta$ of $\theta(x_\theta)=\theta$ globally maximizes
$M_\theta(x)$. For $\theta=0$, the maximum occurs at $x_0=0$.
Let $x_1$ be the unique solution of $\theta(x_1)=1$. Thus
$0<x<x_1$ parametrizes the interior support parameters $0<\theta<1$.

The remaining objective is strictly convex in $q$, with minimizer
\[
 q_\theta=
 \left[1+\exp\left\{
 \frac{2M_\theta(x_\theta)-c\log(9/100)}{c}
 \right\}\right]^{-1}.
\]
To see that $q_\theta$ lies in $(0,1/2)$, evaluate the quantity in the exponent at the $z$ axis.
$M_\theta(\bm e_z)-\frac c2\log(9/100)
=\log3-\theta\log(5/3)>0$.
Substitution gives the full highest-weight support
\[
 h_{\mathrm{hw}}(\theta)
 =-c\log\left(e^{M_\theta(x_\theta)/c}
       +\frac9{100}e^{-M_\theta(x_\theta)/c}\right).
\]
At $\theta=0$, these formulas give $x_0=0$, $q_0=1/10$, and
$h_{\mathrm{hw}}(0)=0$.  The endpoint $\theta=1$ is obtained by continuity
of the minimum over the compact cell-parameter space.
The difference between the two Stein thresholds is
$D(\rho\Vert\sigma)-D_{\mathrm R}(\rho\Vert\sigma)
=\frac25\log9-\frac45\log\frac95=0.4086604990\ldots$.
Both supports are minima of $a-\theta b$ over compact cell-rate
regions. For $h_{\wt}$ this is the single-block weight-seed region
in Theorem~\ref{thm:exact-weight-support}. Since $a-\theta b$ is affine
in $\theta$ for each cell-rate pair $(a,b)$, both supports are concave.
The full-rank sandwiched formula makes $h_{\wt}$ smooth for $0<\theta<1$.
The formula for $h_{\mathrm{hw}}$ is also smooth there, since
$\theta'(x)>0$ gives a smooth inverse $x_\theta$.

For either support $h$, put $b(\theta)=-h'(\theta)$. At an interior
solution of $b(\theta)=r$, the derivative of $\theta r+h(\theta)$
is zero. By concavity its graph lies below the horizontal tangent at
that parameter, so the solution maximizes the objective, including
over its continuous extension to the endpoints. To evaluate
$E_{\mathrm{hw}}$ at $D$, solve $b_{\mathrm{hw}}(\theta)=D$.
In the parametrization $\theta=\theta(x)$ this gives
$x_D=0.4409573516050230080\ldots$, so the maximum at rate $D$
is attained at $\theta(x_D)$.

At a stationary rate of the gap where both exponents are
differentiable and their maximizing parameters are interior, the
two slopes are equal. By Lemma~\ref{lem:compact-rate-envelope-derivatives},
the slope of each exponent is its maximizing parameter $\theta$.
Their alternative-state rates therefore agree at that parameter. Solving $b_{\mathrm{hw}}(\theta(x))=b_{\wt}(\theta(x))$
gives the candidate $x_G=0.7954748006282451311\ldots$.
Write $\theta_G=\theta(x_G)$ and
$r_G=b_{\mathrm{hw}}(\theta_G)=b_{\wt}(\theta_G)$. At this rate
both objectives have zero derivative at $\theta_G$, so both attain
their maxima there. The common term $\theta_G r_G$ cancels, giving
$E_{\mathrm{hw}}(r_G)-E_{\wt}(r_G)
 =h_{\mathrm{hw}}(\theta_G)-h_{\wt}(\theta_G)$.
Numerical evaluation gives
\begin{align}
 E_{\mathrm{hw}}(D)&=0.04441162310634218777\ldots,\label{eq:numerical-Stein-gap}\\
 h_{\mathrm{hw}}(\theta_G)-h_{\wt}(\theta_G)
 &=0.09908232171254030878\ldots,\\
 r_G&=1.4914748002527326827\ldots.\label{eq:numerical-rmax}
\end{align}
The reduction above gives the full support function $h_{\mathrm{hw}}$.
The numerical calculation samples the signs of
$D-b_{\mathrm{hw}}(\theta(x))$ and
$b_{\wt}(\theta(x))-b_{\mathrm{hw}}(\theta(x))$ at 5,000 midpoints in
$0<x<x_1$, comparing the samples on each side of the corresponding
root. These checks leave global maximality over rates uncertified.

\begin{table}[ht]
\centering
\begin{tabular}{rrrr}
\toprule
$r$ & $E_{\mathrm{hw}}(r)$ & $E_{\mathrm{sc}}^{*}(r)=E_{\wt}(r)$ & difference\\
\midrule
1.00&0.068732&0.004326&0.064407\\
1.20&0.116298&0.029123&0.087175\\
1.40&0.172247&0.074313&0.097934\\
1.50&0.203447&0.104374&0.099072\\
1.70&0.273278&0.180116&0.093162\\
1.90&0.355669&0.280411&0.075258\\
2.10&0.457398&0.422133&0.035265\\
2.20&0.523682&0.519300&0.004382\\
2.25&0.569300&0.569300&0\\
\bottomrule
\end{tabular}
\caption{Exponents for concatenated highest-weight cells and unrestricted measurements for the pair \eqref{eq:pair}. The unrestricted values use \eqref{eq:exact-weight-seed-curve}. Entries are rounded independently, and the highest-weight values are numerical estimates without rigorous error bounds.}\label{tab:num}
\end{table}

\begin{figure}[!t]
\centering
\includegraphics[width=.78\linewidth]{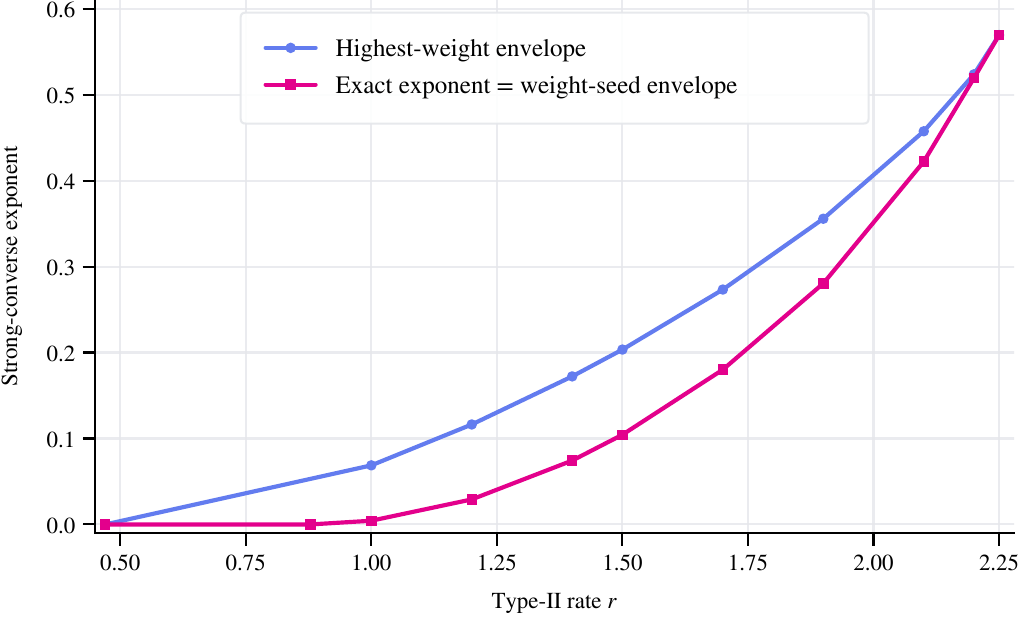}
\caption{Strong-converse exponents for concatenated highest-weight cells and unrestricted measurements for the pair \eqref{eq:pair}. The unrestricted value also equals $E_{\wt}$ by Corollary~\ref{thm:qubit-weight-completeness}. The plotted points give the values in Table~\ref{tab:num} and the exact zero-exponent thresholds. The segments guide the eye. Figure~\ref{fig:intro-qubit-exponents} instead uses arbitrary decision rules for Keyl's measurement on the whole sample.}\label{fig:envelopes}
\end{figure}

The fixed-error Stein estimates in
Figure~\ref{fig:intro-stein-boundary} are discussed in
Appendix~\ref{subsubsec:stein-qubit-arc-reduction}. The Hoeffding and
strong-converse curves in Figure~\ref{fig:intro-qubit-exponents} are
numerical evaluations of the exact variational formulas summarized in
Table~\ref{tab:exponent-summary}.

\FloatBarrier
\section{Proof of the fixed-error Stein theorem}\label{app:merge-stein-proof}
\subsection{Mixture rates and uniform testing}

Fix a nonempty compact set \(K\subset H_\rho\) and a probability measure
\(\nu\) with support exactly \(K\). Write
\[
 \sigma_{K,n}=\int_K(h^\dagger\sigma h)^{\otimes n}\dd\nu(h).
\]
The set \(K\) and the measure \(\nu\) are fixed as \(n\to\infty\).
The state \(\rho\) specifies the group \(H_\rho\), while the state
\(\eta\) in the lemma specifies the input used to test the mixture.
Write \(S(\eta)=-\Tr\eta\log\eta\) for its entropy.

\Needspace{5\baselineskip}
\begin{lemma}[Relative-entropy rate of a rotated mixture]\label{lem:mixture-rate}
For every faithful state \(\eta\), the limit
\begin{equation}\label{eq:E-definition}
 E_\eta(K)=\lim_{n\to\infty}\frac1n
                  D(\eta^{\otimes n}\Vert\sigma_{K,n})
\end{equation}
exists and is independent of the full-support measure \(\nu\).
It is continuous in \(\eta\) and in \(K\), with the Hausdorff metric on
nonempty compact subsets of \(H_\rho\). Moreover, for every \(\delta>0\),
\begin{equation}\label{eq:spectral-concentration}
 \Tr\eta^{\otimes n}
 \one\left\{\left|-\frac1n\log\sigma_{K,n}
                     -S(\eta)-E_\eta(K)\right|>\delta\right\}
 \longrightarrow0.
\end{equation}
The convergence is uniform when \(\eta\) ranges over a compact set of
faithful states. Here \(\one\{\cdot\}\) denotes a spectral projection.
\end{lemma}

\begin{proof}
\textbf{Step 1: Existence and independence of mixture weights.}
Put \(a=\lambda_{\min}(\sigma)>0\),
\(b=\lambda_{\max}(\sigma)\), and
\(\kappa(t)=\log(1+2bt/a)\). For \(h,k\in U(d)\), the operator-norm difference between the rotated
states is at most \(2b\|h-k\|_\infty\). Both states are bounded below
by \(aI\), so
\begin{equation}\label{eq:rotation-comparison}
 e^{-\kappa(t)}k^\dagger\sigma k
 \le h^\dagger\sigma h
 \le e^{\kappa(t)}k^\dagger\sigma k
 \qquad\text{if }\|h-k\|_\infty\le t.
\end{equation}
For each \(t>0\), compactness and full support imply
\(\inf_{h\in K}\nu\{k:\|k-h\|_\infty<t\}>0\).
Indeed, take a finite cover by balls of radius \(t/2\) centered in
\(K\); the minimum of their positive measures is a lower bound.
Integrating \cref{eq:rotation-comparison} therefore proves that there
are numbers \(c_n\ge0\), with \(c_n/n\to0\), such that
\begin{equation}\label{eq:mixture-uniform}
 (h^\dagger\sigma h)^{\otimes n}
       \le e^{c_n}\sigma_{K,n}\qquad(h\in K).
\end{equation}
To see the sublinear growth explicitly, the least admissible \(c_n\)
is at most \(n\kappa(t)-\log\inf_{h\in K}\nu(B(h,t))\).
First let \(n\to\infty\), then \(t\downarrow0\).

Apply \cref{eq:mixture-uniform} to the last \(m\) tensor factors inside
the integral defining \(\sigma_{K,n+m}\). This gives
\[
 \sigma_{K,n+m}\le e^{c_m}\sigma_{K,n}\otimes\sigma_{K,m}.
\]
Operator monotonicity of the logarithm implies
\begin{align*}
 D(\eta^{\otimes(n+m)}\Vert\sigma_{K,n+m})
 \ge{}&D(\eta^{\otimes n}\Vert\sigma_{K,n})
       +D(\eta^{\otimes m}\Vert\sigma_{K,m})-c_m.
\end{align*}
Fix \(m\) and iterate this inequality over blocks of length \(m\).
The remaining relative entropy is nonnegative, so
\[
 \liminf_{n\to\infty}\frac1nD(\eta^{\otimes n}\Vert\sigma_{K,n})
 \ge\frac{D(\eta^{\otimes m}\Vert\sigma_{K,m})-c_m}{m}.
\]
The sequence on the left is bounded, since \(\sigma_{K,n}\ge a^nI\).
Taking \(m\) along a subsequence that realizes the limsup of the
normalized relative entropy, and using \(c_m/m\to0\), shows that
the liminf and limsup agree. If another measure has the same support, integrate
\cref{eq:mixture-uniform} against it and then reverse the roles of the
two measures. Their logarithms differ in operator order by \(o(n)I\),
which proves independence of \(\nu\).

\textbf{Step 2: Concentration and continuity in the null state.}
For a state \(\omega\) and a Hermitian operator \(X\), write
\(\operatorname{Var}_\omega(X)=\Tr\omega X^2-(\Tr\omega X)^2\).
We bound the variance of \(\log\sigma_{K,n}\). Set
\(L=\tfrac12\log(b/a)\). Removing tensor factor \(i\)
from the integral gives
\[
 a(I_i\otimes\sigma_{K,n-1})\le\sigma_{K,n}
       \le b(I_i\otimes\sigma_{K,n-1}),
\]
where the other factors retain their order and \(\sigma_{K,0}=1\).
Taking logarithms yields the uniform bound
\begin{equation}\label{eq:log-local-bound}
 \left\|\log\sigma_{K,n}
   -I_i\otimes\log\sigma_{K,n-1}
   -\tfrac12\log(ab)I\right\|_\infty\le L.
\end{equation}

The product-state variance inequality of
\cite[Theorem 6.6]{DePalmaFanizzaMowryODonnell} now gives
\begin{equation}\label{eq:log-variance}
 \operatorname{Var}_{\eta^{\otimes n}}(\log\sigma_{K,n})\le nL^2.
\end{equation}
The cited inequality bounds the variance by the sum, over tensor factors,
of the squared distances to operators independent of that factor, in the
inner product $\langle A,B\rangle=\Tr\eta^{\otimes n}A^\dagger B$.
Equation~\eqref{eq:log-local-bound} supplies such an approximation with
operator-norm error at most $L$ for each factor. Each squared distance is
therefore at most $L^2$, which gives Eq.~\eqref{eq:log-variance}.

Write \(g_n(\eta)=-n^{-1}\Tr\eta^{\otimes n}\log\sigma_{K,n}\).
The limit already proved gives
\(g_n(\eta)\to S(\eta)+E_\eta(K)\).
Replacing the factors of \(\eta^{\otimes n}\) one at a time by those
of \(\zeta^{\otimes n}\), and using \cref{eq:log-local-bound}, gives
\begin{equation}\label{eq:mean-continuity}
 |g_n(\eta)-g_n(\zeta)|\le L\|\eta-\zeta\|_1.
\end{equation}
The term independent of the replaced factor vanishes because
\(\Tr(\eta-\zeta)=0\). The remaining operator has norm at most \(L\).
The bound gives continuity of the limit and uniform convergence on
compact sets of faithful states by a finite-net argument.
Chebyshev's inequality and \cref{eq:log-variance} give
\[
 \Tr\eta^{\otimes n}
 \one\left\{\left|-\frac1n\log\sigma_{K,n}-g_n(\eta)\right|>\delta\right\}
 \le\frac{L^2}{n\delta^2}.
\]
Together these observations prove \cref{eq:spectral-concentration}.

\textbf{Step 3: Continuity in the rotation set.}
If the Hausdorff distance between compact sets \(K,K'\subset H_\rho\)
is at most \(t\), choose \(k\in K'\) within distance \(t\) of each
\(h\in K\). Then \cref{eq:rotation-comparison,eq:mixture-uniform} give
\[
 (h^\dagger\sigma h)^{\otimes n}
 \le e^{n\kappa(t)}(k^\dagger\sigma k)^{\otimes n}
 \le e^{n\kappa(t)+c_n(K')}\sigma_{K',n},
 \qquad c_n(K')=o(n).
\]
After integration, \(\sigma_{K,n}\le
 e^{n\kappa(t)+c_n(K')}\sigma_{K',n}\).
The reverse comparison also holds. Dividing the resulting relative
entropy bounds by \(n\) proves
\[
 |E_\eta(K)-E_\eta(K')|\le\kappa(t).
\]
This is uniform in \(\eta\); combined with
\cref{eq:mean-continuity} and continuity of entropy, it gives the
claimed continuity.
\end{proof}

The spectral concentration above gives the thresholds for testing the
mixture. The direct bound also applies uniformly to nearby inputs.

\begin{lemma}[Testing against the rotated mixture]\label{lem:stabilizer-testing}
Fix \(K\subset H_\rho\) as above and a faithful state \(\eta\).

\noindent\textbf{(i) Direct estimate.}
For every \(R<E_\eta(K)\), there is a sequence of effects
\(0\le T_n\le I\) such that
\[
 \Tr\eta^{\otimes n}T_n\longrightarrow1,\qquad
 \sup_{h\in K}\Tr(h^\dagger\sigma h)^{\otimes n}T_n
       \le e^{-nR}
\]
for all sufficiently large \(n\).

\noindent\textbf{(ii) Converse estimate.}
For each \(\delta>0\), every
effect \(0\le T_n\le I\), without any covariance requirement, satisfies
\begin{equation}\label{eq:stabilizer-converse}
 \Tr\eta^{\otimes n}T_n
 \le e^{n(E_\eta(K)+\delta)}\Tr\sigma_{K,n}T_n+o(1),
\end{equation}
where the remainder is independent of \(T_n\).

\noindent\textbf{(iii) Direct estimate on neighborhoods.}
For every \(R<E_\rho(K)\), there are a closed neighborhood \(V\) of
\(\rho\) consisting of faithful states, an open neighborhood \(O\) of
\(K\) in \(U(d)\), and one sequence of effects \(0\le T_n\le I\) such
that
\[
 \inf_{\eta\in V}\Tr\eta^{\otimes n}T_n\longrightarrow1,\qquad
 \sup_{h\in O}\Tr(h^\dagger\sigma h)^{\otimes n}T_n
       \le e^{-nR}
\]
for all sufficiently large \(n\). The neighborhoods and the test sequence
may depend on \(R\), but the neighborhoods do not depend on \(n\).
\end{lemma}

\begin{proof}
Fix a small \(\delta>0\). An entropy-typical projection \(\Pi_n\) obeys
\[
 \Tr\eta^{\otimes n}\Pi_n\to1,\qquad
 \rank\Pi_n\le e^{n(S(\eta)+\delta)},\qquad
 \Pi_n\eta^{\otimes n}\Pi_n
       \le e^{-n(S(\eta)-\delta)}\Pi_n.
\]
These follow from the law of large numbers for the eigenvalues of
\(\eta\).

\noindent\textbf{(i) Direct estimate.}
Use \(\Pi_n\) to bound the rank and a spectral projection \(Q_n\)
to restrict acceptance to small eigenvalues of the mixture. Set
\[
 Q_n=\one\{\sigma_{K,n}\le
                     e^{-n(S(\eta)+E_\eta(K)-\delta)}\},\qquad
 T_n=Q_n\Pi_nQ_n.
\]
By \cref{eq:spectral-concentration}, \(\Tr\eta^{\otimes n}Q_n\to1\).
The gentle-projection bound \cite{Winter}, with the constant in
\cite[Appendix~C, Lemma~5]{OgawaNagaoka2002Gentle},
\[
 \|\omega-P\omega P\|_1
       \le2\sqrt{\Tr\omega(I-P)},
\]
shows that \(\Tr\eta^{\otimes n}T_n\to1\). Also,
\[
 \Tr\sigma_{K,n}T_n
 \le e^{-n(S(\eta)+E_\eta(K)-\delta)}\rank\Pi_n
 \le e^{-n(E_\eta(K)-2\delta)}.
\]
Equation \eqref{eq:mixture-uniform} converts this to the same
asymptotic bound uniformly over \(h\in K\).

\noindent\textbf{(ii) Converse estimate.}
Use \(Q_n=\one\{\sigma_{K,n}\ge
 e^{-n(S(\eta)+E_\eta(K)+\delta)}\}\) and put
\(A_n=\Pi_n\eta^{\otimes n}\Pi_n\). Write \(\eta_n=\eta^{\otimes n}\).
On the complementary spectral event \(I-Q_n\),
\(-n^{-1}\log\sigma_{K,n}>S(\eta)+E_\eta(K)+\delta\).
Thus \cref{eq:spectral-concentration} gives
\(\Tr\eta_n(I-Q_n)\to0\), and typicality gives
\(\Tr\eta_n(I-\Pi_n)\to0\). Inserting \(Q_n\eta_nQ_n\) and using
trace-norm contraction under compression gives
\begin{align*}
 \|\eta_n-Q_nA_nQ_n\|_1
 &\le \|\eta_n-Q_n\eta_nQ_n\|_1+\|\eta_n-A_n\|_1\\
 &\le 2\sqrt{\Tr\eta_n(I-Q_n)}+2\sqrt{\Tr\eta_n(I-\Pi_n)}
 \longrightarrow 0.
\end{align*}
The second line applies the gentle-projection bound from part (i)
to the original state \(\eta_n\) with each projection separately.
Also,
\[
 Q_nA_nQ_n\le e^{-n(S(\eta)-\delta)}Q_n
       \le e^{n(E_\eta(K)+2\delta)}\sigma_{K,n}.
\]
For every effect \(T_n\), the trace-norm difference bounds the error
in replacing \(\Tr\eta_n T_n\) by \(\Tr Q_nA_nQ_n T_n\), since
\(\|T_n\|_\infty\le1\). Taking the trace of the operator bound against
\(T_n\) and choosing the proof margin to be half the tolerance in the
statement gives \cref{eq:stabilizer-converse}, with a remainder
independent of \(T_n\).
In particular, a type-II rate strictly above \(E_\eta(K)\) forces
success to tend to zero, even if only a fixed positive success
probability was required.

\noindent\textbf{(iii) Direct estimate on neighborhoods.}
Fix \(R<E_\rho(K)\) and choose
\(\zeta>0\) so that \(R+4\zeta<E_\rho(K)\).
Fix an eigenbasis \((e_i)_{i=1}^d\) of \(\rho\) and put
\(p_{\eta,i}=\langle e_i,\eta e_i\rangle\).
By continuity of Shannon entropy at \(p_\rho\), choose \(\delta_f>0\)
so that every frequency vector within sup-norm distance
\(2\delta_f\) of \(p_\rho\) has entropy at most \(S(\rho)+\zeta/2\). Let \(\Pi_n\) project onto the basis strings
whose empirical frequencies lie in this closed ball. There are at
most \((n+1)^d\) types, each having at most
\(e^{n(S(\rho)+\zeta/2)}\) strings in the ball. Hence
\(\rank\Pi_n\le e^{n(S(\rho)+\zeta)}\) for all sufficiently large \(n\).

Choose a small closed neighborhood \(V\) of \(\rho\) consisting of
faithful states with \(\|p_\eta-p_\rho\|_\infty\le\delta_f\) for every
\(\eta\in V\). The diagonal of \(\eta^{\otimes n}\) in the product
basis gives the law \(p_\eta^{\otimes n}\). A rejected string has
an empirical frequency differing from \(p_{\eta,i}\) by at least
\(\delta_f\) in some coordinate \(i\). Each coordinate frequency has
variance \(p_{\eta,i}(1-p_{\eta,i})/n\le 1/(4n)\). Chebyshev's
inequality and the union bound thus give
\[
 \sup_{\eta\in V}\Tr\eta^{\otimes n}(I-\Pi_n)
 \le \frac{d}{4n\delta_f^2}\longrightarrow0 .
\]

For the mixture threshold, set
\[
 Q_n=\one\{\sigma_{K,n}\le
   e^{-n(S(\rho)+E_\rho(K)-\zeta)}\},\qquad
 T_n=Q_n\Pi_n Q_n.
\]
Shrink \(V\), if necessary, so that for every \(\eta\in V\),
\[
 \bigl|S(\eta)+E_\eta(K)-S(\rho)-E_\rho(K)\bigr|<\zeta/2.
\]
The threshold for \(-n^{-1}\log\sigma_{K,n}\) lies at least
\(\zeta/2\) below \(S(\eta)+E_\eta(K)\) for every \(\eta\in V\).
Uniform spectral concentration in
\cref{eq:spectral-concentration}, with deviation \(\zeta/4\), therefore
gives \(\inf_{\eta\in V}\Tr\eta^{\otimes n}Q_n\to1\).
Apply the gentle-projection bound from part (i) with
\(\omega=\eta^{\otimes n}\) and \(P=Q_n\) to obtain
\[
 \left|\Tr\eta^{\otimes n}T_n-\Tr\eta^{\otimes n}\Pi_n\right|
 \le 2\sqrt{\Tr\eta^{\otimes n}(I-Q_n)} .
\]
The right-hand side tends uniformly to zero on \(V\), so
\(\inf_{\eta\in V}\Tr\eta^{\otimes n}T_n\to1\).
The rank and threshold bounds give
\(\Tr\sigma_{K,n}T_n\le e^{-n(E_\rho(K)-2\zeta)}\).
Equation \eqref{eq:mixture-uniform} transfers this bound uniformly to
\(K\), with a subexponential loss. Finally choose a fixed open neighborhood \(O\) of \(K\) so small that
the further loss from
\cref{eq:rotation-comparison} is less than \(\zeta\) in exponent.
The resulting uniform type-II rate is at least
\(E_\rho(K)-3\zeta-o(1)>R\) for all sufficiently large \(n\).
\end{proof}

\subsection{Fixed-error exponent for phase-invariant observations}
\label{subsec:phase-invariant-stein-proof}

\begin{proof}[Proof of Proposition~\ref{prop:phase-invariant-fixed-error-Stein}]
For the lower bound, use the Keyl test of
Proposition~\ref{prop:keyl-stein-direct-reorganized}. It accepts when
the reported state lies in a trace-norm ball centered at $\rho$.
Every $g\in G_\rho$ fixes $\rho$ and preserves the trace norm, so the
ball is invariant. By covariance the acceptance effect is invariant as well.

For the upper bound let $\rho_n=\rho^{\otimes n}$ and let $\mathcal P_n$
be the pinching obtained by averaging conjugation by $g^{\otimes n}$
over $G_\rho$. Each block specifies the number of copies in each
eigenspace of $\rho$. Thus $\rho_n$ is scalar on every block, also
for repeated eigenvalues. In particular, it commutes with
$\overline\sigma_n=\mathcal P_n(\sigma^{\otimes n})$.

Fix $0<\alpha<1$, set $t=1-\alpha$, and write
$A=\rho^{\alpha/(2t)}\sigma\rho^{\alpha/(2t)}$. Since powers of $\rho_n$
are scalar on each occupation subspace,
\[
 \Tr\rho_n^\alpha\overline\sigma_n^t
 =\Tr\left[(\mathcal P_n(A^{\otimes n}))^t\right]
 \ge\Tr[(A^{\otimes n})^t]
 =(\Tr A^t)^n.
\]
The inequality follows from concavity of $X\mapsto\Tr X^t$ and the
unitary average defining $\mathcal P_n$. Taking logarithms and dividing
by $\alpha-1<0$ gives
\[
 D^{\rm P}_\alpha(\rho_n\Vert\overline\sigma_n)
 \le nD^{\rm RS}_\alpha(\rho\Vert\sigma).
\]
Letting $\alpha\uparrow1$ at fixed $n$ yields
$D(\rho_n\Vert\overline\sigma_n)\le nD_{\mathrm R}(\rho\Vert\sigma)$
by the order-one limit of the reverse-sandwiched divergence
from Section~\ref{subsec:ucov-hoeffding-renyi-converse}.

A phase-invariant effect has the same alternative acceptance under
$\sigma^{\otimes n}$ and $\overline\sigma_n$. The latter is the mixture
with $K=G_\rho\subset H_\rho$ and $\nu=\mu_\rho$. Its rate
$\lim_{n\to\infty}n^{-1}D(\rho_n\Vert\overline\sigma_n)$ exists by
Lemma~\ref{lem:mixture-rate}, and Lemma~\ref{lem:stabilizer-testing}(ii)
forces null acceptance to zero at every type-II rate above this limit. This gives the upper bound at
every fixed $0<\epsilon<1$.
\end{proof}

\subsection{Compact-set formulation, converse, and achievability}

\subsubsection{Optimizing the retained orientations}\label{app:stein-mass-optimization}

We first prove the compact-set form
\begin{equation}\label{eq:stein-compact}
 B_\epsilon^{\mathrm{ucov}}
 =\max_{\substack{K\subset H_\rho\ {\rm compact}\\
                       \mu(K)\ge1-\epsilon}} E_\rho(K).
\end{equation}
The measurable-set formula will then follow by approximation.

For \(0<s<1\), the maximum of \(E_\rho(K)\) over compact
\(K\subset H_\rho\) of measure at least \(s\) exists. Indeed, the space
of nonempty compact subsets of the compact group \(H_\rho\) is compact
in the Hausdorff metric. If \(K_j\to K\), then
\(\limsup_j\mu(K_j)\le\mu(K)\), since \(K_j\) is eventually contained
in every neighborhood of \(K\). Thus the constraint is closed, and
Lemma~\ref{lem:mixture-rate} gives continuity of the objective.

To see continuity as the required mass increases to \(s\), take a
convergent subsequence of maximizers and use upper semicontinuity of
Haar measure. For continuity as the required mass decreases to \(s\),
start with a maximizing set \(K\) at \(s\). If \(\mu(K)=s\), replace
\(K\) by a small closed metric neighborhood. The group \(H_\rho\) is
a product of unitary groups and is connected. A boundary point of \(K\)
has points of its complement arbitrarily close to it, so any such
neighborhood contains an open set outside \(K\) with positive Haar
measure. This gives a strict increase in mass. The enlarged sets
converge to \(K\) in Hausdorff distance, and their exponents converge to
\(E_\rho(K)\). If \(\mu(K)>s\), the original set already meets every
nearby mass constraint. The same argument shows that using
\(\mu(K)>s\) gives the same supremum.

\subsubsection{Converse for arbitrary covariant measurements}

Fix \(0<\delta\) and \(0<2\gamma<s:=1-\epsilon\) independently
of the blocklength.  We will first let \(n\to\infty\) while the cell
partition, \(\delta\), and \(\gamma\) remain fixed; only then do we
refine the partition and let \(\gamma\downarrow0\).
Fix a covariant procedure with \(\alpha_n\le\epsilon\). Let
\(P_{\rho,n}(\dd x)=\Tr\rho^{\otimes n}M_n(\dd x)\).
Covariance and invariance of \(\rho\) under \(H_\rho\) give
\begin{align}
 1-\alpha_n
 &=\int\int_{H_\rho}f_n(hx)\dd\mu(h)\,P_{\rho,n}(\dd x),\label{eq:averaged-success}\\
 \beta_n
 &=\int_{H_\rho}\int f_n(hx)
          \Tr(h^\dagger\sigma h)^{\otimes n}M_n(\dd x)\dd\mu(h).
          \label{eq:averaged-beta}
\end{align}
These identities use only the measurable group action and Fubini's theorem.

Partition \(H_\rho\), up to a null set, into finitely many Borel cells
\(C_1,\ldots,C_N\) with positive measure, null boundaries, and diameter
at most \(\delta\). Such a partition follows from a finite cover by
small balls whose boundaries have measure zero.
At outcome \(x\), mark cell \(j\) if
\[
 \frac1{\mu(C_j)}\int_{C_j}f_n(hx)\dd\mu(h)\ge\gamma.
\]
Let \(m(x)\) be the total measure of the marked cells. The average of
\(f_n(hx)\) is at most \(m(x)+\gamma\), so
\(\int m\,\dd P_{\rho,n}\ge s-\gamma\).
Because \(0\le m\le1\), writing
\(p=P_{\rho,n}\{m\ge s-2\gamma\}\) gives
\[
 s-\gamma\le\int m\,\dd P_{\rho,n}
 \le(s-2\gamma)+(1-s+2\gamma)p,
 \qquad p\ge\frac{\gamma}{1-s+2\gamma}\ge\gamma.
\]
There are at most \(2^N\) possible sets of marked cells. For each \(n\),
one such set with total mass at least \(s-2\gamma\) occurs on a
measurable event of probability at least \(\gamma/2^N\).
Call the selected outcome event \(B_n\), set \(T_n=M_n(B_n)\), and
let \(K\) be the union of the closures of its marked cells. Then
\begin{equation}\label{eq:pattern-success}
 \Tr\rho^{\otimes n}T_n\ge\gamma/2^N,\qquad
 \mu(K)\ge s-2\gamma.
\end{equation}
Fix a marked cell $C_j$ and $h\in\overline C_j$. Write
$\nu_h(\dd x)=\Tr(h^\dagger\sigma h)^{\otimes n}M_n(\dd x)$.
For $k\in C_j$, the distance to $h$ is at most $\delta$.
Equation~\eqref{eq:rotation-comparison}, tensorized and traced against
the POVM, gives $\nu_k\ge e^{-n\kappa(\delta)}\nu_h$ as measures.
Restricting \eqref{eq:averaged-beta} to $C_j$ and $B_n$ therefore gives
\begin{align*}
 \beta_n
 &\ge\int_{C_j}\int_{B_n} f_n(kx)\,\nu_k(\dd x)\dd\mu(k)\\
 &\ge e^{-n\kappa(\delta)}
      \int_{B_n}\left(\int_{C_j} f_n(kx)\dd\mu(k)\right)\nu_h(\dd x)\\
 &\ge e^{-n\kappa(\delta)}\gamma\mu(C_j)
      \Tr(h^\dagger\sigma h)^{\otimes n}T_n .
\end{align*}
The last inequality uses that $C_j$ is marked for every $x\in B_n$.
Taking the supremum over $h\in K$ and the minimum marked-cell mass gives
\begin{equation}\label{eq:pattern-beta}
 \sup_{h\in K}\Tr(h^\dagger\sigma h)^{\otimes n}T_n
 \le
 \frac{e^{n\kappa(\delta)}\,\beta_n}
      {\gamma\min_j\mu(C_j)}.
\end{equation}

The converse estimate \cref{eq:stabilizer-converse} applies to every
effect, including the extracted \(T_n\). The estimate is uniform in the
effect, and only finitely many sets \(K\) can occur.
Integrating \cref{eq:pattern-beta} over the mixture on \(K\) and
using \cref{eq:pattern-success,eq:stabilizer-converse} gives, for each
\(\xi>0\),
\[
 \frac{\gamma}{2^N}
 \le \frac{e^{n(E_\rho(K)+\xi+\kappa(\delta))}\beta_n}
           {\gamma\min_j\mu(C_j)}+o(1).
\]
The remainder tends to zero uniformly over the fixed finite family
of sets \(K\). Since \(\xi>0\) is arbitrary, the limsup type-II
exponent is at most
\[
 \max_{\substack{K\subset H_\rho\ {\rm compact}\\
                  \mu(K)\ge s-2\gamma}}
 E_\rho(K)+\kappa(\delta).
\]
This also applies to effects chosen within a fixed factor of the
finite-copy optimum. Let \(\delta\downarrow0\) and then
\(\gamma\downarrow0\), using continuity in the mass constraint.
This proves the required upper bound on
$\overline B_\epsilon^{\mathrm{ucov}}(\rho\Vert\sigma)$.

\subsubsection{Achievability}
\label{app:stein-achievability}

\textbf{Step 1: Estimating an orientation.}
Diagonalize \(\rho\) in decreasing eigenvalue order. Fix a compact
\(K\subset H_\rho\) with \(\mu(K)>s\), and fix \(R<E_\rho(K)\).
Use Keyl's measurement on \(m\) copies, where \(m\to\infty\) and \(m/n\to0\),
retaining its full unitary coordinate \(U\). This is a legitimate
covariant refinement even when a Young diagram has repeated rows. In the
Schur block \(\lambda\), one can realize the joint label \((\lambda,U)\)
by the POVM density
\[
 d_\lambda\,|\pi_\lambda(U)v_\lambda\rangle
 \langle\pi_\lambda(U)v_\lambda|\otimes I_{\mathcal V_\lambda}\,dU,
\]
where \(\mathcal V_\lambda\) is the multiplicity space from
Appendix~\ref{app:keyl-finite-output-law} and \(dU\) is normalized
Haar measure on \(U(d)\). The factor \(d_\lambda=\dim\mathcal U_\lambda\)
normalizes the Haar integral by Schur's lemma. Left translation of
\(U\) gives covariance. Its
pushforward is the original physical partial-flag observation: conditional
on a physical orbit outcome, the redundant group coordinates are filled
by a state-independent invariant kernel. The full coordinate therefore
adds no information about the input (see
Appendix~\ref{app:keyl-finite-output-law}). Under \(\rho\) the distribution
of \(U\) is left \(H_\rho\)-invariant. The reported state
\(\widehat\tau_m=U\diag(\lambda/m)U^\dagger\) converges to \(\rho\) in
probability. The ordered eigenvalues then converge as well, and
\[
 \|U\rho U^\dagger-\rho\|_\infty
 \le\|\rho-\diag(\lambda/m)\|_\infty
       +\|\widehat\tau_m-\rho\|_\infty\longrightarrow0
\]
in probability. Since \(U(d)\) is compact, the distance from \(U\)
to \(H_\rho\) tends to zero in probability. Every weak limit of the
law of \(U\) is supported on \(H_\rho\) and is left invariant under
\(H_\rho\). The law therefore converges to Haar measure on \(H_\rho\),
including when \(\rho\) has repeated eigenvalues.

\textbf{Step 2: Testing the remaining copies.}
Part~(iii) of Lemma~\ref{lem:stabilizer-testing} supplies a test \(T_{n-m}\)
on the remaining copies. It accepts all null states sufficiently close
to \(\rho\) with success tending uniformly to one, and has type-II error
at most \(e^{-(n-m)R}\) for all rotations in a neighborhood \(O\) of \(K\).
Shrink \(O\) so that \(U^\dagger\rho U\) is in the required null
neighborhood whenever \(U\in O\). Choose a continuous function
\(w:U(d)\to[0,1]\), equal to one on \(K\) and supported in \(O\).
Its integral over \(H_\rho\) is at least \(\mu(K)>s\).

Given the recorded label \((\lambda,U)\), apply the \(U\)-rotated
binary POVM to the remaining \(n-m\) copies:
\[
 \{U^{\otimes(n-m)}T_{n-m}(U^\dagger)^{\otimes(n-m)},\
      I-U^{\otimes(n-m)}T_{n-m}(U^\dagger)^{\otimes(n-m)}\}.
\]
The combined measurement records \((\lambda,U,\text{binary outcome})\).

\textbf{Step 3: Covariance and error bounds.}
Under a joint rotation \(g\), the first record transforms as
\(U\mapsto gU\), and the conditional binary effect transforms into the
one selected by \(gU\), so the combined POVM is covariant. The decision
rule accepts with probability \(w(U)\) when the binary outcome is \(1\),
and rejects when it is \(0\).
Under either state \(X\in\{\rho,\sigma\}\), the copies used by the
second measurement are independent of the first-stage record. The
rotated effect gives the probability of outcome \(1\), and the rule
then gives the conditional probability of final acceptance
\[
 \Pr_X(\mathrm{accept}\mid\lambda,U)
 = w(U)\Tr\big[(U^\dagger XU)^{\otimes(n-m)}T_{n-m}\big].
\]
Under \(\rho\), the trace tends uniformly to one on the support of
\(w\). The final acceptance probability thus differs from
\(\mathbb E_\rho w(U)\) by a quantity tending to zero. By weak
convergence of \(U\) and continuity of \(w\), its limit is
\(\int_{H_\rho}w\,\dd\mu\ge\mu(K)>s\).
Under \(\sigma\), the trace is at most \(e^{-(n-m)R}\) on the support of
\(w\), and final acceptance vanishes elsewhere because \(w(U)=0\).
Since \(0\le w\le1\), the same upper bound holds after averaging over
the first-stage record.
Thus \(\alpha_n\le\epsilon\) for all sufficiently large \(n\) and the
type-II rate is at least \(R\). Arbitrary admissible choices at the
finitely many remaining blocklengths do not affect the limit.

Taking the supremum over \(R,K\) proves Eq.~\eqref{eq:stein-compact}, including
existence of the limit. The mass-continuity argument proves continuity
in \(\epsilon\); monotonicity follows either from the formula or
from the operational definition.

\subsubsection{Measurable sets}

For measurable \(A\) of positive measure, let \(K\) be the support of
\(\mu(\,\cdot\,\cap A)/\mu(A)\). This measure gives zero mass to
\(A\setminus K\), so \(\mu(K)\ge\mu(A)\). By \cref{eq:E-definition},
the mixture has exponent \(E_\rho(K)\).
Conversely, take open metric neighborhoods of a compact
\(K\subset H_\rho\) of mass at least \(s<1\), with radii decreasing
to zero. Their masses exceed \(s\) by the same neighborhood argument
used for \cref{eq:stein-compact}. The support of Haar measure restricted
to an open neighborhood is its closure, and these closures converge
to \(K\) in Hausdorff distance. Continuity of \(E_\rho\)
therefore proves Eq.~\eqref{eq:merge-ucov-fixed-stein}.

\subsubsection{Qubit phase arcs and numerical evaluation}
\label{subsubsec:stein-qubit-arc-reduction}

The fixed-error formula in Theorem~\ref{thm:merge-ucov-fixed-interior-stein}
optimizes over subsets of the stabilizer $H_\rho$.
For the diagonal $\rho$ in Figure~\ref{fig:intro-stein-boundary}, these
rotations have the form $h=\operatorname{diag}(e^{i\phi_0},e^{i\phi_1})$.
The rotated alternative $h^\dagger\sigma h$ depends only on the relative
phase $\phi=\phi_0-\phi_1$ modulo $2\pi$, represented by
$g_\phi=\operatorname{diag}(e^{i\phi},1)$.
Normalized Haar measure $\mu$ on $H_\rho$ induces uniform measure on
the phase circle. Fix $0<\epsilon<1$. Let $A$ be one closed arc or the union of two equal
closed arcs centered $\pi$ apart, of total length $|A|=2\pi(1-\epsilon)$.
Its inverse image $K_A$ in $H_\rho$ has measure $1-\epsilon$. The
normalized restriction of $\mu$ to $K_A$ has full support on $K_A$ and
induces $d\phi/|A|$ on $A$.

\Needspace{5\baselineskip}
Substituting this measure in \eqref{eq:merge-mixture-rate}, with the same
phase on every copy, and using the compact-set maximum
\eqref{eq:merge-ucov-fixed-stein} at $\mu(K_A)=1-\epsilon$ gives
\[
 E_\rho(K_A)=\lim_{n\to\infty}\frac1n
 D\!\left(\rho^{\otimes n}\,\middle\Vert\,
   \frac1{|A|}\int_A (g_\phi^\dagger\sigma g_\phi)^{\otimes n}\,d\phi\right)
 \le B_\epsilon^{\mathrm{ucov}}(\rho\Vert\sigma).
\]

We now reduce this rate to a pure-state relative-entropy rate.
For the pair in Figure~\ref{fig:intro-stein-boundary} write $p=9/10$,
$q=1/10$, and $\psi=|0\rangle\langle0|$. Put
$\sigma_\phi=g_\phi^\dagger\sigma g_\phi$ and
$\sigma_{A,N}=|A|^{-1}\int_A\sigma_\phi^{\otimes N}\,d\phi$.
The pure-state rate for the same arc set $A$ is
\[
 e_0(A)=\lim_{N\to\infty}\frac1N D(\psi^{\otimes N}\Vert\sigma_{A,N}).
\]
Step~1 of the proof of Lemma~\ref{lem:mixture-rate} also gives existence
for $\psi$: the mixture comparison uses faithfulness of $\sigma$, and
the entropy sequence is nonnegative and bounded per copy for a pure
null as well.

Let $j=\lfloor nq\rfloor$ and $N=n-2j$. Under
\eqref{eq:covhoe-product-Schur}, the $j$th block of $\sigma_{A,n}$ is
$(\det\sigma)^j M_{A,N}\otimes I_{m_{n,j}}$, where $M_{A,N}$ is the
restriction of $\sigma_{A,N}$ to $\Sym^N(\mathbb C^2)$.
Let $\omega_n$ be the normalized projector onto the subspace of this
block with symmetric factor $|0\rangle^{\otimes N}$ and the full
multiplicity space. The corresponding weight in $\rho^{\otimes n}$ is
\[
 w_n=m_{n,j}p^{n-j}q^j
 =\frac{n-2j+1}{n-j+1}\binom nj p^{n-j}q^j .
\]
In particular, $\rho^{\otimes n}\ge w_n\omega_n$.
The ratio in front of the binomial mass has a positive limit.
Stirling's formula at $j=\lfloor nq\rfloor$ gives $w_n$ of order
$n^{-1/2}$.

Set $F_n=-n^{-1}\log\sigma_{A,n}$ and $f_n=\Tr\rho^{\otimes n}F_n$.
The variance bound \eqref{eq:log-variance}, applied with null $\rho$,
gives
\[
 |\Tr\omega_n F_n-f_n|^2
 \le \Tr\omega_n(F_n-f_n I)^2
 \le w_n^{-1}\operatorname{Var}_{\rho^{\otimes n}}(F_n)
 =O(n^{-1/2}).
\]
Here $\operatorname{Var}_{\rho^{\otimes n}}(F_n)$ denotes the variance defined in
the proof of Lemma~\ref{lem:mixture-rate}.
The block formula gives
\[
 \Tr\omega_n F_n
 =-\frac{j}{n}\log\det\sigma
  -\frac1n\langle 0^{\otimes N}|\log M_{A,N}|0^{\otimes N}\rangle.
\]
The last term tends to $(p-q)e_0(A)$ because the null has zero entropy,
lies in the symmetric block, and $N/n\to p-q$. On the other hand,
$f_n\to S(\rho)+E_\rho(K_A)$ by \eqref{eq:E-definition}. Consequently,
\[
 E_\rho(K_A)
 =-S(\rho)-q\log\det\sigma+(p-q)e_0(A)
 =\frac45\bigl(e_0(A)+\log(9/10)\bigr).
\]

For a direct finite-degree calculation, let $d=k-\ell$ for Dicke
indices $k,\ell$. In the Dicke basis, the $(k,\ell)$ entry of $M_{A,N}$
is the corresponding entry of $\Sym^N(\sigma)$ multiplied by
$|A|^{-1}\int_A e^{id\phi}\,d\phi$. A common shift of the arc centers
conjugates the matrix by a diagonal unitary and preserves the
expectation in $|0\rangle^{\otimes N}$. With $s=|A|/(2\pi)$, a single
arc centered at zero gives the multiplier $\operatorname{sinc}(\pi s d)$.
Two equal arcs centered at zero and $\pi$ give
$\tfrac12(1+(-1)^d)\operatorname{sinc}(\pi s d/2)$.
Here $\operatorname{sinc}(x)=\sin(x)/x$ for $x\ne0$ and
$\operatorname{sinc}(0)=1$.

\medskip\noindent\textbf{Numerical evaluation.}
For Figure~\ref{fig:intro-stein-boundary}, we evaluate the rates for
both arc families at each sampled interior tolerance and use the larger
of the two numerical estimates. Each family's estimates are linearly
interpolated between the sampled tolerances. The plotted curve follows
the pointwise maximum of these interpolants, with an additional vertex
at their crossing. The open endpoints use the exact one-sided limits
$D_{\mathrm R}(\rho\Vert\sigma)$ as $\epsilon\downarrow0$ and
$D(\rho\Vert\sigma)$ as $\epsilon\uparrow1$.

The empirical absolute-error estimate for these candidate evaluations,
based on mesh refinement, extrapolation, and interpolation checks,
is about $2\times10^{-5}$ nats. This is an empirical assessment of the
numerical evaluations within the two chosen families, rather than a
rigorous error bound. It does not bound the gap to the optimal covariant
exponent, and optimality of these families remains open.

\section{Proof of the qubit Hoeffding exponent}
\label{app:merge-qubit-proof}
\subsection{Keyl's measurement on qubits}

Given a covariant qubit test, we construct one Keyl decision rule
whose type-I and type-II errors are at most polynomial multiples
of the respective original errors. We first reduce the measurement
to rank-one seeds in each Schur block. A spherical comparison
supplies one rotation for which both densities are large. Integrating
over nearby rotations bounds each weighted sum of the errors. Convex
separation then produces one rule satisfying both bounds.

Write $\Proj$ for the projective line of qubit rays,
identified with the unit Bloch sphere.  Let $\omega$ be normalized
sphere area, and use normalized Haar measure on $G=U(2)$.  For an
integer $n\ge1$ and $0\le j\le\lfloor n/2\rfloor$ define
$N=n-2j$, $d_N=N+1$ and
$m_{n,j}=\binom nj-\binom n{j-1}$, where $\binom n{-1}=0$.
The qubit Schur--Weyl decomposition gives, for every positive qubit
operator $X$,
\begin{equation}
 X^{\otimes n}\simeq\bigoplus_j
   (\det X)^j\Sym^N(X)\otimes I_{m_{n,j}}.
 \label{eq:covhoe-product-Schur}
\end{equation}
The coherent-state resolution on the spin-$N/2$ factor is
\begin{equation}
 d_N\int_{\Proj}|u^{\otimes N}\rangle\langle u^{\otimes N}|
                 \dd\omega(u)=I_{d_N}.
 \label{eq:covhoe-coherent-resolution}
\end{equation}
For the $N=0$ block this is the one-dimensional identity.
Keyl's qubit measurement therefore has outcome density, relative to
counting measure in $j$ and $\omega$ in $u$,
\begin{equation}
 p^{\mathrm K}_{X,n}(j,u)=m_{n,j}d_N(\det X)^j
                          \langle u|X|u\rangle^N.
 \label{eq:covhoe-Keyl-density}
\end{equation}
Summing over $j$ and integrating over $u$ gives one for a state $X$.
When $N=0$, the uniform coordinate $u$ carries no information.

\subsection{Reduction to rank-one seeds}
\label{covhoe:sec:reduction}

Permutation averaging preserves the outcome law under both product
states. The refinement records the Schur label $j$, an index $(y,\ell)$
selecting the seed ray in that block, and a rotation $g$ of the seed.
Applying the original rule to the recovered outcome $gy$ preserves
both errors.

For a unit $v\in\Sym^N(\mathbb C^2)$ write $U_N(g)=\Sym^N(g)$ and set
\[
 A_j=(\det\rho)^j\Sym^N(\rho),\qquad
 B_j=\lambda(\det\sigma)^j\Sym^N(\sigma).
\]
Write $\Gamma_{j,v}(\lambda)$ for the minimum weighted error contributed
by seed $v$ in block $j$ before multiplying by the factor $m_{n,j}d_N$.
\begin{equation}
 \Gamma_{j,v}(\lambda)
 =\int_G\min\{\langle v|U_N(g)^\dagger A_jU_N(g)|v\rangle,
                 \langle v|U_N(g)^\dagger B_jU_N(g)|v\rangle\}\dd g.
 \label{covhoe:eq:Gamma}
\end{equation}
Haar measure is normalized. Determinant phases may be suppressed in
$U_N(g)$, since they cancel in the matrix coefficients.

\begin{proposition}[Exact pure-orbit risk formula]
\label{covhoe:prop:pure-orbit}
For every $n$ and $\lambda>0$,
\begin{equation}
 R_n^{\ucov}(\lambda)=\sum_jm_{n,j}d_N
                          \min_{\|v\|=1}\Gamma_{j,v}(\lambda).
 \label{covhoe:eq:pure-risk}
\end{equation}
The minima are attained. The Keyl risk $R_n^{\mathrm K}(\lambda)$ is the same sum with a coherent
seed $v=e^{\otimes N}$ in each block.
\end{proposition}
\begin{proof}
\textbf{Step 1. Permutation averaging.}
For every outcome event $A$, average the effect $M(A)$ over
permutations of the tensor factors to obtain $\overline M(A)$.
Permutations fix both product states and commute with $g^{\otimes n}$.
Thus the average preserves both outcome laws and covariance.
Schur--Weyl duality gives
\[
 \overline M(A)=\bigoplus_j E_j(A)\otimes I_{m_{n,j}},
\]
where $E_j$ is a normalized covariant POVM on $\Sym^N(\mathbb C^2)$.
Recording the block $j$ refines $\overline M$.

\textbf{Step 2. Recording a group element.}
Fix $j$ and define the probability measure
$\nu_j(A)=\Tr E_j(A)/d_N$.
Finite-dimensional Radon--Nikodym differentiation gives a measurable
positive matrix field $S_j(y)$ with
\[
 E_j(\dd y)=S_j(y)\nu_j(\dd y),\qquad
 \Tr S_j(y)=d_N\quad\nu_j\text{-almost everywhere}.
\]
For example, differentiate its matrix entries and use positivity on a
countable dense set of vectors to choose a common positive version.
Covariance of $E_j$ also implies that the scalar measure $\nu_j$ is $G$-invariant:
$\nu_j(gA)=\nu_j(A)$ for each Borel set $A$.
Define on $\Omega\times G$ the block POVM
\begin{equation}
 \widetilde E_j(\dd y,\dd g)
 =U_N(g)S_j(y)U_N(g)^\dagger\,
                         \nu_j(\dd y)\dd g.
 \label{covhoe:eq:Borel-lift}
\end{equation}
The action is $h(y,g)=(y,hg)$; the space and action are
standard Borel. The irreducible Haar twirl of $S_j(y)$ equals
$I_{d_N}$, so this is normalized and covariant.
Its pushforward under $\pi(y,g)=gy$ is exactly $E_j$:
\begin{align}
 \widetilde E_j(\pi^{-1}(A))
 &=\int_G U_N(g)E_j(g^{-1}A)U_N(g)^\dagger\dd g
 =E_j(A).
 \label{covhoe:eq:lift-pushforward}
\end{align}
The last equality is covariance. The map $\pi$ intertwines the actions
because $\pi(y,hg)=h\pi(y,g)$. The joint Borel action makes the integrands
measurable, and the finite traces justify Fubini's theorem.

\textbf{Step 3. Recording the seed.}
Choose a measurable spectral resolution
$S_j(y)=\sum_\ell s_\ell(y)
                   |v_\ell(y)\rangle\langle v_\ell(y)|$.
Such a choice exists in finite dimension: order the measurable
eigenvalues and choose bases in degenerate spectral subspaces by
projecting the fixed standard basis and applying Gram--Schmidt with
the first nonzero projection at each step.
The weights
\[
 \eta_j(\dd y,\ell)
     =\nu_j(\dd y)\frac{s_\ell(y)}{d_N}
\]
form a probability measure, independent of the input state. With
$(y,\ell)$ recorded, \eqref{covhoe:eq:Borel-lift} is its mixture of
normalized pure-orbit POVMs
$d_N U_N(g)|v_\ell(y)\rangle\langle v_\ell(y)|
U_N(g)^\dagger\dd g$.
Under the map $(j,y,\ell,g)\mapsto gy$, this refinement recovers
$\overline M$. The rule $f(gy)$ therefore has the same two errors as
$f$ applied to $M$. Allowing the rule to use the additional coordinates
can only decrease the optimal weighted error.

For any observation with densities $p,q$, minimizing its rule gives
\begin{equation}
 \inf_{0\le f\le1}\int[(1-f)p+\lambda f q]
                =\int\min\{p,\lambda q\}.
 \label{covhoe:eq:binary-minimum}
\end{equation}
The likelihood rule is measurable, so the equality is attained.
By \eqref{eq:covhoe-product-Schur}, a branch with seed $v$ contributes
at least $m_{n,j}d_N\Gamma_{j,v}(\lambda)$ to the weighted error before
averaging over $(y,\ell)$. This average is at least the minimum over
unit seeds times $m_{n,j}d_N$. Summing over $j$ gives the lower bound
in \eqref{covhoe:eq:pure-risk}.

\textbf{Step 4. Attainment.}
The unit seed sphere is compact. The integrand in \eqref{covhoe:eq:Gamma} is
continuous in $v$ and uniformly bounded in finite dimension, so
$\Gamma_{j,v}$ is continuous and attains its minimum.
Choose a minimizing seed for each of the finitely many blocks, record
$(j,g)$, and use its likelihood rule. The resulting direct-sum pure
orbit POVM is admissible and attains \eqref{covhoe:eq:pure-risk}.
For a coherent seed the coefficients are
$(\det X)^j\langle ge|X|ge\rangle^N$; pushing Haar measure forward to
the ray $ge$ gives \eqref{eq:covhoe-Keyl-density}. Conversely, averaging any
group rule over the fibers of $g\mapsto ge$ is state-independent and
preserves both errors, because both densities are fiber-constant.
This proves the last assertion, including the scalar $N=0$ block.
\end{proof}

\subsection{Spherical comparison for every finite-degree seed}
\label{covhoe:sec:sphere}

For $N\ge1$ and a unit seed $v$, define its squared coherent overlap
and its normalized $N$th root by
\begin{equation}
 p_v(z)=|\langle v|z^{\otimes N}\rangle|^2,\qquad
 M_v=\max_z p_v(z),\qquad
 f_v(z)=(p_v(z)/M_v)^{1/N}.
 \label{covhoe:eq:seed-functions}
\end{equation}
Equation~\eqref{eq:covhoe-coherent-resolution} gives
$1/d_N\le M_v\le1$.
The homogeneous coherent amplitude factors as
$c\prod_{\ell=1}^N\langle u_\ell|z\rangle$, where $u_\ell$ are
unit representatives of the factor rays. Define
\[
 \mu_v=\frac1N\sum_{\ell=1}^N\delta_{u_\ell},\qquad
 V_\mu(z)=\int_{\Proj}\log|\langle a|z\rangle|^2\,\dd\mu(a).
\]
Then $\log f_v=V_{\mu_v}-\max V_{\mu_v}$; the zeros are the antipodes
of the factor rays. This fixes the Majorana convention used here.
For positive definite $Y$ define
\begin{equation}
 \Phi_Y(\mu_v)=\log\max_{z\in\Proj}
                  \frac{f_v(z)}{\langle z|Y^{-1}|z\rangle}.
 \label{covhoe:eq:Phi}
\end{equation}
This agrees with \eqref{eq:majorana-phi} since
$\log f_v=V_{\mu_v}-\max V_{\mu_v}$. We write
$\mathcal A(R)=(1-\cos R)/2$ for the normalized area of a unit-sphere cap
of angular radius $R\in[0,\pi]$.

For each state we need an axis with a large response. The two chosen
axes must make the same angle as the original pair, so one rotation
can produce both. We build two connected sets of candidate axes that
contain the same maximizing ray. The first step is to bound the area
of the component of $\{f_v>t\}$ containing that ray.

\begin{lemma}[Area of the component containing a maximum]
\label{covhoe:lem:component}
Choose any maximum point $e$ of $f_v$. For every regular value
$0<t<1$, the component $E_t$ of $\{f_v>t\}$ containing $e$ obeys
\begin{equation}
 \omega(E_t)\ge1-t.
 \label{covhoe:eq:component-area}
\end{equation}
\end{lemma}
\begin{proof}
\textbf{Step 1: The Laplacian and the boundary flux.}
In a stereographic chart the coherent amplitude is a polynomial divided
by $(1+|z|^2)^{N/2}$, so
\[
 \log p_v(z)=\log|P(z)|^2-N\log(1+|z|^2)+\text{constant}.
\]
The unit-sphere Laplacian is $(1+|z|^2)^2\Delta_{\mathbb R^2}/4$.
The function $\log|P|^2$ is harmonic where $P\ne0$, so away from the
finitely many zeros,
\begin{equation}
 \Delta_{S^2}\log f_v=-1.
 \label{covhoe:eq:lap}
\end{equation}
For $t>0$ the closure of $E_t$ avoids the zeros. At a regular level it
is a smooth domain, with all boundary components included. Its outward
normal points toward decreasing $f_v$; Green's formula gives
\begin{equation}
 \int_{\partial E_t}|\nabla\log f_v|\dd\ell=4\pi\omega(E_t).
 \label{covhoe:eq:component-flux}
\end{equation}

\textbf{Step 2: Coarea and isoperimetry at regular levels.}
Put $x=-\log t$ and $a(x)=\omega(E_{e^{-x}})$.
Around a regular $x_0>0$, choose a compact level band avoiding the
zeros and $e$, with $\nabla\log f_v\ne0$ throughout.  The normalized
gradient flow $-\nabla\log f_v/|\nabla\log f_v|^2$, smoothly cut off
outside that band, fixes $e$ and carries the components containing
$e$ between nearby levels.  Thus no merger occurs in the band, and
the boundary-speed formula gives
\[
 a'(x)=\frac1{4\pi}\int_{\partial E_t}
                         |\nabla\log f_v|^{-1}\dd\ell.
\]
The classical unit-sphere isoperimetric inequality is
$|\partial E_t|^2\ge16\pi^2a(1-a)$, valid also for domains with holes.
Cauchy--Schwarz and \eqref{covhoe:eq:component-flux} give
\[
 16\pi^2a(1-a)\le|\partial E_t|^2
                       \le16\pi^2a a'(x).
\]
Since $a>0$, this proves $a'(x)\ge1-a(x)$.
This is the spherical level-set argument also used for the full
superlevel set in Frank \cite[Lemma 12, Case 2]{Frank}; here it is
applied to the specified single component.

\textbf{Step 3. All levels.}
The components containing $e$ grow as $x$ increases. Hence
$B(x)=e^x a(x)$ is nondecreasing. By Sard's theorem almost every level
is regular, and there $B'(x)=e^x(a(x)+a'(x))\ge e^x$. The increment
of a nondecreasing function is at least the integral of its derivative.
In particular, upward jumps when components merge can only increase
this increment. Thus $B(x)-B(y)\ge e^x-e^y$ for $0<y<x$. It follows that
$1-a(x)\le e^{y-x}(1-a(y))\le e^{y-x}$. Letting $y\downarrow0$ gives
\eqref{covhoe:eq:component-area}, including levels at which components merge.
\end{proof}

\begin{lemma}[Area of angular neighborhoods]
\label{covhoe:lem:neighborhood}
Let $A\subset S^2$ be compact with positive area. For $r\ge0$, let
$A_r=\{z:\dist(z,A)\le r\}$. If
$\omega(A)\ge\mathcal A(R)$ for $0\le R\le\pi$, then
\begin{equation}
 \omega(A_r)\ge\mathcal A(\min\{\pi,R+r\}).
 \label{covhoe:eq:neighborhood}
\end{equation}
\end{lemma}
\begin{proof}
Let $C$ be a spherical cap with $\omega(C)=\omega(A)$ and radius
$R_0\ge R$.  Spherical neighborhood isoperimetry
\cite[Proposition~1]{BarnesOzguWu} applies to arbitrary measurable
subsets and their closed angular neighborhoods.  After dividing the
surface-area measure there by $4\pi$, it gives
\[
 \omega(A_r)\ge\omega(C_r)
 =\mathcal A(\min\{\pi,R_0+r\})
 \ge\mathcal A(\min\{\pi,R+r\}),
\]
which proves the claim, including the full-area case.
\end{proof}

\begin{lemma}[Two connected sets realize the required angle]
\label{covhoe:lem:two-sets}
Let compact connected $A,B\subset S^2$ contain a common point.
Suppose their normalized areas are at least
$\mathcal A(R_A)$ and $\mathcal A(R_B)$, where $R_A,R_B>0$.
Then for every $0\le\gamma\le\min\{\pi,R_A+R_B\}$ there are
$a\in A,b\in B$ with $\dist(a,b)=\gamma$.
\end{lemma}
\begin{proof}
The maximum distance $D$ is attained. If $D<\pi$, then for every
$0<r<\pi-D$, $B$ is disjoint from $(-A)_r$: the triangle inequality
would otherwise give a pair of points of $A,B$ at distance greater
than $D$. Lemma~\ref{covhoe:lem:neighborhood} implies
\[
 \mathcal A(R_B)+\mathcal A(\min\{\pi,R_A+r\})\le1.
\]
Let $r\uparrow\pi-D$. Positive area of $B$ excludes
$R_A+\pi-D\ge\pi$, and monotonicity of $\mathcal A$ then gives
$R_B\le D-R_A$. Thus $D\ge R_A+R_B$ if $D<\pi$; if the radius sum
is at least $\pi$, necessarily $D=\pi$.
The continuous distance image of the connected set $A\times B$ is
an interval. It contains zero because of the common point, and $D$
by compactness. This proves every required intermediate angle.
\end{proof}

Here $\pauli=(\sigma_x,\sigma_y,\sigma_z)$ is the vector of Pauli matrices. For $0<b<1$ and an axis $a\in S^2$ write
$X_a=(I+b\,a\cdot\pauli)/2$, $d_X=(1-b^2)/4$,
$s_X=(1-b)/2$, $p_X=(1+b)/2$. In these coordinates
\begin{equation}
 \Phi_{X_a}(\mu_v)
    =\log\max_z\frac{2d_X f_v(z)}{1-b\,a\cdot z}.
 \label{covhoe:eq:Phi-axis}
\end{equation}
Fix a ray $e$ with $f_v(e)=1$. The axes satisfying
$\langle e|X_a|e\rangle\ge w$, with $s_X<w<p_X$, form a cap of radius
\begin{equation}
 R_X(w)=\arccos((2w-1)/b).
 \label{covhoe:eq:cap-radius}
\end{equation}

\begin{lemma}[A connected set of large responses]
\label{covhoe:lem:response}
Fix $e$ with $f_v(e)=1$. For $s_X<w<p_X$, the component containing
$e$ of the set $\{a:\Phi_{X_a}(\mu_v)\ge\log w\}$ is compact and
has area at least $\mathcal A(R_X(w))$.
\end{lemma}
\begin{proof}
The response $\Phi_{X_a}(\mu_v)$ is continuous in $a$ since the
ratio in \eqref{covhoe:eq:Phi-axis} is continuous on the product of
the two spheres and its denominator is bounded away from zero.
For a regular $t\in(0,1)$ with $-1<(1-2d_Xt/w)/b<1$, put
$r(t)=\arccos((1-2d_Xt/w)/b)$.
Let $E_t$ be the component through $e$ from
Lemma~\ref{covhoe:lem:component}. The closed $r(t)$-neighborhood of
$\overline E_t$ is compact and connected and contains $e$. Every axis
$a$ in this neighborhood has a point $z\in\overline E_t$ within
angle $r(t)$. Then
$f_v(z)\ge t$ and $a\cdot z\ge\cos r(t)$, giving
$2d_X f_v(z)/(1-b\,a\cdot z)\ge w$.
Lemmas~\ref{covhoe:lem:component} and \ref{covhoe:lem:neighborhood} bound its area
below by that of a cap with radius
$\min\{\pi,\arccos(2t-1)+r(t)\}$.

To choose $t$, take an axis $a$ on the boundary of the coherent cap
of radius $R_X(w)$, so $\langle e|X_a|e\rangle=w$. For a coherent seed
at $e$, Cauchy--Schwarz gives
\[
 |\langle e|z\rangle|^2
 \le \langle e|X_a|e\rangle\,\langle z|X_a^{-1}|z\rangle,
\]
with equality at the ray of $z=X_ae/\|X_ae\|$.
Thus this ray reaches the coherent response threshold $w$ in the
ratio in \eqref{covhoe:eq:Phi}. Its squared overlap with $e$ gives the
level $t_*=w^2/(w-d_X)$ used in the neighborhood construction.
The identity $(w-p_X)(w-s_X)<0$ gives $0<t_*<1$.
The two radii at $t_*$ add up to the desired radius,
\begin{equation}
 \arccos(2t_*-1)+r(t_*)=R_X(w).
 \label{covhoe:eq:tangent}
\end{equation}
In a basis of top and bottom eigenvectors of this $X_a$, a unit representative of $e$ is
$(\cos(R_X(w)/2),\sin(R_X(w)/2))$ after phases are chosen.
The ray $z=X_ae/\|X_ae\|$ is at distance
\[
 r_*=2\arctan\left(\frac{s_X}{p_X}
                           \tan\frac{R_X(w)}2\right)
\]
from $a$, strictly between $0$ and $R_X(w)$. Thus $z$ is on the
short meridian between $a$ and $e$.
Since $\langle e|X_a|e\rangle=w$ and $X_a^2=X_a-d_XI$,
\[
 |\langle e|z\rangle|^2=\frac{w^2}{w-d_X}=t_*,\qquad
 \langle z|X_a^{-1}|z\rangle=\frac{w}{w-d_X}.
\]
These identities give
$\dist(e,z)=\arccos(2t_*-1)$ and
$\cos r_*=(1-2d_Xt_*/w)/b$, proving \eqref{covhoe:eq:tangent}.
In particular $0<r(t_*)<\pi$.
The compact component of the response set containing $e$ contains
all these neighborhoods. Taking regular $t$ tending to $t_*$ gives
the area bound $\mathcal A(R_X(w))$ by continuity of the two radii.
\end{proof}

We now choose two axes at the same angle as the original state axes.
One rotation then takes the original pair to the chosen pair.

\begin{proposition}[One rotation dominates both coherent responses]
\label{covhoe:prop:joint}
For every $N\ge1$, every unit seed $v$, every faithful qubit pair,
and every qubit ray $u$, some $g\in U(2)$ satisfies simultaneously
\begin{equation}
 \Phi_{g^\dagger\rho g}(\mu_v)\ge\log\langle u|\rho|u\rangle,
 \qquad
 \Phi_{g^\dagger\sigma g}(\mu_v)\ge\log\langle u|\sigma|u\rangle.
 \label{covhoe:eq:joint}
\end{equation}
\end{proposition}
\begin{proof}
Fix a maximizer $e$ of $f_v$ and write $w_X=\langle u|X|u\rangle$.
For a nonscalar $X$, evaluating at $e$ gives
$\Phi_{X_a}(\mu_v)\ge\log s_X$ for every axis $a$. Aligning the
larger-eigenvalue axis with $e$ gives response $\log p_X$, since
$f_v\le1$ and $X^{-1}\ge p_X^{-1}I$. A scalar state always gives
response $-\log 2$. Thus if one state is scalar or if $w_X=s_X$,
its constraint holds under every rotation. For the other state,
align its larger-eigenvalue axis with $e$ if it has one. If both
states are nonscalar and share that axis, align it with $e$.

Now let $\gamma>0$ be the angle between the largest-eigenvalue axes
of two nonscalar states, and suppose both $w_X>s_X$.
The radii $R_\rho(w_\rho)$ and $R_\sigma(w_\sigma)$, extended to
allow $R_X(p_X)=0$, have sum at least $\gamma$ by the triangle
inequality through $u$. For two interior thresholds,
Lemma~\ref{covhoe:lem:response} gives two compact connected sets
through $e$ with these area bounds. By
Lemma~\ref{covhoe:lem:two-sets}, we can select one axis from each
with angle $\gamma$. A proper rotation takes the original ordered
pair to the selected pair. For $0<\gamma<\pi$ this follows by
matching the oriented orthonormal frames. For $\gamma=\pi$, it
suffices to match the first axis, as the second is its antipode.
The appropriate lift to $SU(2)$ realizes the selected axes as
those of $g^\dagger\rho g$ and $g^\dagger\sigma g$.

If either threshold is $p_X$, first use smaller interior thresholds,
whose radii can only be larger. Let the thresholds increase to their
original values and take a convergent subsequence of the rotations
in $SU(2)$. Continuity of $\Phi_{X_a}(\mu_v)$ gives both inequalities
at the limiting rotation.
\end{proof}

\subsection{Uniform finite-block comparison of weighted errors}
\label{covhoe:sec:risk}

A single favorable rotation has Haar measure zero. We bound the
change in the response under small rotations so that a whole
neighborhood contributes to the integral. The constants may depend
on the state pair and are independent of the block, seed, and
weight $\lambda$.

\begin{lemma}[An exact norm bound and a rotation modulus]
\label{covhoe:lem:norm}
For every $N\ge1$, unit seed $v$, and $Y>0$,
\begin{equation}
 \langle v|\Sym^N(Y)|v\rangle
 \ge M_v e^{N\Phi_Y(\mu_v)}
 \ge d_N^{-1}e^{N\Phi_Y(\mu_v)}.
 \label{covhoe:eq:norm-bound}
\end{equation}
For a fixed faithful qubit state $X$ with eigenvalues $p_X\ge s_X$,
a relative rotation of angle at most $\delta$ changes
$\Phi_{g^\dagger Xg}(\mu_v)$ by at most $(p_X-s_X)\delta/(2s_X)$.
\end{lemma}
\begin{proof}
The first inequality in Eq.~\eqref{covhoe:eq:norm-bound} is the
exact coherent-norm estimate established in the proof of
Theorem~\ref{thm:majorana-asymptotic-pure-seeds}: the squared norm of
$\Sym^N(Y^{1/2})v$ is at least its maximum squared coherent overlap
$M_v e^{N\Phi_Y(\mu_v)}$.  The bound $M_v\ge d_N^{-1}$ gives the
second inequality.

For the rotation bound, the inverse expectation of $X_a$ is
$(1-(p_X-s_X)a\cdot z)/(2\det X)$. It is at least $1/p_X$, and
moving the axis through angle at most $\delta$ changes it by at most
$(p_X-s_X)\delta/(2\det X)$. The mean-value theorem for its
logarithm gives the bound $(p_X-s_X)\delta/(2s_X)$, which is zero
for a scalar state. Taking the maximum in \eqref{covhoe:eq:Phi} preserves
this bound.
\end{proof}

\begin{proposition}[The uniform polynomial risk comparison]
\label{covhoe:prop:risk-bound}
For every $n\ge1$ and $\lambda>0$,
\[
 D_n^{-1}R_n^{\mathrm K}(\lambda)
 \le R_n^{\mathrm{ucov}}(\lambda)
 \le R_n^{\mathrm K}(\lambda),
\]
where, as in Eq.~\eqref{eq:covhoe-Dn},
\[
 L=\max_{X=\rho,\sigma}\frac{p_X-s_X}{2s_X},\qquad
 D_n=16\pi e^L(n+1)^4,
\]
and $p_X\ge s_X>0$ are the eigenvalues of $X$.
Thus Eq.~\eqref{eq:covhoe-risk} holds with constants independent of
the seed, the Schur block, and the weight $\lambda$.
\end{proposition}
\begin{proof}
Fix a block $j$ with $N\ge1$ and write
\[
 T_j(\lambda)=\max_{u\in\Proj}\min\{
       (\det\rho)^j\langle u|\rho|u\rangle^N,
       \lambda(\det\sigma)^j\langle u|\sigma|u\rangle^N\}.
\]
Choose a ray $u$ attaining $T_j(\lambda)$. For any seed $v$,
Proposition~\ref{covhoe:prop:joint} applied to $u$ gives a rotation $g_0$.
Within rotation angle $\delta=1/(n+1)$ of $g_0$,
Lemma~\ref{covhoe:lem:norm} bounds both terms of the minimum in
\eqref{covhoe:eq:Gamma}. On this neighborhood the integrand is at least
\begin{equation}
 d_N^{-1}e^{-NL\delta}T_j(\lambda)
 \ge d_N^{-1}e^{-L}T_j(\lambda).
 \label{covhoe:eq:rotation-integrand}
\end{equation}
One $SU(2)$ cap lifting this neighborhood has normalized Haar mass
\[
 \frac{\delta-\sin\delta}{2\pi}\ge\frac{\delta^3}{16\pi}
 \quad(0<\delta\le1).
\]
To check the normalization, identify $SU(2)$ with the unit $S^3$;
rotation angle $\delta$ corresponds to geodesic cap radius $\delta/2$.
Integrating $4\pi\sin^2 t\,\dd t$ and dividing by $2\pi^2$ gives the
displayed mass. The bound follows from
$\delta-\sin\delta\ge\delta^3/6-\delta^5/120\ge\delta^3/8$.
Integrals over normalized $U(2)$ equal the corresponding $SU(2)$
integrals here, because central phases do not affect conjugation.
Thus
\[
 \Gamma_{j,v}(\lambda)
 \ge\frac{e^{-L}T_j(\lambda)}{16\pi d_N(n+1)^3}
 \ge D_n^{-1}\Gamma_{j,\mathrm{coh}}(\lambda),
\]
where $\Gamma_{j,\mathrm{coh}}\le T_j$ and $d_N\le n+1$ were used.
For $N=0$ both costs are the same scalar cost, so the inequality still
holds. Haar invariance makes the bound independent of the chosen
center $g_0$. Thus the same $D_n$ applies separately to every seed and
block; the favorable rotations need not agree between blocks.
Multiply by $m_{n,j}d_N$, take the seed minima, and sum in
Proposition~\ref{covhoe:prop:pure-orbit}. This gives the first inequality
in \eqref{eq:covhoe-risk}. The second follows because the Keyl POVM is
an admissible covariant observation. The comparison is finite-block
and uniform in $\lambda$, so it includes any $n$-dependent weights.
\end{proof}

\subsection{A single Keyl rule and the Hoeffding exponent}
\label{covhoe:sec:completion}

We use the weighted-error bounds to find one Keyl rule that controls
both $\alpha_n$ and $\beta_n$. Applying this rule at each $n$ preserves
the required exponent of $\beta_n$ and achieves at least the original
exponent of $\alpha_n$.

\begin{lemma}[Simultaneous comparison of error pairs]
\label{covhoe:lem:pair}
For each covariant test at blocklength $n$, a Keyl rule exists satisfying
\eqref{eq:covhoe-pair}.
\end{lemma}
\begin{proof}
Let $\mathcal E_n^{\mathrm K}\subset[0,1]^2$ be the set of all Keyl error pairs.
Relative to the finite reference measure used in
\eqref{eq:covhoe-Keyl-density}, the set of rules $0\le f\le1$ is a weak-*
compact convex subset of $L^\infty$. The two errors are affine,
weak-* continuous functionals with $L^1$ densities. Thus
$\mathcal E_n^{\mathrm K}$ is compact and convex. Its upward closure
$\mathcal U_n=\mathcal E_n^{\mathrm K}+\mathbb R_+^2$ is closed and convex.
Rules are identified up to null sets in this compactness argument;
a Borel version in $[0,1]$ can always be chosen on the standard Borel
output space.

For a given covariant pair $(\alpha,\beta)$,
Proposition~\ref{covhoe:prop:risk-bound} implies
\[
 \min_{(x,y)\in\mathcal E_n^{\mathrm K}}(x+\lambda y)
   =R_n^{\mathrm K}(\lambda)\le D_n(\alpha+\lambda\beta),\qquad\lambda>0.
\]
If $z=(D_n\alpha,D_n\beta)$ were outside $\mathcal U_n$, strict
finite-dimensional separation would give a nonzero vector $c$ with
$c\cdot z<\inf_{w\in\mathcal U_n}c\cdot w$.
Both coordinates of $c$ must be nonnegative: otherwise the infimum
over the upward closure would be $-\infty$.
If both are positive, dividing by the first coordinate contradicts
the displayed inequality with $\lambda=c_2/c_1$.
If one is zero, the infimum is zero, since the always-accept rule has
$\alpha=0$ and the always-reject rule has $\beta=0$; nonnegativity of
$z$ again contradicts strict separation.
Hence $z\in\mathcal U_n$. By definition there is a single Keyl pair
$(\alpha^{\mathrm K},\beta^{\mathrm K})$ with $\alpha^{\mathrm K}\le D_n\alpha$ and
$\beta^{\mathrm K}\le D_n\beta$, as claimed.
\end{proof}

\begin{proof}[Proof of parts (i) and (ii) of Theorem~\ref{thm:merge-qubit-full-hoeffding}]
Proposition~\ref{covhoe:prop:risk-bound} and
Lemma~\ref{covhoe:lem:pair} give the two finite-block comparisons.
For any sequence of covariant tests, choose the rule from the lemma
at each $n$. Since $\log D_n/n\to0$, its errors satisfy
\[
 \liminf_n-\frac1n\log\alpha_n^{\mathrm K}
       \ge\liminf_n-\frac1n\log\alpha_n,\qquad
 \liminf_n-\frac1n\log\beta_n^{\mathrm K}
       \ge\liminf_n-\frac1n\log\beta_n.
\]
These bounds also hold for zero errors by \eqref{eq:covhoe-pair}.
Every covariant sequence feasible at rate $r$ thus gives a feasible
Keyl sequence with at least the same type-I exponent. The reverse
inequality follows from class inclusion. The exact Keyl exponent
in Theorem~\ref{thm:keyl-hoeffding-exact-reorganized} now gives
\eqref{eq:merge-qubit-H}.
\end{proof}

\begin{remark}[Scope of the error comparison]
Both errors are bounded by factors of $D_n$ with $\log D_n=o(n)$,
which suffices for comparing the two decay exponents in the Hoeffding
problem. The fixed-error allowance and the
decay exponent of $1-\alpha_n$ require separate arguments.
\end{remark}

\section{Scalar evaluation of the exact qubit exponent}\label{app:merge-qubit-scalar}
\subsection{Eliminating the eigenvalue and verifying the duality}

The qubit reverse-divergence formula used below is, for
$\tau=t|u\rangle\langle u|+(1-t)|u^\perp\rangle\langle u^\perp|$,
$1/2\le t\le1$ and faithful $X$,
\begin{equation}
 D_{\mathrm R}(\tau\Vert X)
 =t\log t+(1-t)\log(1-t)
 -(2t-1)\log(u^\dagger Xu)-(1-t)\log\det X.
 \label{eq:merge-qubit-DR}
\end{equation}
\begin{proof}[Proof of Theorem~\ref{thm:merge-qubit-full-hoeffding}(iii)]
First minimize the weighted sum
\[
 C_\alpha
 =\min_\tau\{\alpha\DR(\tau\Vert\rho)
                       +(1-\alpha)\DR(\tau\Vert\sigma)\},
 \qquad 0<\alpha<1.
\]
As the weight varies, the minimizing divergence pairs attain every
required interior value of the constraint, which yields the dual
formula.

\textbf{Step 1. Minimize over the eigenvalue.}
Recall from Eq.~\eqref{eq:merge-Kalpha} that
\[
 K_\alpha=\max_{\|u\|=1}
 (u^\dagger\rho u)^\alpha(u^\dagger\sigma u)^{1-\alpha}.
\]
For a fixed unit \(u\), put
\[
 k=(u^\dagger\rho u)^\alpha(u^\dagger\sigma u)^{1-\alpha},
 \qquad z=(\det\rho)^\alpha(\det\sigma)^{1-\alpha}.
\]
By Eq.~\eqref{eq:merge-qubit-DR}, the quantity to minimize over \(1/2\le t\le1\) is
\[
 t\log t+(1-t)\log(1-t)-(2t-1)\log k-(1-t)\log z.
\]
If \(k\ge\sqrt z\), the minimizer is \(t=k^2/(k^2+z)\) and its value
is \(-\log(k+z/k)\). If \(k<\sqrt z\), the minimizer is \(t=1/2\)
and the value is \(-\log(2\sqrt z)\).
A Bloch direction perpendicular to both input Bloch vectors gives
\(u^\dagger\rho u=u^\dagger\sigma u=1/2\). Hence
\(K_\alpha\ge1/2\).
The two eigenvalues of a qubit state sum to one, so its determinant
is at most \(1/4\), with equality only for \(I/2\). Both weights
in \(z\) are positive. If at least one state is nonscalar, then
\(z<1/4\) and \(K_\alpha>\sqrt z\). On \(k>\sqrt z\),
\(k+z/k\) is strictly increasing and exceeds \(2\sqrt z\).
Hence every minimizer uses \(k=K_\alpha\) and
\(t=K_\alpha^2/(K_\alpha^2+z)\), giving \eqref{eq:merge-Calpha}.
If both states are \(I/2\), all expectation pairs are
\((1/2,1/2)\) and \(k=K_\alpha=\sqrt z=1/2\). The minimizing
eigenvalue is \(t=1/2\), so the minimizing state is \(I/2\)
and \(C_\alpha=0\), again as in \eqref{eq:merge-Calpha}.

\textbf{Step 2. Attain the required constraint value.}
The set of expectation pairs
\((x,y)=(u^\dagger\rho u,u^\dagger\sigma u)\) is the compact convex
joint numerical range of the two qubit matrices. In Bloch coordinates,
it is the image of the unit sphere under an affine map whose linear part
has rank at most two, hence also the image of the unit ball. For
\(0<\alpha<1\), \(\alpha\log x+(1-\alpha)\log y\) is strictly
concave on the positive quadrant. Its maximizing \emph{pair} \((x,y)\)
is therefore unique. As the weight tends to any interior value,
compactness and continuity of the objective imply that every limit
point of maximizing pairs maximizes the limiting objective.
Uniqueness then gives continuity of the pair in \(\alpha\).
Different rays can realize the same pair. When at least one state is
nonscalar, Step~1 gives \(t\) as a continuous function of \(K_\alpha\)
and \(z\). Equation~\eqref{eq:merge-qubit-DR} gives both divergence values from
\(t\) and the maximizing pair. When both states are \(I/2\), the
minimizing divergence pair is the constant \((0,0)\).
Thus the divergence pair is unique and continuous in \(\alpha\)
on \((0,1)\).

For any feasible state with $\DR(\tau\Vert\sigma)\le r$,
\[
 \DR(\tau\Vert\rho)
 \ge\frac{C_\alpha-(1-\alpha)r}{\alpha}.
\]
To attain equality for $0<r<\DR(\rho\Vert\sigma)$, let $\tau_\alpha$
be a minimizing state in the definition of $C_\alpha$. Compactness,
continuity and the unique zeros of the divergences imply that
$\tau_\alpha$ converges to $\sigma$ as $\alpha\downarrow0$ and to
$\rho$ as $\alpha\uparrow1$. Thus the continuous value
$\DR(\tau_\alpha\Vert\sigma)$ takes every value strictly between
zero and $\DR(\rho\Vert\sigma)$. Choose $\alpha$ so that it equals $r$.
Then $\tau_\alpha$ attains equality in the displayed bound.
For $r\ge\DR(\rho\Vert\sigma)$ the constrained minimum is zero,
attained by $\rho$. The displayed lower bound is then nonpositive
for every $\alpha$ and tends to zero as $\alpha\uparrow1$.
This proves \eqref{eq:merge-Hdual}.

The scalar value $C_\alpha$ and its derivative also recover the two
divergences of a minimizing state. Bound the difference quotient between
two parameter values by evaluating each objective at the other's
minimizer. Continuity of the two divergence values then gives
\begin{equation}\label{eq:tilted}
 \DR(\tau_\alpha\Vert\sigma)=C_\alpha-\alpha C_\alpha',
 \qquad
 \DR(\tau_\alpha\Vert\rho)=C_\alpha+(1-\alpha)C_\alpha'.
\end{equation}
For $0<r<\DR(\rho\Vert\sigma)$, the condition
$\DR(\tau_\alpha\Vert\sigma)=r$ for choosing an interior weight thus
becomes the scalar equation $C_\alpha-\alpha C_\alpha'=r$.

\textbf{Step 3. Reduce \(K_\alpha\) to a scalar minimization.}
The weighted arithmetic--geometric mean inequality gives, for
every \(s>0\) and every unit \(u\),
\[
 (u^\dagger\rho u)^\alpha(u^\dagger\sigma u)^{1-\alpha}
 \le u^\dagger\left(\alpha s\rho+
          (1-\alpha)s^{-\alpha/(1-\alpha)}\sigma\right)u.
\]
At the maximizing expectation pair $(x,y)$, take
$s=(y/x)^{1-\alpha}$, so $sx=s^{-\alpha/(1-\alpha)}y$ and the
bound is an equality there. The coefficients of the linear expression
on the right are proportional to the gradient of
$\alpha\log x+(1-\alpha)\log y$ at this pair. The first-order condition
on the convex set of expectation pairs therefore makes $(x,y)$
a maximizer of that linear expression. Its maximum is the largest
eigenvalue on the right of \eqref{eq:merge-Kscalar}, so this $s$ attains
the minimum.
Each Rayleigh quotient there is convex in \(s\), so their supremum is
convex. Faithfulness makes it diverge at both zero and infinity.
\end{proof}

\end{document}